\documentclass[apj,numberedappendix,iop,twocolappendix]{emulateapj}

\pdfoutput=1

\usepackage%
[
pagebackref, 
colorlinks=true,
linkcolor=red,filecolor=red,
citecolor=blue,
urlcolor=cyan,
pdftitle={Capture and Evolution of Planetesimals in Circumjovian Disks},
pdfauthor={D'Angelo and Podolak},
pdfsubject={Capture and Evolution of Planetesimals in Circumjovian Disks},
pdfcreator={KAZEditor},
pdfproducer={Kazzimiei},
pdfkeywords={Accretion, accretion disks; Hydrodynamics; Methods: numerical; Planet-disk interactions; Planets and satellites: formation; Protoplanetary disks},
bookmarks=true,
breaklinks=true,
hypertexnames=false,
pdfstartview=FitV,
hyperfootnotes=true,
bookmarksopen=false,
pdfpagemode=None]{hyperref}
\usepackage{breakurl}
\newcommand{\Mearth}{\mbox{$M_{\mathrm{E}}$}}        
\newcommand{\Mp}{\mbox{$M_{p}$}}                     
\newcommand{\Ms}{\mbox{$M_{\star}$}}                 
\newcommand{\dMp}{\dot{M}_{p}}                       
\newcommand{\Rhill}{\mbox{$R_\mathrm{H}$}}           
\newcommand{\K}{\mbox{$\mathrm{K}$}}                 
\newcommand{\AU}{\mbox{AU}}                          
\newcommand{\figlen}{0.97\linewidth}
\newcommand{\figlew}{1.00\linewidth}
\slugcomment{The Astrophysical Journal, in press}

\shorttitle{Capture and Evolution of Planetesimals in Circumjovian Disks}
\shortauthors{D'Angelo \& Podolak}

\begin{document}

\title{Capture and Evolution of Planetesimals in Circumjovian Disks}

\author{Gennaro D'Angelo\altaffilmark{1,2} and Morris Podolak\altaffilmark{3}}
\altaffiltext{1}{NASA Ames Research Center, MS 245-3, Moffett Field, CA 94035, USA 
(\href{mailto:gennaro.dangelo@nasa.gov}{gennaro.dangelo@nasa.gov})}
\altaffiltext{2}{SETI Institute, 189 Bernardo Avenue, Mountain View, CA 94043, USA}
\altaffiltext{3}{Department of Geosciences, Tel Aviv University, Ramat Aviv 69978, Israel
(\href{mailto:morris@post.tau.ac.il}{morris@post.tau.ac.il})}

\begin{abstract} 
We study the evolution of planetesimals in evolved gaseous disks, which
orbit a solar-mass star and harbor a Jupiter-mass planet at $a_{p}\approx 5\,\AU$.  
The gas dynamics is modeled with a three-dimensional hydrodynamics code 
that employes nested-grids and achieves a resolution of one Jupiter's radius 
in the circumplanetary disk. The code models solids as individual particles.
Planetesimals are subjected to gravitational forces by the star and the planet,
drag force by the gas, disruption via ram pressure, and mass loss through ablation.
The mass evolution of solids is calculated self-consistently with their
temperature, velocity, and position. We consider icy and icy/rocky bodies
of radius $0.1$--$100\,\mathrm{km}$, initially deployed on orbits 
around the star within a few Hill radii ($\Rhill$) of the planet's orbit.
Planetesimals are scattered inward, outward, and toward disk regions 
of radius $r\gg a_{p}$.
Scattering can relocate significant amounts of solids, provided that regions 
$|r-a_{p}|\sim 3\,\Rhill$ are replenished with planetesimals. 
Scattered bodies can be temporarily captured on planetocentric orbits. 
Ablation consumes nearly all solids at gas temperatures $\gtrsim 220\,\K$.
Super-keplerian rotation around and beyond the outer edge of the gas gap 
can segregate $\lesssim 0.1\,\mathrm{km}$ bodies, producing solid gap edges 
at size-dependent radial locations.
Capture, break-up, and ablation of solids result in a dust-laden circumplanetary
disk with low surface densities of $\mathrm{km}$-size planetesimals, 
implying relatively long timescales for satellite formation.
After a giant planet acquires most of its mass, accretion of solids is unlikely 
to alter significantly its heavy-element content. 
The luminosity generated by solids' accretion can be of a similar order 
of magnitude to the contraction luminosity.
\end{abstract}


\section{Introduction}
\label{sec:Intro}

\defcitealias{podolak1988}{PPR88}

Planetesimal accretion is a key process in the formation of a giant planet.  
In the Core-Nucleated Accretion (CNA) scenario \citep{pollack1996}, 
planetesimal accretion accounts for the formation of the initial core (Phase~1).  
Additional planetesimal capture during the slow accretion of the gaseous 
envelope (Phase~2) releases gravitational energy, which must be radiated 
so that the envelope can contract.  As the captured planetesimals pass 
through the envelope and ablate, they leave behind solid grains which 
affect the opacity.  Both the energy release by sinking solids and 
the opacity affect the planet luminosity and help determine 
the duration of Phase~2 \citep{hubickyj2005,naor2008,naor2010}.  
This ablated material will also affect the subsequent composition 
in the envelope \citep{iaroslavitz2007,mousis2014}.
After the rapid gas accretion and ensuing contraction (Phase~3), 
when the planet has acquired most of its mass, 
additional accretion of solids occurs both onto the planet and 
in the subdisk surrounding the planet (the circumplanetary disk).
This additional accretion adds relatively little to the planet itself, 
but can have important consequences for the formation of the regular 
satellites \citep[e.g.,][]{canup2009,estrada2009} and, possibly, 
for the occurrence of the irregular satellites.

As computing power has increased, studies of planetesimal accretion 
have become more detailed. \citet{pollack1996} calculated the accretion 
rate assuming that the planetesimals were uniformly distributed over 
the feeding zone. \citet{inaba2003b} used a statistical model, which 
simulated such effects as scattering and collisions between planetesimals. 
Modifications and improvements of this idea have been
employed over the past few years \citep{kobayashi2010,bromley2011}.
More recently, \citet{gennaro2014} revisited the problem combining 
detailed calculations for the evolution of a swarm of planetesimals 
and for the structure of the planet's envelope, up to the beginning 
of Phase~2.

The last stage of accretion, after the planet has undergone its rapid 
contraction, presents special difficulties. The planet is massive enough 
to open a gap in the gas density distribution of the circumstellar disk, 
and this constrains the motion of the gas flowing toward the planet
\citep[e.g.,][and references therein]{lissauer2009}.  
The inflowing gas and the associated circumplanetary disk, in turn, 
affect how planetesimals are delivered to and captured by the subdisk 
and by the planet itself.  
Additionally, planet-induced perturbations on the gas may impact 
the redistribution of planetesimals in the circumstellar disk.
In this work, we present the results of calculations that combine the 
three-dimensional (3D) gas dynamics with an N-body module to study 
these processes in detail, including the determination of temperature,
ablation, and fragmentation of planetesimals.
We consider the case of a giant planet of one Jupiter's mass and
model solids in the (initial) size range $0.1$--$100\,\mathrm{km}$.
Results are presented also for smaller bodies, down to $1\,\mathrm{cm}$
in radius.

In what follows, we describe how the thermodynamical evolution of the
disk's gas and of the solids is calculated in Sections~\ref{sec:TED} and 
\ref{sec:TEP}, respectively.
The numerical procedures are outlined in Section~\ref{sec:NM}.
Results for the disk evolution are presented in Section~\ref{sec:DDD},
and those for the evolution of planetesimals are presented in 
Sections~\ref{sec:PCD} and \ref{sec:PCP}. 
We conclude discussing our findings in Section~\ref{sec:DC}.
Finally, further details and numerical tests are given in 
Appendix~\ref{sec:CD} and \ref{sec:TPE}.

\section{Thermodynamical Evolution of the Disk}
\label{sec:TED}

\begin{deluxetable}{cl}
\tablecolumns{2}
\tablewidth{0pc}
\tablecaption{List of Symbols\label{table:qq}}
\tablehead{
\colhead{Symbol}&\colhead{Definition}
}
\startdata
$\{r,\theta,\phi\}$&Stellocentric spherical polar coordinates \\
$r_{\mathrm{mn,mx}}$& Min/Max grid radius \\
$\theta_{\mathrm{mn,mx}}$& Min/Max grid co-latitude angle \\
$\Ms$              & Stellar mass \\
$\Omega_{f}$& Frame rotation rate \\
$\Omega_{p}$& Planet rotation rate \\
$\Mp$              & Planet mass \\
$R_{p}$           & Planet radius \\
$\Rhill$            & Planet's Hill radius \\
$a_{p}$           & Planet semi-major axis \\
$\mathbf{r}_{p}$   & Planet position \\
$\mathbf{v}_{g}$& Gas velocity \\
$\rho_{g}$          & Gas volume density \\
$\Sigma_{g}$     & Gas surface density \\
$P_{g}$           & Gas pressure \\
$c_{g}$           & Gas sound speed \\
$\nu_{g}$           & Turbulence kinematic viscosity \\
$\alpha_{g}$       & Turbulence viscosity parameter \\
$\{\rho_{0},\Sigma_{0}\}$ & Circumstellar disk reference densities \\
$H$                & Circumstellar disk pressure scale-height\\
$T_{n}$           & Circumstellar disk temperature \\
$T_{g}$           & Gas temperature \\
$\mu_{g}$       & Gas mean molecular weight \\
$\eta_{g}$          & Gas molecular dynamical viscosity \\
$\{L_{r},L_{\theta},L_{\phi}\}$ & Particle specific linear/angular momenta \\
$\{\mathbf{r}_{s}, \mathbf{v}_{s},\mathbf{a}_{s}\}$ & Particle position, velocity, and acceleration \\
$\{\mathbf{F}_{D},\mathbf{a}_{D}\}$ & Aerodynamic drag force and acceleration \\
$C_{D}$             & Drag coefficient \\
$\mathcal{M}$& Mach number \\
$\mathcal{R}$& Reynolds number \\
$\mathcal{B}$& Biot number \\
$\rho_{s}$          & Particle density \\
$M_{s}$              & Particle mass \\
$R_{s}$          & Particle radius \\
$\epsilon_{s}$ & Particle emissivity \\
$L_{s}$           & Particle specific vaporization energy \\
$C_{s}$           & Particle specific heat \\
$T_{s}$           & Particle temperature \\
$\lambda_{s}$ & Particle thermal conductivity \\
$\delta_{s}$     & Particle isothermal depth \\
$\mu_{s}$       & Particle mean molecular weight \\
$\sigma_{s}$   & Nominal material compressive strength \\
$P_{v}$          & Particle vapor pressure \\
$T_{\mathrm{cr}}$ & Particle critical temperature \\
$P_{\mathrm{dy}}$ & Dynamical pressure \\
$R_{\mathrm{dy}}$ & Particle break-up radius \\
$\Psi$             & Rock volume fraction of mixed medium \\
$\{\Phi_{l},\Phi_{u}\}$ & Conductivity efficiency factors of mixed medium \\
$\{x,y,z\}$                & Planetocentric cartesian coordinates \\
$\tilde{r}$                 & Distance from the planet \\
$T_{e}$                   & Circumplanetary disk effective temperature \\
$\tilde{H}$               & Circumplanetary disk local thickness \\
$\kappa_{\mathrm{R}}$ & Rosseland mean opacity
\enddata
\end{deluxetable}
We work in a reference frame whose origin is fixed to the star and 
which rotates about the origin at a rate $\Omega_{f}$, equal to 
the angular velocity of the planet around the star, $\Omega_{p}$. 
For a planet on a circular orbit, $\Omega_{p}$ is equal to the mean-motion:
\begin{equation}
\Omega_{p}=\sqrt{\frac{G(\Ms+\Mp)}{a_{p}^{3}}},
\label{eq:omega}
\end{equation}
where $\Ms$ and $\Mp$ are the star's and planet's mass,
respectively, and $a_{p}$ is the planet's semi-major axis.
The planet-to-star mass ratio is $\Mp/\Ms=9.8\times 10^{-4}$.
Table~\ref{table:qq} contains a list of the main symbols used 
in this paper. 

The circumstellar disk is represented by a spherical sector 
with an inner hole.
Consider a spherical polar coordinate system $\{\mathcal{O};r,\theta,\phi\}$, 
where $r$ indicates the polar distance from the origin, $\mathcal{O}$, 
the angle $\theta$ is the co-latitude ($\theta=0$ is the north pole,
$\theta=\pi/2$ is the mid-plane, and $\pi/2-\theta$ is the latitude), 
and the angle $\phi$ is the azimuth. The disk volume is within the
range given by $[r_{\mathrm{mn}},r_{\mathrm{mx}}]\times%
[\theta_{\mathrm{mn}},\theta_{\mathrm{mx}}]\times 2\pi$,
where $r_{\mathrm{mn}}=0.4\,a_{p}$ and $r_{\mathrm{mx}}=4\,a_{p}$.
We assume that the planet's orbit lies in the disk's equatorial plane,
$\theta=\pi/2$, and that the disk is symmetric with respect to this
plane. Consequently, only half of the disk volume needs to be simulated.
Therefore, we set $\theta_{\mathrm{mn}}\simeq 2\pi/5$ and 
$\theta_{\mathrm{mx}}=\pi/2$.

The disk's gas is approximated to a viscous fluid of constant kinematic
viscosity $\nu_{g}$, volume density $\rho_{g}$, and velocity $\mathbf{v}_{g}$.
In the following, all gas-related quantities will bear the subscript `$g$'.
The viscosity $\nu_{g}$ is typically assumed to arise from turbulence 
(of unspecified origin)
within the gas, and needs not to be confused with the molecular viscosity,
introduced below, which is much smaller in magnitude.
We set $\nu_{g}=10^{-5}\,a^{2}_{p}\Omega_{p}$, which corresponds 
to a turbulence parameter \citep{S&S1973} $\alpha_{g}=0.004$ for our
choice of the disk thickness.

We generally assume that the circumstellar gas is locally isothermal 
(the temperature depends only on $r$) and that the pressure is
\begin{equation}
P_{g}=c_{g}^{2}\rho_{g}.
\label{eq:pg}
\end{equation}
The gas sound speed is $c_{g}=(H/r)v_{\mathrm{K}}$, and 
$v_{\mathrm{K}}$ is the local Keplerian velocity. The relative
thickness of the disk above the equatorial plane, $H/r$, 
is taken to be constant and equal to $0.05$.  
Therefore, the simulated disk volume extends in the vertical
direction for more than $5.2$ scale-heights, $H$. 
By assuming the equation of state for an ideal gas, the temperature 
in the circumstellar disk becomes
\begin{equation}
T_{n}=\left(\frac{\mu_{g} m_{\mathrm{H}}}{k_{\mathrm{B}}}\right)c_{g}^{2},
\label{eq:Tg}
\end{equation}
where $\mu_{g}$ is the mean molecular weight of the gas, 
$m_{\mathrm{H}}$ is the hydrogen mass, and $k_{\mathrm{B}}$ is 
the Boltzmann constant. 
Since $c_{g}^{2}\propto 1/r$, the gas temperature 
in the circumstellar disk is proportional to $1/r$ as well.

Given the thermal state of the gas, the disk region in which
the gravity of the planet dominates over that of the star has a 
linear size on the order of the Hill radius,
$\Rhill=a_{p}\left[\Mp/(3\Ms)\right]^{1/3}$, which strictly speaking 
represents the distance of the Lagrange point L$_{1}$ from the 
planet 
\citep[to leading order in $\Mp/\Ms$, e.g.,][]{kopal1978,murray2000}. 
In absence of gas (i.e., neglecting pressure and viscosity effects), 
this region is a solid of revolution (around the planet-star axis) whose 
volume is only about a third of that occupied by the Hill sphere. 
Thus, the effective (volumetric mean) radius of the region is 
$\approx 2\Rhill/3$ \citep{kopal1959,Paczynski1971,kopal1978,eggleton1983}.

\begin{deluxetable}{cccccc}
\tablecolumns{6}
\tablewidth{0pc}
\tablecaption{Disk's Gas Constants\label{table:dgc}}
\tablehead{
\colhead{$H/r$}&\colhead{$\nu_{g}$\tablenotemark{a}}&\colhead{$\mu_{g}$}&%
\colhead{$\gamma_{g}$}&\colhead{$\Sigma_{0}$\tablenotemark{b}}&%
\colhead{$\rho_{0}$\tablenotemark{b}}
}
\startdata
$0.05$ & $10^{-5}$ & $2.39$ & $1.4$ & $10$--$100$ & $10^{-12}$--$10^{-11}$
\enddata
\tablenotetext{a}{In units of $a^{2}_{p}\Omega_{p}$.}
\tablenotetext{b}{Unperturbed $\Sigma_{g}$ and $\rho_{g}$, in cgs units, at $5.2\,\AU$.}
\end{deluxetable}
Table~\ref{table:dgc} summarizes the disk's gas parameters, assuming $\Ms=M_{\odot}$. 
The disk's reference densities at $5.2\,\AU$ are derived from evolution 
models discussed in Section~\ref{sec:DDD}.
The mid-plane temperature of the disk at $5.2\,\AU$, 
for the choice of parameters in Table~\ref{table:dgc}, is $\approx 120\,\K$.
As explained in Section~\ref{sec:CPD}, the gas temperature distribution
is given by Equation~(\ref{eq:Tg}) but, in the restricted region of the 
circumplanetary disk, it is modified
according to simple arguments based on local viscous heating,
black-body heating by background radiation, and radiative cooling.

Orbital migration of the planet is neglected. At the higher $\Sigma_{0}$ 
considered here, the planet would drift inward at a speed $\sim \nu_{g}/a_{p}$ 
\citep{gennaro2008}. 
At the lower $\Sigma_{0}$, migration would be inertia-limited and slower.

\section{Thermodynamical Evolution of Planetesimals}
\label{sec:TEP}

Here we describe the physical model for the thermodynamical evolution 
of the planetesimals. Since the model generally applies to any solid
particle, regardless of the size, in this section we shall refer to the 
planetesimals simply as particles. All particle-related quantities will bear 
the subscript `$s$'. Sometimes, for ease of notation, this subscript is 
dropped.

\subsection{Particle Dynamics}
\label{sec:PD}

Let us introduce the linear momentum per unit mass in the radial
direction $L_{r}=v_{r}$, the meridional angular momentum per unit 
mass $L_{\theta}=r v_{\theta}$, and the azimuthal angular momentum 
per unit mass $L_{\phi}=r\sin{\theta}v^{A}_{\phi}$, all defined in 
an inertial frame of reference. The velocity $v^{A}_{\phi}$ is the absolute 
azimuthal velocity: $v^{A}_{\phi}=v_{\phi}+(r\sin{\theta})^{2}\Omega_{f}$.
In terms of these momenta, the equations of motion of a particle can
be written as
\begin{eqnarray}
\frac{dL_{r}}{dt}&=&a_{r}+\frac{1}{r}\left[\left(\frac{L_{\theta}}{r}\right)^{2}%
                                                     +\left(\frac{L_{\phi}}{r\sin{\theta}}\right)^{2}\right]%
                                                     \label{eq:momeqr}\\
\frac{dL_{\theta}}{dt}&=&a_{\theta}r+\frac{\cos{\theta}}{\sin{\theta}}%
                                    \left(\frac{L_{\phi}}{r\sin{\theta}}\right)^{2}%
                                                     \label{eq:momeqtet}\\
\frac{dL_{\phi}}{dt}&=&a_{\phi}r\sin{\theta} \label{eq:momeqphi},
\end{eqnarray}
where $a_{r}$, $a_{\theta}$, and $a_{\phi}$ are the spherical
components of the gravitational acceleration imparted to the 
particle. 
Notice that the subscript `$s$' associated to the coordinates 
and momenta of the particle is dropped. 

In our case, the acceleration in Equations~(\ref{eq:momeqr}),
(\ref{eq:momeqtet}), and (\ref{eq:momeqphi})
arises from the gravitational forces exerted by the star and 
the planet, from non-inertial forces, and from the drag force 
exerted by the gas
\begin{eqnarray}
\mathbf{a}_{s}&=&\frac{G\Mp(\mathbf{r}_{p}-\mathbf{r}_{s})}{|\mathbf{r}_{p}-\mathbf{r}_{s}|^{3}}%
                      -\frac{G\Ms}{|\mathbf{r}_{s}|^{3}}\mathbf{r}_{s}%
                      +\mathbf{a}_{D}\nonumber\\%
                       & &%
                       -\frac{G\Mp}{|\mathbf{r}_{p}|^{3}}\mathbf{r}_{p}%
                       -\frac{G M_{s}}{|\mathbf{r}_{s}|^{3}}\mathbf{r}_{s}\nonumber\\%
                       & &%
                      -\mathbf{\Omega}_{f}\mathbf{\times}%
                        (\mathbf{\Omega}_{f}\mathbf{\times}\mathbf{r}_{s})%
                      -2\,\mathbf{\Omega}_{f}\mathbf{\times}\mathbf{v}_{s}%
\label{eq:acc}
\end{eqnarray}
where $\mathbf{r}_{p}$ is the position vector of the planet, 
$\mathbf{r}_{s}$, $\mathbf{v}_{s}$, and $M_{s}$ are position, 
velocity, and mass of the particle.
The rotation rate vector, $\mathbf{\Omega}_{f}$, is parallel 
to the direction of the north pole ($\theta=0$), i.e., 
$\mathbf{\Omega}_{f}=\Omega_{f}\mathbf{\hat{z}}$.
The components $a_{r}$, $a_{\theta}$, and $a_{\phi}$ are 
found by projecting $\mathbf{a}_{s}$ along the spherical polar 
unit vectors.

In Equation~(\ref{eq:acc}), the third term on the right-hand side,
$\mathbf{a}_{D}$, is the drag acceleration.
The fourth and fifth terms are the non-inertial accelerations
imparted to the star (the origin) by the planet and the particle, 
respectively.
The last two terms are the centrifugal and Coriolis accelerations.
Additional terms may be included, such as the gravitational 
force per unit mass exerted by the disk on the particle, which we 
ignore here, and that exerted on the star, another non-inertial term, 
which is ignored as well.

Notice that Equations~(\ref{eq:momeqr}), (\ref{eq:momeqtet}), and 
(\ref{eq:momeqphi}) use absolute linear and angular momenta, 
hence they apply regardless of whether the vector 
$\mathbf{\Omega}_{f}$ 
is constant or not. Equation~(\ref{eq:acc}) is valid for 
$\mathbf{\dot{\Omega}}_{f}=\mathbf{0}$ and requires the additional 
non-inertial term 
$-\mathbf{\dot{\Omega}}_{f}\mathbf{\times}\mathbf{r}_{s}$
in case $\mathbf{\dot{\Omega}}_{f}\neq \mathbf{0}$.

Let us indicate with $A_{s}$ the cross section of a particle, then
the drag force experienced by the particle while moving through 
the gas is
\begin{equation}
\mathbf{F}_{D}=\frac{1}{2}C_{D}A_{s}\rho_{g}%
                                   |\mathbf{v}_{g}-\mathbf{v}_{s}|%
                                   (\mathbf{v}_{g}-\mathbf{v}_{s}),
\label{eq:FD}
\end{equation}
where $C_{D}$ is the drag coefficient. For spherical particles of
uniform density $\rho_{s}$ and radius $R_{s}$, the drag force per
unit mass becomes
\begin{equation}
\mathbf{a}_{D}=\frac{3}{8}\frac{C_{D}}{R_{s}}%
                                   \left(\frac{\rho_{g}}{\rho_{s}}\right)%
                                   |\mathbf{v}_{g}-\mathbf{v}_{s}|%
                                   (\mathbf{v}_{g}-\mathbf{v}_{s}).
\label{eq:aD}
\end{equation}
  
In general, the drag coefficient,  $C_{D}$, depends on the 
(relative) Mach number
\begin{equation}
\mathcal{M}=\frac{|\mathbf{v}_{g}-\mathbf{v}_{s}|}{c_{g}},
\label{eq:Mn}
\end{equation}
and on the (relative) Reynolds number, which can be written as
\begin{equation}
\mathcal{R}=2 R_{s}\rho_{g}%
                    \frac{|\mathbf{v}_{g}-\mathbf{v}_{s}|}{\eta_{g}},
\label{eq:Rn}
\end{equation}
in which $\eta_{g}$ represents the molecular dynamical viscosity
of the gas. We use the expression for $C_{D}$ derived by \citet{melosh2008},
which is a continuos function applicable over the full range of
$\mathcal{M}$ and $\mathcal{R}$. In the continuum flow limit, 
that occurs when $\mathcal{M}/\mathcal{R}\ll 1$,
we embed in the coefficient of \citet{melosh2008}
a drag formula proposed by \citet{brown2003}.
The drag coefficient is discussed in more detail 
in Appendix~\ref{sec:CD}.

\subsection{Particle Thermodynamics}
\label{sec:PT}

A particle moving through gas sweeps a mass per unit time 
equal to $A_{g}\rho_{g}|\mathbf{v}_{g}-\mathbf{v}_{s}|$.
Collisions between gas atoms (and/or molecules) and the 
particle transfer some amount of the specific kinetic energy 
of the gas, $|\mathbf{v}_{g}-\mathbf{v}_{s}|^{2}/2$,
to the particle at a rate
$(f_{\mathrm{EK}}/2) A_{g}\rho_{g}|\mathbf{v}_{g}-\mathbf{v}_{s}|^{3}$,
where $f_{\mathrm{EK}}$ is the fraction of the total collisional kinetic
energy transmitted as heat to the particle. This energy exchange 
can also be interpreted in terms of the rate at which work is done
on the particle by drag in the gas frame
$\mathbf{F}_{D}\mathbf{\cdot}(\mathbf{v}_{g}-\mathbf{v}_{s})=%
(C_{D}/2)A_{s}\rho_{g}|\mathbf{v}_{g}-\mathbf{v}_{s}|^{3}$,
hence $f_{\mathrm{EK}}\propto C_{D}$ with a proportionality
factor $\le 1$.
\citealt{podolak1988} (hereafter \citetalias{podolak1988}) 
argued that an upper limit to the proportionality factor is $1/4$, 
though they discussed the possibility for it to be smaller.
Here we take this upper limit and assume that
$f_{\mathrm{EK}}=C_{D}/4$. 
Therefore, the rate at which the particle gains energy due to 
frictional heating with the gas is
$(\pi/8)C_{D}\rho_{g}R_{s}^{2}\,|\mathbf{v}_{g}-\mathbf{v}_{s}|^{3}$.

Another source of heating is represented by the energy 
absorbed from the radiation emitted by the ambient gas at 
temperature $T_{g}$,
$4\pi R_{s}^{2}\, \epsilon_{s} \sigma_{\mathrm{SB}} T_{g}^{4}$
(assuming black-body emission), 
where $\epsilon_{s}$ is the thermal emissivity of the particle 
(here assumed a perfect black-body radiator, 
$\epsilon_{s}=1$\footnote{%
This is a very good approximation for ice (and water), and
typically a reasonable approximation for silicates.})
and $\sigma_{\mathrm{SB}}$ is the Stefan-Boltzmann constant.
Similarly, energy is lost via radiation emitted through the particle 
surface,
$4\pi R_{s}^{2}\, \epsilon_{s} \sigma_{\mathrm{SB}} T_{s}^{4}$.
Loosely speaking, $T_{s}$ represents the particle temperature.
More precisely, as we clarify below, it is the temperature
of an outer isothermal layer of the particle.

Finally, there is energy involved in the phase transition of
the particle's material. If $dM_{s}/dt$ is the rate of change 
of the particle's mass and all of $dM_{s}$ is involved in the
phase transition, $L_{s}dM_{s}/dt$ is the energy per unit 
time absorbed (or released, depending on the sign of $dM_{s}/dt$) 
in the process, where $L_{s}$ is the energy per unit mass required 
to vaporize the substance.  

Accounting for all heating and cooling sources presented above, 
the energy balance equation takes the form
\begin{eqnarray}
\frac{4}{3}\pi R_{s}^{3}\rho_{s} C_{s}\frac{dT_{s}}{dt}&=&%
\frac{\pi}{8}C_{D}\rho_{g}R_{s}^{2}\,|\mathbf{v}_{g}-\mathbf{v}_{s}|^{3}\nonumber\\
                                                                                    &+&%
4\pi R_{s}^{2}\, \epsilon_{s} \sigma_{\mathrm{SB}}%
\left(T_{g}^{4}-T_{s}^{4}\right)\nonumber\\%
                                                                                    &+&%
L_{s}\frac{dM_{s}}{dt}.
\label{eq:dTsdt0}
\end{eqnarray}
In Equation~(\ref{eq:dTsdt0}), $C_{s}$ is the specific heat of 
the particle. \citet{whipple1950} estimated that, for typical
meteoritic material, the the left-hand side may be ignored 
for particles smaller than $0.01\,\mathrm{cm}$ in radius.

In Equation~(\ref{eq:dTsdt0}), the variation of the particle's 
internal energy (the left-hand side), assumes that the 
temperature $T_{s}$ is uniform throughout the volume of 
the body.
Such assumption requires that there be no temperature 
gradient inside the body, i.e., that internal heat conduction
be infinite.
This may indeed be the case for small particles but, as the
particle radius increases, the presence of a temperature 
gradient within the body becomes increasingly non-negligible. 
For example, \citet{love1991} concluded that a significant
temperature gradient may begin to appear across particles 
with a diameter larger than $\sim 0.1\,\mathrm{cm}$ when
$T_{s}\approx 1500\,\K$. 
Therefore, the isothermality assumption advocated in
Equation~(\ref{eq:dTsdt0})
may be justified only in an outer shell of the body.

In order to evaluate the thickness of the surface layer of 
a particle, which may be approximated as isothermal at 
temperature $T_{s}$, we follow the approach of 
\citet{mcauliffe2006}, based on the work of \citet{love1991}. 
A measure of whether or not a temperature gradient develops
inside a heated body can be derived from the Biot number,
which is defined as
\begin{equation}
\mathcal{B}\equiv\frac{h_{c}l_{c}}{\lambda_{s}},
\label{eq:Bi}
\end{equation}
where $h_{c}$ is a characteristic heat transfer coefficient, 
with the units of an energy flux per unit temperature, $l_{c}$ 
is a characteristic  length, and $\lambda_{s}$ is the thermal 
conductivity of the material.
The quantity $h_{c}$ is intended to represent the rate of heat 
exchange between the body and the surrounding environment, 
as a function of the difference of temperature between them.
The characteristic length $l_{c}$ is typically defined 
as the volume-to-surface ratio. For a sphere, $l_{c}$ is
a third of the radius.

It is customary to assume that temperature gradients inside 
a given substance are negligible for $\mathcal{B}\le 0.1$
\citep[e.g.,][]{lienhard2008}.
\citet{love1991} approximated the characteristic heat transfer 
coefficient $h_{c}$ of a layer at temperature $T_{s}$ as 
$\sigma_{\mathrm{SB}}T^{3}_{s}$. Thus, from 
Equation~(\ref{eq:Bi}), one can approximate the maximum 
thickness of the isothermal layer $\delta_{s}$ to
\begin{equation}
\delta_{s}=0.3%
\left(\frac{\lambda_{s}}{\sigma_{\mathrm{SB}}T^{3}_{s}}\right).
\label{eq:diso}
\end{equation}
Obviously, $\delta_{s}$ has an upper bound at $R_{s}$, in which
case the body can be considered as fully isothermal.

Therefore, in general, we will assume that heating and cooling 
processes affect only a surface layer of the body, of thickness 
$\delta_{s}$, rather than its entire volume. In this approximation, 
Equation~(\ref{eq:dTsdt0}) can be re-written as
\begin{eqnarray}
\frac{4}{3}\pi\!\left[R_{s}^{3}-(R_{s}-\delta_{s})^3\right]\!\rho_{s}%
C_{s}\frac{dT_{s}}{dt}&=&%
\frac{\pi}{8}C_{D}\rho_{g}R_{s}^{2}\,|\mathbf{v}_{g}-\mathbf{v}_{s}|^{3}\nonumber\\
                                                                                    &+&%
4\pi R_{s}^{2}\, \epsilon_{s} \sigma_{\mathrm{SB}}%
\left(T_{g}^{4}-T_{s}^{4}\right)\nonumber\\%
                                                                                    &+&%
L_{s}\frac{dM_{s}}{dt}.
\label{eq:dTsdt}
\end{eqnarray}
Note that, in the above equation, the particle radius may vary 
with time due to ablation.
At $T_{s}=100\,\K$, the maximum isothermal depth, $\delta_{s}$, 
of an icy particle is few tens of meters, and somewhat less than 
ten meters at $150\,\K$. 
Since $\lambda_{s}$ varies by a factor less than $2$ between the
two temperatures (see Table~\ref{table:mapar}), this change in 
$\delta_{s}$ is mainly dictated by the increased heat exchange with
the surroundings.
The quoted depths become larger for rocky (quartz) bodies by a factor 
of $\approx 3$ (see Table~\ref{table:mapar}),
but they are broadly in accord with the estimate of 
\citetalias{podolak1988}, who concluded that heating and cooling 
would affect only a relatively thin layer of large, km-size bodies.
This approach, however, is rendered necessary by the fact that
planetesimals may spend most of their time in the cool circumstellar 
and circumplanetary disk environments.

In Equations~(\ref{eq:diso}) and (\ref{eq:dTsdt}), both the thermal
conductivity $\lambda_{s}$ and the specific heat $C_{s}$ are 
functions of $T_{s}$\footnote{Here, the material is assumed to be
compact, and possible effects due porosity, inhomogeneity and 
impurity of the substance are neglected.}. 
We use piece-wise fits to the data reported
by \citet{CRC92} and \citet{jensen1980} for ice and by
\citet{powell1966} and \citet{chase1998} for quartz (SiO$_{2}$).
The specific energy of vaporization, $L_{s}$, is instead approximated 
as constant (see discussion in \citetalias{podolak1988}). 
Here, rather than rocks, we consider a mixture in which rocks
are embedded in an icy matrix.
Indicating the mass fractions of ice and rock with $\chi_{\mathrm{ice}}$ and 
$\chi_{\mathrm{rock}}$ (so that $\chi_{\mathrm{ice}}+\chi_{\mathrm{rock}}=1$),
the specific heat of the mixture is given by
\begin{equation}
C_{s}=\chi_{\mathrm{ice}}C_{s}^{\mathrm{ice}}+\chi_{\mathrm{rock}}C_{s}^{\mathrm{rock}}.
\label{eq:Csmix}
\end{equation}
The thermal conductivity of the mixture is approximated as
\begin{equation}
\lambda_{s}=\lambda_{s}^{\mathrm{ice}}\left[(1-\Psi)\Phi_{u}+\Psi\Phi_{l}\right],
\label{eq:lsmix}
\end{equation}
where $\Psi=\chi_{\mathrm{rock}}\rho_{s}^{\mathrm{ice}}/%
                   (\chi_{\mathrm{rock}}\rho_{s}^{\mathrm{ice}}+%
                   \chi_{\mathrm{ice}}\rho_{s}^{\mathrm{rock}})$
is the fraction of the volume occupied by rock.
The quantities $\Phi_{u}$ and $\Phi_{l}$ represent efficiency factors 
for the thermal conductivity of a mixed medium, composed of a matrix 
of one material embedding grains of a second material
\citep[see discussion in][]{prialnik2004}.
Rock would constitute the matrix of the mixed medium for $\Psi>0.5$.
In Equation~(\ref{eq:lsmix}), $\Phi_{u}$ and $\Phi_{l}$ are both functions 
of $\Psi$ and $\lambda_{s}^{\mathrm{rock}}/\lambda_{s}^{\mathrm{ice}}$,
and are respectively given by Equations~(25) and (26) of \citet{prialnik2004}. 
Equation~(\ref{eq:lsmix}) converges to
$\lambda_{s}^{\mathrm{ice}}$ for $\Psi\rightarrow 0$ 
(both $\Phi_{u}$ and $\Phi_{l}\rightarrow 1$)
and to $\lambda_{s}^{\mathrm{rock}}$ for $\Psi\rightarrow 1$
(both $\Phi_{u}$ and $\Phi_{l}\rightarrow \lambda_{s}^{\mathrm{rock}}/\lambda_{s}^{\mathrm{ice}}$).
We use an ice mass fraction $\chi_{\mathrm{ice}}=0.6$, hence $\Psi\simeq 0.334$.
The medium remains mixed throughout the evolution and possible effects 
of differentiation \citep{mosqueira2010} are ignored.
A summary of some material's properties is listed in Table~\ref{table:mapar},
including values of $C_{s}$ and $\lambda_{s}$ at three representative 
temperatures.

\begin{deluxetable}{lccc}
\tablecolumns{4}
\tablewidth{0pc}
\tablecaption{Material's Properties\label{table:mapar}}
\tablehead{
\colhead{Symbol}&\colhead{Ice}&\colhead{Rock}
&\colhead{Ice+Rock}
}
\startdata
$\rho_{s}\tablenotemark{a}$ [$\mathrm{g\,cm^{-3}}$] &  $1.00$ & $2.65$ & $1.33$\\
$\varepsilon_{s}$\tablenotemark{b}&  $1.00$ & $1.00$ & $1.00$\\
$L_{s}$\tablenotemark{c} [$\mathrm{erg\,g^{-1}}$] &$2.83\times 10^{10}$ & $8.08\times 10^{10}$&$2.83\times 10^{10}$\\
$L_{s}$\tablenotemark{d} [$\mathrm{erg\,g^{-1}}$] &$2.50\times 10^{10}$ & $7.92\times 10^{10}$&$2.50\times 10^{10}$\\
$\mu_{s}$\tablenotemark{e}&  $18.0$ & $60.1$ & $25.0$ \\
$\sigma_{s}$\tablenotemark{f} [$\mathrm{dyne\,cm^{-2}}$] &$10^{6}$ & $10^{7}$ & $10^{6}$ \\ 
\multicolumn{2}{l}{$C_{s}$\tablenotemark{g} [$\mathrm{erg\,g^{-1}\,K^{-1}}$]} &                 &    \\
\multicolumn{1}{r}{at $50\,\mathrm{K}$}&%
                                            $4.35\times 10^{6}$ & $9.56\times 10^{5}$ & $2.99\times 10^{6}$ \\
\multicolumn{1}{r}{at $100\,\mathrm{K}$}&%
                                            $8.30\times 10^{6}$ & $2.67\times 10^{6}$ & $6.05\times 10^{6}$ \\
\multicolumn{1}{r}{at $200\,\mathrm{K}$}&%
                                            $1.58\times 10^{7}$ & $5.43\times 10^{6}$ & $1.17\times 10^{7}$ \\
\multicolumn{2}{l}{$\lambda_{s}$\tablenotemark{h} [$\mathrm{erg\,s^{-1}\,cm^{-1}\,K^{-1}}$]}%
                                                                                                                           &                 &    \\
\multicolumn{1}{r}{at $50\,\mathrm{K}$}&%
                                            $1.33\times 10^{6}$ & $5.89\times 10^{6}$ & $1.85\times 10^{6}$ \\
\multicolumn{1}{r}{at $100\,\mathrm{K}$}&%
                                            $6.41\times 10^{5}$ & $2.09\times 10^{6}$ & $8.33\times 10^{5}$ \\
\multicolumn{1}{r}{at $200\,\mathrm{K}$}&%
                                            $3.10\times 10^{5}$ & $9.55\times 10^{5}$ & $3.98\times 10^{5}$
\enddata
\tablenotetext{a}{Density.}
\tablenotetext{b}{Thermal emissivity.}
\tablenotetext{c}{Specific vaporization energy of the solid phase.}
\tablenotetext{d}{Specific vaporization energy of the liquid phase.}
\tablenotetext{a}{Mean molecular weight.}
\tablenotetext{f}{Nominal compressive strength at $R_{s}=10^{5}\,\mathrm{cm}$.}
\tablenotetext{g}{Specific heat.}
\tablenotetext{h}{Thermal conductivity.}
\end{deluxetable}

\subsection{Particle Ablation}
\label{sec:PA}

The heat deposited in the outer layer of a body can cause phase 
transitions of its material, and hence mass loss. Here we consider 
that mass loss is caused by transition to the gas phase.
The rate at which vaporization removes mass from a solid body
can be approximated by the Hertz-Knudsen-Langmuir equation 
\citep[e.g.,][and references therein]{blottner1971,campbell2004}.
Indicating with $P_{v}$ and $\mu_{s}$, respectively, the vapor 
pressure and the mean molecular weight of the material, 
arguments from the kinetic theory of gases imply that the flux 
of atoms/molecules leaving the surface of a body is 
$\mu_{s}m_{\mathrm{H}} P_{v}/(k_{\mathrm{B}}T_{s})\bar{\mathcal{V}}_{s}/4$,
where $\bar{\mathcal{V}}_{s}$ is the average thermal speed of 
atoms/molecules in the vapor \citep[e.g.,][]{m&m}:
\begin{equation}
\bar{\mathcal{V}}_{s}=\sqrt{\frac{8}{\pi}%
                             \frac{k_{\mathrm{B}}T_{s}}{\mu_{s}m_{\mathrm{H}}}}.
\label{eq:vvap}
\end{equation}
Integrating the flux over the surface of the (spherical) body, 
we have that the mass loss rate is
\begin{equation}
\frac{dM_{s}}{dt}=-4 \pi R_{s}^{2}%
         P_{v}\sqrt{\frac{\mu_{s}m_{\mathrm{H}}}{2\pi k_{\mathrm{B}}T_{s}}},
\label{eq:dmsdt}
\end{equation}
and $P_{v}=P_{v}(T_{s})$.
Equation~(\ref{eq:dmsdt}) assumes that the vapor is rapidly carried 
away from the body's surface, 
i.e., the partial pressure of the vapor
in the gas is unimportant.
In case of the ice-rock mixture, vapor 
carries away the icy matrix first (due to higher vapor pressure), 
but we assume that the rocky material embedded in the matrix is also 
lost by appropriately modifying $\mu_{s}$.

\begin{deluxetable*}{crrrrrr}
\tablecolumns{7}
\tablewidth{0pc}
\tablecaption{Constants in Vapor Pressure Formulas\tablenotemark{a}\label{table:pvc}}
\tablehead{
\colhead{Equation}&\colhead{$a_{0}$}&\colhead{$a_{1}$}&\colhead{$a_{2}$}&
\colhead{$a_{3}$}&\colhead{$a_{4}$}&\colhead{$a_{5}$}
}
\startdata
(\ref{eq:pvwash}) &$-2445.5646$ &$8.2312$ &%
                              $-0.01677006$                  &$0.0000120514$ &%
                              $-3.632266$    & \\
(\ref{eq:pvwag})   &$-7.85951783$                         &$1.84408259$&%
                               $-11.7866497$&$22.6807411$&%
                               $-15.9618719$&$1.80122502$  \\
(\ref{eq:pvquartz})&$31.82319964$&$46071.4304$   & $58.883$ & & &
\enddata
\tablenotetext{a}{The pressure is in units of dyne/cm$^{2}$.}
\end{deluxetable*}
The vapor pressure can be obtained by integrating the 
Clausius-Clapeyron equation. The resulting function depends
on a number of constants that are fixed using physical 
arguments and experimental data.
For icy bodies, at temperatures below the melting point ($T_{s}=273.16\,\K$), 
we use the formula of  \citet{washburn1924}, which reads
\begin{equation}
\log{P_{v}}=a_{0}+a_{1}T_{s}+a_{2}T^{2}_{s}+\frac{a_{3}}{T_{s}}%
                 +a_{4}\log{T_{s}}.
\label{eq:pvwash}
\end{equation}
For $P_{v}$ expressed in units of dyne/cm$^{2}$ ($=0.1\,\mathrm{Pa}$), 
the constants $a_{i}$ are given in Table~\ref{table:pvc}. 
Although \citeauthor{washburn1924}'s 
formula dates back $90$ years, it agrees very well with the
2011 release of the sublimation pressure of ordinary water ice from
The International Association for the Properties of Water and Steam.
\citet{washburn1924} found that Equation~(\ref{eq:pvwash})
satisfactorily reproduced the experimental data accessible to him  
(above $\sim 170\,\K$). We find that this formula actually gives 
a good fit to all the values of the vapor pressure of ice reported 
by \citet{CRC92}, which extend down to $50\,\K$.

Above the melting point and below the critical temperature,
$T_{\mathrm{cr}}=647.096\,\K$, 
we use the fitting function from
\citet{wagner2002}
\begin{eqnarray}
\ln{\!\left(\frac{P_{v}}{P_{\mathrm{cr}}}\right)\!}=%
\!\left(\frac{T_{\mathrm{cr}}}{T_{s}}\right)\!%
\left(\right.
\!a_{0}\vartheta_{s}&+&a_{1}\vartheta_{s}^{1.5}+a_{2}\vartheta_{s}^{3}%
                                   +a_{3}\vartheta_{s}^{3.5}\nonumber\\
                               &+&a_{4}\vartheta_{s}^{4}+a_{5}\vartheta_{s}^{7.5}%
\!\left.\right),
\label{eq:pvwag}
\end{eqnarray}
where $\vartheta_{s}=(1-T_{s}/T_{\mathrm{cr}})$,
$P_{\mathrm{cr}}=2.2064\times 10^{8}\,\mathrm{dyne/cm}^{2}$
is the vapor pressure at the critical temperature, and the constants 
$a_{i}$ can be found in Table~\ref{table:pvc}.
We find that the vapor pressure from Equation~(\ref{eq:pvwag})
becomes larger than that from Equation~(\ref{eq:pvwash}) for
$T_{s}>272.84\,\K$, hence we use this temperature value
for the transition between the two formulas, 
Equations~(\ref{eq:pvwash}) and (\ref{eq:pvwag}).

Alternatively, the vapor pressure can be derived by fitting expreimental
data. For quartz, we adopt the fitting function published on the Chemistry 
WebBook of the National Institute of Standards and Technology (NIST)
\begin{equation}
\ln{P_{v}}=a_{0}-\frac{a_{1}}{T_{s}+a_{2}}.
\label{eq:pvquartz}
\end{equation}
The constants $a_{i}$ are displayed in Table~\ref{table:pvc} for 
$P_{v}$ expressed in units of dyne/cm$^{2}$. 
This NIST fit applies over a limited range of temperatures, and it should
be considered as an extrapolation at lower temperatures and
up to the critical temperature, $T_{\mathrm{cr}}=4500\,\K$. 
However, since the vapor pressure of the icy matrix is much higher,
the mass loss of mixed-composition particles is also governed by 
Equations~(\ref{eq:pvwash}) and (\ref{eq:pvwag}).

At temperatures greater than $T_{\mathrm{cr}}$,  there is no 
distinction between the vapor and the liquid phase and
the mass loss rate is energy limited (see \citetalias{podolak1988}).
For $T_{s}\ge T_{\mathrm{cr}}$, the mass vaporization rate is
\begin{eqnarray}
\frac{dM_{s}}{dt}&=&\frac{1}{L_{s}}\!\left[%
4\pi R_{s}^{2}\, \epsilon_{s} \sigma_{\mathrm{SB}}%
\left(T^{4}_{\mathrm{cr}}-T_{g}^{4}\right)\phantom{\frac{1}{1}}\right.\nonumber\\%
                                                                                    &-&%
\left.
\frac{\pi}{8}C_{D}\rho_{g}R_{s}^{2}\,|\mathbf{v}_{g}-\mathbf{v}_{s}|^{3}%
\right].
\label{eq:dmsdt_crit}
\end{eqnarray}
When $T_{s}=T_{\mathrm{cr}}$, a particle evaporates at a constant 
temperature \citep{hood1991}. At and beyond the critical temperature,
any net energy input is used for ablation.
If there is a net energy output (i.e., when radiative cooling 
becomes larger than the sum of frictional and radiative heating) 
the vaporization rate is set to zero.

Equations~(\ref{eq:dmsdt}) and (\ref{eq:dmsdt_crit}) are applied together
with the Equations of motion (\ref{eq:momeqr}), (\ref{eq:momeqtet}), and
(\ref{eq:momeqphi}) under the hypothesis of isotropic mass loss, where 
the isotropy is with respect to the center of mass of the moving body. 
In other words, it is assumed that the absolute momenta of the escaping 
mass are equal to those said mass would have if it was attached to 
the moving body \citep{kopal1978}.

\subsection{Fracturing and Break-up of Planetesimals}
\label{sec:FBP}

A solid body acted upon by external forces is stressed to some degree.
In case of a spherical body, if the stress overcomes the 
compressive strength of the material, the body can fracture.
\citet{pollack1979} \citep[see also][]{baldwin1971} approximated 
the differential force (per unit surface area) across a body traveling 
though gas as the dynamical pressure
\begin{equation}
P_{\mathrm{dy}}=\frac{1}{2}\rho_{g}|\mathbf{v}_{g}-\mathbf{v}_{s}|^{2}.
\label{eq:pdy}
\end{equation}

Non-spherical bodies are also subject to bending, hence they can 
fracture at stresses lower by about an order of magnitude, i.e., 
once the tensile strength is exceeded 
\citep[see discussion in][]{baldwin1971}. The compressive strength
has typically an inverse dependence on the body size, the body temperature, 
and the material porosity \citep{petrovic2003}. 
Material strengths are also sensitive to the rate of strain
\citep[e.g.,][]{lange1983}, i.e., the rate at which the external force is 
applied. 
This may be  especially important for large bodies.
Simulations and laboratory experiments suggest that the strength 
of rocky, iron, and icy bodies is proportional to $1/R_{s}$ to a 
some power, which is typically between $0.3$ and $0.5$ 
\citep{housen1999,benz1999}.

A fractured body can quickly break apart, unless it is held together
by its own gravity, which occurs if the radius exceeds
\begin{equation}
R_{\mathrm{dy}}=\sqrt{\frac{5}{4\pi}%
\frac{P_{\mathrm{dy}}}{G\rho^{2}_{s}}}.
\label{eq:rdy}
\end{equation}
\citep{pollack1979,pollack1986}. Here we assume
that if $R_{s}<R_{\mathrm{dy}}$ and the dynamical pressure 
exceeds the compressive strength of the particle's material, 
the body is completely disrupted and the fragments quickly 
dissolve (which is probably a good approximation if the 
fragments are sufficiently small).
If $R_{s}>R_{\mathrm{dy}}$ the body \textit{does not} break apart,
independently of $P_{\mathrm{dy}}$.

The compressive strengths of planetesimals are largely unknown.
Data obtained from the fragmentation of stony and iron meteorites
in the Earth's atmosphere imply strengths within the range from
$10^{6}$ to $10^{9}\,\mathrm{dyne/cm}^{2}$
\citep{ceplecha1993,petrovic2001,popova2011}. 
The compressive strength of solid ice is on the order of 
$10^{7}\,\mathrm{dyne/cm}^{2}$ \citep{petrovic2003},
though it is expected to be lower for porous ice \citep{cox1985}.
The inferred compressive strength of primitive icy bodies
in the solar system, such as comets, is much smaller, 
$\lesssim 10^{4}\,\mathrm{dyne/cm}^{2}$ \citep{toth2006}. 
\citetalias{podolak1988} argued that the old age of comets may have 
significantly altered their mechanical properties through outgassing.
\citet{biele2009} also pointed out that strengths measured from 
comets may be affected by pre-existing faulting.
Thus, the compressive strength of ``young'' icy planetesimals may 
as well be in the range from $\sim 10^{5}$ to 
$\sim10^{6}\,\mathrm{dyne/cm}^{2}$.

Given the large uncertainties, we set the material compressive 
strength to $\sigma_{s}\sqrt{1\,\mathrm{km}/R_{s}}$
\citep{holsapple2009}. 
The nominal strength, $\sigma_{s}$, at the $1\,\mathrm{km}$-scale
size is $10^{6}$ and $10^{7}\,\mathrm{dyne/cm}^{2}$
for icy and rocky planetesimals, respectively
(see Table~\ref{table:mapar}).
This choice of the compressive strengths implies that,
according Equation~(\ref{eq:rdy}), only icy (rocky) bodies whose radius
is smaller than $\approx 10\,\mathrm{km}$ ($\approx 20\,\mathrm{km}$) 
can fragment, 
if the dynamical pressure is just marginally larger than the compressive
strength. Larger bodies can fracture more easily, but are held together 
by their own gravity. However, the maximum radius for break-up
increases as $P_{\mathrm{dy}}$ increases.

\section{Numerical Methods}
\label{sec:NM}

\subsection{Solution for the Disk Hydrodynamics}
\label{sec:SDH}

The disk's gas is described as a continuum viscous fluid via
the Navier-Stokes equations \citep[see, e.g.,][]{m&m}, 
written in terms of the specific linear momentum and the specific 
total angular momenta of the gas in a rotating frame
\citep[see][]{gennaro2005}. 
These equations are solved in a stepwise fashion by means 
of a finite-difference code. The solution of the advection term, 
referred to as the transport step, applies the monotonic transport 
of \citet{vanleer1977} and uses an operator-splitting technique 
\citep[see][]{stone1992a} to cope with the three dimensions.
In the source step, the other terms of the equations are taken
into account, namely the apparent forces,
the gradients of pressure and gravity, and the viscous stresses.
Numerical stability is ensured by constraining the integration 
time step, $\Delta t$, according to the Courant-Friedrichs-Lewy 
condition \citep[see][]{stone1992a}.
Overall, the algorithm is second-order accurate in 
space and effectively second-order accurate in time 
\citep[e.g.,][]{boss1992}.
The code was compared against other fluid dynamics codes 
in studies involving problems of tidal interactions between 
planets and disks \citep{devalborro2006,masset2006a,devalborro2007}.

\begin{deluxetable}{crrrc}
\tablecolumns{5}
\tablewidth{0.9\linewidth}
\tablecaption{Grid Structure\label{table:grids}}
\tablehead{
\colhead{Level}&\colhead{$N_{r}$}&\colhead{$N_{\theta}$}&\colhead{$N_{\phi}$}&
\colhead{Volume\tablenotemark{a}}
}
\startdata
 $1$ & $243$  &  $22$ & $423$&  disk\\
 $2$ &  $84$  &   $24$ &  $84$ & $9.13\times 2.53\times 9.15$\\
 $3$ & $104$  &  $34$ & $104$& $5.65\times 1.79\times 5.66$\\
 $4$ & $124$  &  $44$ & $124$& $3.37\times 1.16\times 3.37$\\
 $5$ & $164$  &  $64$ & $164$& $2.23\times 0.84\times 2.23$\\
 $6$ & $244$  &  $84$ & $244$& $1.66\times 0.55\times 1.66$\\
 $7$ & $404$  & $104$& $404$& $1.37\times 0.34\times 1.37$\\
 $8$ & $724$  & $144$& $724$& $1.23\times 0.24\times 1.23$
\enddata
\tablenotetext{a}{Volume is in units of $R^{3}_{\mathrm{H}}$, except for grid level $1$.}
\end{deluxetable}
The Navier-Stokes momentum equations are discretized over 
a spherical polar grid with constant spacing in all three coordinate 
directions.
The code allows for grid refinements by means of  a nested-grid
technique \citep{gennaro2002,gennaro2003b}. The increase
of volume resolution is a factor of $2^{3}$ for any level added
to the grid system.
In this study, we employ a grid system with $8$ levels, the details
of which are given in Table~\ref{table:grids}.
The first level encloses the entire disk, whereas additional levels
enclose smaller and smaller disk portions around the planet.
In Table~\ref{table:grids}, $N_{r}$, $N_{\theta}$, and $N_{\phi}$
indicate the number of grid points along the correspondent 
coordinate directions.
The last column gives the volume occupied by each grid level,
where the lengths are in units of \Rhill.
Overall, the grid system contains about $103$ million grid elements.
The region of the wider circumplanetary disk, typically taken as
$\sim \Rhill/4$ around the planet, is discretized over the 
$8^{\mathrm{th}}$ grid with more than $12$ million gird elements.

The spatial resolution on the first grid level is such that
$\Delta r/a_{p}\simeq a_{p}\,\Delta \phi/(r\sin{\theta})\simeq 0.015$
and $a_{p}\,\Delta \theta/r\simeq 0.013$.
On the $8^{\mathrm{th}}$ grid level, the linear resolution around the planet
is $\approx 10^{-4}\,a_{p}\approx 1.4\times 10^{-3}\,\Rhill$, 
which is about equal to Jupiter's current radius, $R_{J}$, at $5.2\,\AU$. 
Note that the actual radius of the planet, $R_{p}$, at these
late stages of accretion (i.e., when it is no longer accreting substantial
quantities of gas compared to its mass) is likely $\gtrsim 1.3\,R_{J}$ and 
$\lesssim 1.8\,R_{J}$ \citep{lissauer2009}.
We adopt the value $R_{p}=1.6\,R_{J}$.

We apply boundary conditions at the inner and outer disk radii,
$r_{\mathrm{mn}}$ and $r_{\mathrm{mx}}$, using the procedures 
of \citet{devalborro2006}. Boundaries at the disk surface 
($\theta=\theta_{\mathrm{mn}}$) and at the equatorial plane are 
handled as in \citet{masset2006a}.
In these calculations, we do not account for accretion on
the central star which, in conjunction with accretion on the planet,
can alter the density in the disk interior of the planet's orbit
\citep{lubow2006}.

\subsection{Solution for the Planetesimal Thermodynamics}
\label{sec:SPE}

The set of Equations~(\ref{eq:momeqr})--(\ref{eq:momeqphi}) is completed by the
equations to obtain the spherical polar coordinates of a particle
\begin{eqnarray}
\frac{dr}{dt}        &=&L_{r}                                                               \nonumber\\
\frac{d\theta}{dt}&=&\frac{L_{\theta}}{r^{2}}                                    \label{eq:rtpeq}\\
\frac{d\phi}{dt}   &=&\frac{L_{\phi}}{(r\sin{\theta})^{2}}-\Omega_{f}\nonumber.
\end{eqnarray}

The system of first order ordinary differential equations (ODE),  
represented by Equations~(\ref{eq:rtpeq}),
(\ref{eq:momeqr}), (\ref{eq:momeqtet}), (\ref{eq:momeqphi}), 
(\ref{eq:dTsdt}), and (\ref{eq:dmsdt}) or (\ref{eq:dmsdt_crit}),
is solved numerically by means of a variable (arbitrarily high) order 
and variable step-size 
Gragg-Bulirsch-Stoer extrapolation algorithm \citep{hairer1993}.
Indicating with $\Delta t$ the time step of the hydrodynamical
calculation at time $t$ (see Section~\ref{sec:SPE}), 
the system of ODE is integrated three times, according to the 
step-size sequence $(\Delta t/4,\Delta t/2,\Delta t/4)$, using 
gas field distributions centered at $(t, t+\Delta t/2, t+\Delta t)$, 
respectively.

The ODE solver chooses automatically the order of the algorithm
and a series of appropriate internal time intervals so to advance 
the solution to the required end time. The algorithm's order and 
the length of each time interval are constrained by 
user-supplied 
tolerances on the local truncation error of the solution, which is
estimated from the comparison of solutions at different orders. 
Here we apply tolerances in the range from $2\times10^{-16}$ to $10^{-10}$. 

The 3D gas field distributions of $\rho_{g}$, $T_{g}$, and $\mathbf{v}_{g}$, 
are interpolated in space at the position of the particles by means of a 
second-order accurate algorithm, based on monotonic harmonic means
\citep{vanleer1977}, according to the approach of \citet{gennaro2002}, 
extended to three dimensions. The advantage of this method rests on its
capability of handling discontinuities and shock-like conditions in the gas.
The spatial as well as temporal interpolations are performed on 
the gas field distributions with the highest available resolution, which 
are those calculated on the most refined grid level where the particle 
is located.

Several tests of the planetesimal thermodynamics solver are presented 
in Appendix~\ref{sec:TPE}. These include standard two- and three-body 
problems, drag-induced orbital decay and free-fall of particles, and various 
thermal evolution problems.

\subsection{Gas and Particle Accretion}
\label{sec:GPA}
 
The accretion of gas onto a gap-opening planet is a complex problem.
It was suggested by \citet{gennaro2003a} and \citet{bate2003}, and later
confirmed 
\citep[see the recent studies by][and references therein]{tanigawa2012,ayliffe2012a,gressel2013,szulagyi2014}, 
that gas mostly proceeds off the mid-plane (at and above the surface) 
of the disk around the planet, prior to accreting on its
envelope. We do not model the planet's envelope here and adopt 
a prescription for gas accretion along the lines of \citet{gennaro2003a} 
and \citet{gennaro2008}, a procedure that is directionally unbiased.
The spherical volume around the planet from which gas is removed
to mimic accretion extends for  $1.4\,R_{p}$ in radius ($\approx 2.2\,R_{J}$), 
which makes the procedure independent of the mode of gas delivery
to the planet's envelope. 
At such short distances,
the thermal energy of the gas is much smaller than the gravitational 
energy binding the gas to the planet \citep[][]{bodenheimer1986}, 
hence gas cannot escape.
Planet formation calculations do indicate that the rates of accretion 
calculated in this manner correspond to the actual rates of envelope 
growth \citep{lissauer2009}.

The accretion of planetesimals is a simpler problem,
since arbitrarily close encounters with the planet are allowed.
During close approaches, the gravitational potential of the planet 
is always used, and no regularization is applied 
\citep[see, e.g.,][]{bodenheimer2006}.
We adopt two criteria for accretion: if the particle approaches the planet 
within a distance $\le R_{p}$ ($1.6\,R_{J}$), a head-on impact 
is assumed;
otherwise, and if the distance of approach is $\le 2.2\,R_{p}$
($3.5\,R_{J}$), the particle is deemed as accreted if its relative velocity 
is less than the escape velocity from the planet at that distance.

\section{Disk Density and Dynamics}
\label{sec:DDD}

\citet{gennaro2012} performed calculations of circumstellar 
disk evolution driven by viscous diffusion and photoevaporation, 
exploring ranges of stellar EUV luminosity, initial disk mass, 
initial mass distribution, and gas kinematic viscosity representative 
of the protosun and the early solar nebula. 
These parameters can be constrained by the requirements that 
the disk's gas lifetime be shorter than $20\,\mathrm{Myr}$ 
\citep[e.g.,][]{haisch2001,pascucci2006,roberge2011,williams2011,bell2013} 
and longer 
than the formation time of a Jupiter-mass planet at $\approx 5\,\AU$, 
which is $\gtrsim 1\,\mathrm{Myr}$ 
\citep{hubickyj2005,alibert2005b,lissauer2009,naor2010,mordasini2011}.

\begin{figure}[t!]
\centering%
\resizebox{\linewidth}{!}{%
\includegraphics{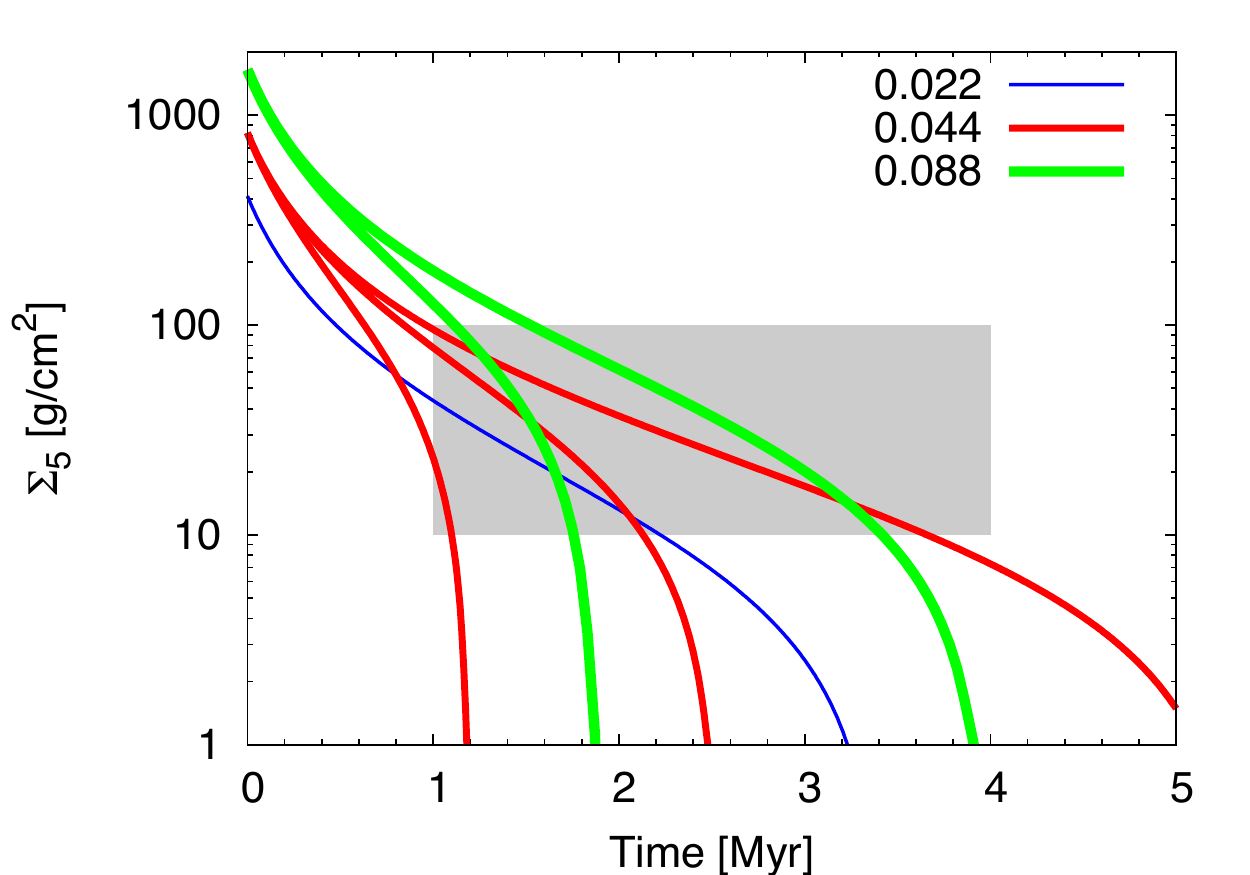}}
\caption{%
             Surface density at $5\,\AU$ versus time in disk models
             whose evolution is driven by viscous diffusion and
             photoevaporation. The numbers in the legend indicate
             the initial disk mass, in units of $\Ms$, within $40\,\AU$ 
             of the star. The models also employ different gas kinematic 
             viscosities and EUV fluxes emitted by the star.
             See text for an explanation of the shaded area.
             Data from the models of \citet{gennaro2012}.
              }
\label{fig:S5}
\end{figure}
In Figure~\ref{fig:S5}, we plot results from some models
of \citet{gennaro2012}, in particular the surface density at $5\,\AU$
versus time. The ratio of the initial disk mass to the stellar
mass is indicated in the top-right corner.
The curves indicate that, after $\sim 1\,\mathrm{Myr}$,
the density is at most $\approx 200\,\mathrm{g\,cm}^{-2}$,
and
is typically smaller than $100\,\mathrm{g\,cm}^{-2}$
(but could be much smaller, $\approx 20\,\mathrm{g\,cm}^{-2}$). 
After
$\sim 2\,\mathrm{Myr}$, the surface density ranges from
$\sim10$ to $\sim 50\,\mathrm{g\,cm}^{-2}$.
In all the
models in the figure, the initial surface density inside of about 
$10\,\AU$ is proportional to $1/\sqrt{r}$, consistent with that
used in the hydrodynamical calculations.

Therefore, given the range of possible disk densities at the time
of Jupiter's formation (between, say, $\sim 1$ and $\sim 4\,\mathrm{Myr}$), 
we consider two values for the unperturbed 
surface density, $\Sigma_{0}$, at $5.2\,\AU$
of $\approx 10$ and  $\approx 100\,\mathrm{g\,cm}^{-2}$, 
which correspond to mass densities $\rho_{0}\approx 10^{-12}$
and $\rho_{0}\approx 10^{-11}\,\mathrm{g\,cm}^{-3}$.
The shaded area in Figure~\ref{fig:S5} represents the 
area covered by our choice of $\Sigma_{0}$.
The disk mass in units of $\Ms$, inside of $\approx 21\,\AU$, 
is $\approx 8\times 10^{-5}\,\Sigma_{0}$, where $\Sigma_{0}$ 
is in units of $\mathrm{g\,cm}^{-2}$.

\begin{figure*}
\centering%
\resizebox{\figlen}{!}{%
\includegraphics{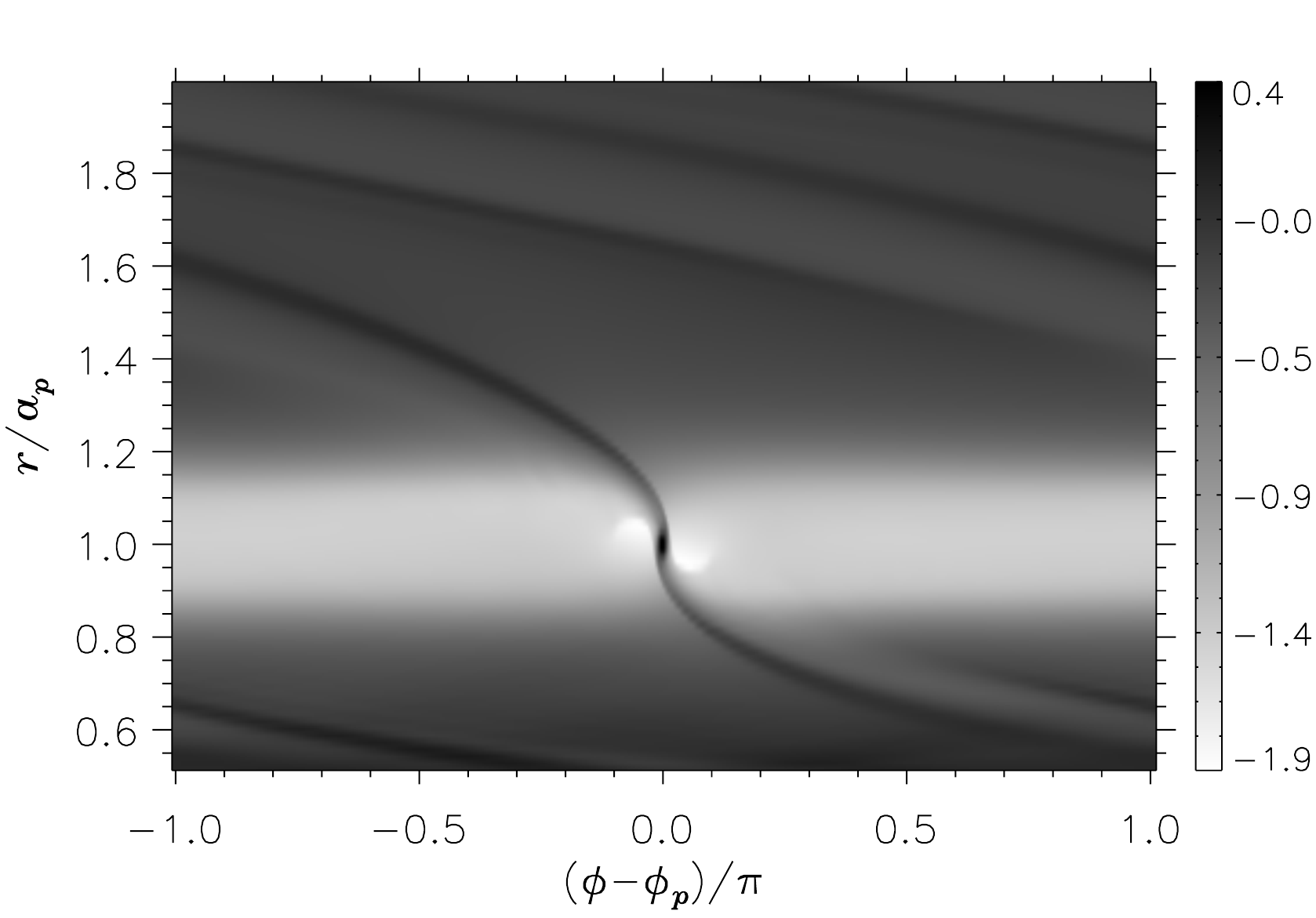}%
\includegraphics{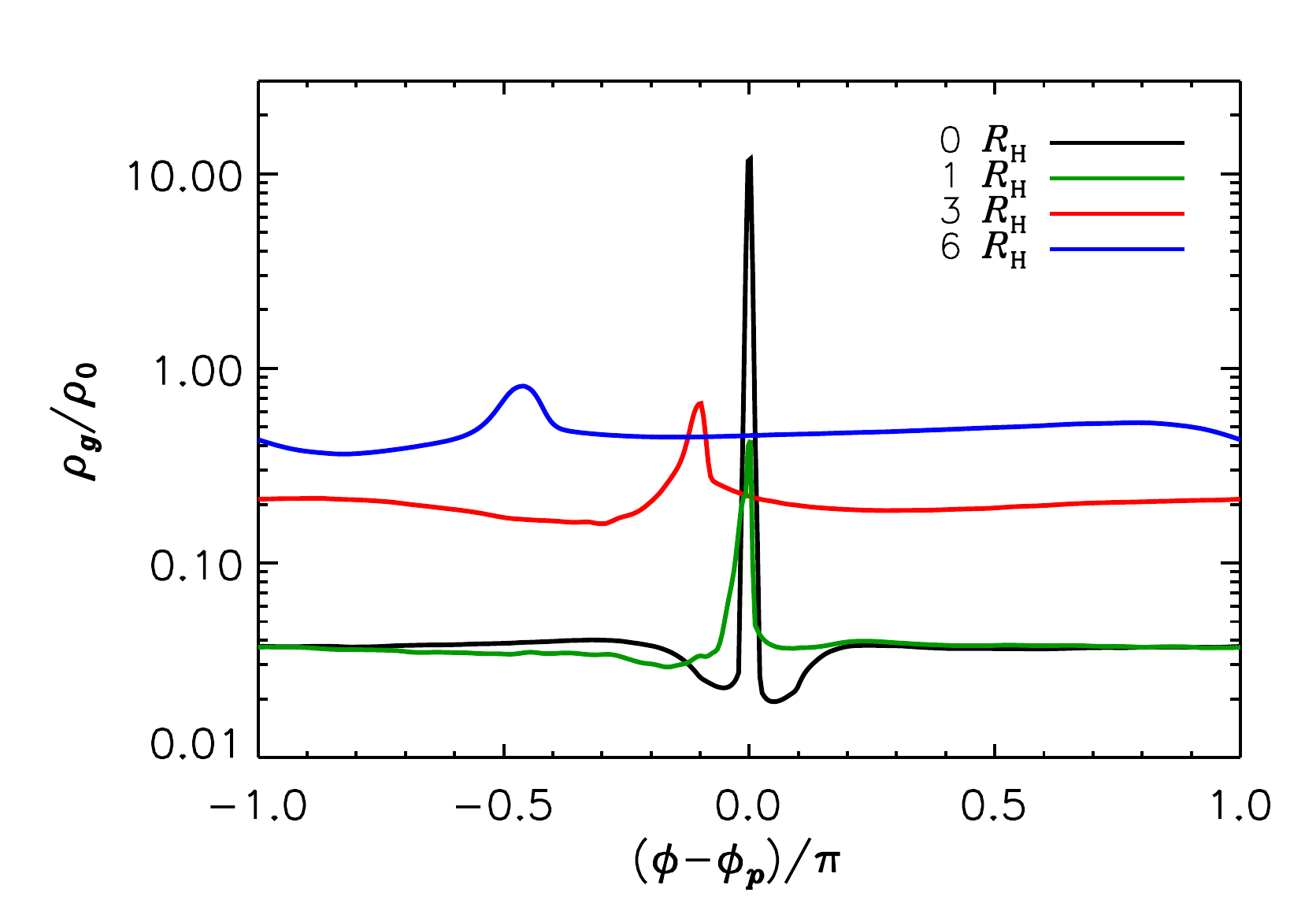}}
\resizebox{\figlen}{!}{%
\includegraphics[clip]{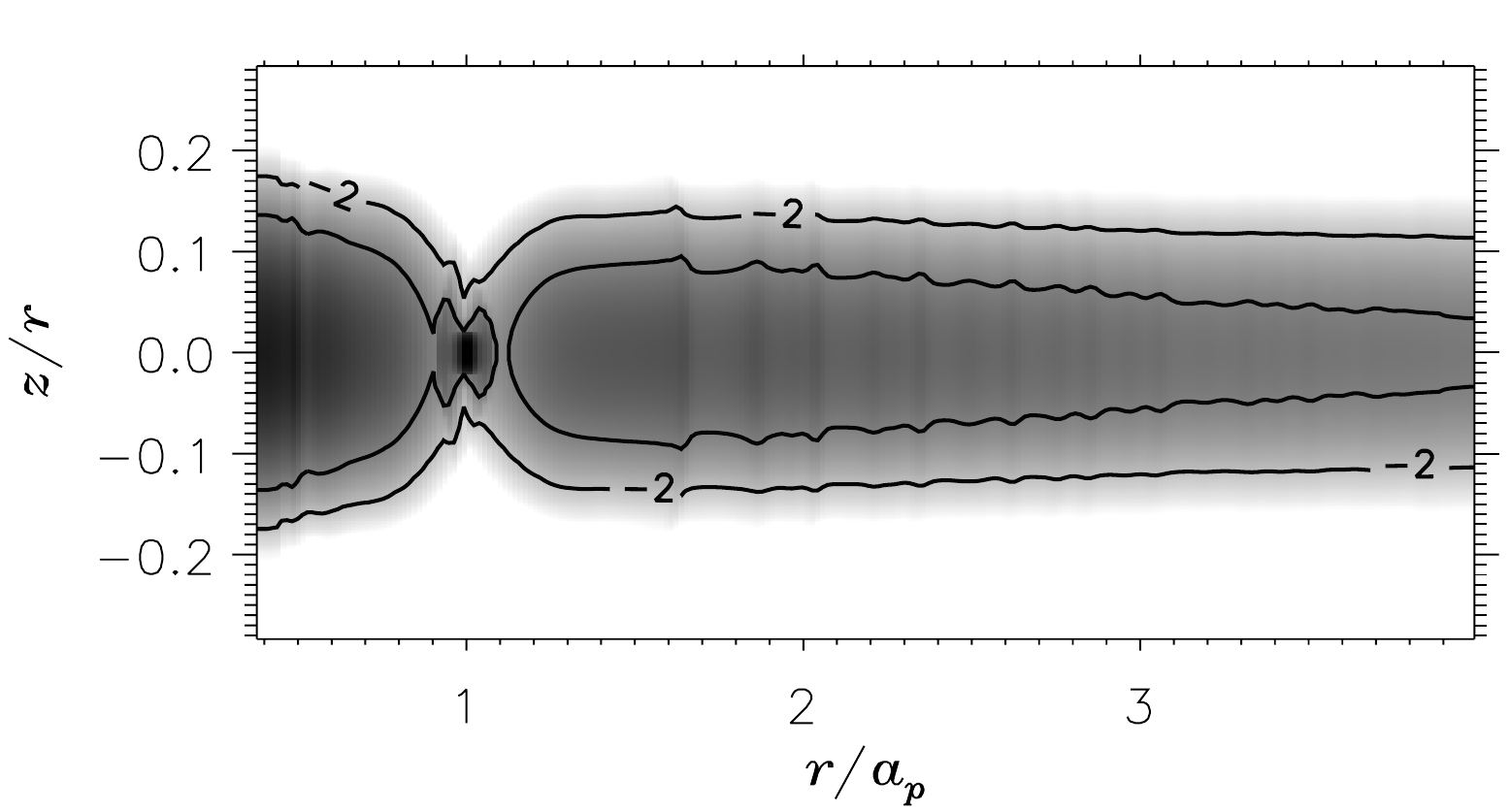}%
\includegraphics[clip]{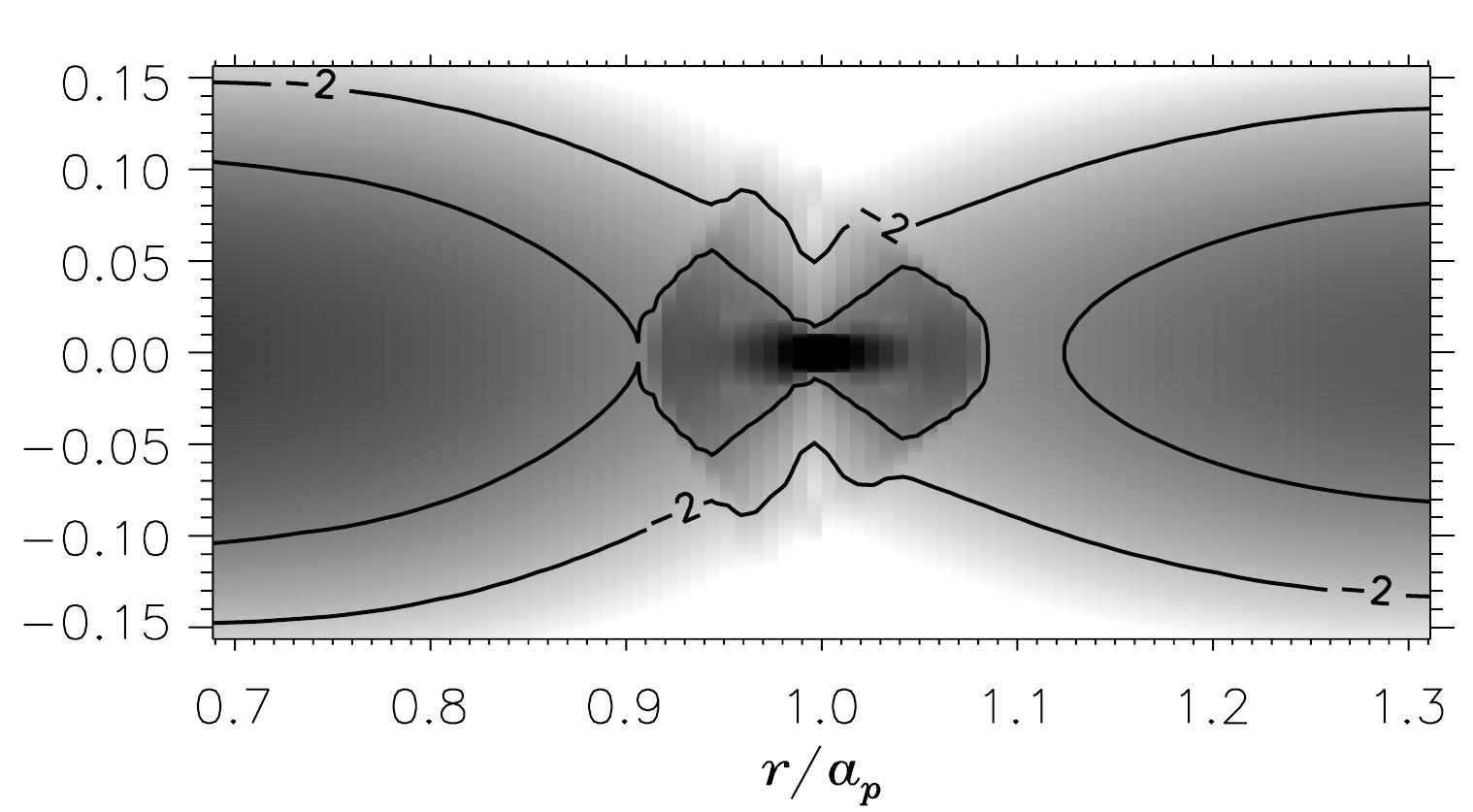}\includegraphics[clip]{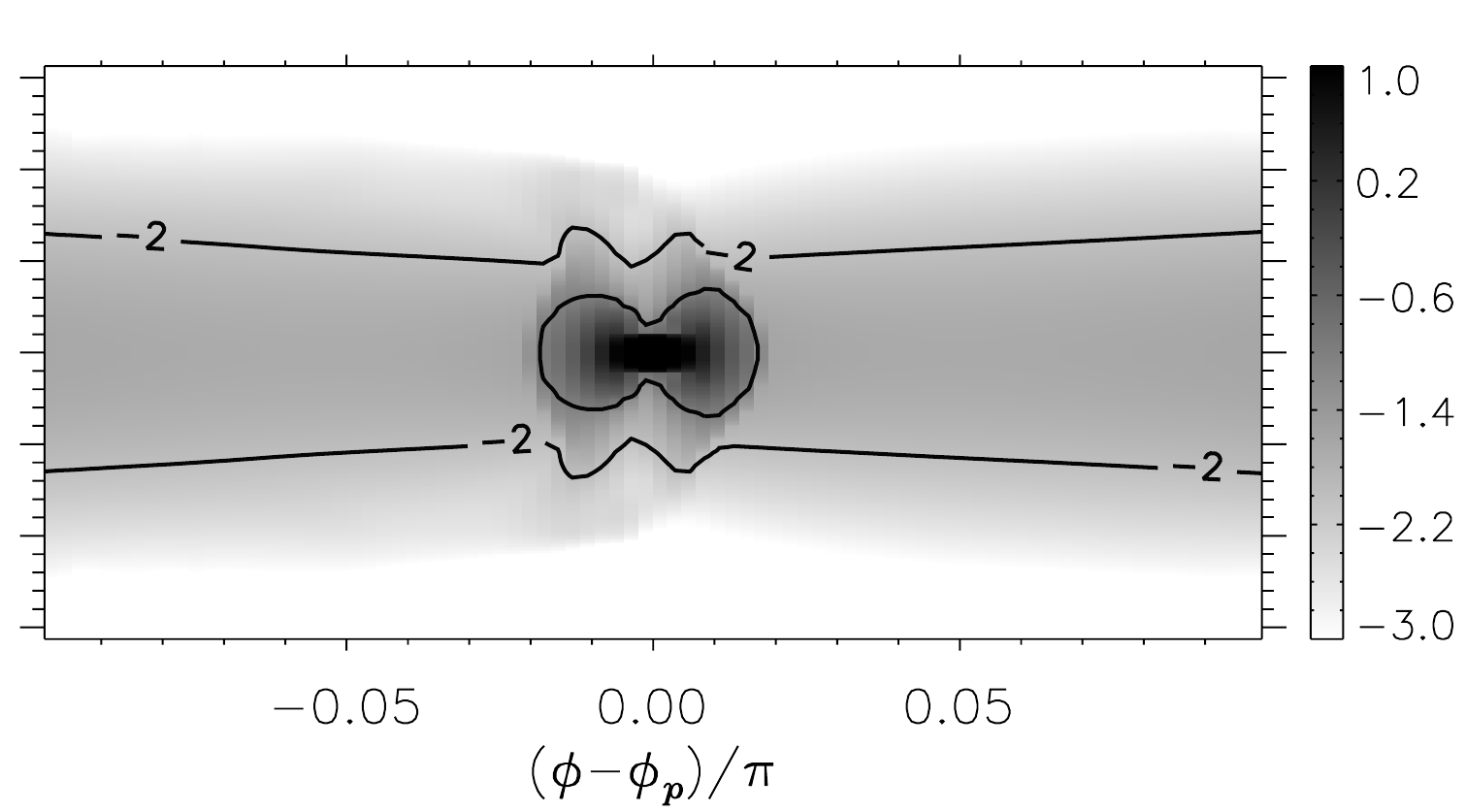}}
\resizebox{\figlen}{!}{%
\includegraphics{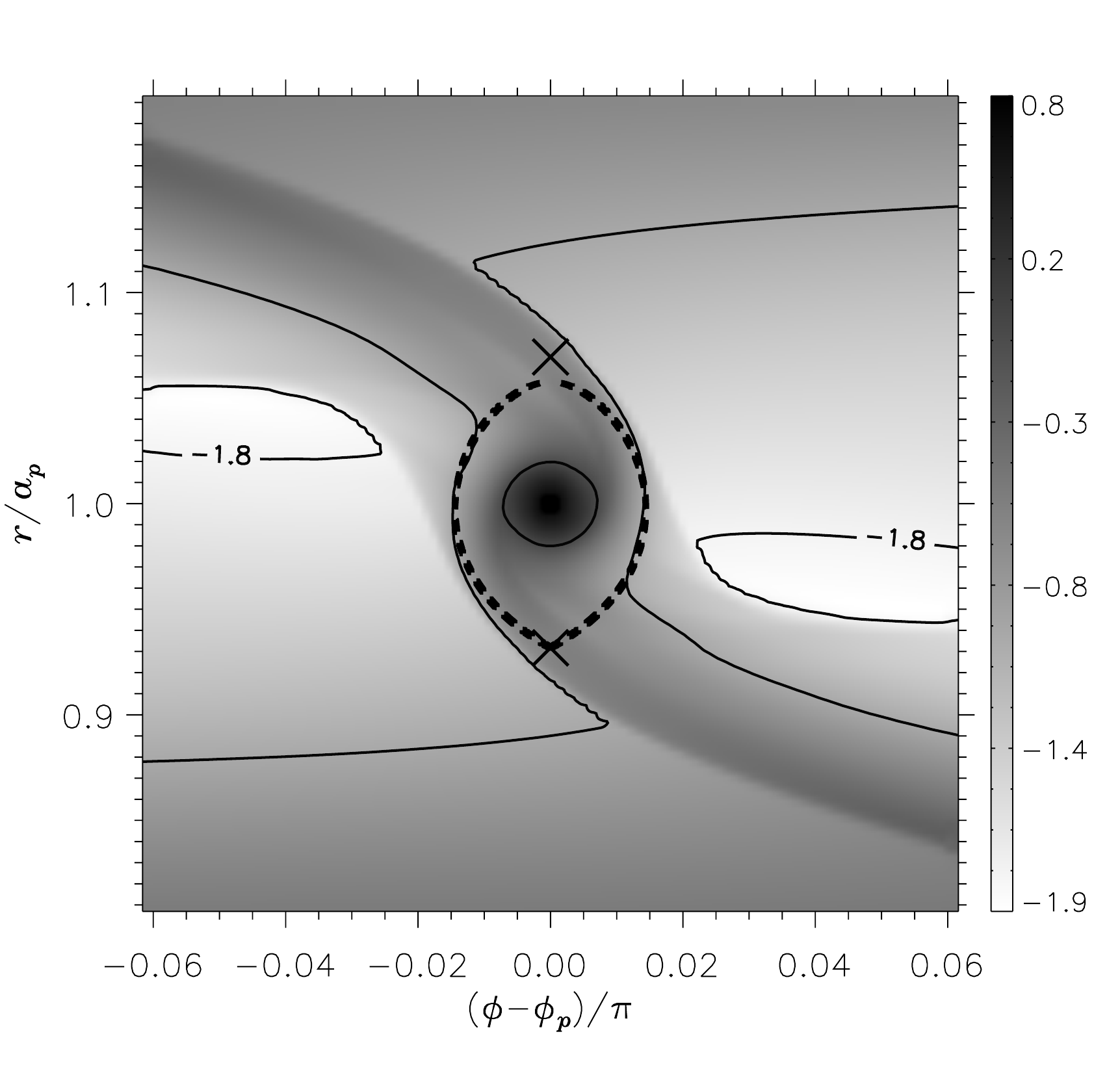}%
\includegraphics{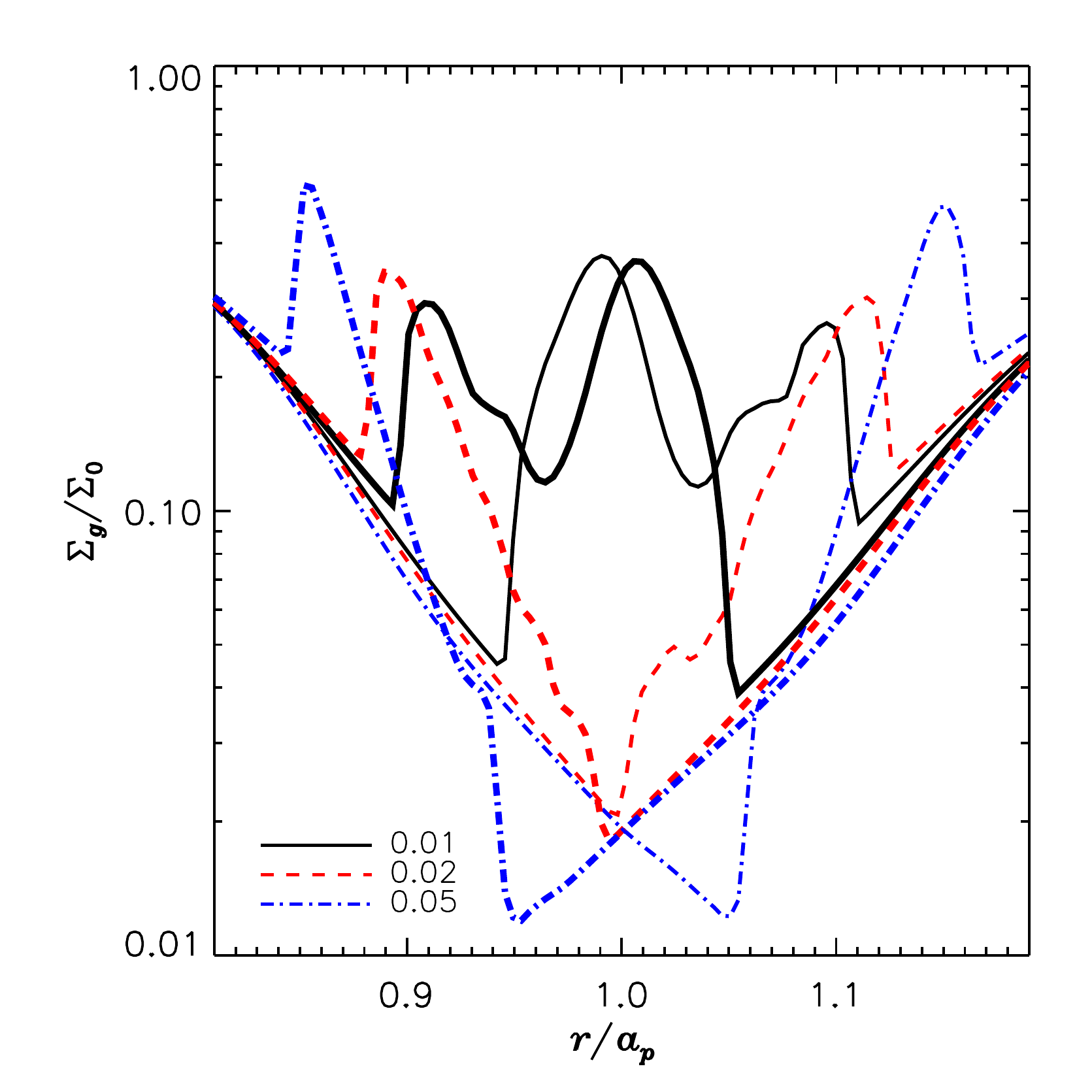}}
\caption{%
             Top-left. Color scale rendering of $\log{(\Sigma_{g}/\Sigma_{0})}$ as a function
             of the azimuthal angle from the planet and the distance from the star in units of 
             $a_{p}$.
             Top-right. Volume density, $\rho_{g}$ in the disk's mid-plane, normalized to 
             $\rho_{0}$ as a function of 
             the azimuth at various distances from the planet's orbit, as indicated in the legend.
             Center. Color scale rendering of $\log{(\rho_{g}/\rho_{0})}$ on disk slices passing
             through the planet. The latitude ($\approx z/r$) is on the vertical axis and the 
             linear distance from the star (left and middle) or angular distance from the planet 
             (right) is on the horizontal axis.
             Bottom. Color scale rendering of $\log{(\Sigma_{g}/\Sigma_{0})}$ in the proximity
             of the planet's Roche lobe (left) and $\Sigma_{g}/\Sigma_{0}$ versus radial distance 
             (right), at various separation angles, $|\phi-\phi_{p}|/\pi$, as indicated. 
             Thicker lines are for $\phi>\phi_{p}$.
             }
\label{fig:img_disk}
\end{figure*}
Tidal interactions between the disk's gas and the planet excite density 
waves at Lindblad resonances \citep{goldreich1980} and deplete
the gas within a few $\Rhill$ from the planet's orbit, where tidal torques 
exceed viscous torques \citep{lin1986a}. Residual gas is still present
in the tidal gap region (as shown below), even at a viscosity 
much smaller than that adopted here.

The main features of the surface and volume density, on a global
disk scale ($\gtrsim a_{p}$), are illustrated in Figure~\ref{fig:img_disk}. 
In all cases, densities are normalized to
either $\Sigma_{0}$ or $\rho_{0}$. Both spiral density waves and 
the tidally-produced gap are visible  in the top-left panel, while the
plot on the right shows, more quantitatively, the volume density
in the disk's equatorial planet, at several distances from the planet's
orbit. The residual gas in the tidal gap region is also visible as a
function of the azimuthal angle.
The center panels illustrate the density on orthogonal disk slices 
passing through the planet's position, on different length scales. 
The surface density of the region around the planet's Roche lobe 
is shown in the bottom panels (the left panel also shows the Roche
lobe trace and the positions of the Lagrange points $L_{1}$ 
-- lower cross -- and $L_{2}$ -- upper cross). 
Radial cuts of $\Sigma_{g}/\Sigma_{0}$ at various azimuthal angles 
are plotted in the right panel. All images are saturated in order
to improve the contrast between low and high density regions.

\begin{figure}
\centering%
\resizebox{\linewidth}{!}{%
\includegraphics{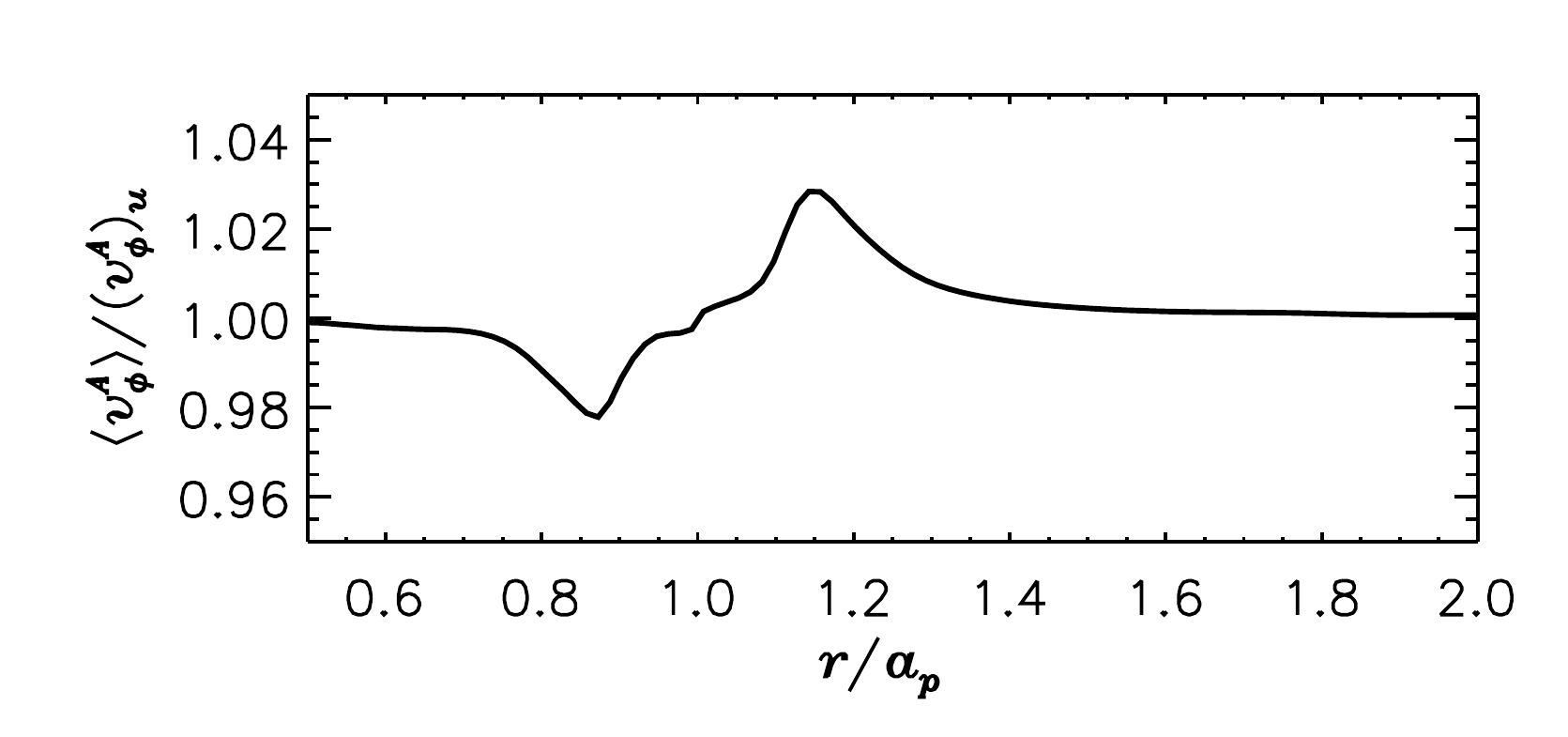}}
\caption{%
              Absolute azimuthal velocity of the gas, averaged over $2\pi$ in
              azimuth around the star and normalized to the unperturbed 
              velocity of Equation~(\ref{eq:rotacu}). The radial pressure gradient
              of the gas induces sub/super-Keplerian rotation at the inner/outer
              gap edge.
              }
\label{fig:vphiu}
\end{figure}
The rotation curve of the unperturbed disk is affected by the gas
pressure gradient, which depends on both density and
temperature gradients. In terms of the Keplerian velocity,
$v_{\mathrm{K}}$,
the (absolute) azimuthal velocity of the gas, in absence of the 
planet and in the mid-plane of the disk, would be \citep[e.g.,][]{tanaka2002}
\begin{equation}
\left(v^{A}_{\phi}\right)_{u}=v_{\mathrm{K}}%
\left[1-\frac{5}{2}\left(\frac{H}{r}\right)^{2}\right]^{1/2},
\label{eq:rotacu}
\end{equation}
which accounts for the fact that the unperturbed density
is $\rho_{g}\propto 1/r^{3/2}$ and $T_{g}\propto 1/r$.
The subscript ``$u$'' stresses the fact that this expression
does not account for the perturbation induced by the planet.
Velocity $(v^{A}_{\phi})_{u}$ differs by less than $1$\% from
the Keplerian velocity.
The ratio of the perturbed velocity, $v^{A}_{\phi}$, to that in 
Equation~(\ref{eq:rotacu}) is shown in Figure~\ref{fig:vphiu}.
The largest deviations from the unperturbed rotation curve
occur around the edges of the gap, where the magnitude of
the density gradient is the largest 
(see top-left panel in Figure~\ref{fig:img_disk}).
The negative/positive pressure (radial) gradient triggers a 
sub/super-Keplerian rotation around the inner/outer edge 
of the gap.
In fact, centrifugal balance requires that 
$(v^{A}_{\phi})^{2}=v^{2}_{\mathrm{K}}+(r/\rho_{g})\partial P_{g}/\partial r$
or, using Equation~(\ref{eq:pg}),
\begin{equation}
v^{A}_{\phi}=v_{\mathrm{K}}%
\left[1-\left(\frac{H}{r}\right)^{2}\!\left(1-\frac{\partial\ln{\rho_{g}}}{\partial\ln{r}}\right)\right]^{1/2},
\label{eq:rotacp}
\end{equation}
which reduces to 
Equation~(\ref{eq:rotacu}) for the unperturbed disk case.
Figure~\ref{fig:vphiu} and Equation~(\ref{eq:rotacp}) suggest that
the magnitude of the gradient $\partial\ln{\rho_{g}}/\partial\ln{r}$
is marginally larger at the outer edge of the gap than it is at the inner edge.
If a particle moved at a Keplerian speed, on average it would 
experience a tail wind when orbiting near the outer edge of the density 
gap and a head wind when orbiting near the inner gap edge.

\subsection{Circumplanetary Disk Thermodynamics}
\label{sec:CPD}

Le us introduce a local reference frame $\{\mathcal{O}^{\prime}; x,y,z\}$, 
with origin $\mathcal{O}^{\prime}$ on the planet, coordinate $x$ pointing
away from the star, $y$ pointing toward the direction of orbital motion, 
and $z$ pointing away from the disk's equatorial plane ($\theta=\pi/2$) 
in the direction $\theta=0$.

\begin{figure*}
\centering%
\resizebox{\figlew}{!}{%
\includegraphics[clip]{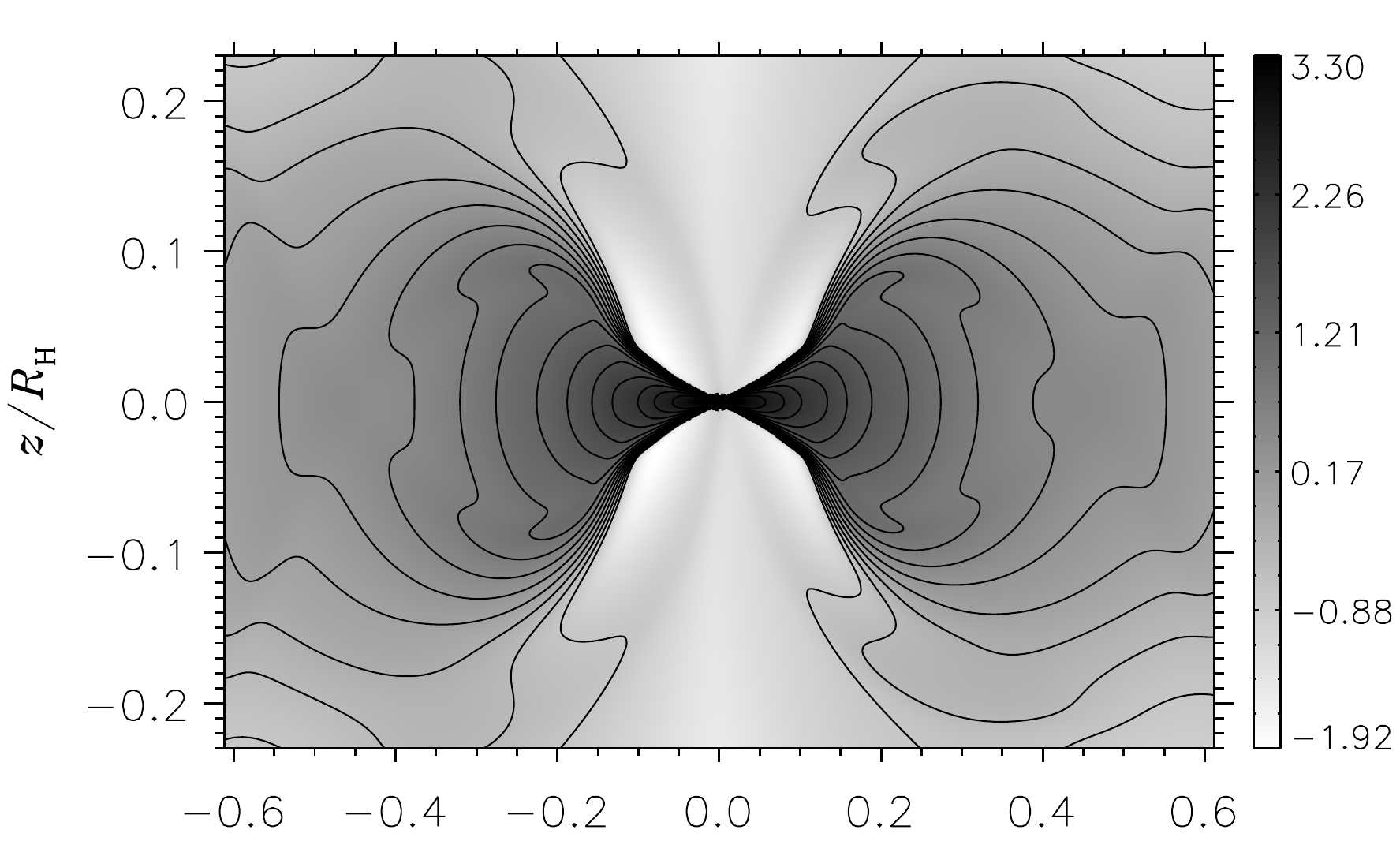}%
\includegraphics[clip]{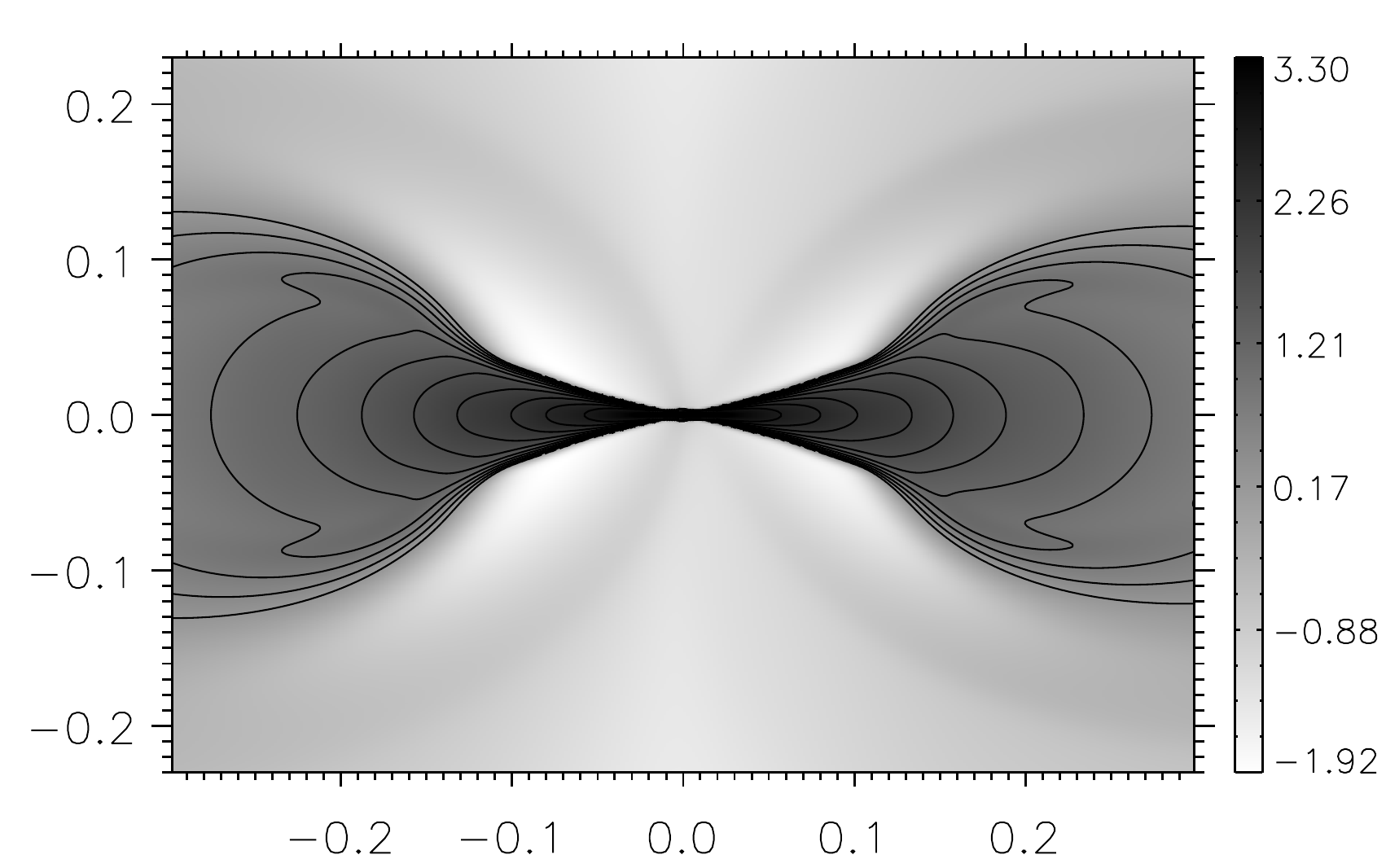}}
\resizebox{\figlew}{!}{%
\includegraphics[clip]{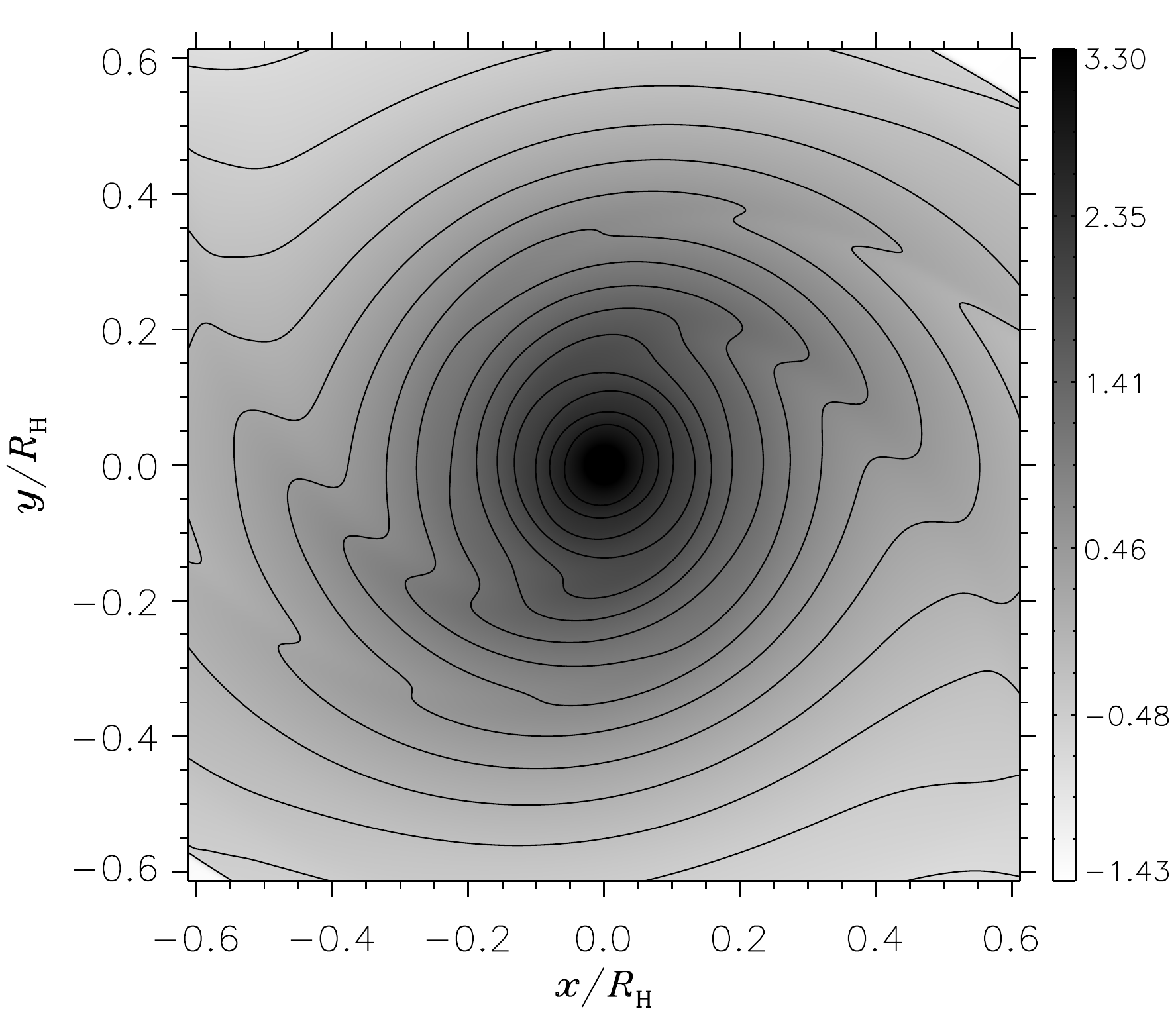}%
\includegraphics[clip]{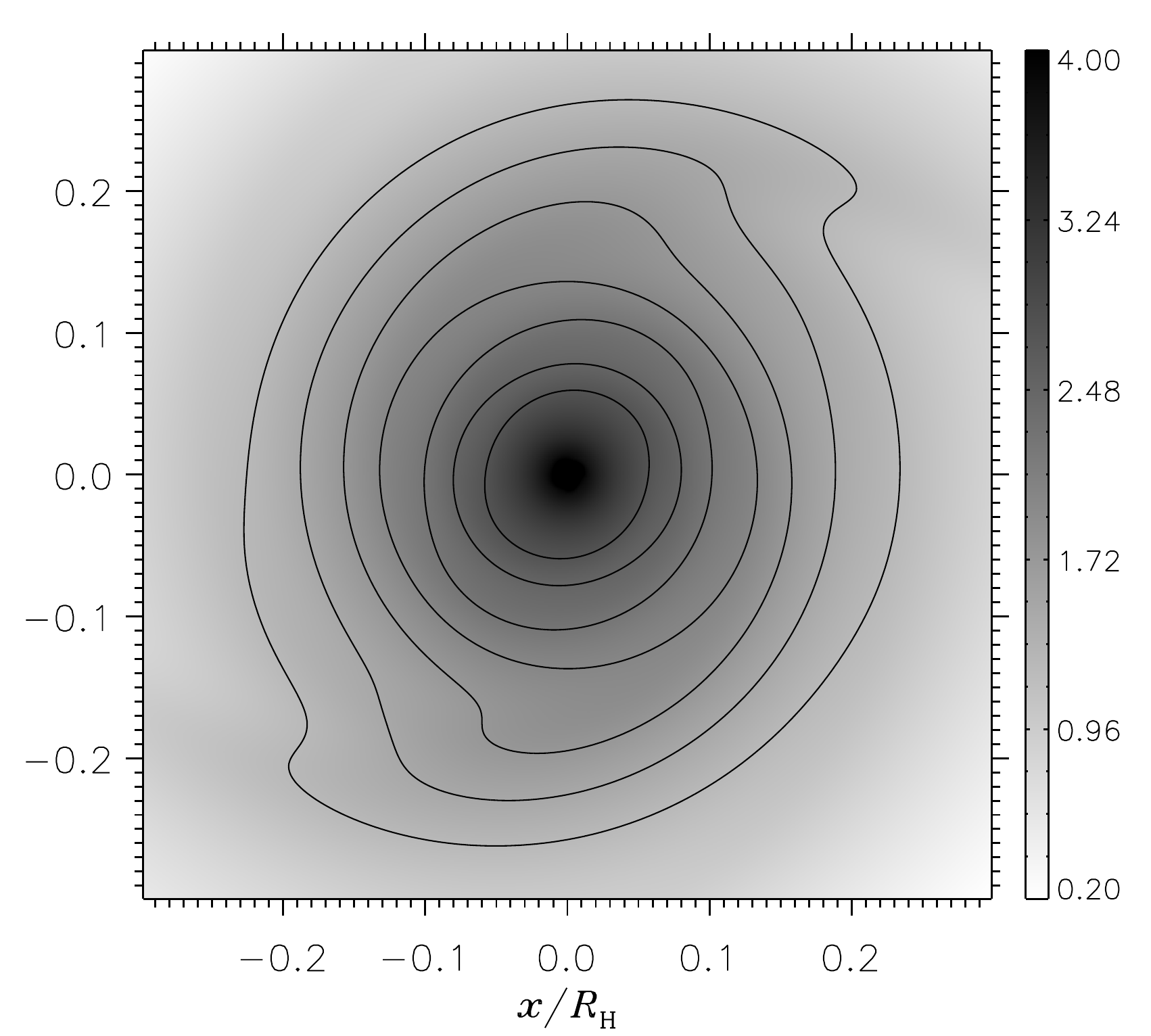}}
\caption{%
              Color scale rendering of $\log{(\rho_{g}/\rho_{0})}$ on disk slices passing
              through the planet's location. Vertical slices are illustrated in the top panels
              whereas slices in the equatorial plane are illustrated in the bottom panels. 
              All distances from the planet are in units of the planet's Hill radius,
              $\Rhill\simeq 0.0689\,a_{p}$.
             }
\label{fig:img_sub}
\end{figure*}
The normalized density, $\rho_{g}/\rho_{0}$, is illustrated in 
Figure~\ref{fig:img_sub}, on length scales $\lesssim \Rhill$, on vertical
(top) and equatorial (bottom) slices passing though the position of the
planet. The density contour levels in the lower panels of the figure 
indicate that a perturbation in the form of a spiral wave propagates 
toward the planet, but it does not propagate in the inner disk, within 
a distance of $0.1\,\Rhill$, or about $75\,R_{J}$ of the planet.
This is in agreement with previous 3D calculations.
Note that while all of Jupiter's regular satellites orbit inside 
$30\,R_{J}$ of the planet, irregular satellites have semi-major axes 
between $\sim 100\,R_{J}\approx 0.13\,\Rhill$ and 
$\sim 425\,R_{J}\approx 0.56\,\Rhill$ \citep[e.g.,][]{jewitt2004,jewitt2007}.
Beyond $\sim240\,R_{J}\approx 0.32\,\Rhill$, these satellites are
all on retrograde orbits.

\begin{figure}
\centering%
\resizebox{\linewidth}{!}{%
\includegraphics[clip]{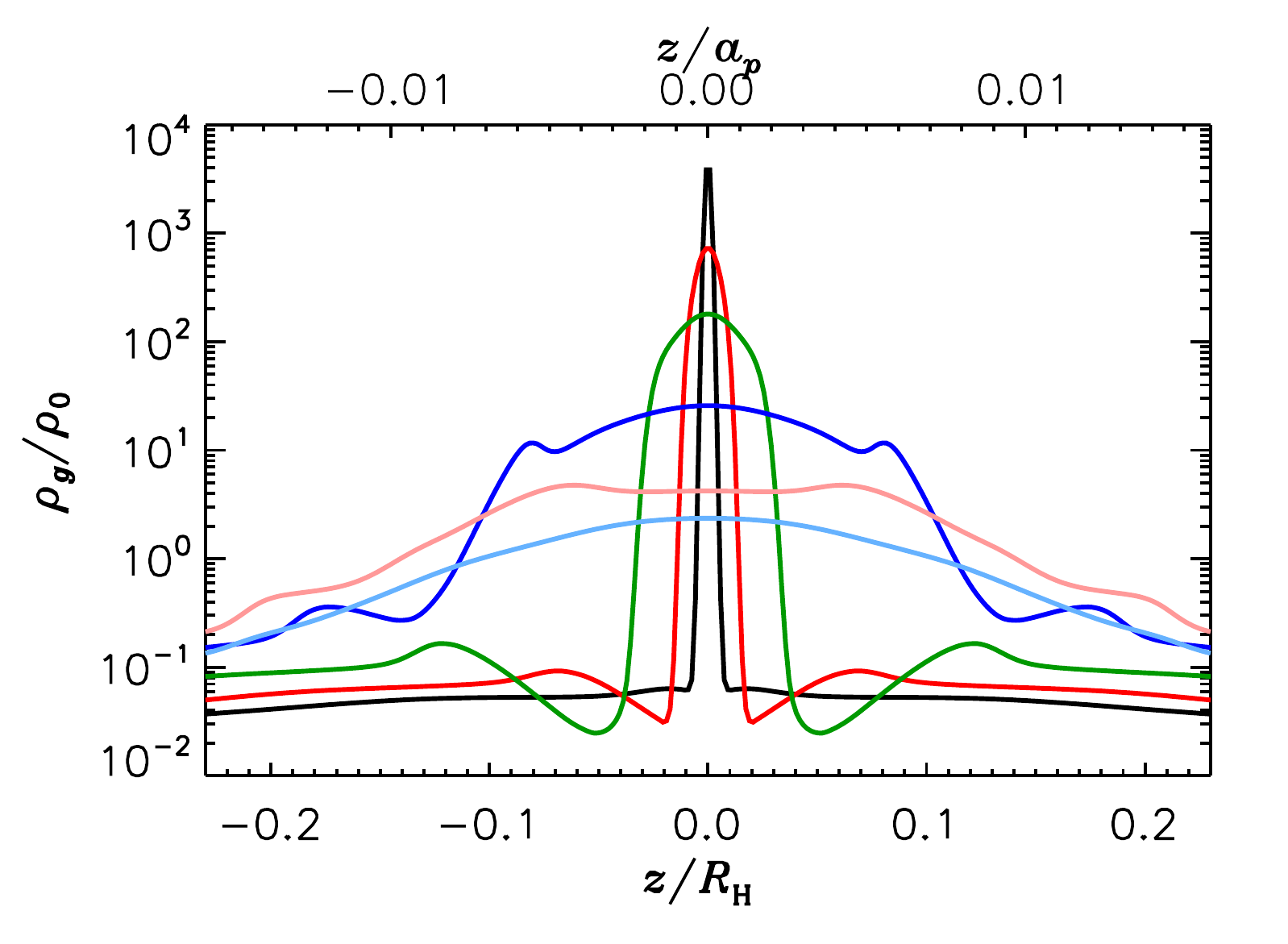}}
\resizebox{\linewidth}{!}{%
\includegraphics[clip]{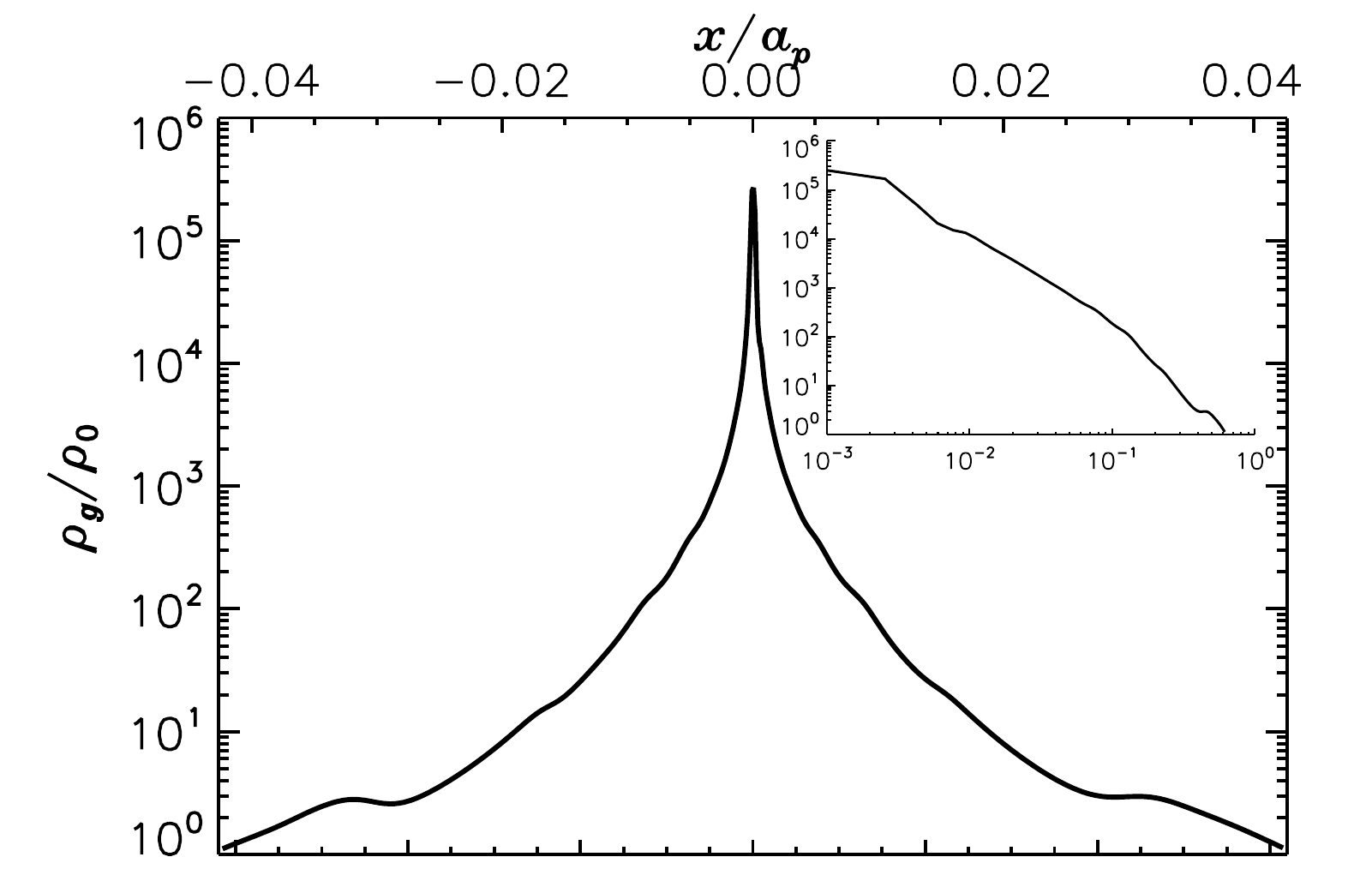}}
\resizebox{\linewidth}{!}{%
\includegraphics[clip]{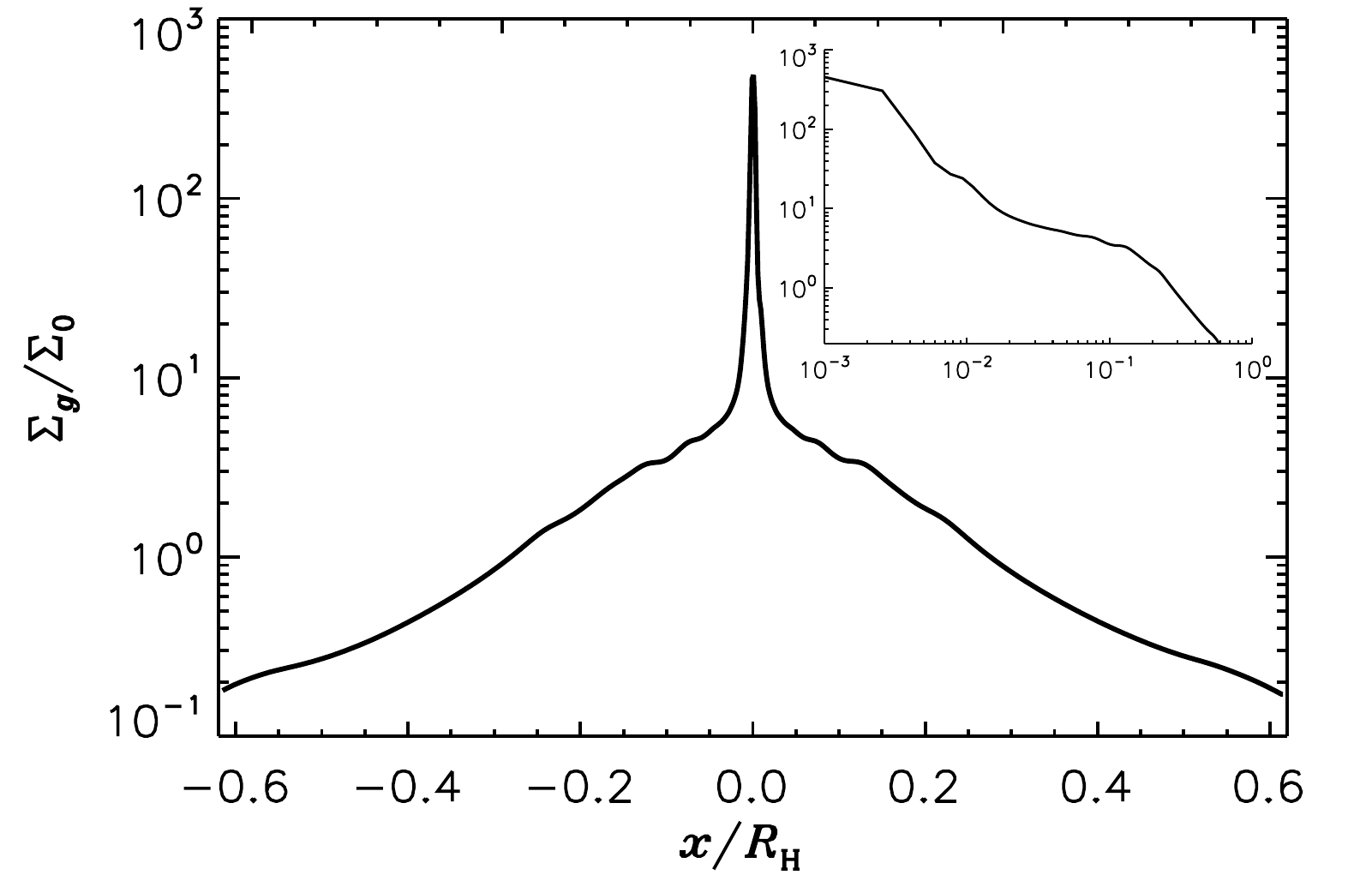}}
\caption{%
             The top panel shows the normalized volume density, 
             $\rho_{g}/\rho_{0}$, along
             the vertical direction at various distances from the planet.
             From profiles with lower to higher peak density, the distance is
             $0.5$, $0.35$, $0.2$, $0.1$, $0.05$, and $0.02\,\Rhill$.
             The middle and bottom panels show, respectively, the
             normalized the (mid-plane) volume and surface 
             density around the planet. The top $x$-axis of each
             panel gives the distance in units of the planet's 
             semi-major axis, $a_{p}$, while the bottom axis is in
             units of $\Rhill$.
             }
\label{fig:prof_sub}
\end{figure}
Line plots of the vertical density distribution are shown in the top 
panel of Figure~\ref{fig:prof_sub}, at various distances from the
planet's location (see figure caption for details).  The volume and 
surface densities within the inner Roche lobe are shown in the 
middle and bottom panels, respectively. A close-up of these
quantities around the planet can be seen in  the insets.
The surface density between $\approx 0.02$ and $\approx 0.15\,\Rhill$
of the planet roughly declines as a $-0.3$ power of the distance. 
This slope becomes approximately $-1$ between $\approx 0.15$ 
and $\approx 0.5\,\Rhill$.

The two-dimensional models of \citet{gennaro2003b} showed that
circumplanetary disks are typically thick, with an aspect ratio
ranging from $\sim 0.2$ to $\sim 0.4$, depending on the thermal
state of the disk. 
This conclusion was later confirmed by several other studies 
\citep[e.g.,][]{machida2008,ayliffe2009b,martin2011,tanigawa2012}.
The top-right panel of Figure~\ref{fig:img_sub} also predicts
a thick disk, in agreement with previous results. 
A direct measurement from the data in the figure, of the
location where there is a sharp density drop, yields an aspect 
ratio of about $0.36$ within $\approx 0.13\,\Rhill$ of the planet.

Although we use a local isothermal equation of state 
(Equation~(\ref{eq:pg})), which becomes effectively isothermal 
in the small region around the planet occupied by the circumplanetary
disk, we set a gas temperature using simple arguments based on 
local heating via viscous dissipation and black-body radiation from
ambient gas, and vertical radiative cooling.
Indicating with $\tilde{r}^{2}=x^{2}+y^{2}$, the effective disk
temperature, $T_{e}$, is given by \citep{pringle1981}
\begin{equation}
T^{4}_{e}-T^{4}_{n}=\frac{3}{8\pi}%
           \frac{G\Mp\dMp}{\sigma_{\mathrm{SB}}\,\tilde{r}^{3}}%
           \left(1-\sqrt{\frac{R_{p}}{\tilde{r}}}\right),%
\label{eq:Teff}
\end{equation}
where $T_{n}$ refers to the cricumstellar disk temperature
in Equation~(\ref{eq:Tg}) and $R_{p}=1.6\,R_{J}$. 
The maximum of the right-hand side of Equation~(\ref{eq:Teff}) is
$\approx 0.00677\,G\Mp\dMp/(\sigma_{\mathrm{SB}}R_{p}^{3})$
and occurs at $\tilde{r}=(49/36)\,R_{p}$. 
The gas accretion rate on the planet, measured from the calculation 
(see Section~\ref{sec:GPA}), is 
$\dMp\approx 2\times10^{-5}\,\Omega_{p}a_{p}^{3}\rho_{0}$.
However, the accretion rate involved in Equation~(\ref{eq:Teff}) is
actually that through the disk, which is only a small fraction of 
$\dMp$ (as mentioned in Section~\ref{sec:GPA}). Although this
fraction varies with distance from the planet, based on the analysis 
of \citet{tanigawa2012}, we simply approximate it to about $1/6$.
Since there is no physical boundary at $R_{p}$, the interface
between the disk and the planet,
for $\tilde{r}^{2}+z^{2}\le R^{2}_{p}$ we set $T_{e}$
equal to $600\,\K$, the effective temperature of a protojupiter 
right after most of the envelope has been accreted\footnote{%
In the models of \citet{lissauer2009},
the effective temperature of a protojupiter, after most of the
gaseous envelope has been acquired, may initially depend on the 
accretion history of the planet, but it subsequently converges to
$\sim 600\,\K$ within a few $10^{5}$ years (see their Figure~10)
and to $\sim 500\,\K$ by $10\,\mathrm{Myr}$ \citep{marley2007}.} 
\citep{lissauer2009}.
We neglect possible heating effects in the circumplanetary
due to irradiation by the planet.

Following \citet{lunine1982}, the vertical temperature can be derived
by considering energy transfer via radiation in the vertical direction, 
which leads to the the equation
\begin{equation}
\frac{dT_{g}^{4}}{dz}= -3\rho_{g}\kappa_{\mathrm{R}}%
           \left(T^{4}_{e}-T^{4}_{n}\right)
           \left(\frac{z}{\tilde{H}}\right),%
\label{eq:TLS}
\end{equation}
where $\kappa_{\mathrm{R}}$ is a frequency-integrated opacity,
which we approximate as the Rosseland mean opacity,
and $\tilde{H}\approx 0.36\,\tilde{r}$ is the local circumplanetary disk scale 
height (see top panels of Figure~\ref{fig:img_sub}).
By solving Equation~(\ref{eq:TLS}), we have
\begin{equation}
T_{g}^{4}=T^{4}_{e}+\frac{3}{2}\tau_{\mathrm{R}}(z)%
           \left(T^{4}_{e}-T^{4}_{n}\right)
           \left(\frac{\tilde{H}+|z|}{\tilde{H}}\right),%
\label{eq:Tcp}
\end{equation}
with a height-dependent optical depth
$\tau_{\mathrm{R}}=\bar{\rho}_{g}\kappa_{\mathrm{R}}(\tilde{H}-|z|)$,
assumed to be non-negative.
The quantity $\bar{\rho}_{g}$ is a vertically averaged volume density 
between heights $|z|$ and $\tilde{H}$. 
Although the equation above is derived for a vertically constant
opacity, we assume that $\kappa_{\mathrm{R}}$ 
is either constant or a linear function of $T_{g}$. Thus, 
the solution of Equation~(\ref{eq:Tcp}) typically requires a root-finding 
iteration procedure, for which we use an algorithm based on the Brent's 
method \citep{brent1973}.

We recall that, in Equation~(\ref{eq:Tcp}),
$T_{g}$ indicates the temperature in the circumplanetary
disk, $T_{n}$ is the temperature in the circumstellar disk
(Equation~(\ref{eq:Tg})), and $T^{4}_{e}-T^{4}_{n}$ is given
by Equation~(\ref{eq:Teff}).
At $|z|=\tilde{H}$, we have that $T_{g}=T_{e}$ whereas, for $|z|>\tilde{H}$,
we impose an exponential decline over height of $T_{g}$ until 
it matches $T_{n}$.

\begin{figure}
\centering%
\resizebox{\linewidth}{!}{%
\includegraphics[clip]{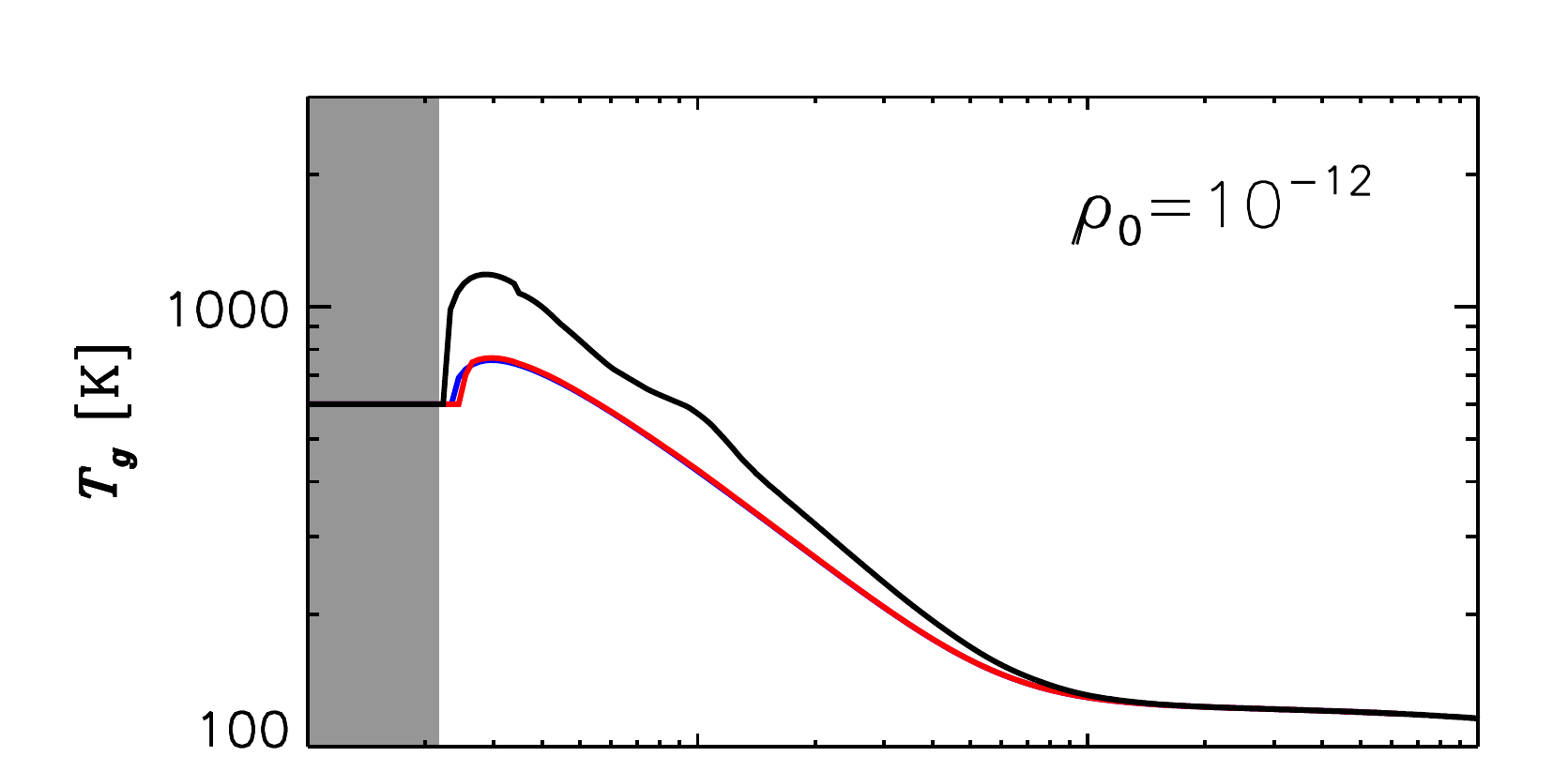}}
\resizebox{\linewidth}{!}{%
\includegraphics[clip]{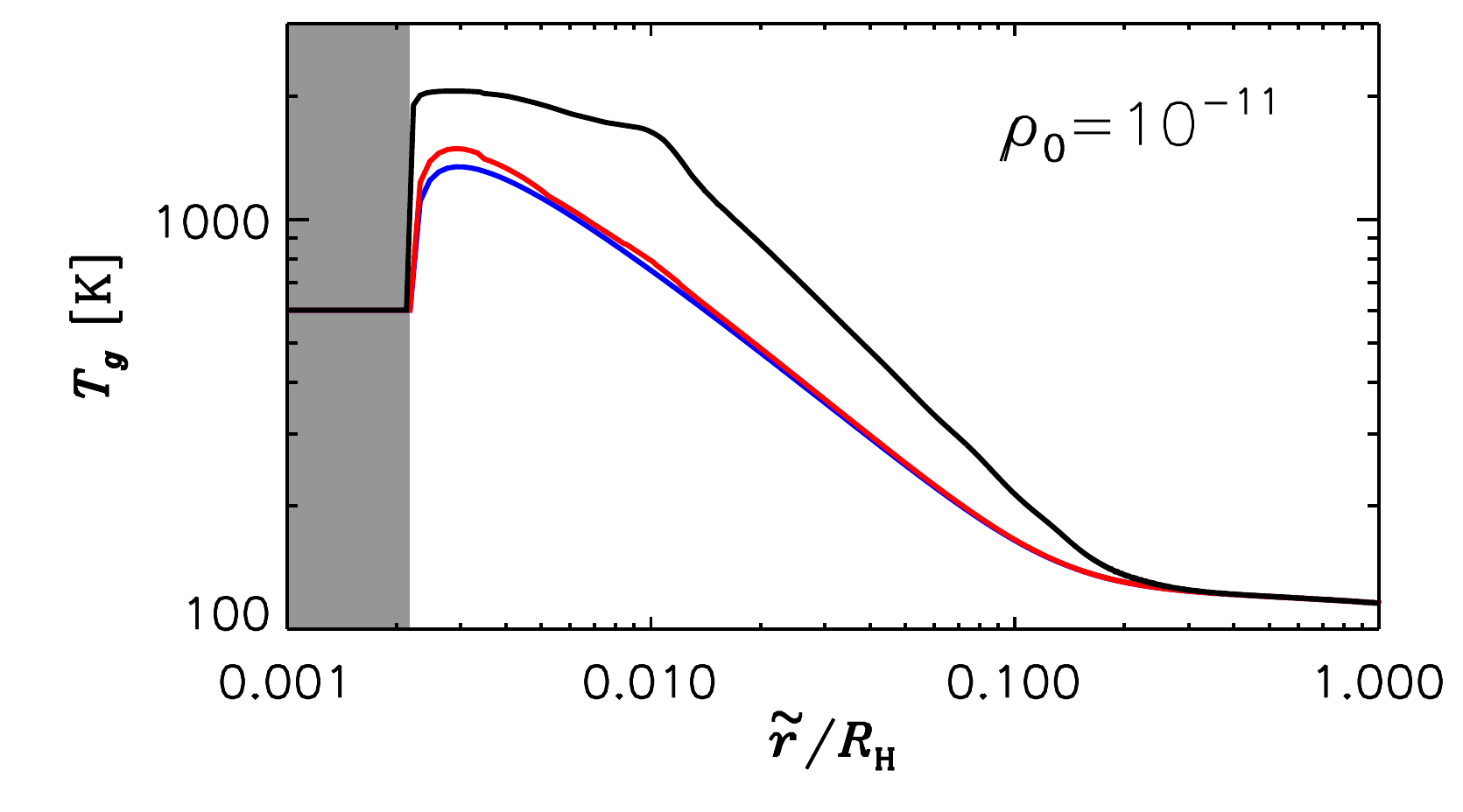}}
\caption{%
              Temperature in the equatorial plane of the circumplanetary
              disk, from Equation~(\ref{eq:Tcp}), as a function of the distance
              from the planet, for background volume density
              $\rho_{0}=10^{-12}\,\mathrm{g\,cm}^{-3}$ (top) and
              $10^{-11}\,\mathrm{g\,cm}^{-3}$ (bottom). 
              The shaded areas indicate distances $\tilde{r}<R_{p}$.
              The black curves use a  Rosseland mean opacity 
              $\kappa_{\mathrm{R}}=10^{-2}\,\mathrm{cm^{2}\,g^{-1}}$
              for $T_{g}<1600\,\K$, which decreases linearly toward 
              a gas-dominated opacity 
              ($\kappa_{\mathrm{R}}=10^{-4}\,\mathrm{cm^{2}\,g^{-1}}$).
              The red curves are for a constant, gas-dominated 
              opacity $\kappa_{\mathrm{R}}=10^{-4}\,\mathrm{cm^{2}\,g^{-1}}$, 
              while the blue curves represent the effective disk temperature, $T_{e}$
              (Equation~(\ref{eq:Teff})).
              }
\label{fig:tcp}
\end{figure}
The underlying local isothermal approximation used in the hydrodynamics 
calculations implies that the opacity of the medium (gas+dust) is low. 
If we apply a gas-dominated opacity, 
$\kappa_{\mathrm{R}}=10^{-4}\,\mathrm{cm^{2}\,g^{-1}}$, 
as contemplated in some of the models of \citet{canup2002}, 
then $T_{g}$ becomes about equal to the effective temperature 
$T_{e}$, as can be seen by comparing the red and blue 
curves in the two panels of Figure~\ref{fig:tcp}.

In Equation~(\ref{eq:Tcp}), as anticipated above, 
we use an opacity that is a piece-wise function of the temperature. 
Below the vaporization temperature of dust grains \citep{pollack1994}, 
we set $\kappa_{\mathrm{R}}=0.01\,\mathrm{cm^{2}\,g^{-1}}$,
which may be the case in an evolved disk
where grains have undergone significant growth \citep[e.g.,][]{dalessio2001}. 
We assume that $\kappa_{\mathrm{R}}$ linearly transitions to a 
gas-dominated opacity, 
$\kappa_{\mathrm{R}}=10^{-4}\,\mathrm{cm^{2}\,g^{-1}}$,
in the temperature interval from $1600\,\K$ to about $2000\,\K$, 
and it remains constant at larger temperatures.
For this opacity law, the equatorial temperature of the
circumplanetary disk is illustrated as a black line in Figure~\ref{fig:tcp},
for both reference background densities 
$\rho_{0}=10^{-12}$ (top) and
$10^{-11}\,\mathrm{g\,cm}^{-3}$ (bottom).
At distances from the planet $\tilde{r}\gtrsim 0.2\,\Rhill$, 
the radial distribution of temperature
merges with the temperature distribution in the circumstellar disk.

Since Equation~(\ref{eq:Tcp}) is not consistent
with the equation of state applied in the hydrodynamics calculations
(Equation~(\ref{eq:pg})), we also consider cases in which the
gas temperature is given everywhere by Equation~(\ref{eq:Tg}).
In these calculations, the circumplanetary disk is basically isothermal 
with a gas temperature 
$T_{n}\simeq (\mu_{g} m_{\mathrm{H}}/k_{\mathrm{B}})(G\Ms/a_{p})(H/a_{p})^{2}$.
The lower temperature close to the planet may affect the ablation history,
and hence the mass evolution, of some planetesimals.

\section{Evolution of Planetesimals in the Circumstellar Disk}
\label{sec:PCD}

We follow the evolution of planetesimals initially equally distributed 
in four size bins with radii $R_{s}=0.1$, $1$, $10$, and $100\,\mathrm{km}$.
The planetesimals are placed on elliptical orbits about the star.
The initial orbital eccentricity, $e_{s}$, and inclination, $i_{s}$,
are randomly selected within the range from $0$ to $0.05$ and 
from $0$ to $0.05$ radiants ($\approx 2.9^{\circ}$). 
The initial argument of periapsis, longitude of the ascending node, and
true anomaly are chosen randomly between $0$ and $2\pi$.
At the beginning, 
three regions in semi-major axis, $a_{s}$, are populated: 
from $0.77\,a_{p}$ to $0.82\,a_{p}$,
from $0.965\,a_{p}$ to $1.035\,a_{p}$, and
from $1.2\,a_{p}$ to $1.25\,a_{p}$.
These regions are all inside $4\,\Rhill$
of the planet's orbit, the classical ``feeding zone'' for accretion of 
solids \citep[e.g.,][]{greenzweig1990,lissauer1993b}.
Planetesimals deployed in the corotation region of the planet,
$a_{p}\mp\Rhill/2$, have near-circular obits and are deployed with 
an azimuth such that $|\phi-\phi_{p}|>\pi/3$, i.e., in between the 
triangular Lagrange points L$_{4}$ (leading) and L$_{5}$ (trailing). 
These three regions are each populated with $132000$ planetesimals
(for a total of $0.65$ or $0.86$ Mars masses of solids, depending on the
material). Initial surface densities of solids are between 
$\approx 0.05$ and $\approx 0.12\,\mathrm{g\,cm}^{-2}$, depending 
on the region and the material.
It is important to stress that, given the equal number densities per
size bin, the solid mass is almost entirely carried by the largest bodies.
Icy and mixed-composition planetesimals are considered in separate 
calculations.
The results presented here can be rescaled by the initial surface density 
of solids, provided that interactions among planetesimals can be neglected
(see Section~\ref{sec:ECR}).

In a planet-less disk, the orbital evolution of planetesimals would be
dictated only by gas drag (and stellar gravity). 
Thus, orbital eccentricity and inclination would be damped on a timescale 
$\tau_{\mathrm{damp}}\sim |\mathbf{v}_{s}-\mathbf{v}_{g}|/|\mathbf{a}_{D}|$.
In a nearly Keplerian disk, with negligible radial velocity and with azimuthal 
velocity given by Equation~(\ref{eq:rotacu}), the approximation
$|\mathbf{v}_{s}-\mathbf{v}_{g}|\sim%
(5 e^{2}_{s}/8+i^{2}_{s}/2+\xi^{4}/4)^{1/2} a_{s}\Omega_{\mathrm{K}}$ 
can be adopted, where $\xi^{2}=(5/2)(H/r)^{2}$ and 
$\Omega_{\mathrm{K}}$ is the Keplerian orbital frequency of the
planetesimal about the star \citep[e.g.,][]{adachi1976,ogihara2009}. 
Therefore,
\begin{equation}
\frac{1}{\Omega_{\mathrm{K}}\tau_{\mathrm{drag}}}\sim%
\frac{3}{8}C_{D}\!\left(\frac{a_{s}}{R_{s}}\right)\!\left(\frac{\rho_{g}}{\rho_{s}}\right)\!%
\sqrt{\frac{5}{8}e^{2}_{s}+\frac{1}{2}i^{2}_{s}+\frac{1}{4}\xi^{4}}.
\label{eq:ada}
\end{equation}
The timescale for the removal of orbital energy, and
hence for the variation of the planetesimal semi-major axis,  
is much longer, of order $\tau_{\mathrm{drag}}/\xi^{2}$ 
(see also Equation~(\ref{eq:dasdtpeale})). 
Our initial conditions ($e_{s}$ and $i_{s}\le 0.05$) would lead to damping 
timescales, at $\sim 5\,\AU$ and for $\rho_{0}=10^{-11}\,\mathrm{g\,cm}^{-3}$, 
$\tau_{\mathrm{drag}}\gtrsim 20$ orbits for 
$\sim 0.1\,\mathrm{km}$-size bodies ($C_{D}\approx 6$) and $\gtrsim 2\times 10^{5}$ 
orbits for $\sim 100\,\mathrm{km}$-size bodies ($C_{D}\approx 0.4$).
It is worth noticing, however, that Equation~(\ref{eq:ada}) assumes that
$\rho_{g}$ is constant along the trajectory of the planetesimal.
The timescales for drag-induced orbital decay would be over two orders
of magnitude as long.
The planetesimal evolution presented here lasts for $\approx 580$ planet's orbits,
or $\approx 7000$ years.
Although the initial orbits of planetesimals are arbitrary,
by the end of the calculations the spatial distributions of the solids are in
a state of quasi-equilibrium, in the sense that they vary slowly over tens 
of orbital periods of the planet. Over much longer timescales, the lack of 
a supply of planetesimals from other regions (besides those considered 
here) likely inhibits a state of true dynamical equilibrium.

\begin{figure*}
\centering%
\resizebox{\figlew}{!}{\includegraphics[clip]{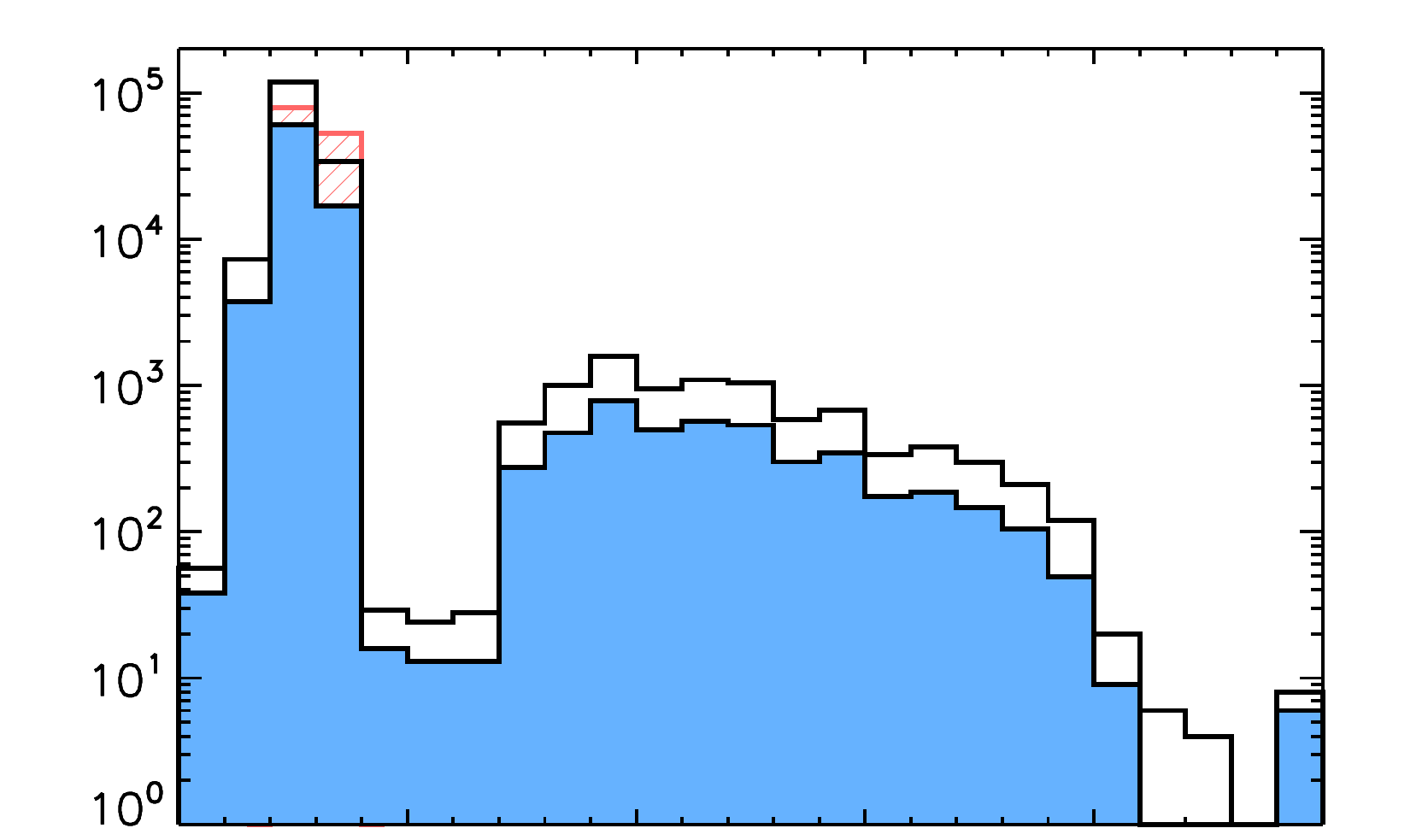}%
                                 \includegraphics[clip]{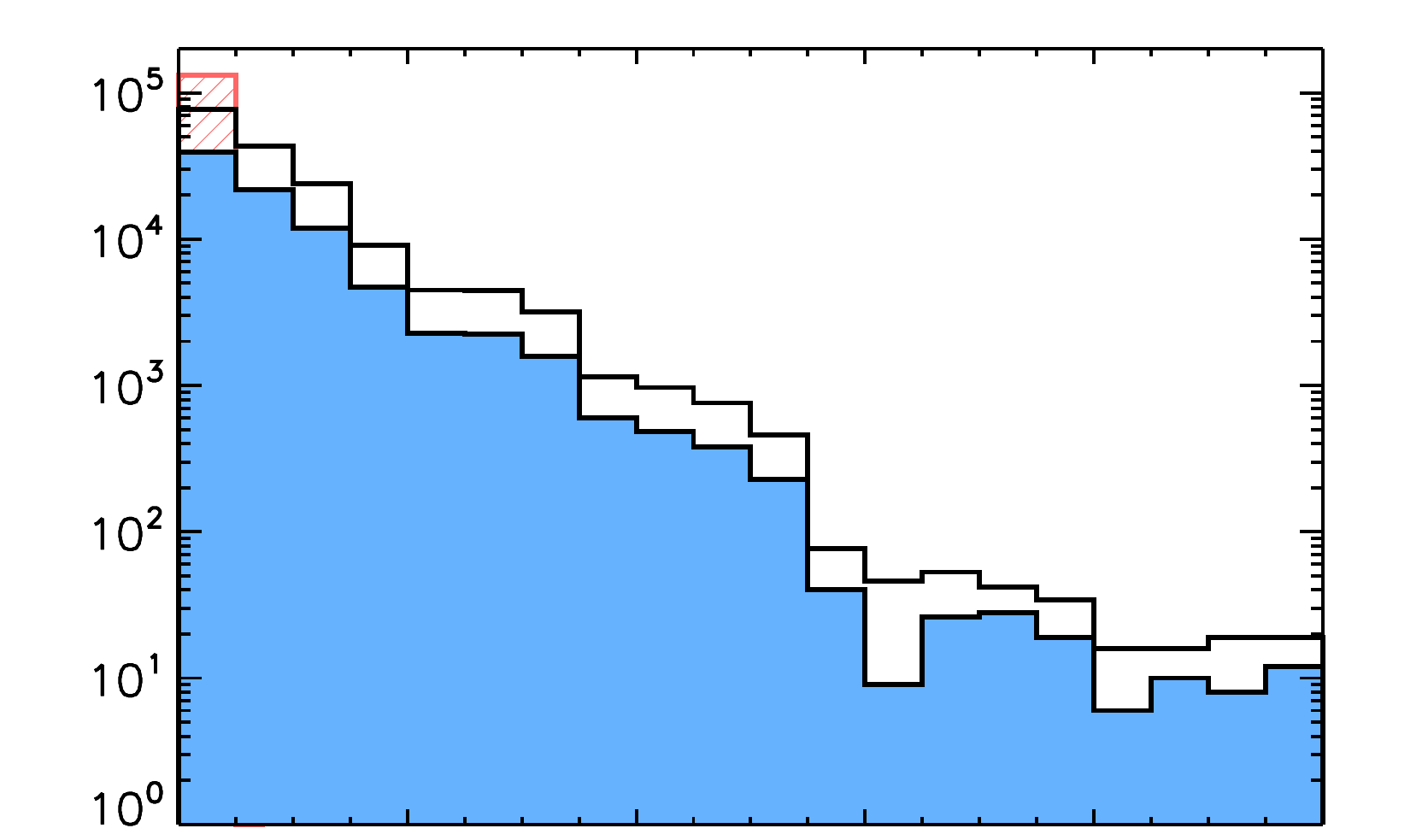}%
                                 \includegraphics[clip]{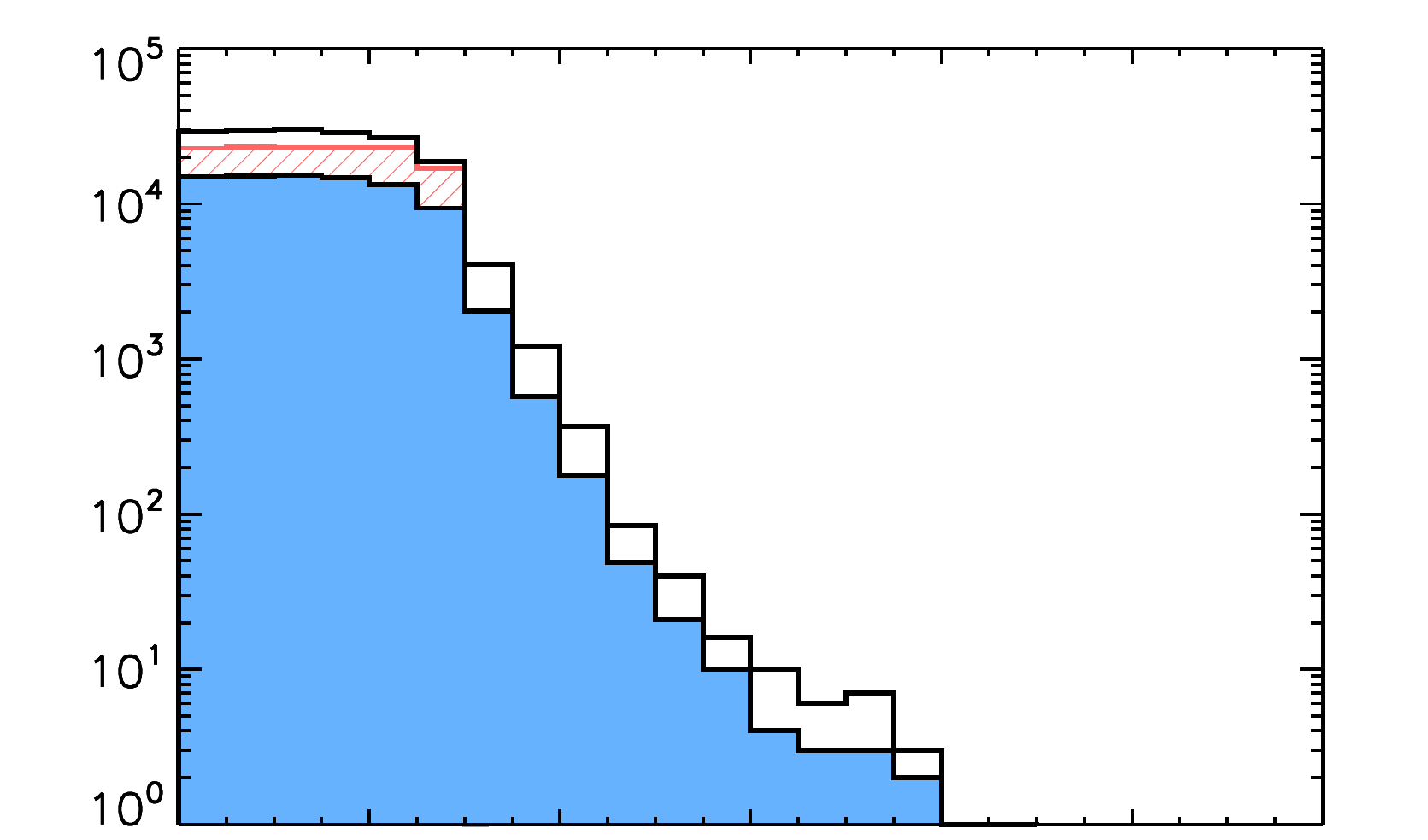}}
\resizebox{\figlew}{!}{\includegraphics[clip]{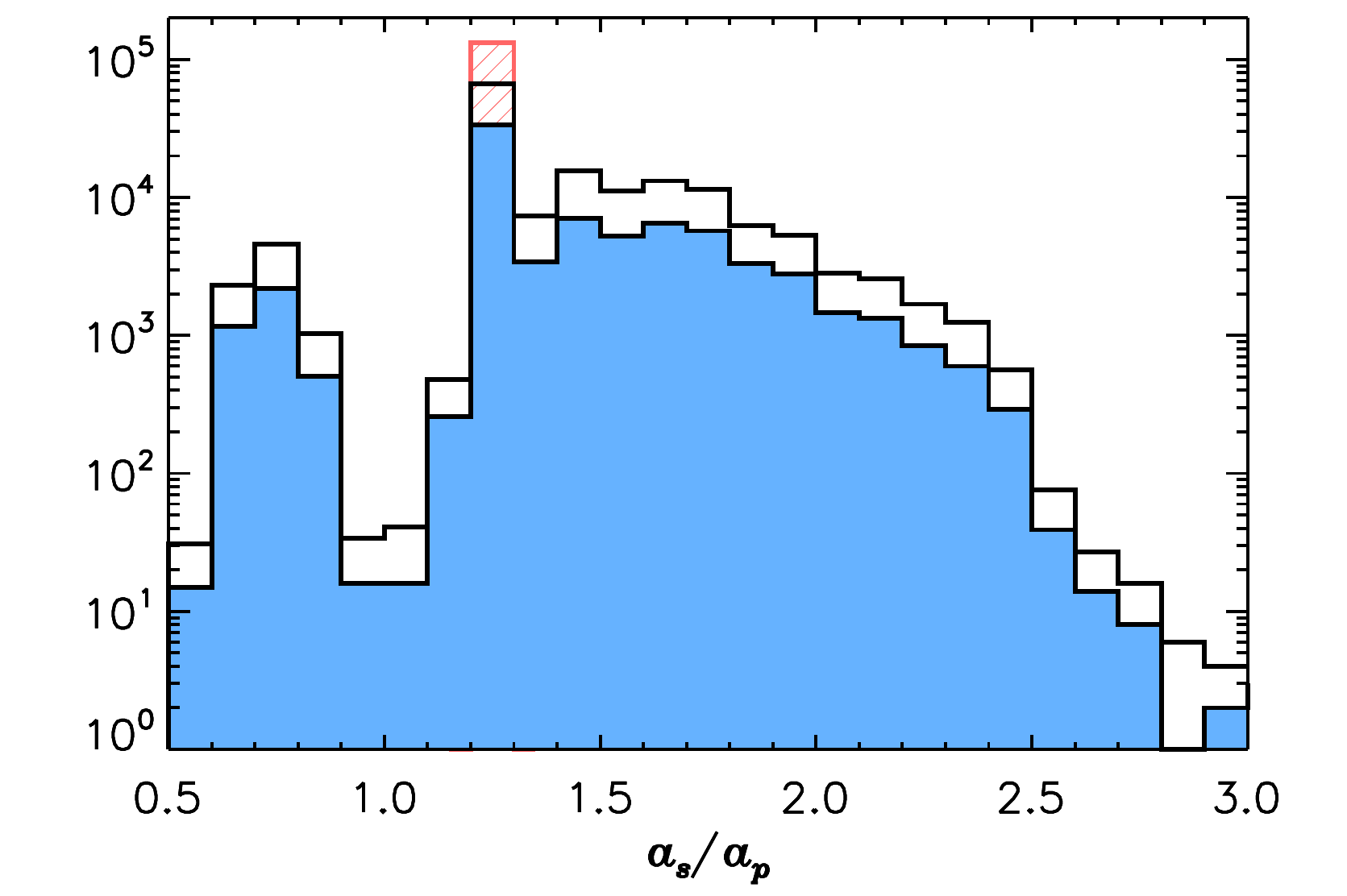}%
                                 \includegraphics[clip]{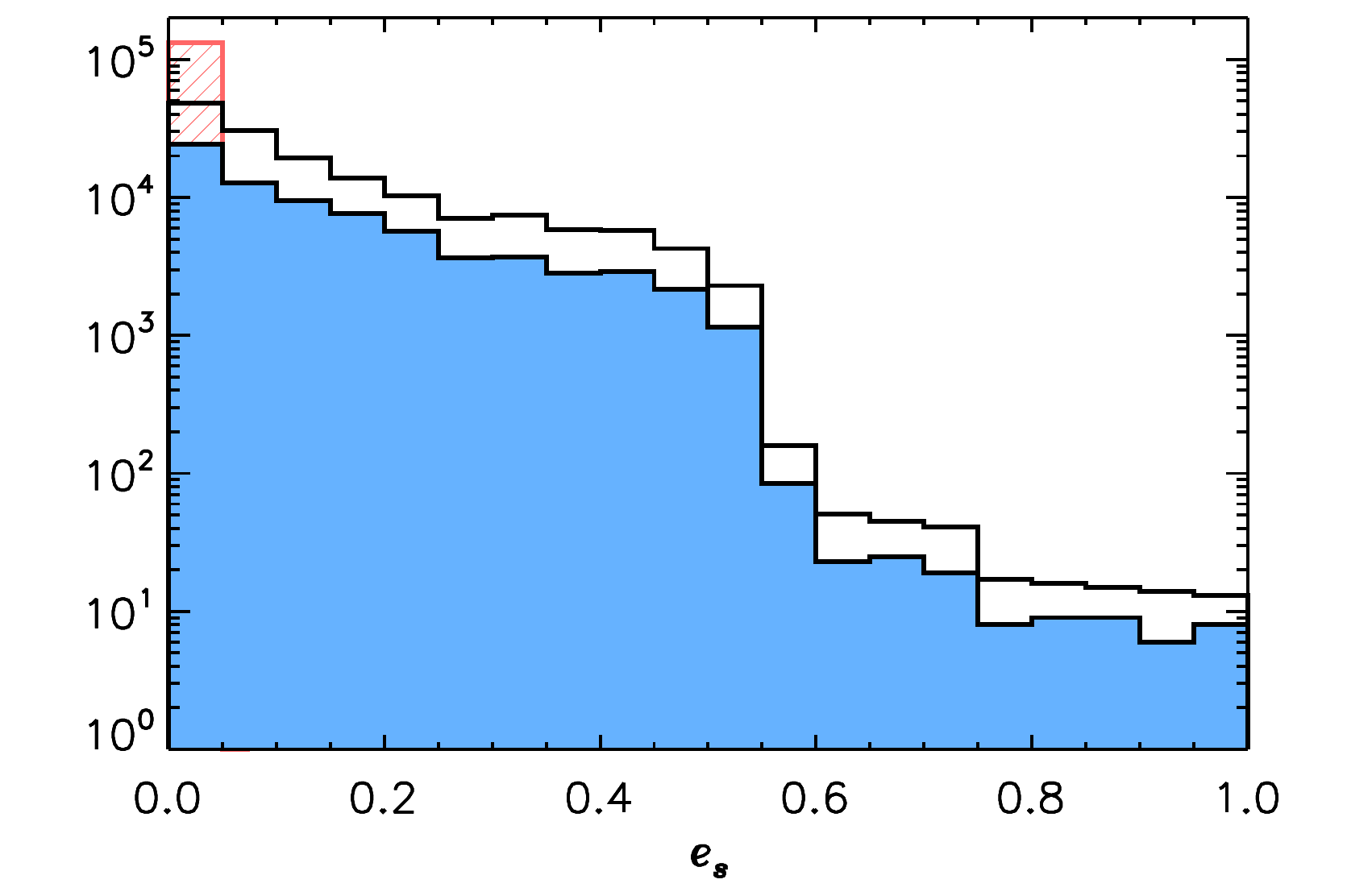}%
                                 \includegraphics[clip]{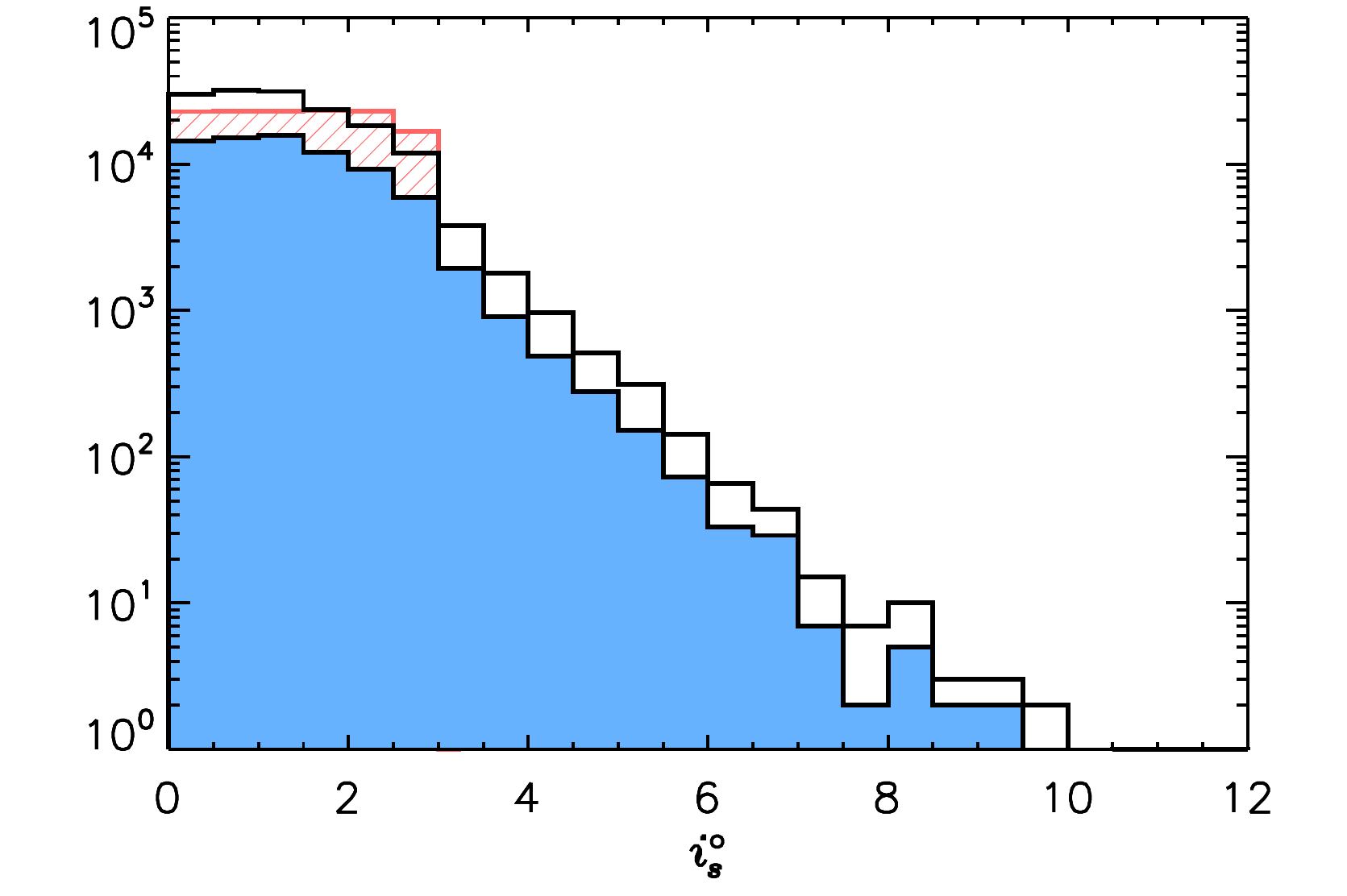}}
\resizebox{\figlew}{!}{\includegraphics[clip]{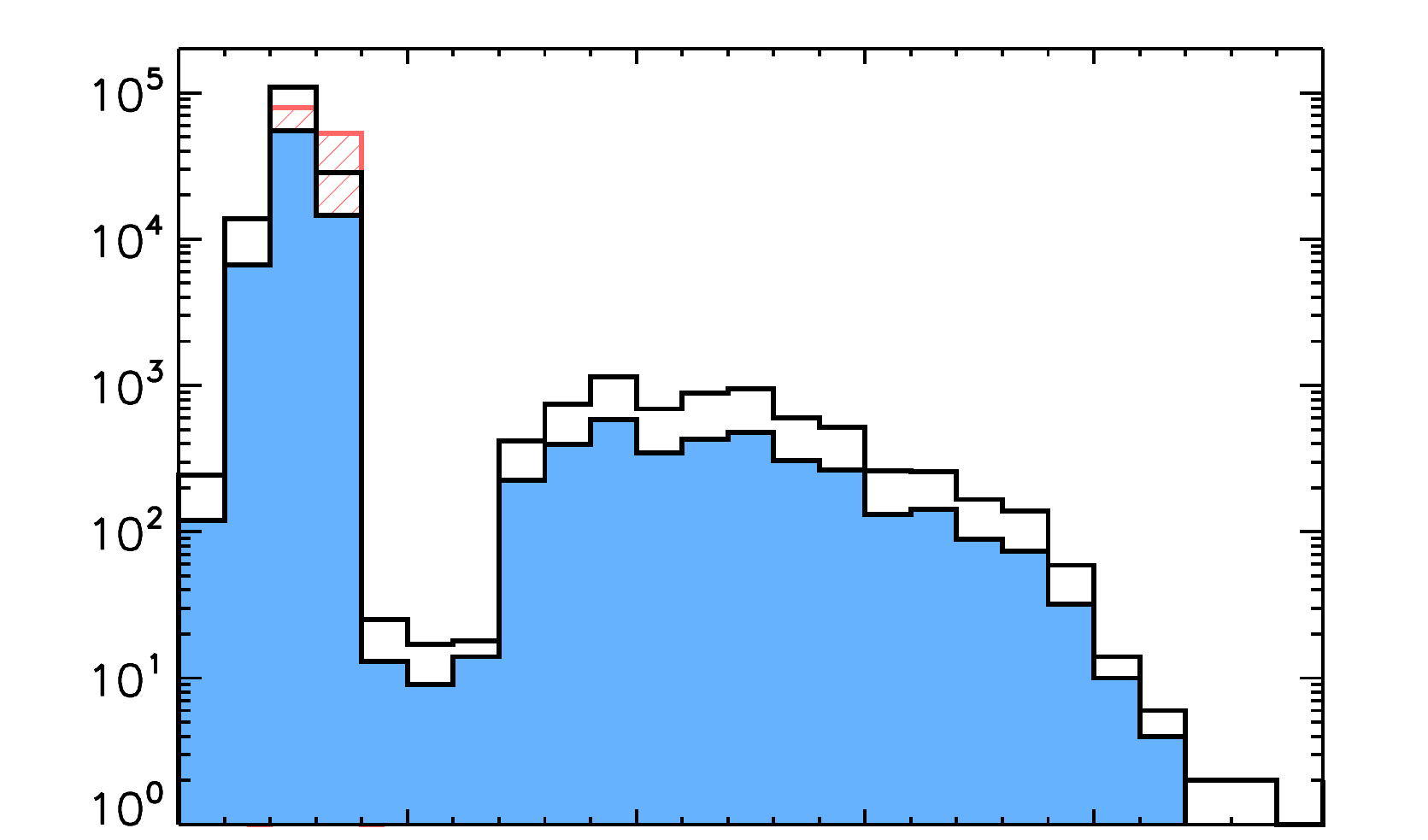}%
                                 \includegraphics[clip]{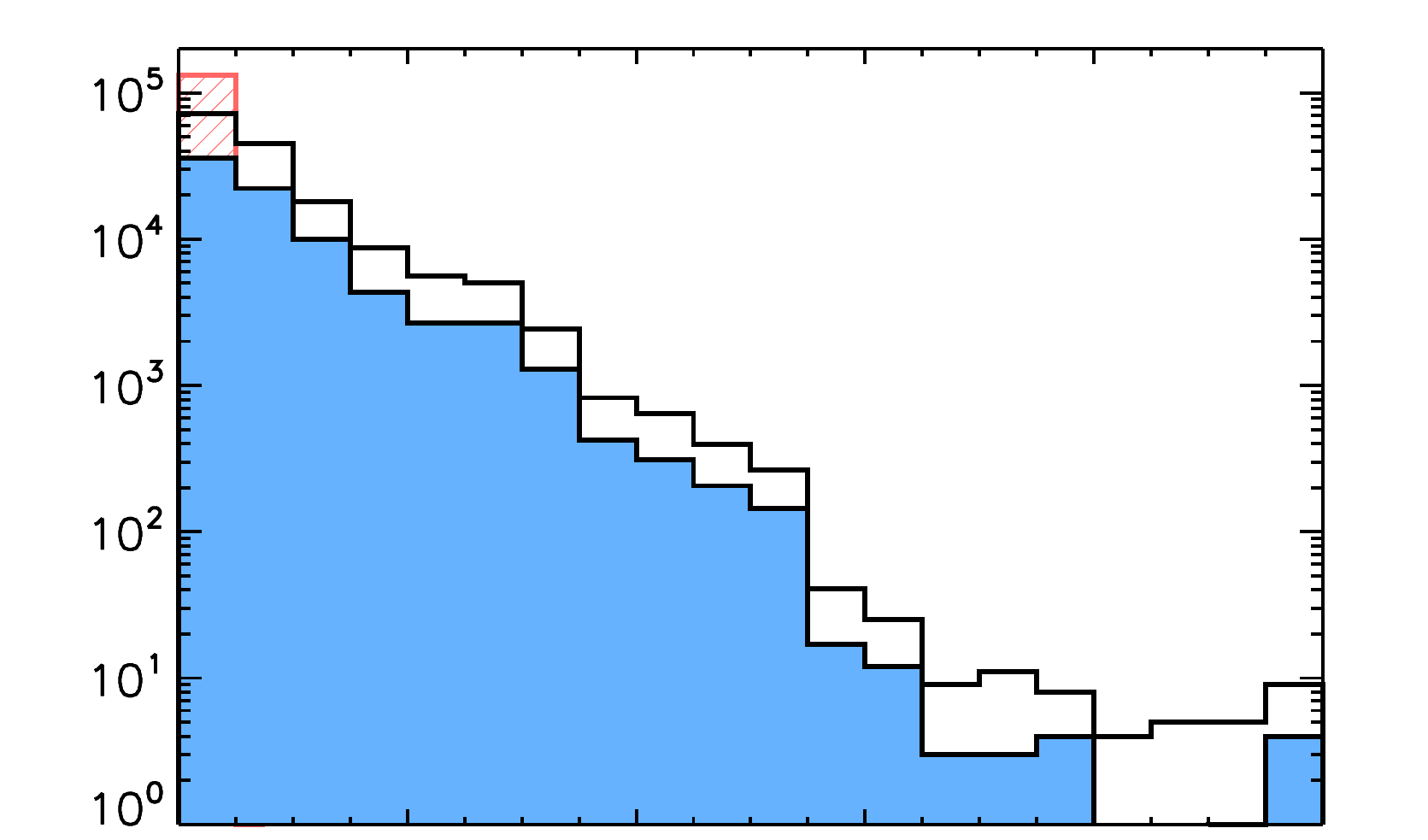}%
                                 \includegraphics[clip]{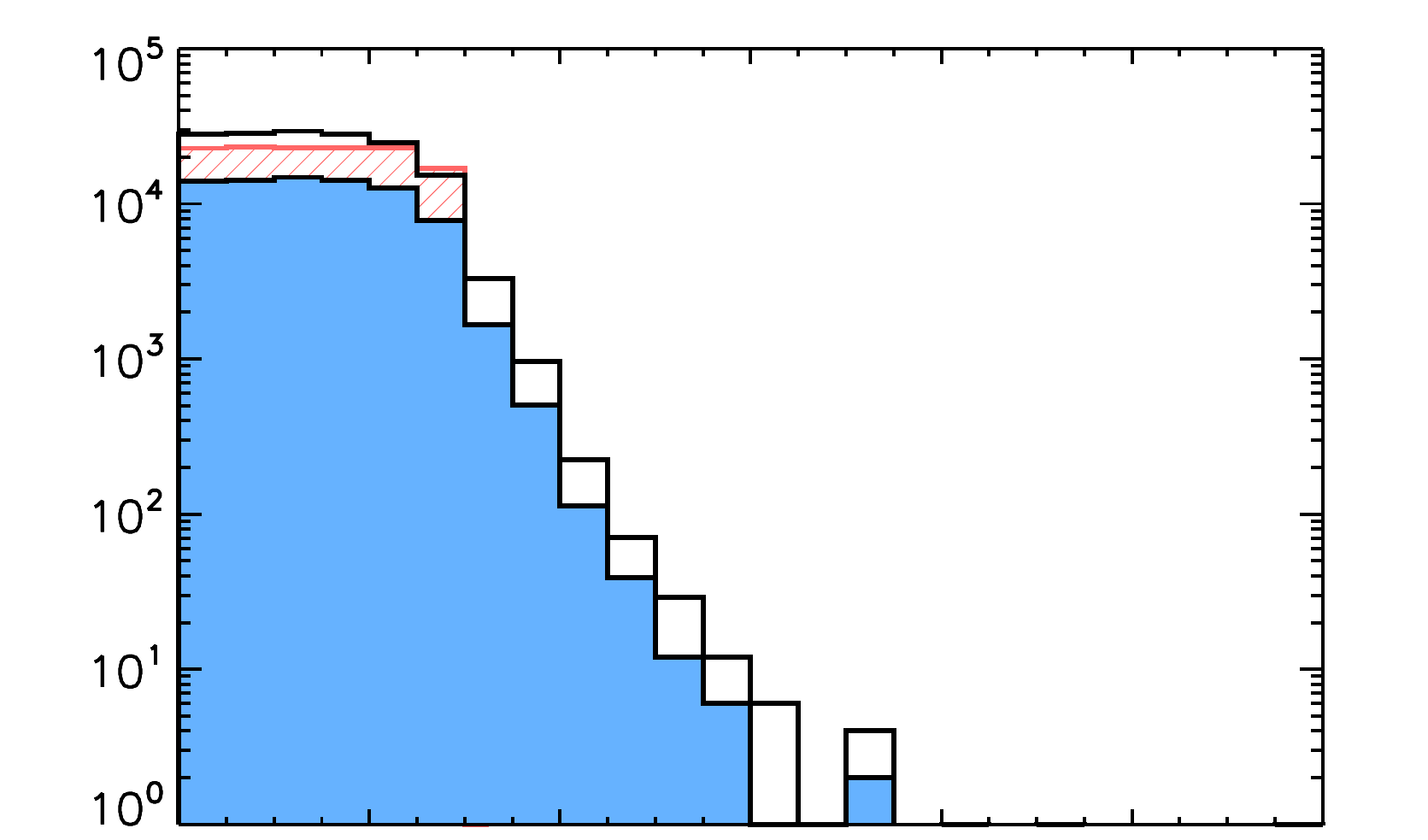}}
\resizebox{\figlew}{!}{\includegraphics[clip]{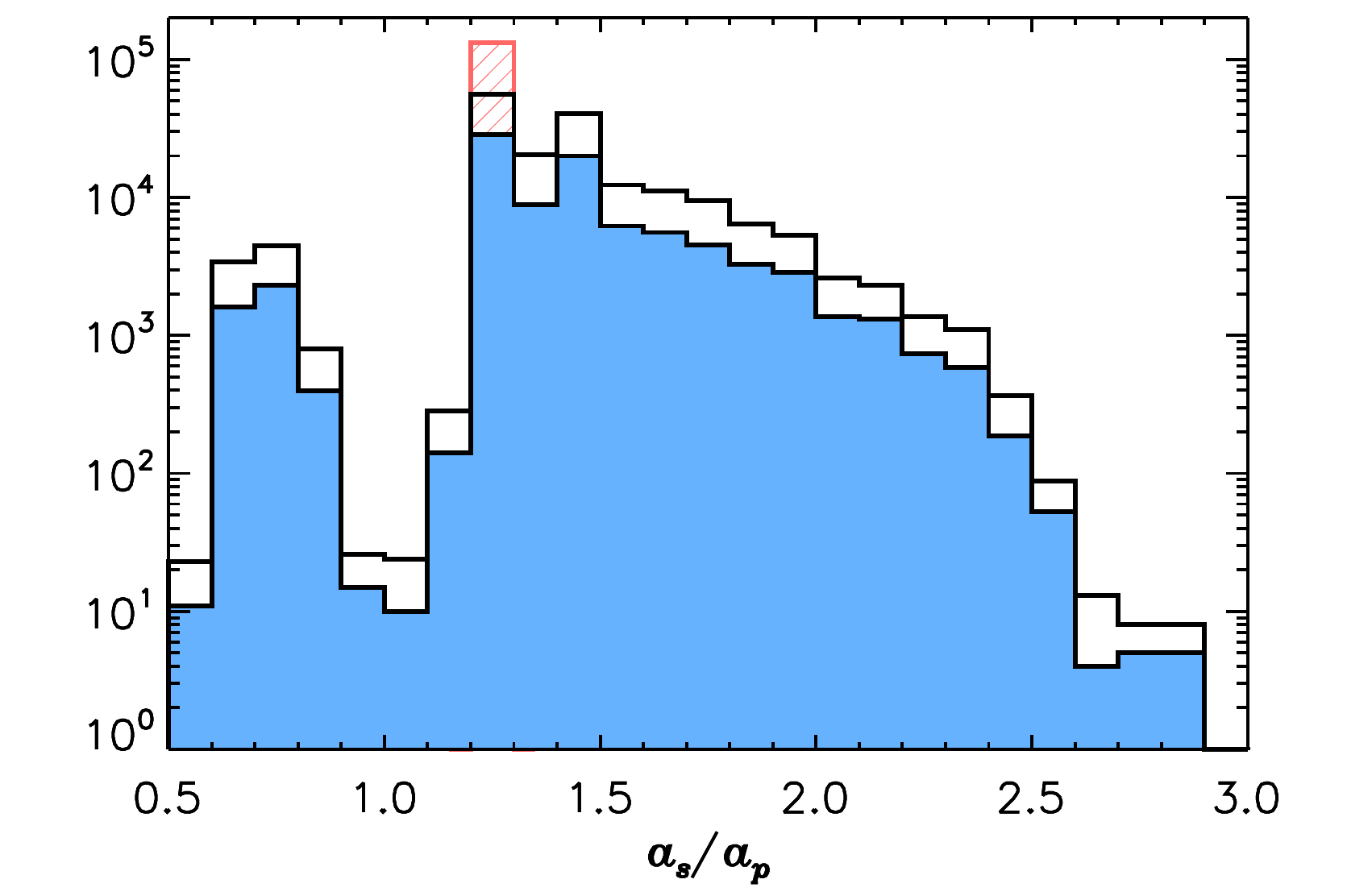}%
                                 \includegraphics[clip]{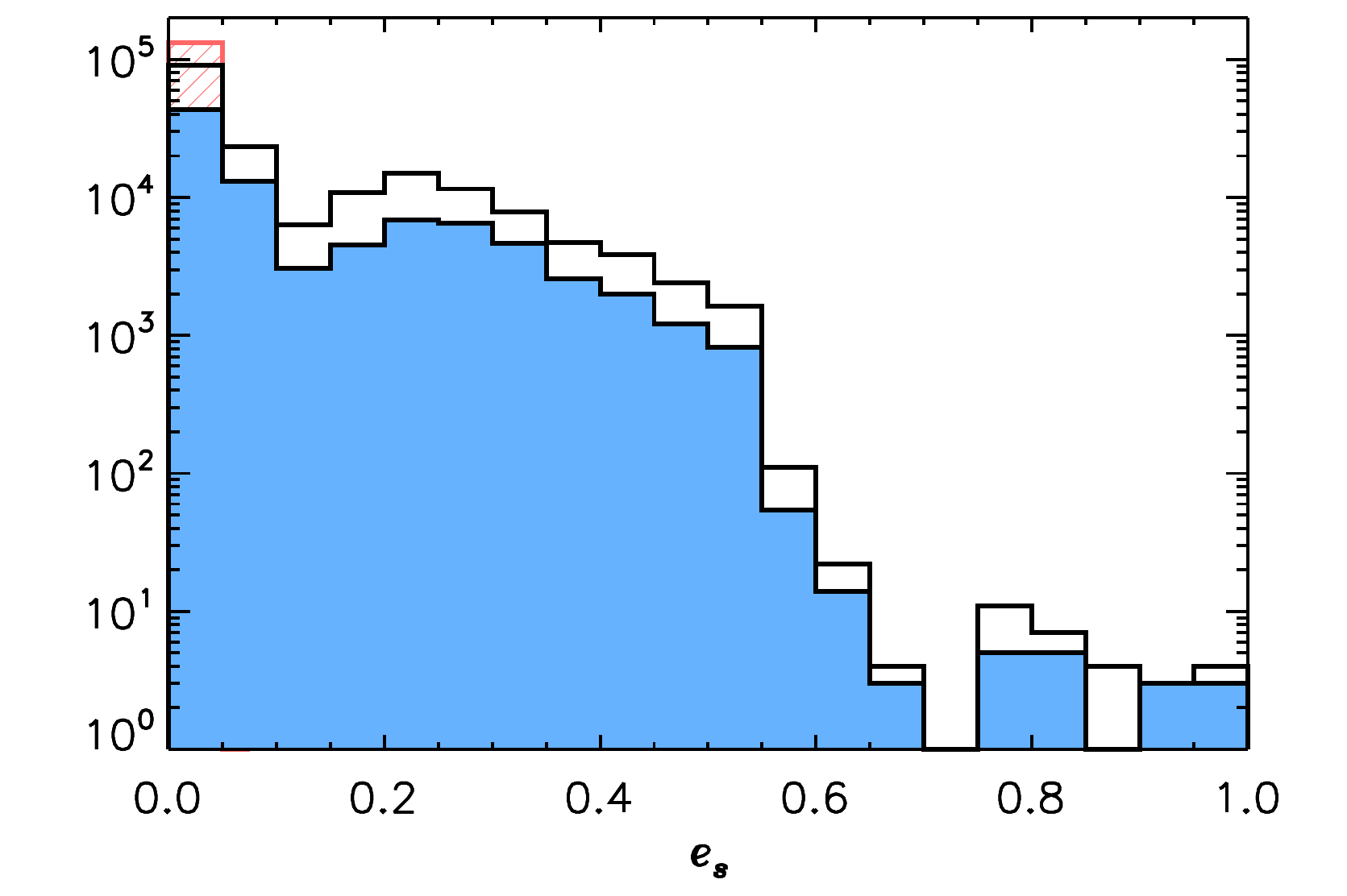}%
                                 \includegraphics[clip]{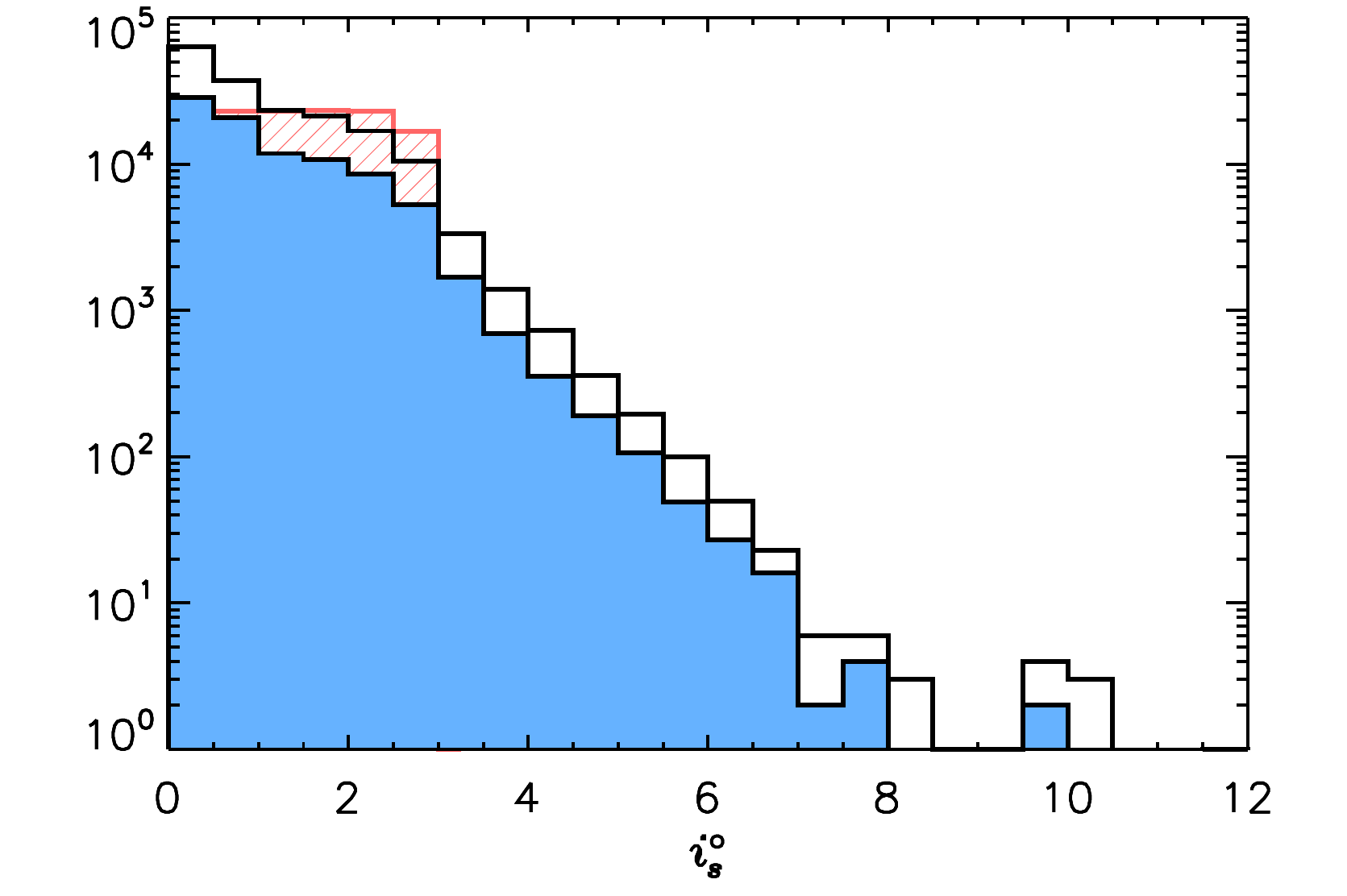}}
\caption{%
             Distributions of semi-major axis (left), eccentricity (center), and inclination
             (right) for $\rho_{0}=10^{-12}$ (upper pair of rows) and $10^{-11}\,\mathrm{g\,cm}^{-3}$ 
             (lower pair of rows). Odd (even) rows illustrate the distributions of planetesimals 
             deployed in the region inside (outside) of the planet's orbit.
             The  cross-hatched histograms show the initial distributions of 
             the mixed-composition bodies.
             The color-shaded and line histograms represent the distributions, respectively,  
             of the mixed-composition population and of both the icy and mixed-composition
             populations, about $580$ orbits after deployment.
             }
\label{fig:hie_disk}
\end{figure*}
\begin{figure*}
\centering%
\resizebox{\figlew}{!}{\includegraphics[clip]{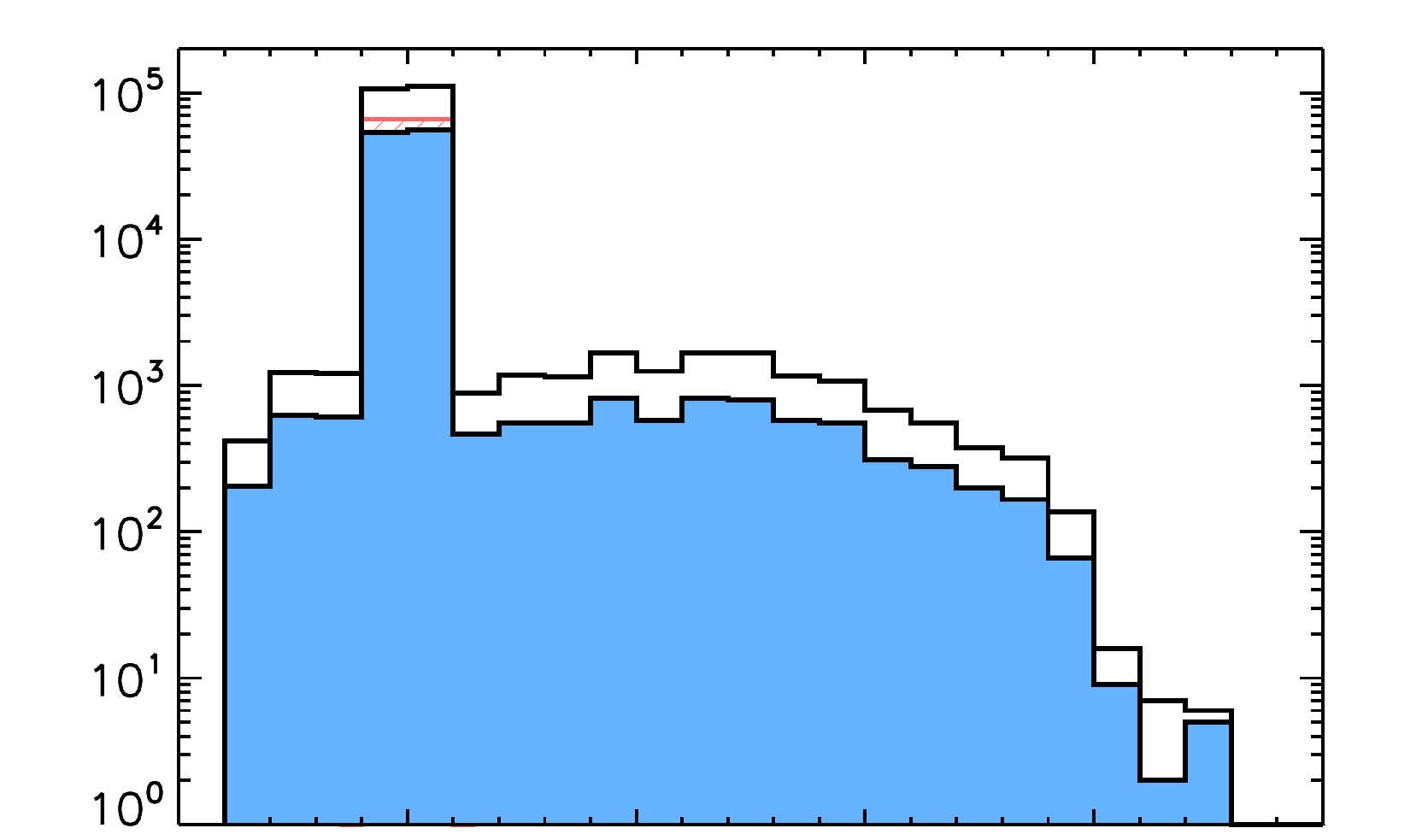}%
                                 \includegraphics[clip]{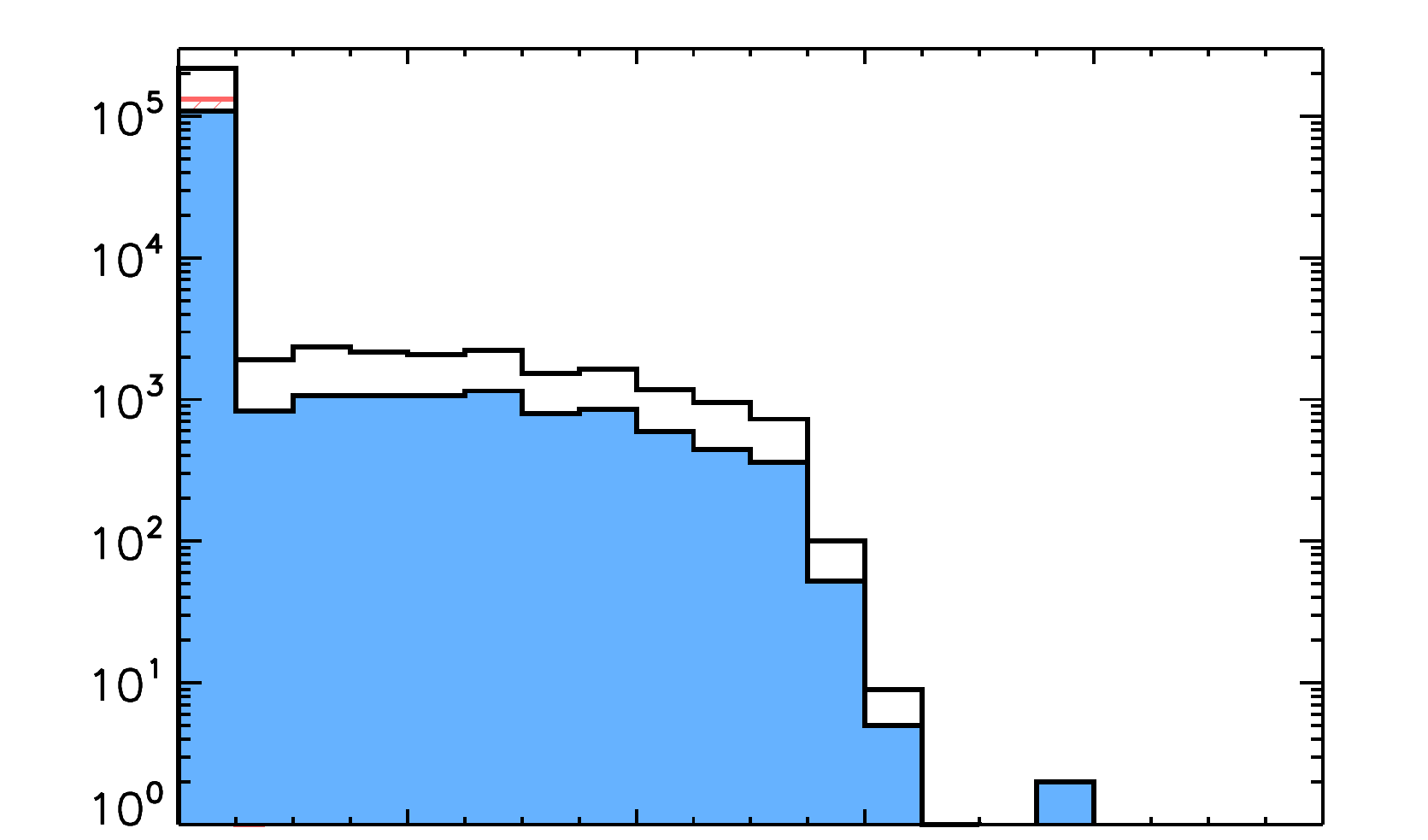}%
                                 \includegraphics[clip]{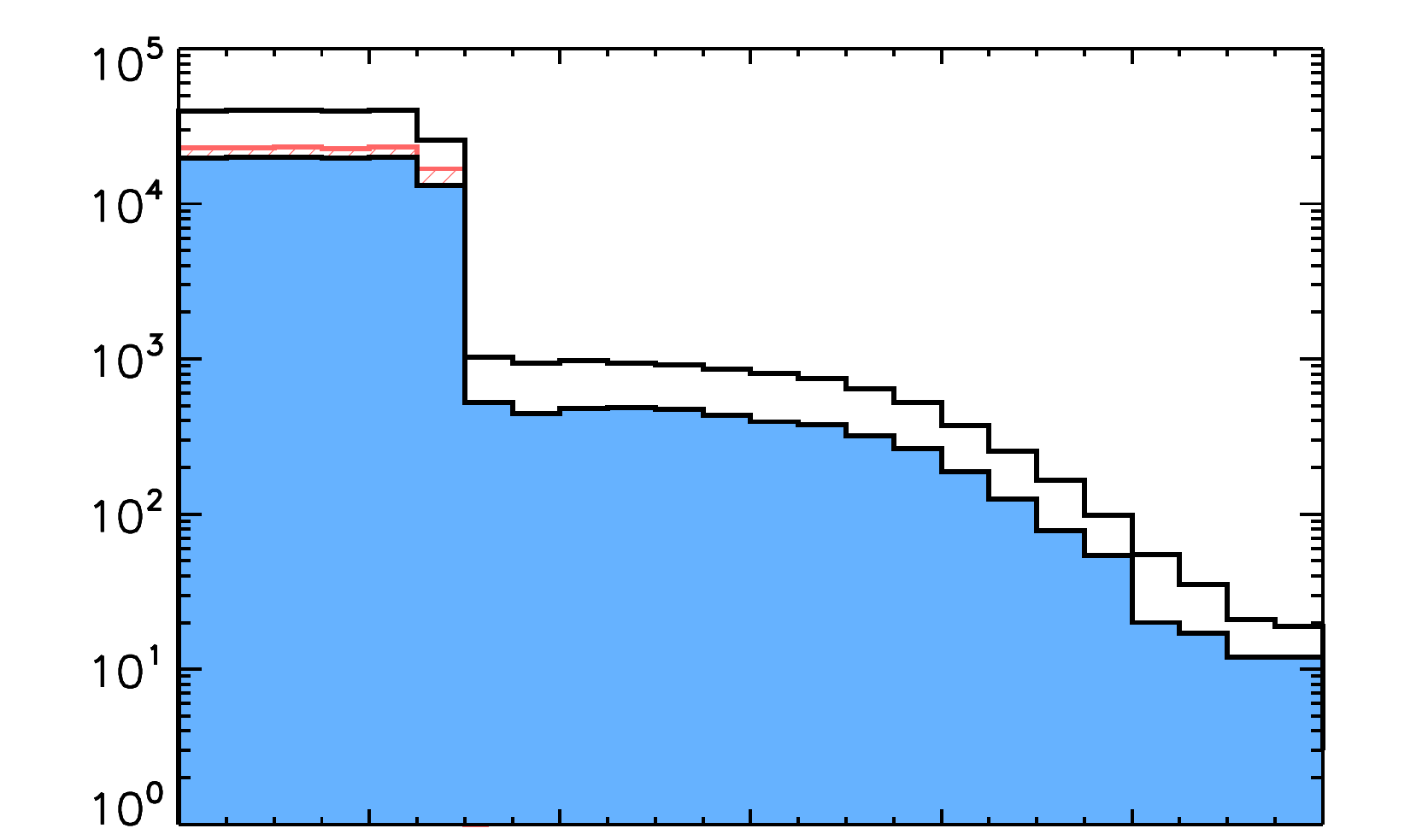}}
\resizebox{\figlew}{!}{\includegraphics[clip]{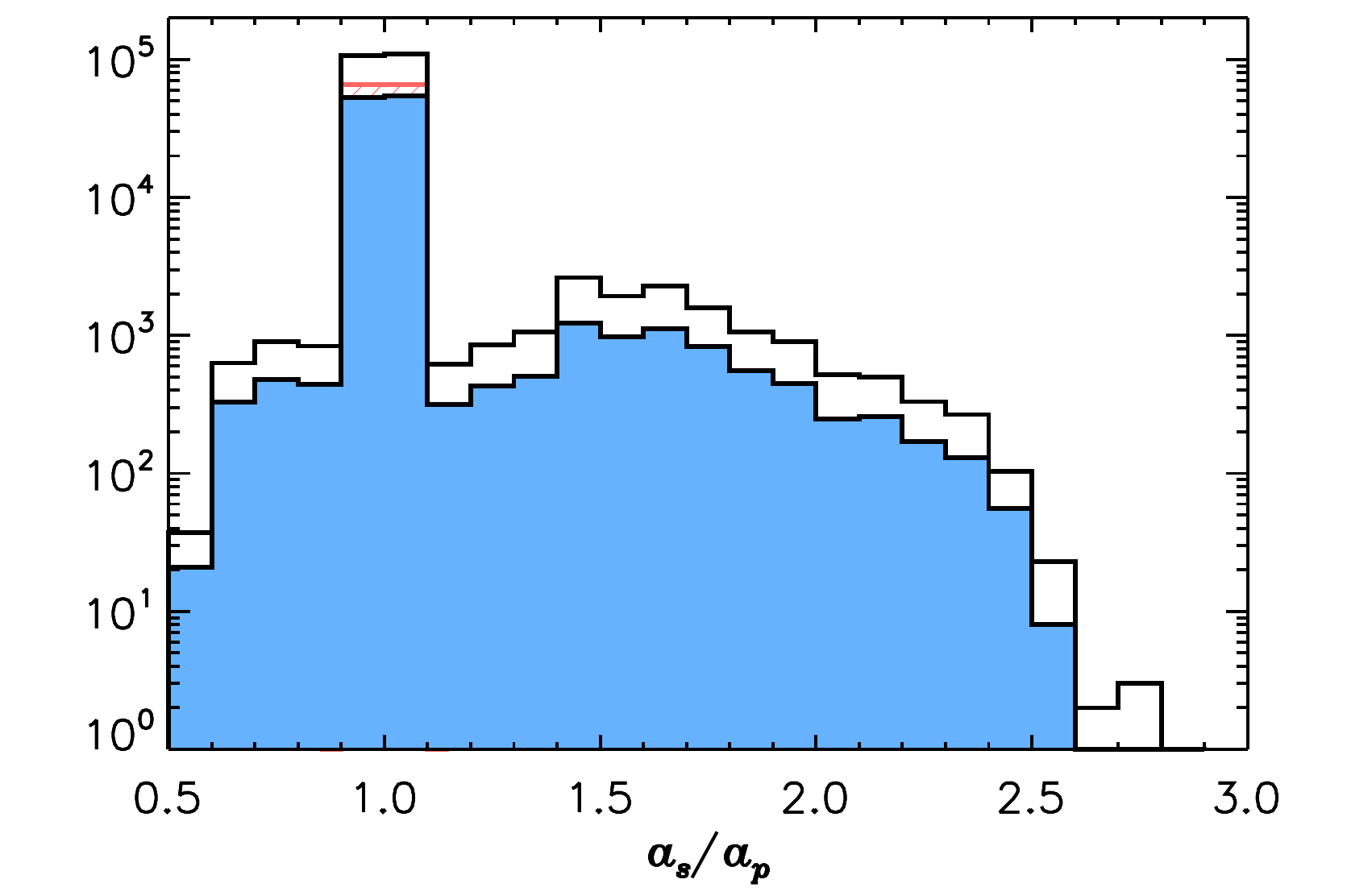}%
                                 \includegraphics[clip]{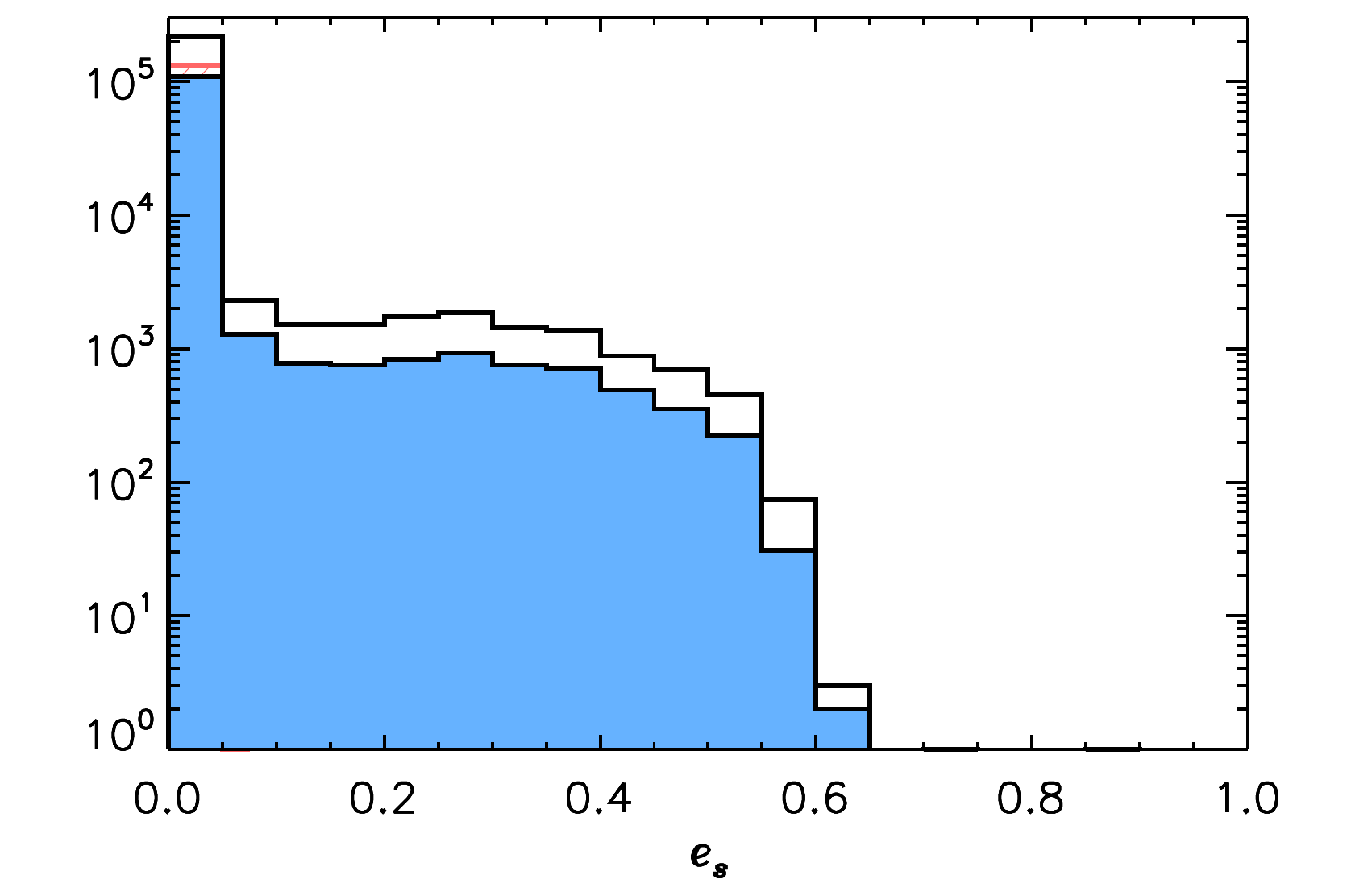}%
                                 \includegraphics[clip]{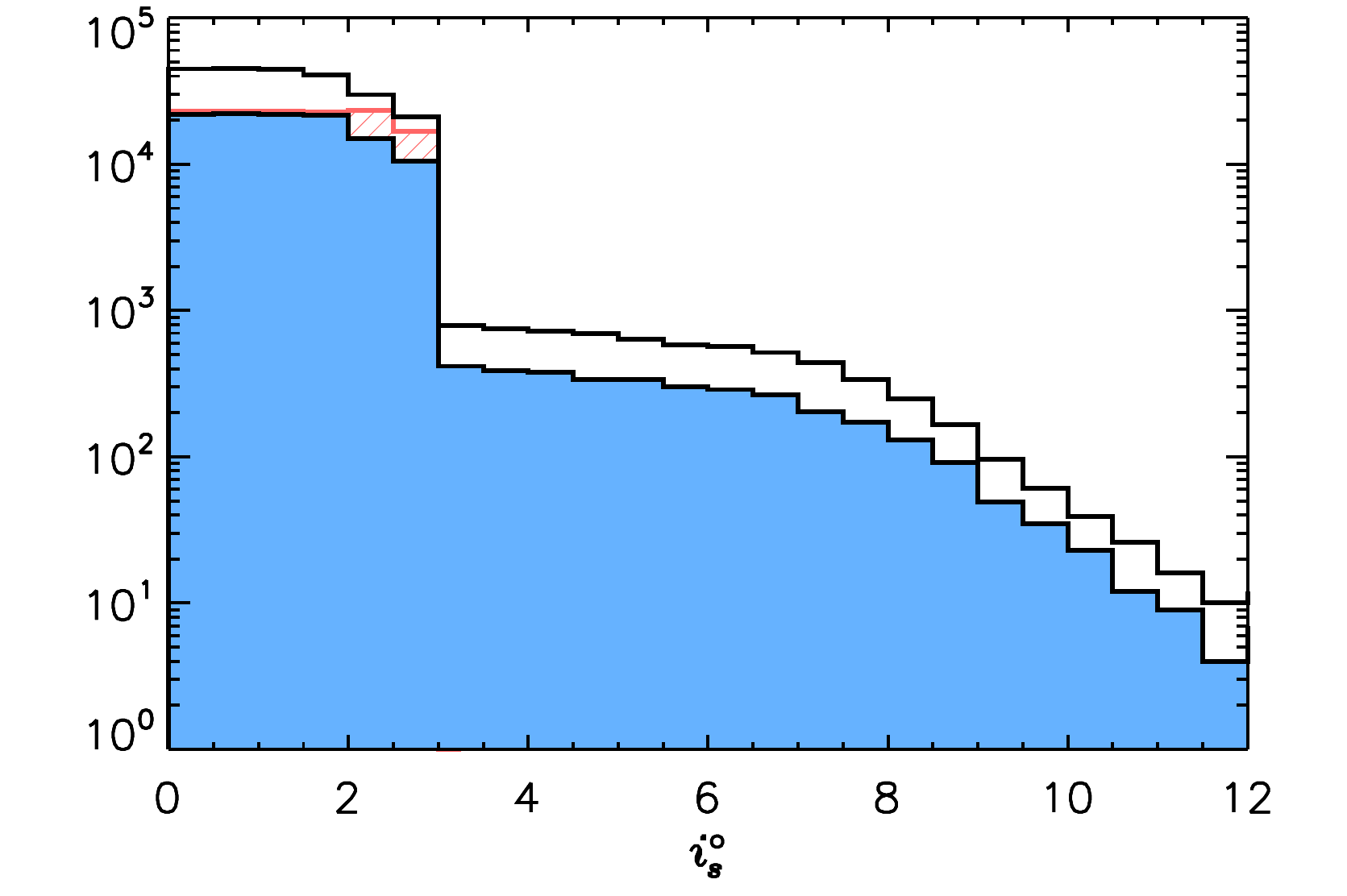}}
\caption{%
             As in Figure~\ref{fig:hie_disk}, but for bodies initially placed in the planet 
             corotation region. Top (bottom) panels refer to $\rho_{0}=10^{-12}$ ($10^{-11}$) 
             $\mathrm{g\,cm}^{-3}$. 
             }
\label{fig:hco_disk}
\end{figure*}
The distributions of some instantaneous (osculating) orbital elements 
are presented in Figures~\ref{fig:hie_disk} and \ref{fig:hco_disk}, 
at the end of the evolution (see the figure captions for further details). 
The histograms refer to bodies deployed in the regions interior 
and exterior of the planet's orbit (Figure~\ref{fig:hie_disk}) and 
in the planet's corotation region (Figure~\ref{fig:hco_disk}).
Bodies that move beyond the disk boundaries (see Section~\ref{sec:TED}) 
are removed from the calculations and excluded from the analysis.
Between $\approx 11$\% and  $\approx 14$\% of the initial mass in
solids moves out of the computational domain by the end of the 
calculations, where the larger percentage is generated by cases
with $\rho_{0}=10^{-12}\,\mathrm{g\,cm}^{-3}$ (icy and mixed-composition
bodies produce similar fractions). 
The average rates of mass loss through the boundaries, between $2$\% 
and $3$\% of the initial mass per $100$ orbits, are consistent with those 
measured over the last $100$ orbits of the calculations.
The probability of ejecting out of boundaries bodies of 
$\sim 10$--$100\,\mathrm{km}$ in radius is similar (within factors of order unity),
and higher or somewhat higher (depending on $\rho_{0}$) than 
the probability of ejecting $R_{s}\lesssim 1\,\mathrm{km}$ bodies.

During the evolution, only about $0.3$\% of the initial mass 
is ablated. The rates of mass lost to ablation are roughly steady during 
the course of  the calculations. 
Higher gas densities produce slightly more ablation than do lower densities.
Models that use everywhere the gas temperature given by 
Equation~(\ref{eq:Tg}), instead of the locally modified temperature 
discussed in Section~\ref{sec:CPD}, yield similar fractions for 
the ablated mass (see Section~\ref{sec:PCPiso}).
The small percentage of ablated mass masks the fact that the total
mass in solids is basically carried by $100\,\mathrm{km}$ bodies, which
shed relatively little mass. 
Smaller bodies, however, are more prone to ablation. In fact,
$R_{s}\approx 10\,\mathrm{km}$ bodies lose $\sim 5$\% of their total initial 
mass and $R_{s}\approx 1\,\mathrm{km}$ planetesimals shed $\sim 20$\% 
of their total initial mass. The fraction becomes $\sim 40$\% for 
$R_{s}\approx 0.1\,\mathrm{km}$ bodies.
This outcome can be understood from Equation~(\ref{eq:dmsdt}), assuming
bodies of equal temperature (and hence vapor pressure, $P_{v})$, which
yields a ratio of the ablated to the total mass proportional to $1/R_{s}$.
Had all size bins contained equal masses, the swarm would have lost
to ablation over $10$\% of its original mass.

Negligible fractions of the initial mass are lost through fracturing
and break-up:
$\sim 10^{-6}$ for  $\rho_{0}=10^{-12}\,\mathrm{g\,cm}^{-3}$ and
$\sim 10^{-5}$ for $\rho_{0}=10^{-11}\,\mathrm{g\,cm}^{-3}$. 
Bodies can break up only if their radius is smaller than $R_{\mathrm{dy}}$
(Equation~(\ref{eq:rdy})) and the body's compressive strength is exceeded
by the dynamical pressure, $P_{\mathrm{dy}}$, which
is proportional to the gas density (Equation~(\ref{eq:pdy})).
Break-up of planetesimals occurs within $0.2\,\Rhill$ of the planet, when
they impact with the dense regions of the circumplanetary disk.

Comparisons of the histograms show only marginal differences between 
the orbital elements of icy and mixed-composition bodies, for the same 
value of the gas density $\rho_{0}$. As expected from the discussion above,
calculations with $\rho_{0}=10^{-12}$ and $10^{-11}\,\mathrm{g\,cm}^{-3}$ 
provide similar distributions of semi-major axes, and distributions of
eccentricities and inclinations differing mainly toward large values 
($e_{s}\gtrsim 0.5$ and $i_{s}\gtrsim 5^{\circ}$).
Regardless of the planetesimal composition and background gas density, 
by the end of the calculations, about $5$\% of the remaining mass of solids
initially placed interior of the planet's orbit is scattered outside the orbit.
Toward the end of the calculations, the average scattering rate is about 
$1$\% of the remaining mass per $100$ orbital periods. 
Around $11$\% of the available mass of solids initially present exterior of 
the planet's orbit is scattered inside, with an average scattering rate 
(toward the end) of about $6$\% of the remaining mass per $100$ orbits.
This would amount to an average of $\sim 10^{24}\,\mathrm{g}$ 
of solids scattered toward the inner disk during a planet's orbital
period, if the initial surface density of planetesimals between $1.2\,a_{p}$ 
and $1.25\,a_{p}$ was $1\,\mathrm{g\,cm}^{-2}$.
Although scattering involves planetesimals of all sizes,
larger size objects ($R_{s}\gtrsim 1\,\mathrm{km}$) are typically scattered
more efficiently, in either radial direction, than are smaller size objects.
Moreover, $R_{s}\sim 0.1\,\mathrm{km}$ bodies are more easily scattered
outward, from inner disk regions, at the lower (rather than at the higher)
value of the reference density, $\rho_{0}$.
While not visible in the histograms of Figure~\ref{fig:hie_disk}, 
because of the large bin size, distributions with finer sampling in 
semi-major axis show several dips, in proximity to the position of 
the 3:2, 5:3, and 2:1 mean-motion resonances with the planet,
and to the corresponding resonant locations exterior of the 
planet's orbit.

Most of the mass deployed in the corotation region ($a_{p}\mp\Rhill/2$) 
remains within $a_{p}\mp\Rhill$ throughout the calculations 
(see Figure~\ref{fig:hco_disk}). 
Only a small fraction, $\sim 0.5$\%, of the initial mass 
is scattered toward the inner disk and a fraction of a few percent 
is scattered outward. Taking into account all possible fates for the
solids, the mass depletion rate of the radial region $a_{p}\mp\Rhill$ is
around $2$\% of the initial mass per hundred orbits of the planet.
Note that gas drag in the region along the planet's orbit is reduced 
(except very close to the planet) due to the density gap 
(see Figure~\ref{fig:img_disk}).
Within factors of order unity, the ejection probability toward the inner 
disk and that toward the outer disk are constant in the size range 
$0.1\,\mathrm{km}\lesssim R_{s}\lesssim 100\,\mathrm{km}$.
Along the planet's orbit, the largest number densities in the frame 
corotating with the planet occur around the L$_{4}$ and L$_{5}$ 
Lagrange points. 
Although the longitude relative to the planet of these points
can be affected by gas drag, they effectively lie $60^{\circ}$ ahead 
(L$_{4}$) and behind (L$_{5}$) the planet (for $R_{s}\gtrsim 0.1\,\mathrm{km}$) 
owing to the presence of the density gap \citep{peale1993}.
In the radial region of tadpole orbits, $a_{p}\mp0.74\Rhill$
\citep[e.g.,][]{murray2000}, the number densities within 
a $15^{\circ}$ longitude of either point are roughly constant
for planetesimals of all sizes.
The smallest number density (in the corotating frame) is 
around the collinear $L_{3}$ point.

\begin{figure}
\centering%
\resizebox{\linewidth}{!}{\includegraphics[clip]{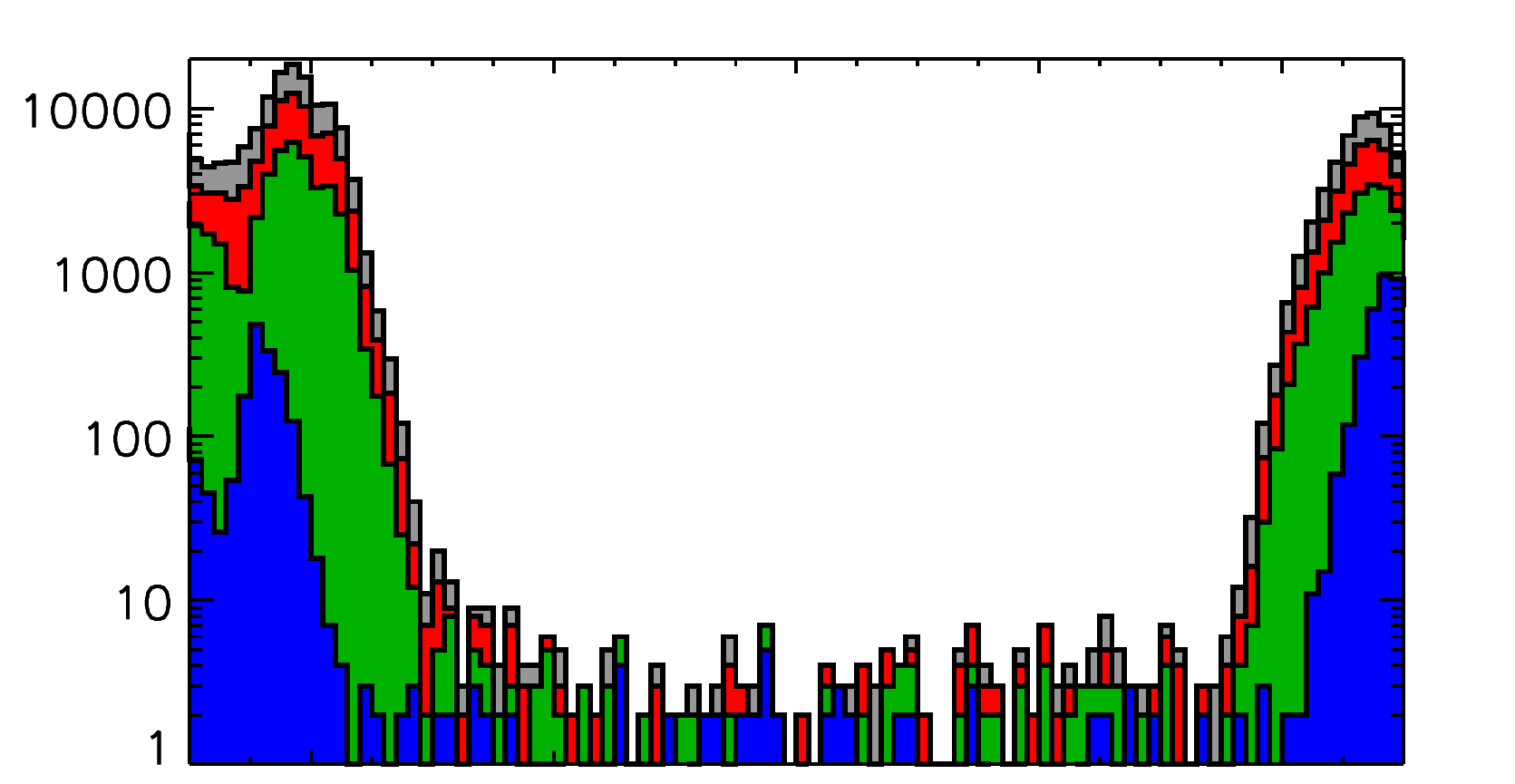}}
\resizebox{\linewidth}{!}{\includegraphics[clip]{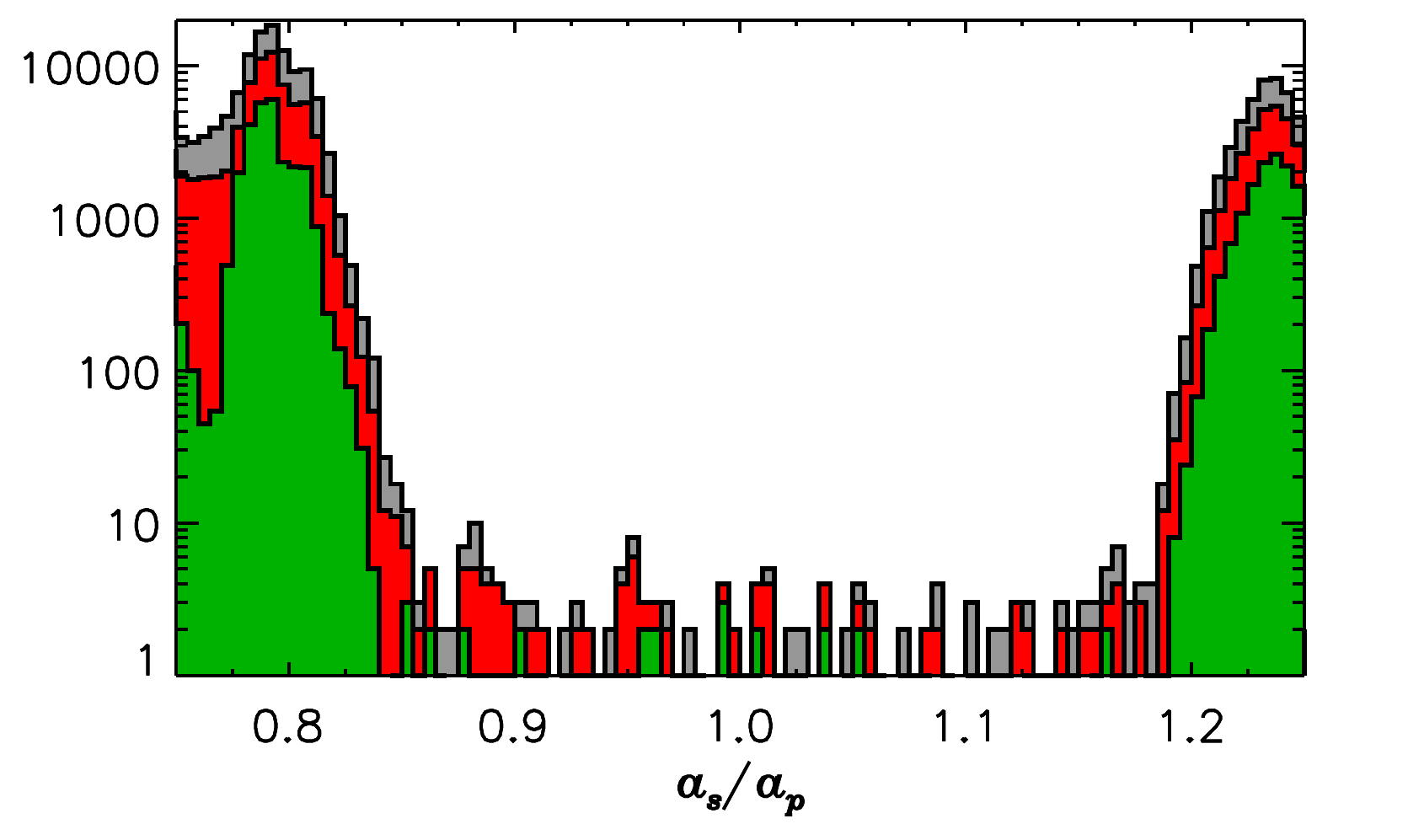}}
\caption{%
             Semi-major axis distributions of planetesimals in proximity of the planet's
             orbit. The histograms include bodies initially placed inside and outside of
             the orbit and of both compositions. The reference gas density is
             $\rho_{0}=10^{-12}$ (top) and $10^{-11}\,\mathrm{g\,cm}^{-3}$ (bottom).
             Histograms of different colors progressively include larger and larger bodies: 
             $R_{s}\le 0.1$ (lowest-count histogram), $1$, $10$, and $100\,\mathrm{km}$ 
             (highest-count histogram). In the bottom panel, $R_{s}\le 0.1\,\mathrm{km}$ 
             bodies do not appear because they are located farther from the planet's 
             orbit.
             }
\label{fig:solid_gap}
\end{figure}
The gap in the planetesimal disk, due to the torques exerted by the planet
on the solids, 
is illustrated in Figure~\ref{fig:solid_gap} as a function of the semi-major
axis, for various planetesimal radii (see figure caption for further details).
The distributions comprise icy and mixed-composition bodies, initially deployed
interior and exterior of the planet's orbit. The positions of the gap edges move 
further away from the planet's orbit as $R_{s}$ reduces because of gas drag effects
(see below).  Differences with respect to the gas density appear marginal 
for radii $R_{s}\gtrsim 1\,\mathrm{km}$, but they become more pronounced
at radii $R_{s}\lesssim 0.1\,\mathrm{km}$ (particles of these sizes do not
appear in the bottom panel of Figure~\ref{fig:solid_gap}).
For $\rho_{0}=10^{-11}\,\mathrm{g\,cm}^{-3}$, the outer gap edge 
of $R_{s}\sim 0.1\,\mathrm{km}$ bodies recedes at $r\approx 1.25\,a_{p}$, 
where the (azimuthally) averaged rotation velocity of the gas 
exceeds the azimuthal velocity of particles because the radial pressure 
gradient of the gas is locally positive
(although only marginally, this effect is already present at $r\sim 1.6\,a_{p}$).
The inner gap edge of $0.1\,\mathrm{km}$ bodies is found at $r\approx 0.67\,a_{p}$.
To address the behavior of small fragments, in Section~\ref{sec:DC}
we discuss some experiments conducted with $\mathrm{cm}$--to--$\mathrm{m}$ 
size particles.

In fact,
the super-Keplerian rotation at the outer edge of the density gap, mentioned in
Section~\ref{sec:DDD} (see Figure~\ref{fig:vphiu}), can lead to planetesimal
segregation, by halting or pushing outward bodies that reach those radial
locations \citep[see, e.g,][and references therein]{ayliffe2012b}.
Neglecting collisions and gravitational encounters among bodies, 
which would redistribute bodies
and likely work against segregation, the efficiency of this process depends 
on the competing effects of the gravitational torque (perpendicular to 
the orbital plane) exerted by the planet and the vertical component 
of the gas drag torque $\mathbf{r}_{s}\mathbf{\times}\mathbf{F}_{D}$.
Considering planetesimals on near-Keplerian orbits,
$\mathbf{r}_{s}\mathbf{\times}(\mathbf{v}_{g}-\mathbf{v}_{s})$
has vertical component $\sim a_{s}(v^{A}_{\phi}-v_{\mathrm{K}})$,
where $v^{A}_{\phi}$ is given by Equation~(\ref{eq:rotacp}) (see also
Figure~\ref{fig:vphiu}), which is positive/negative at the outer/inner gap edge.
For large enough objects ($C_{D}\sim 1$), the gas drag torque is then
$\propto R^{2}_{s}\rho_{g}a_{s}v^{2}_{\mathrm{K}}\propto R^{2}_{s}\rho_{g}$ 
whereas the gravitational torque due to the planet is proportional to 
the body mass and hence to $R^{3}_{s}$, yielding a segregation 
efficiency $\propto \rho_{g}/R_{s}$.
Therefore, for a given value of the local density (and distance from 
the planet's orbit), one can expect segregation of smaller planetesimals 
to be more efficient than that of larger bodies.
As expected, we do not observe strict segregation of planetesimals.
However, it does appear that $R_{s}\sim 0.1\,\mathrm{km}$ bodies, initially
placed in the region exterior of the planet's orbit, are less likely to move toward 
the inner disk than are $R_{s}\sim 1\,\mathrm{km}$ bodies, by a factor of
$\sim 10$ (see Figure~\ref{fig:solid_gap}, top panel). 
Additionally, $R_{s}\sim 0.1\,\mathrm{km}$ bodies are less prone
to cross from the outer to the inner disk at the higher value of $\rho_{0}$ 
than they are at the lower density, again by a factor of order $10$.

\begin{figure}
\centering%
\resizebox{\linewidth}{!}{%
\includegraphics[clip]{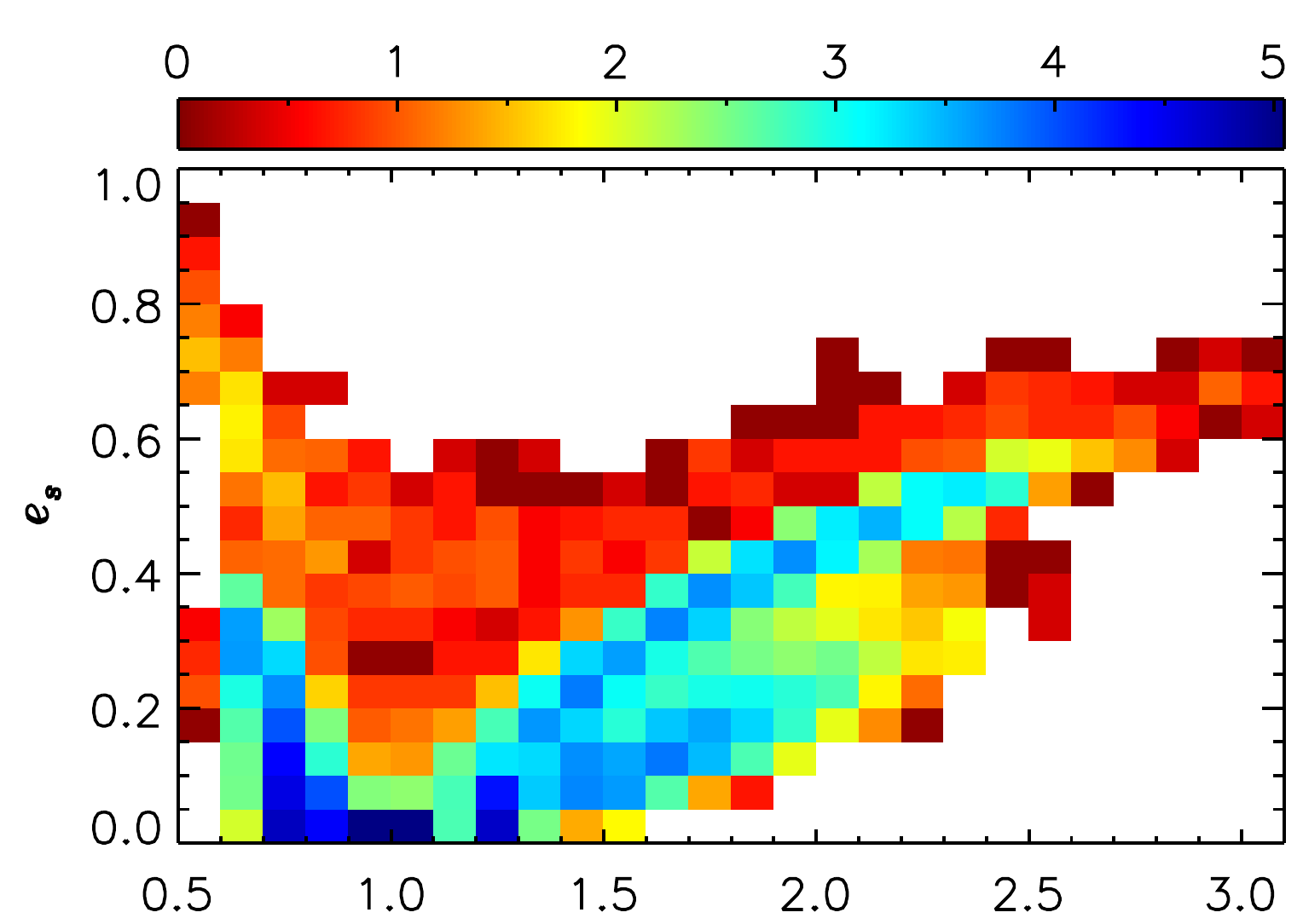}%
\includegraphics[clip]{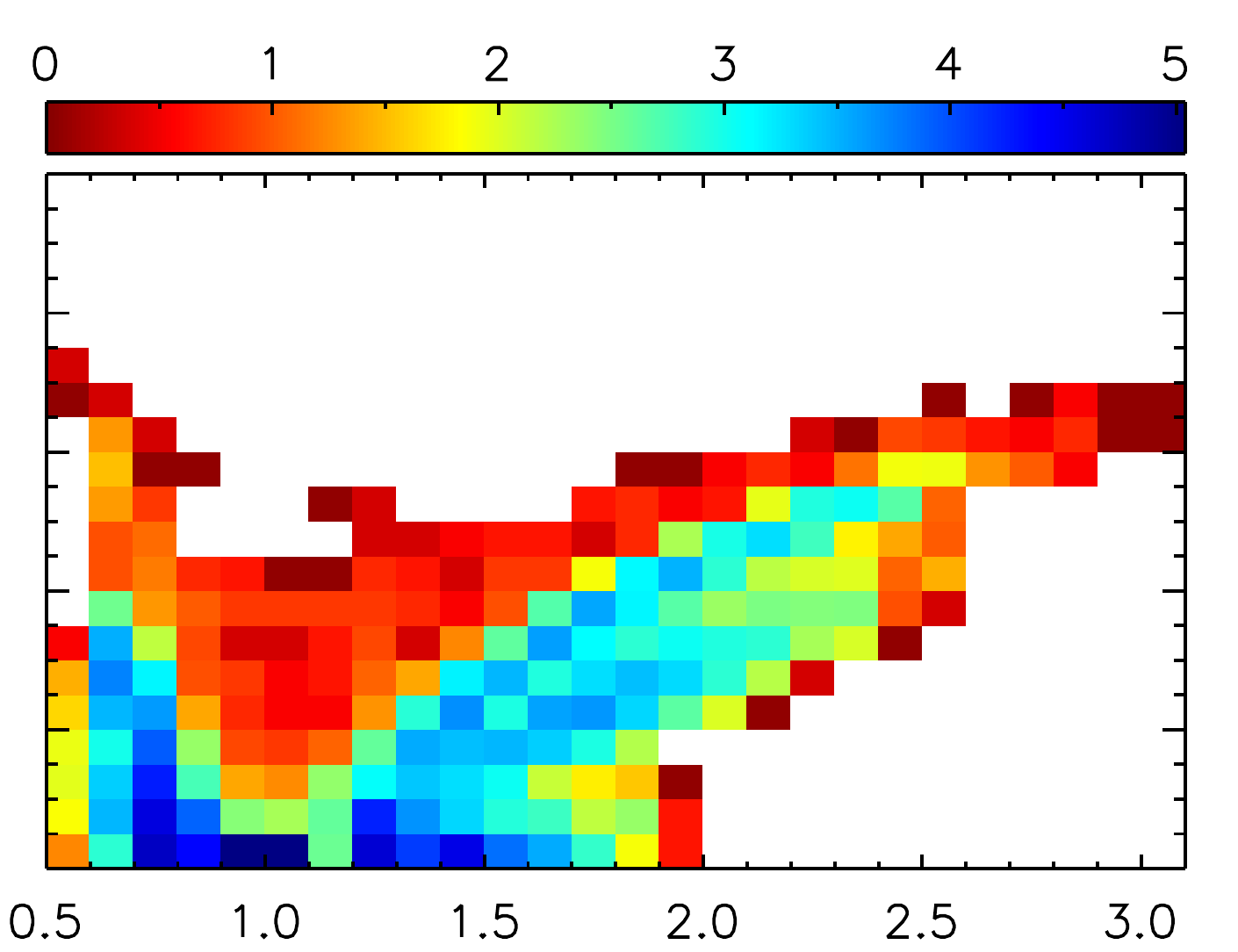}}
\resizebox{\linewidth}{!}{%
\includegraphics[clip]{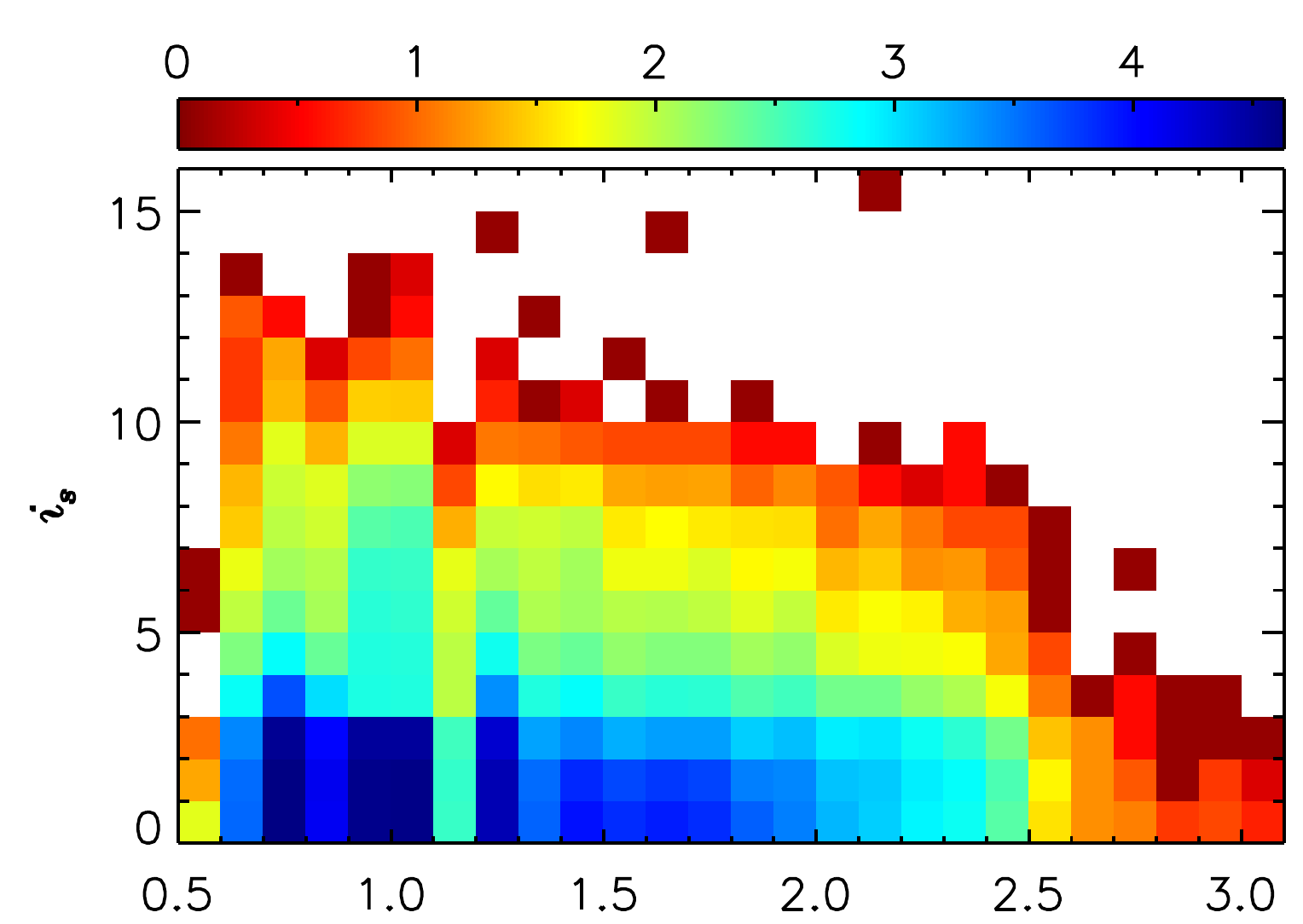}%
\includegraphics[clip]{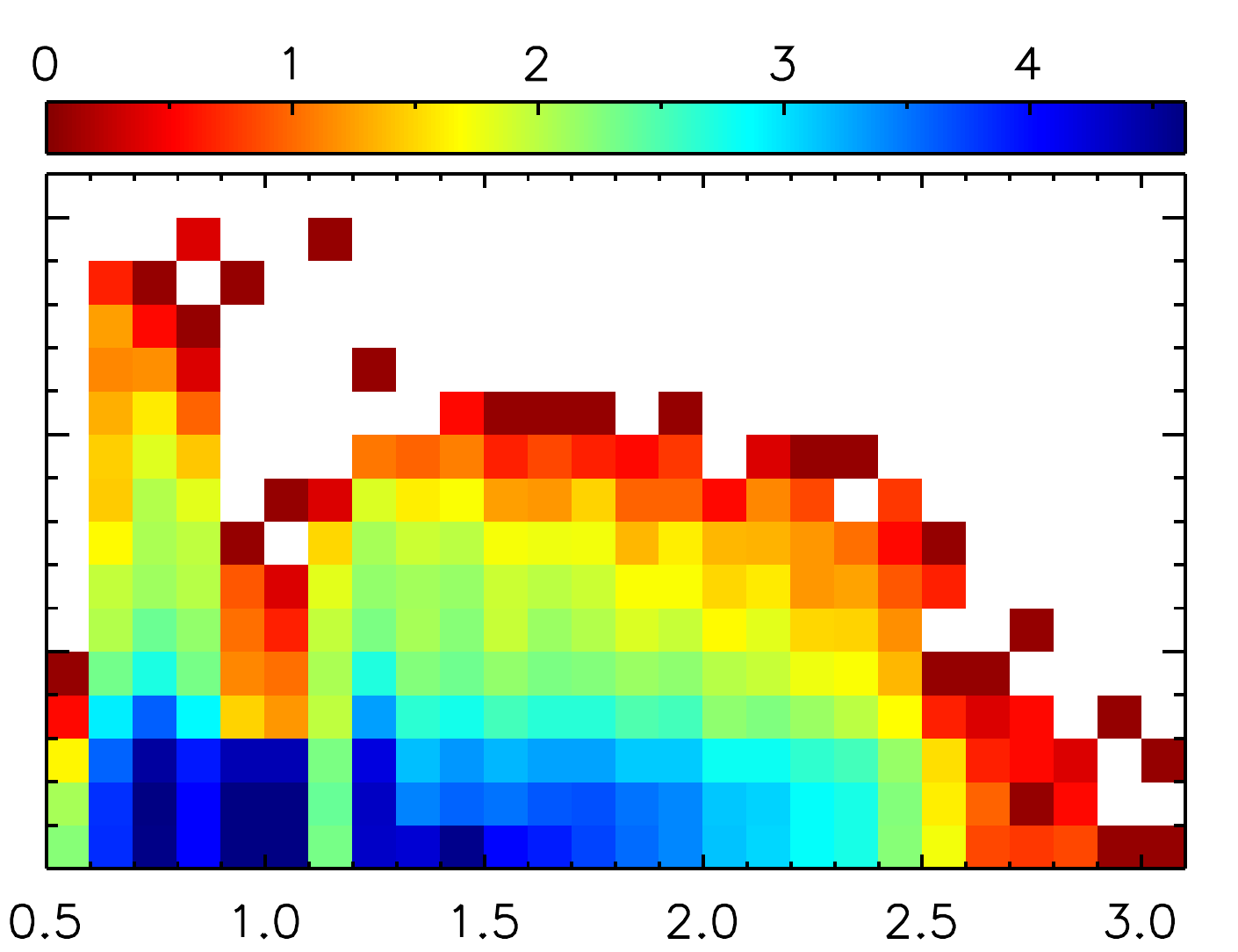}}
\resizebox{\linewidth}{!}{%
\includegraphics[clip]{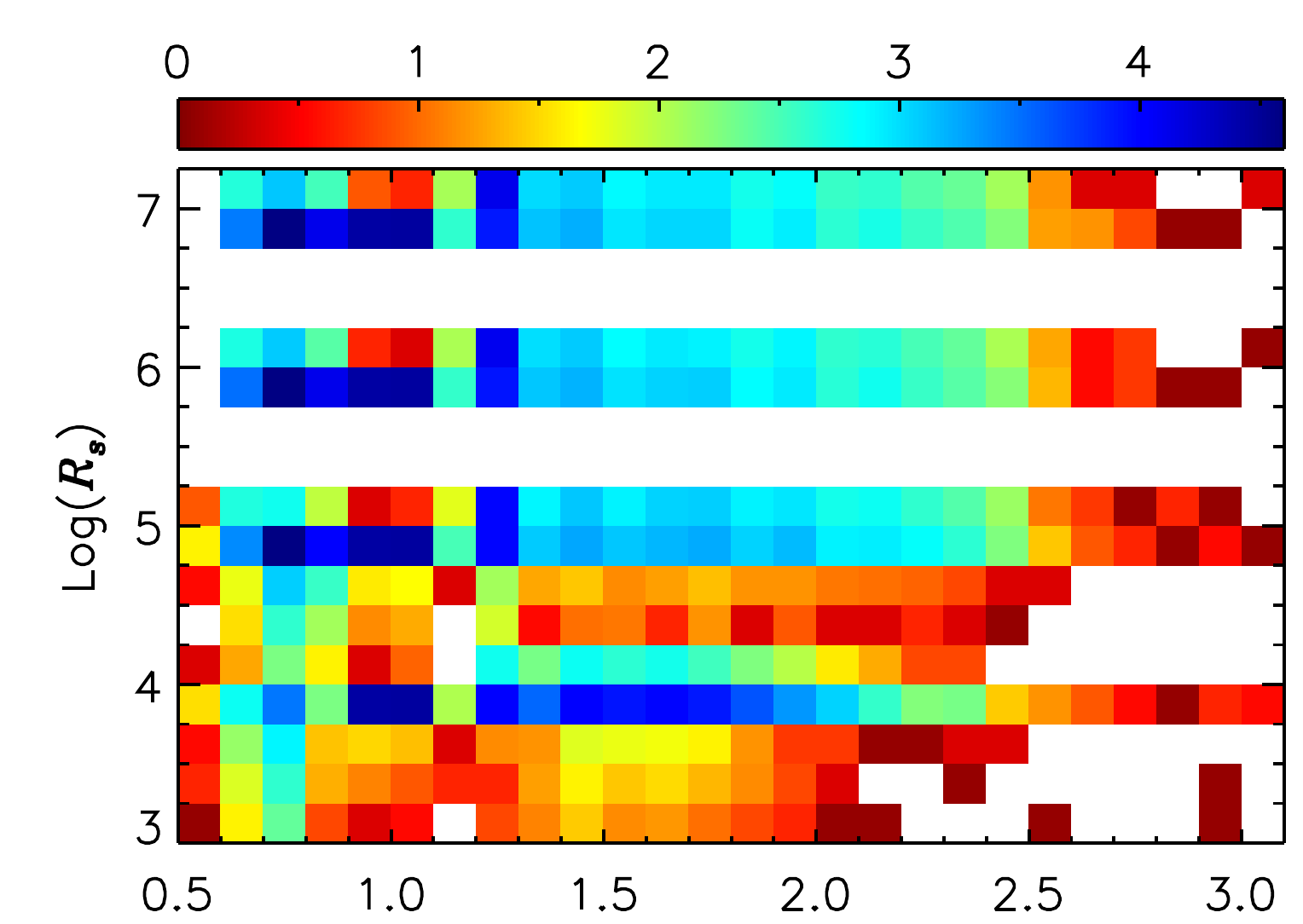}%
\includegraphics[clip]{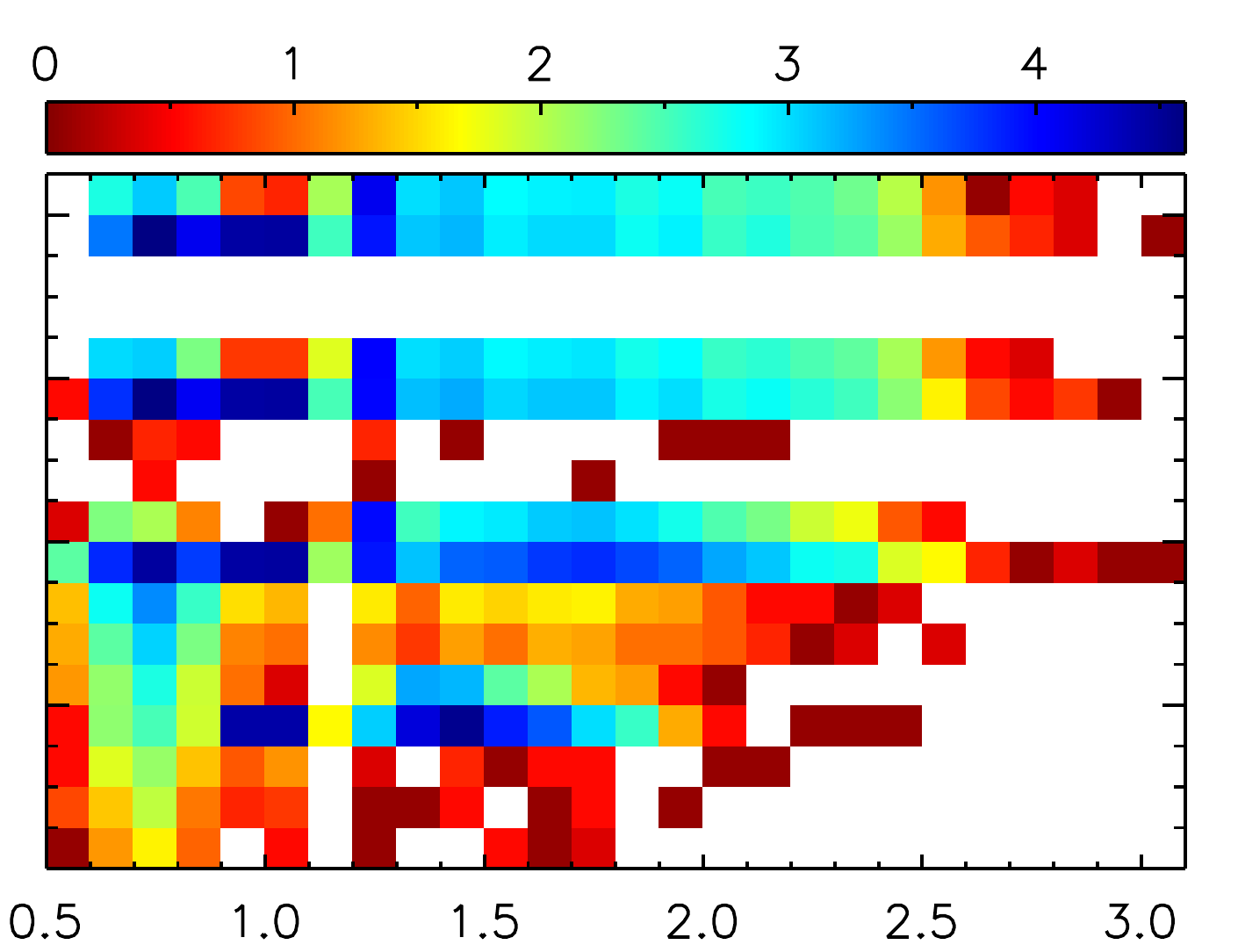}}
\resizebox{\linewidth}{!}{%
\includegraphics[clip]{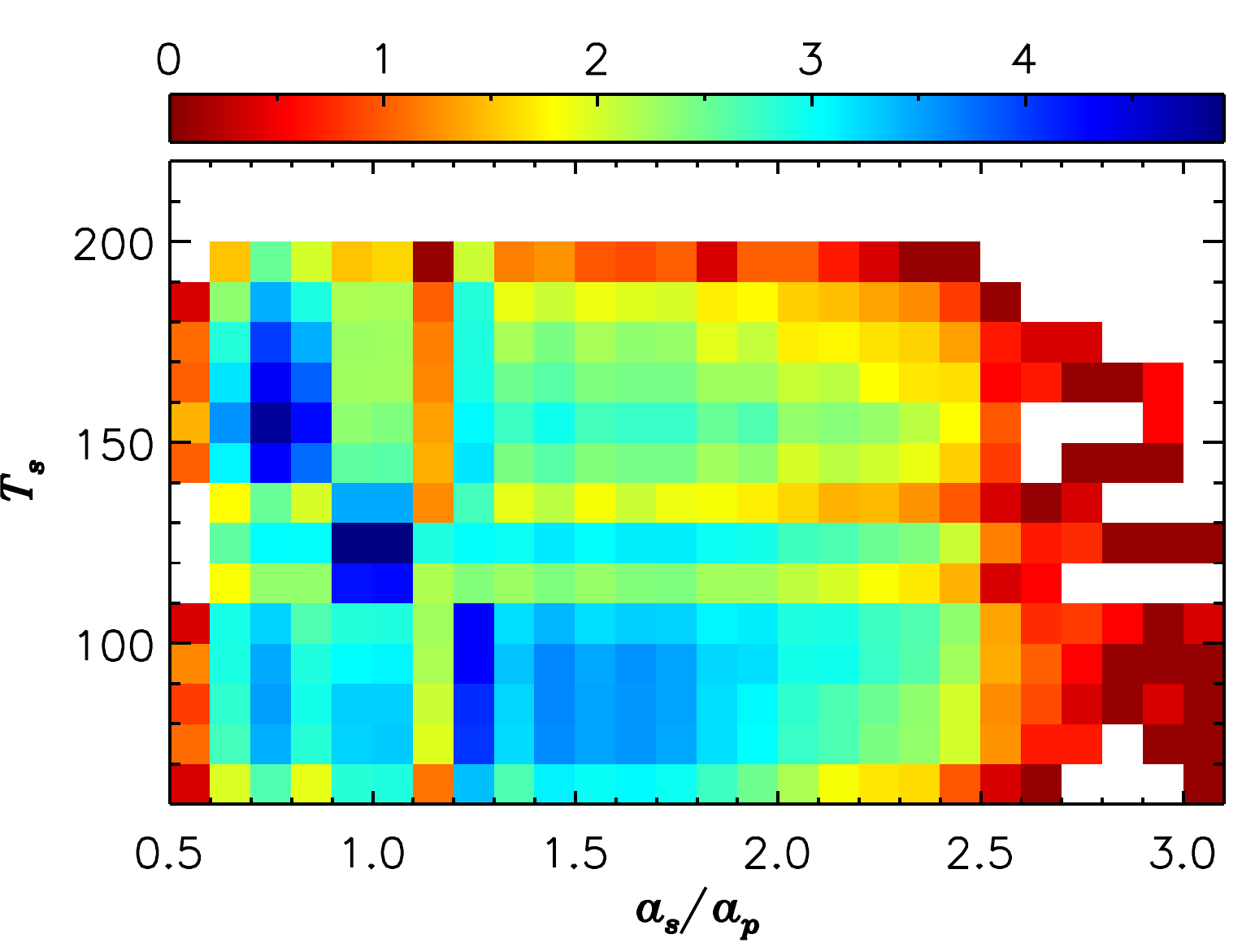}%
\includegraphics[clip]{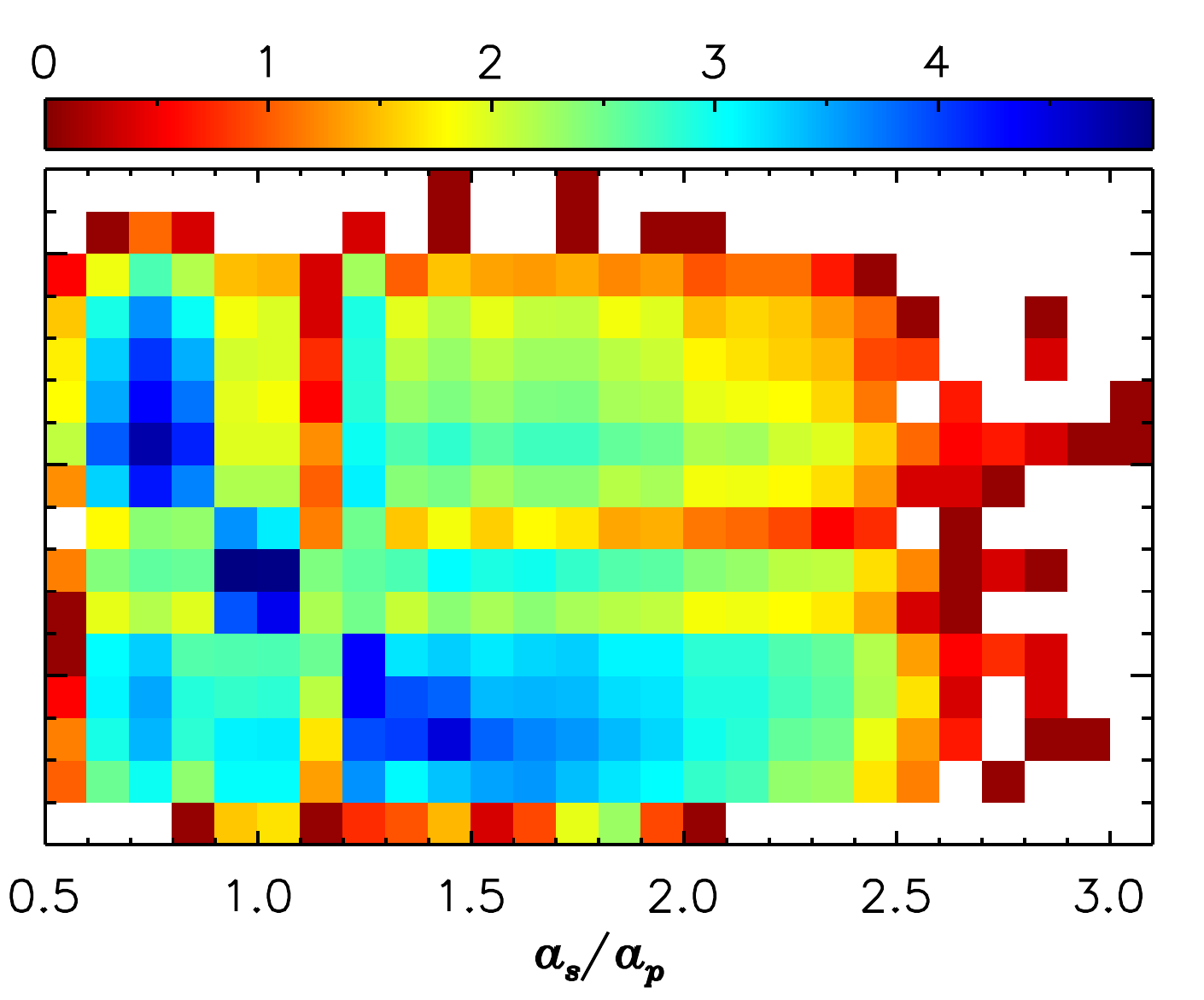}}
\caption{%
              Distribution of icy and mixed-composition particles versus semi-major axis 
              and versus
              orbital eccentricity, orbital inclination, radius, and temperature, as indicated
              on the vertical axes. 
              Bodies initially placed inside and outside of the planet's orbit and in 
              the corotation region are all included.
              Left and right panels refer, respectively, to the lower and 
              higher gas densities ($\rho_{0}=10^{-12}$ and $10^{-11}\,\mathrm{g\,cm}^{-3}$). 
              The color
              scale indicates the logarithm of the number of particles. Inclinations are in
              degrees, radii are in $\textrm{cm}$ and temperatures are in $\textrm{K}$.
              }
\label{fig:disk_2dh}
\end{figure}
Figure~\ref{fig:disk_2dh} shows two-dimensional histograms of 
eccentricity, inclination, size, and temperature of planetesimals
versus semi-major axis. The distributions include both icy and
mixed-composition bodies, for 
$\rho_{0}=10^{-12}$ (left) and $10^{-11}\,\mathrm{g\,cm}^{-3}$ 
(right), initially deployed in all three disk regions mentioned above.
As indicated by Figure~\ref{fig:hie_disk}, orbital eccentricities can 
exceed $0.8$ at both gas reference densities, but these values 
typically occur for $a_{s}\gtrsim 3\,a_{p}$.
Most of planetesimals produced via ablation of initially larger bodies 
have temperatures $T_{s}\gtrsim 150\,\K$, which indicates that ablation
is ongoing. In some instances, temperatures have settled at lower values, 
indicating that these planetesimals orbit in colder disk regions.
At $T_{s}\gtrsim 200\,\K$, bodies are consumed relatively
rapidly. In fact, below the critical temperature ($T_{s}<T_{\mathrm{cr}}$), 
the timescale for ablation of a planetesimal is
\begin{equation}
M_{s}\left|\frac{dM_{s}}{dt}\right|^{-1}=%
                 \frac{1}{3}\frac{R_{s}\rho_{s}}{P_{v}}%
                 \sqrt{\frac{2\pi k_{\mathrm{B}}T_{s}}{\mu_{s}m_{\mathrm{H}}}}.
\label{eq:tabla}
\end{equation}
In the equation above, for mixed-composition planetesimals,
$P_{v}$ is the vapor pressure of ice and $\mu_{s}$ is the mean molecular 
weight of ice modified by the ice mass fraction.
At $T_{s}=150\,\K$, the vapor pressure of ice is 
$\approx 5\times 10^{-5}\,\mathrm{dyne/cm^{2}}$
and the timescale for ablation of a $R_{s}=100\,\mathrm{km}$ body 
would be in excess of $10^{8}$ years (assuming that $T_{s}$ remains
constant and neglecting re-condensation). 
But this timescale rapidly declines as $T_{s}$
(and hence $P_{v}$) rises, becoming $4$--$5\times 10^{3}$ years
at $200\,\K$ ($P_{v}\approx 1.6\,\mathrm{dyne/cm^{2}}$)  
and $\sim 10^{3}$ years at $210\,\K$ 
($P_{v}\approx 7\,\mathrm{dyne/cm^{2}}$).
The distribution of $T_{s}$ versus $a_{s}$ in Figure~\ref{fig:disk_2dh} 
would place the ice line at around $2.8\,\AU$, where the gas temperature
$T_{g}=T_{n}$ is about $220\,\K$ (see Equation~(\ref{eq:Tg}), although 
$T_{s}$ needs not be in equilibrium with $T_{n}$). 
However, global models of evolving disks 
\citep[e.g.,][see also \citealp{gennaro2012}]{dalessio2005} 
predict somewhat lower gas temperatures in those disk regions at
times $\gtrsim 1\,\mathrm{Myr}$. Therefore, the ice line can move
inside $2\,\AU$ as long as the gas remains optically thick to stellar
radiation.

\subsection{An Estimate of Collision Rates}
\label{sec:ECR}

The rates of collisions among planetesimals can be
derived from simple arguments. The average relative velocity 
between two bodies in a swarm is such that \citep[e.g.,][]{stewart1980}
\begin{equation}
\langle v^{2}_{\mathrm{rel}}\rangle=a^{2}_{s}\Omega^{2}_{\mathrm{K}}
\left(\frac{5}{8}\langle e^{2}_{s}\rangle+\frac{1}{2}\langle \sin^{2}{i_{s}}\rangle\right).
\label{eq:v_rel}
\end{equation}
Based on the two-body approximation, the cross section for collisions 
of a target planetesimal of radius $R_{s}$ with those of radius $R_{j}$ 
is
\begin{equation}
S_{j}=\pi\left(R_{s}+R_{j}\right)^{2}%
         \left(1+\frac{v^{2}_{\mathrm{esc}}}{\langle v^{2}_{\mathrm{rel}}\rangle}\right),
\label{eq:Sj}
\end{equation}
where $v^{2}_{\mathrm{esc}}=2GM_{s}/R_{s}$. The rate of collisions 
on the target body is then
\begin{equation}
\frac{dN}{dt}=\sum_{j} \mathcal{N}_{j} S_{j} \langle v^{2}_{\mathrm{rel}}\rangle^{1/2},
\label{eq:dNdt}
\end{equation}
where $\mathcal{N}_{j}$ is the number density of solids, which involves
the planetesimals' surface density, $\Sigma_{s}$, and the swarm
 thickness, $\langle a_{s}\sin{i_{s}}\rangle$. 
The averages $\langle v^{2}_{\mathrm{rel}}\rangle$ and 
$\langle a_{s}\sin{i_{s}}\rangle$, as a function of $a_{s}$, are directly 
evaluated from the calculations. 
In a swarm with equal number densities of $0.1$--$100\,\mathrm{km}$ 
bodies, the largest planetesimals have the highest collision rates.
In the circumstellar disk, at some distance from the planet's orbit
and for $\Sigma_{s}\sim1\,\mathrm{g\,cm}^{-2}$, during the course of 
the calculations the number of collisions on any planetesimal would be negligible.
The same arguments applied to planetocentric orbits in the circumplanetary 
disk (see Section~\ref{sec:PCP}) would yield a collision rate of 
$dN/dt\sim 10^{-4}\,\sqrt{G\Mp/a^{3}_{s}}\,(\Sigma_{s}/1\,\mathrm{g\,cm}^{-2})$
for $a_{s}\lesssim 0.15\,\Rhill$.

\section{Evolution of Planetesimals in the Circumplanetary Disk}
\label{sec:PCP}

\begin{figure*}
\centering%
\resizebox{\figlew}{!}{\includegraphics[clip]{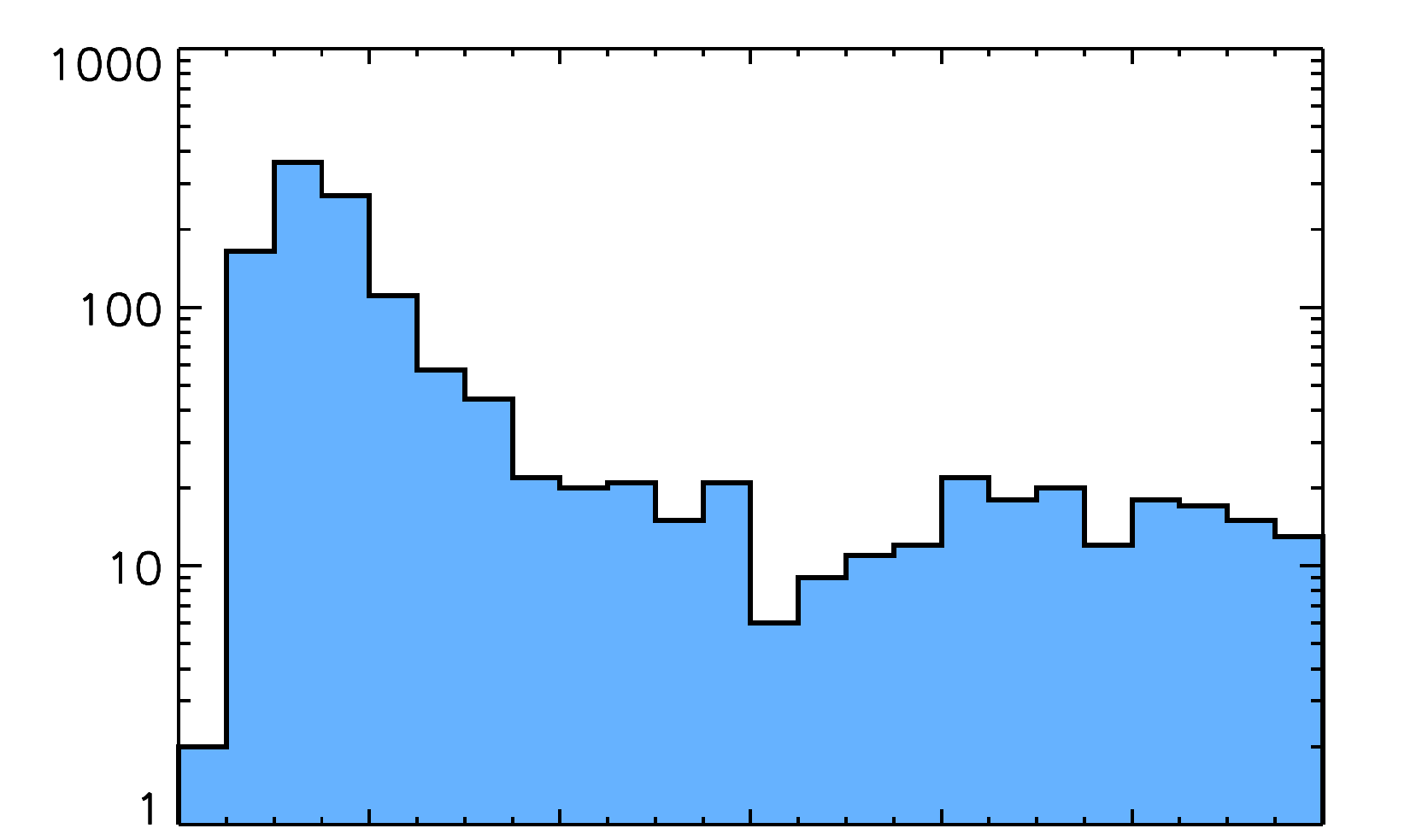}%
                                 \includegraphics[clip]{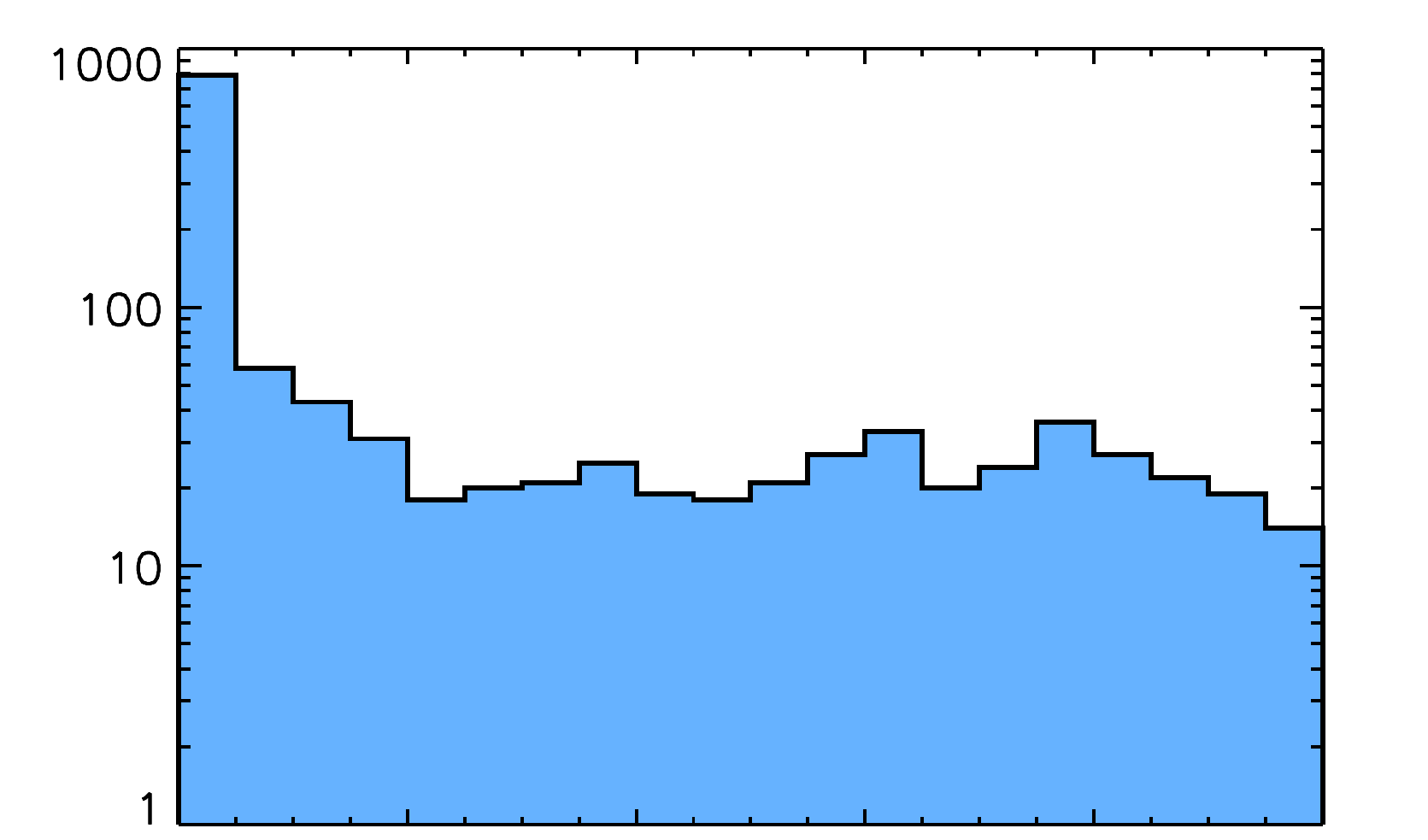}%
                                 \includegraphics[clip]{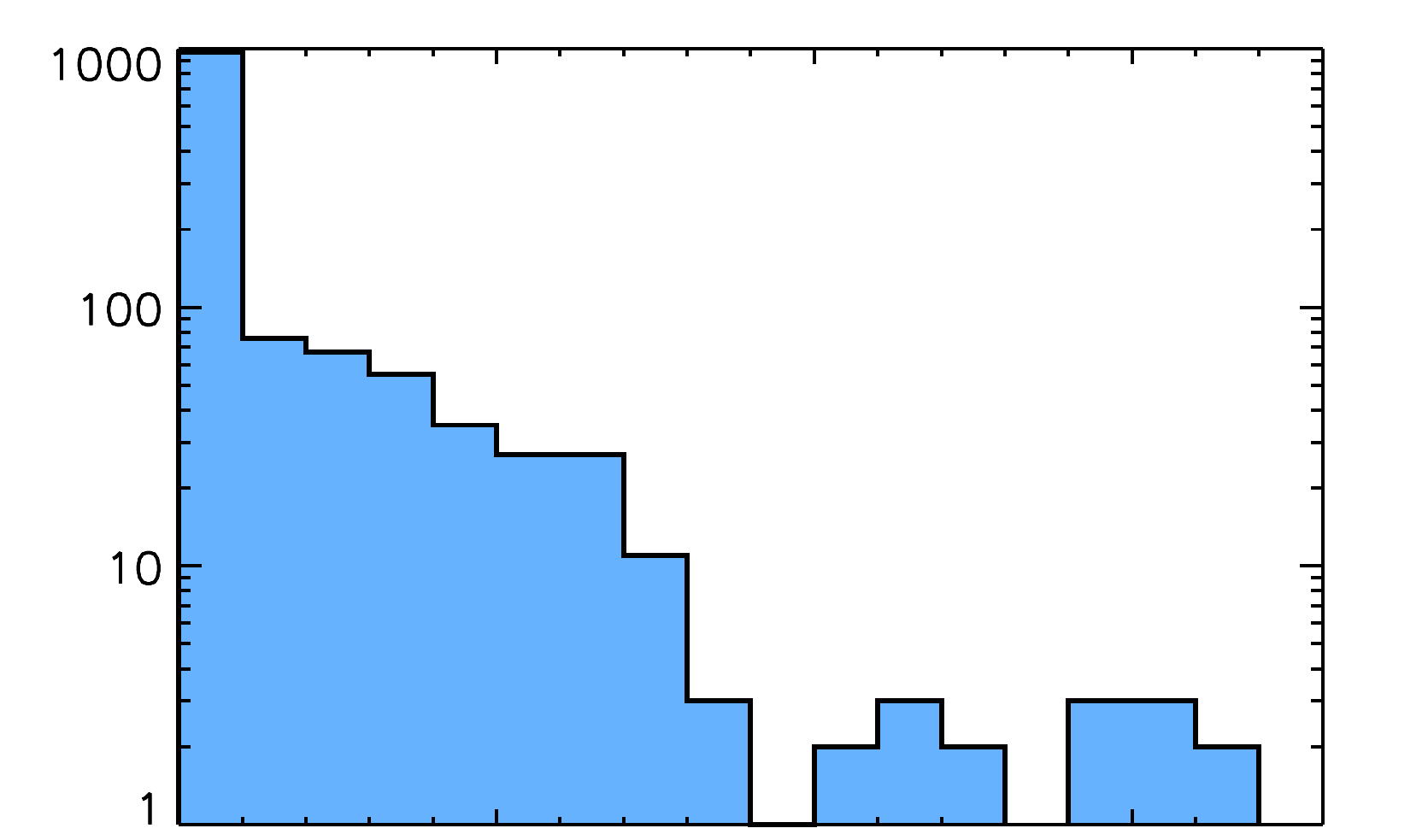}}
\resizebox{\figlew}{!}{\includegraphics[clip]{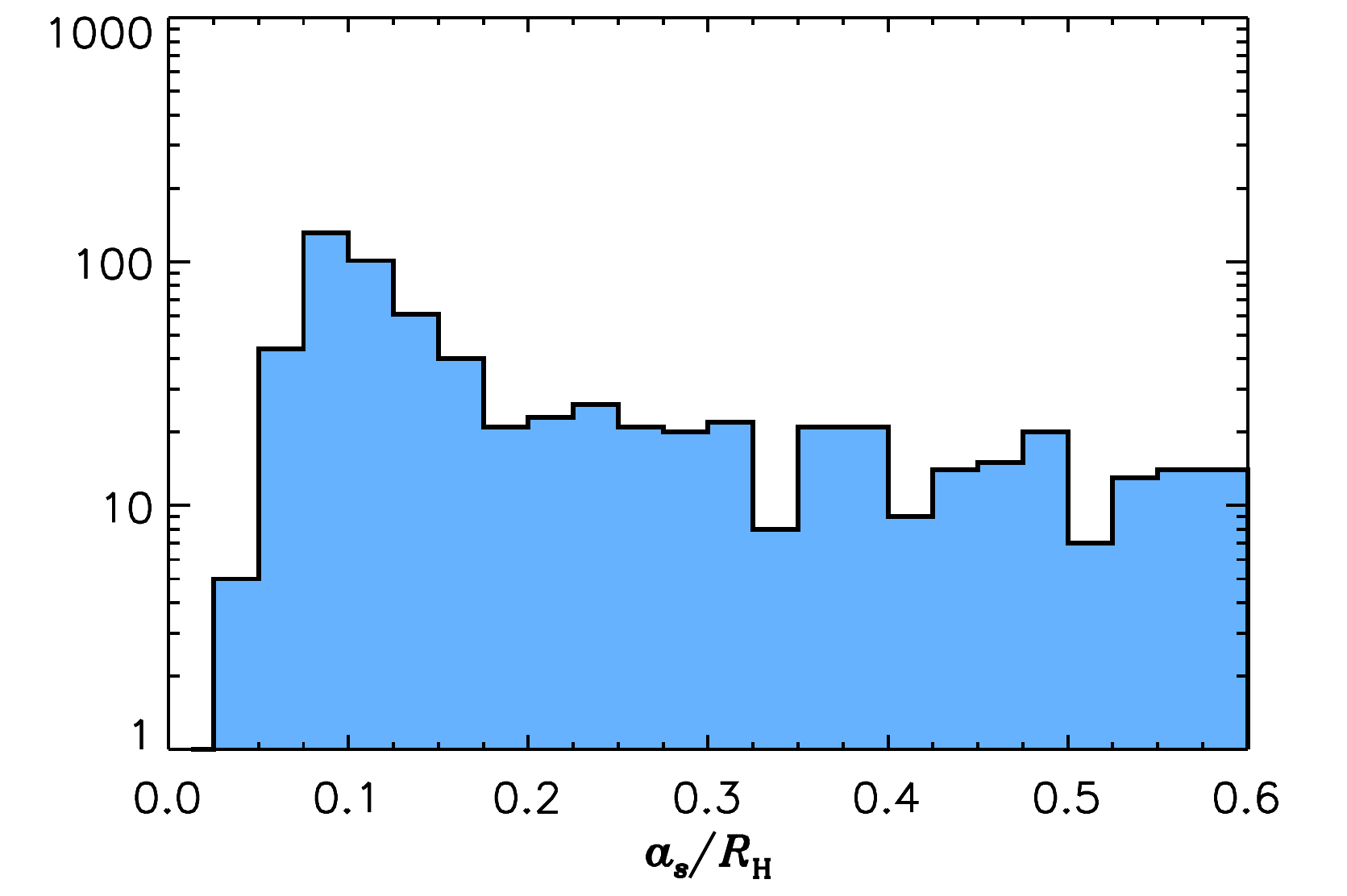}%
                                 \includegraphics[clip]{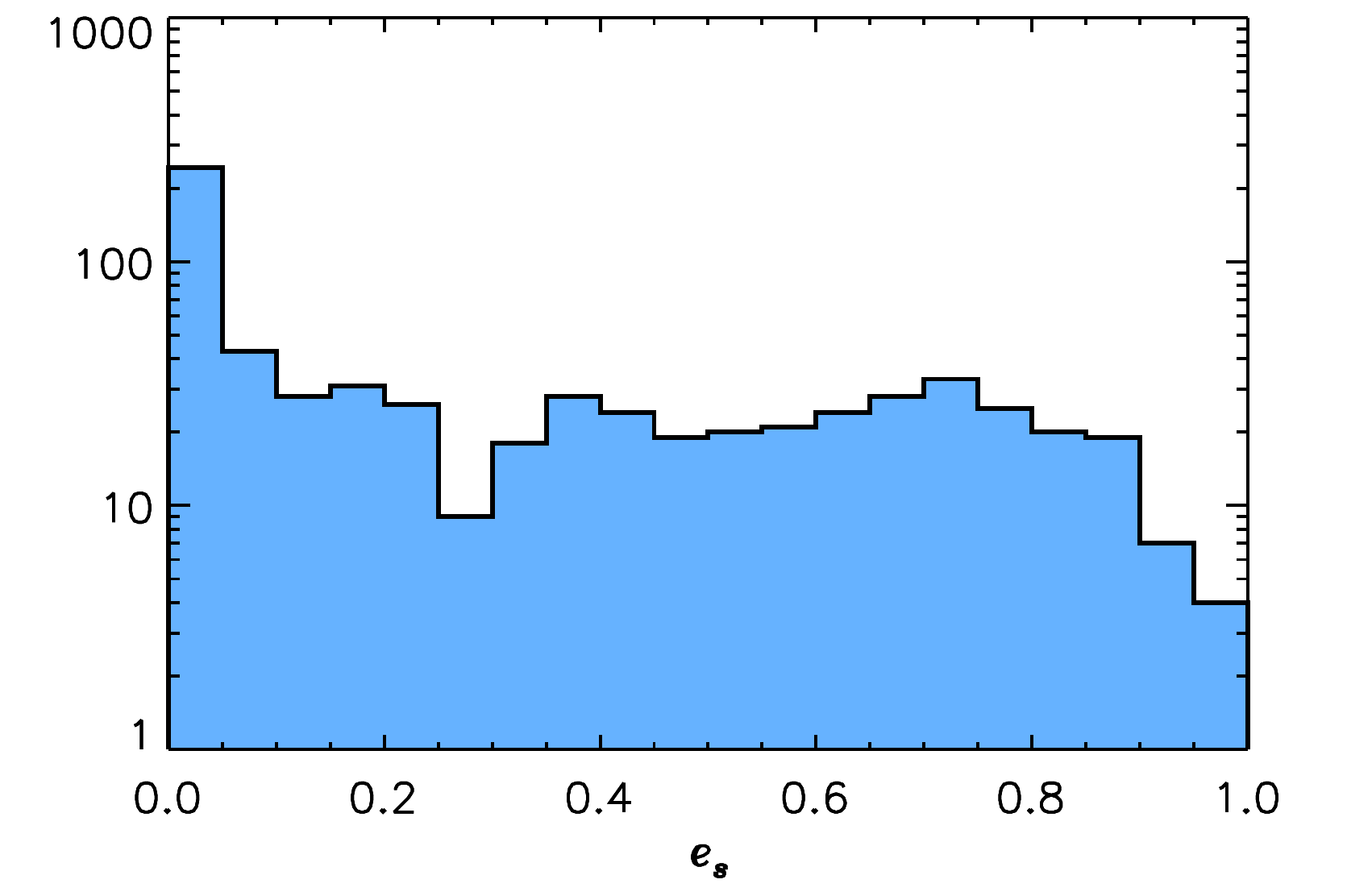}%
                                 \includegraphics[clip]{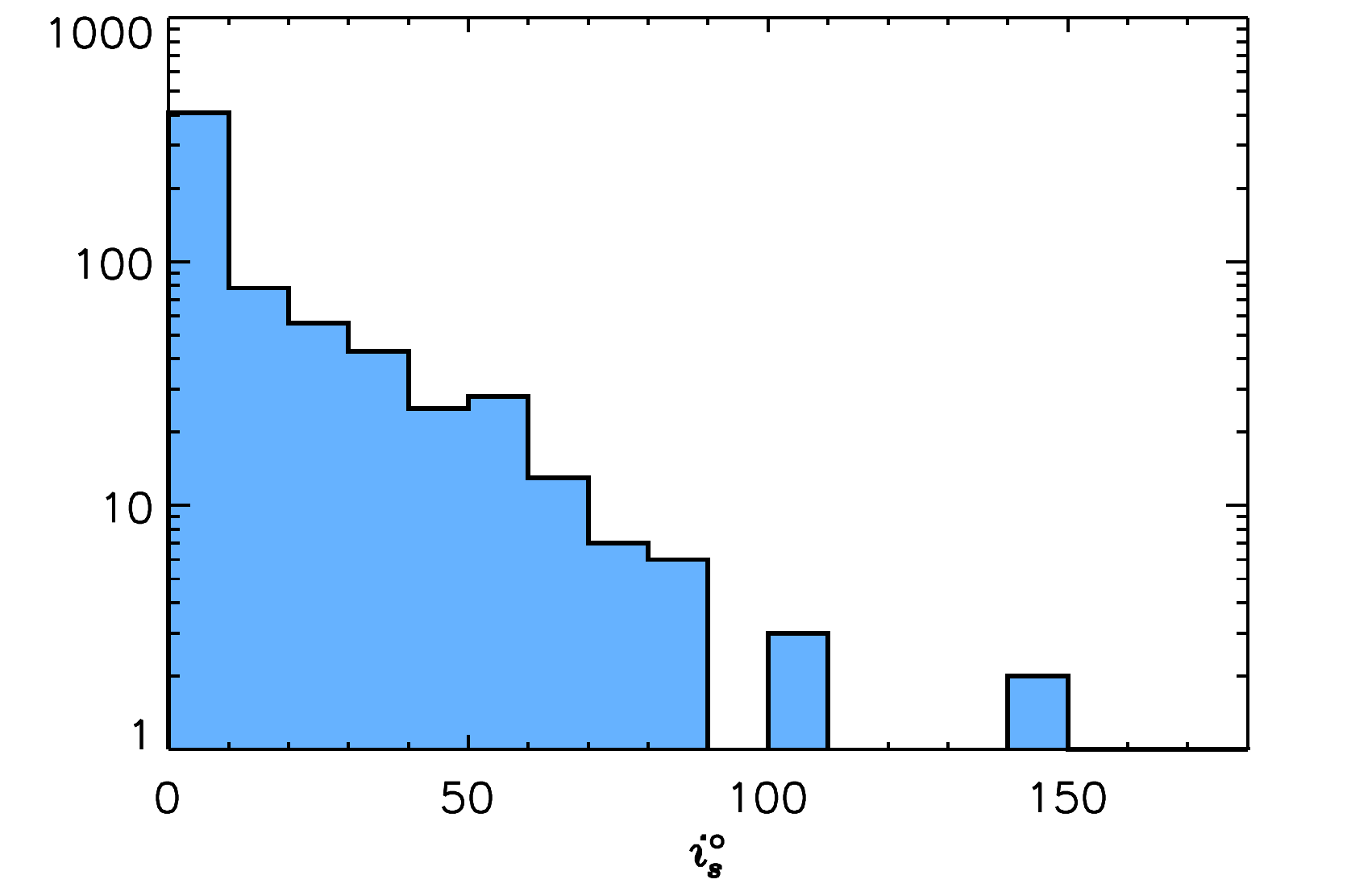}}
\caption{%
             Planetocentric orbital elements, semi-major axis (left), eccentricity (center), 
             and inclination (right), of icy and mixed-composition planetesimals bound 
             to the planet.
             Top (bottom) panels refer to the density $\rho_{0}=10^{-12}$ ($10^{-11}$) 
             $\mathrm{g\,cm}^{-3}$. 
             }
\label{fig:h_cpd}
\end{figure*}
\begin{figure}
\centering%
\resizebox{\linewidth}{!}{%
\includegraphics[clip]{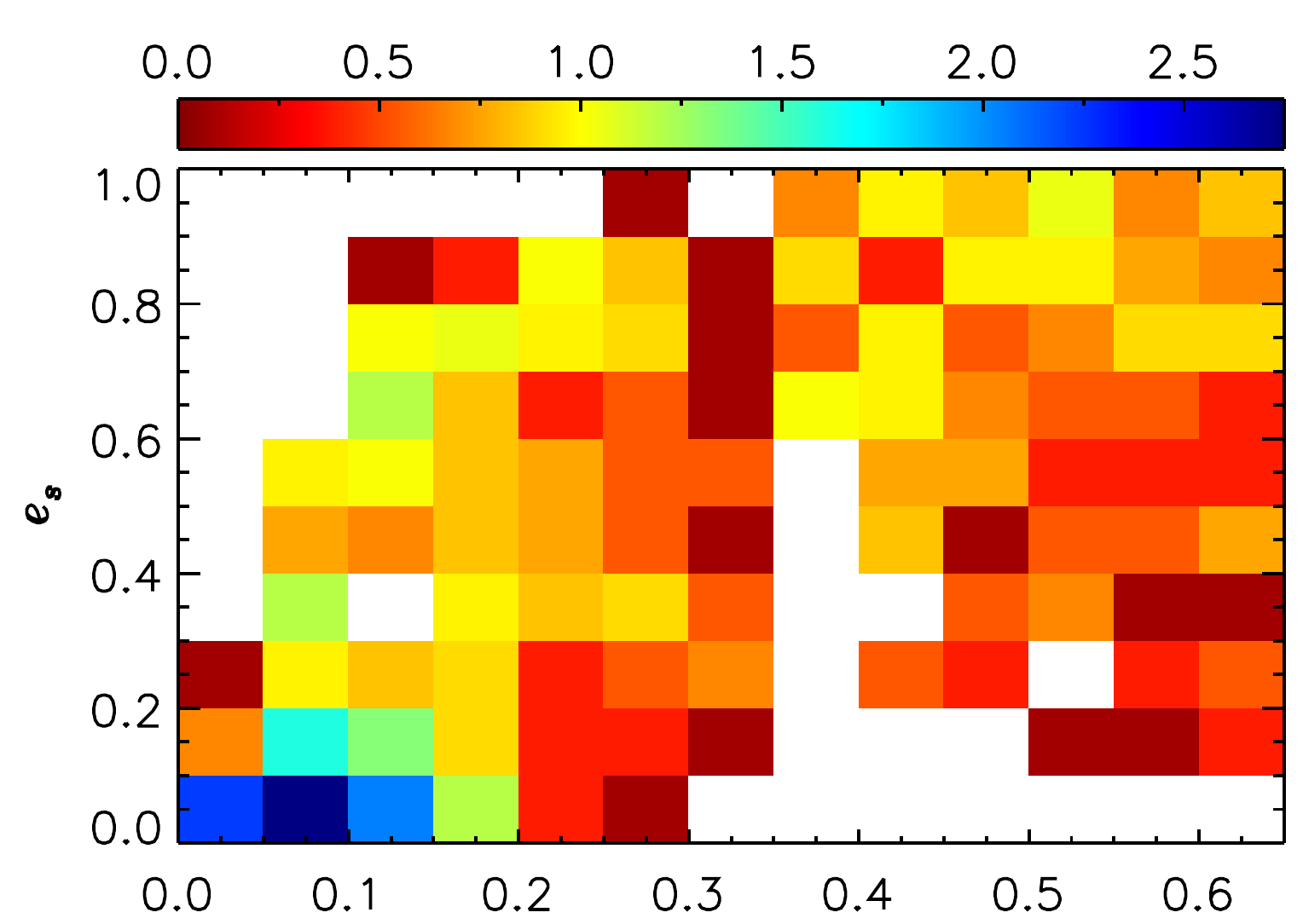}%
\includegraphics[clip]{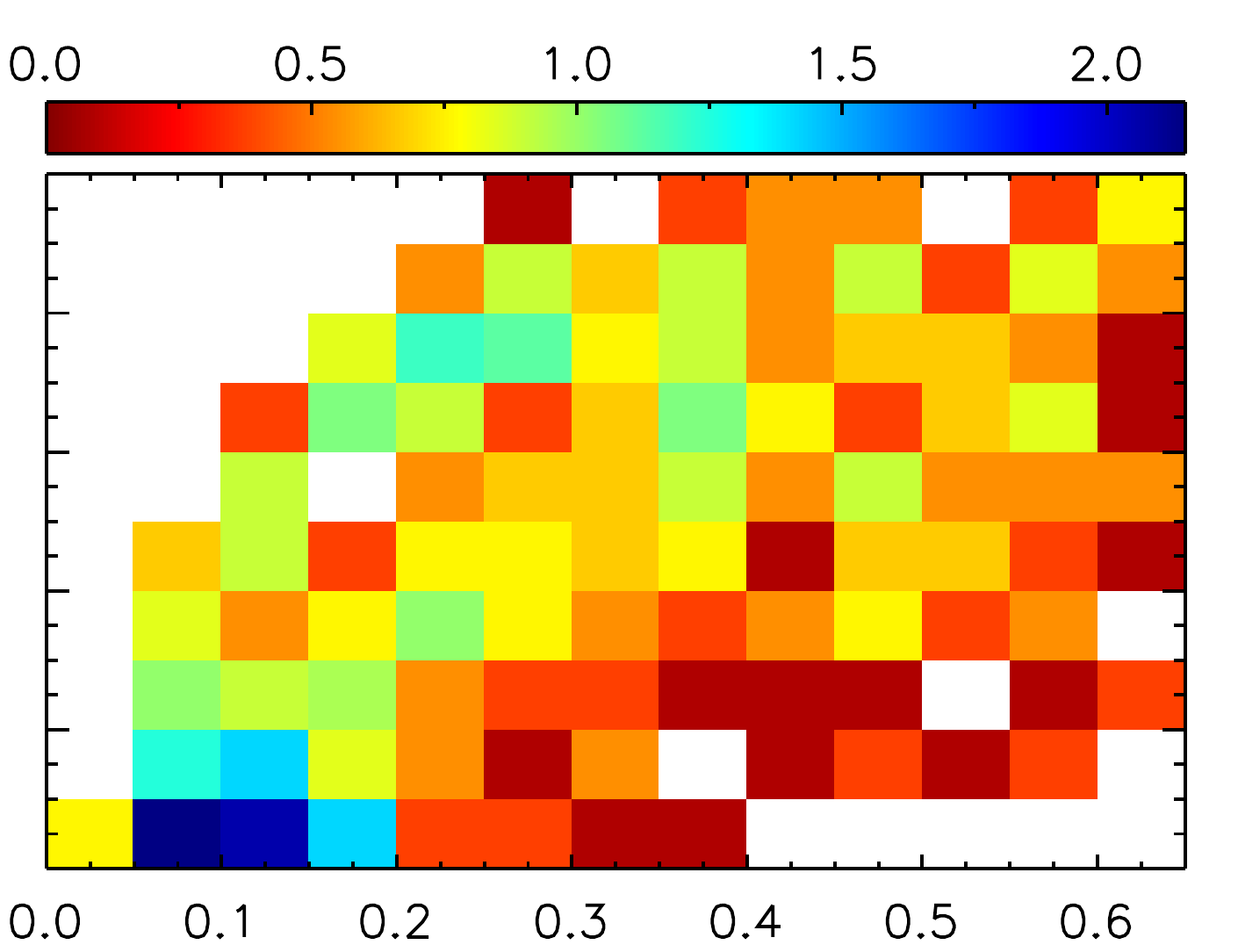}}
\resizebox{\linewidth}{!}{%
\includegraphics[clip]{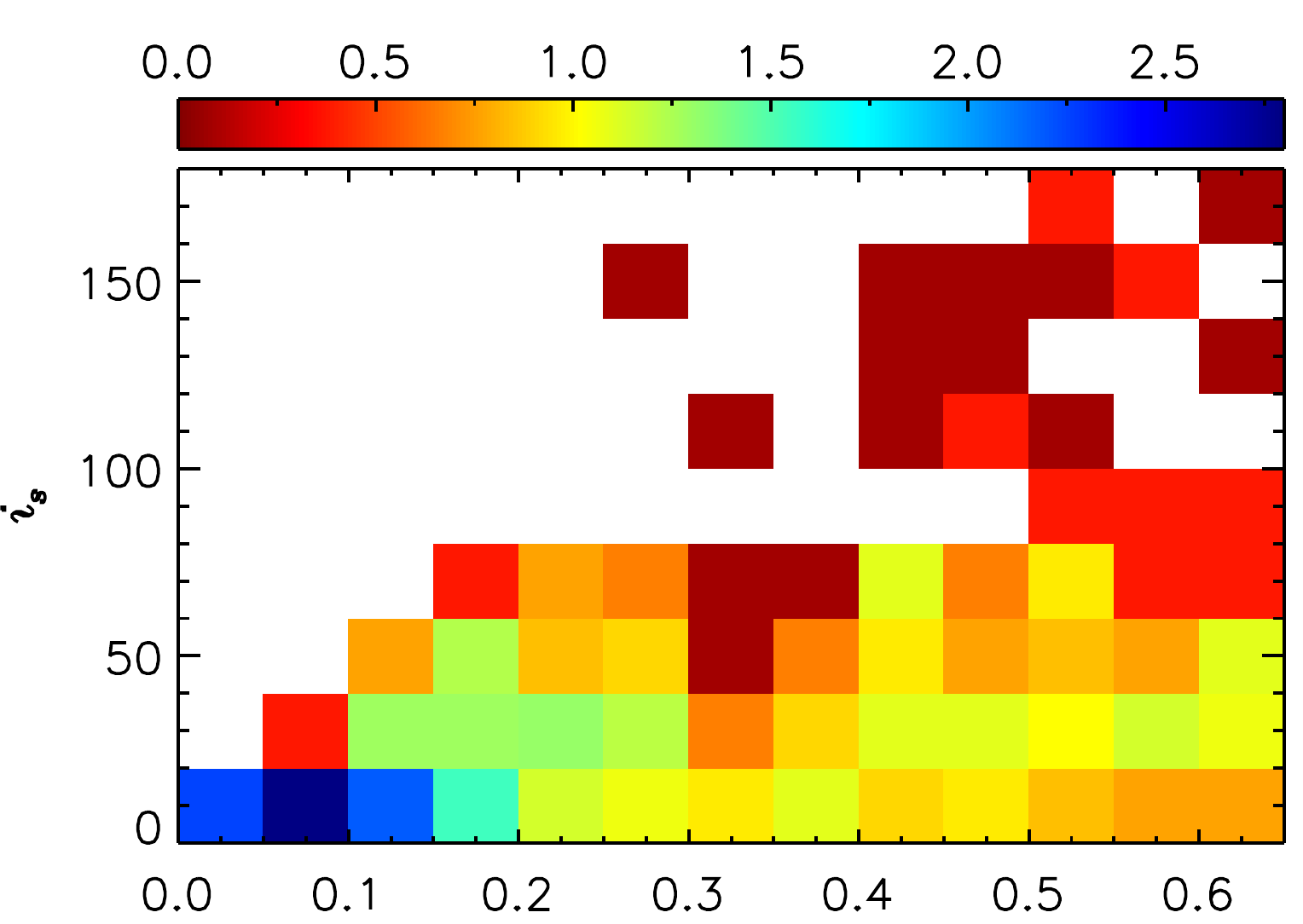}%
\includegraphics[clip]{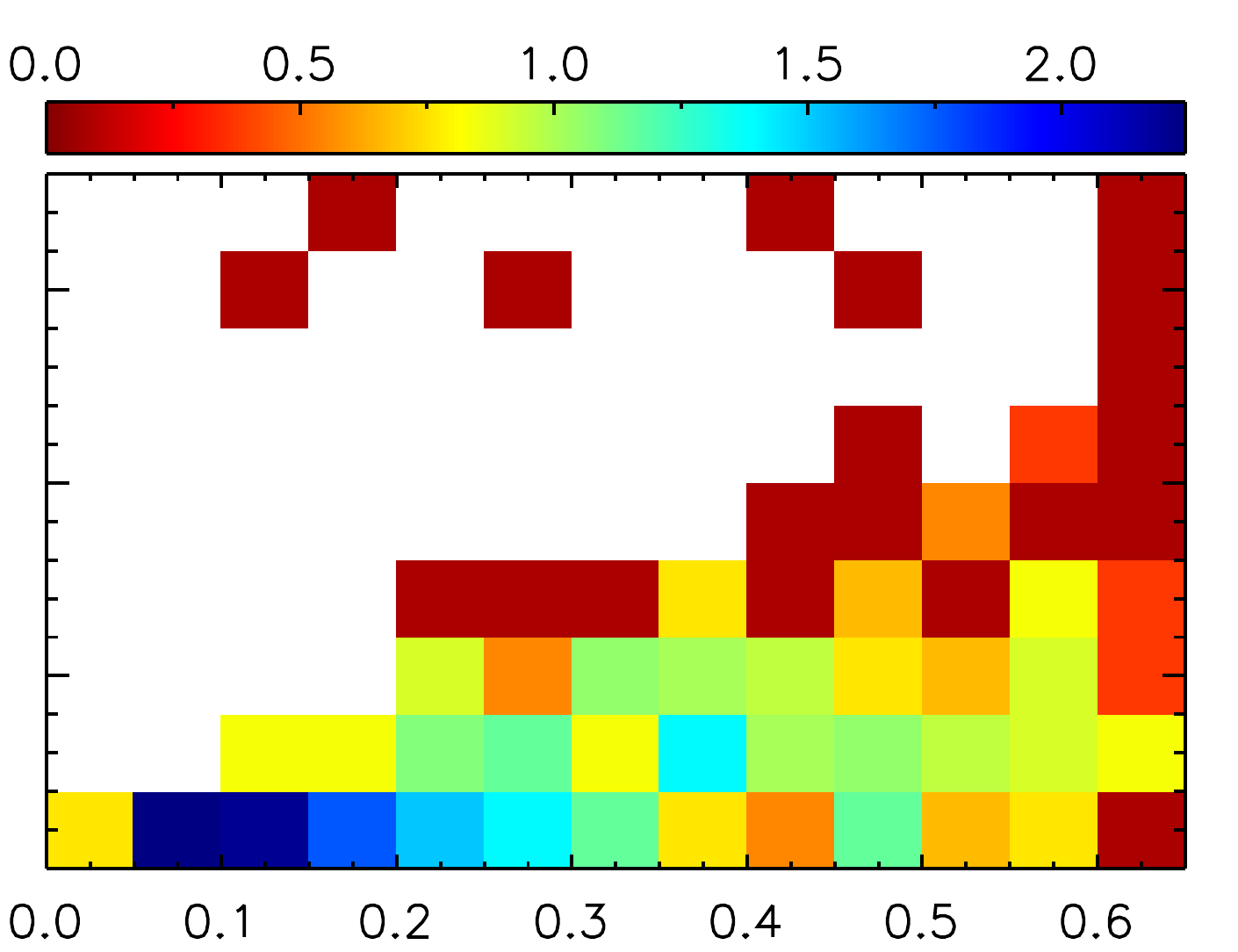}}
\resizebox{\linewidth}{!}{%
\includegraphics[clip]{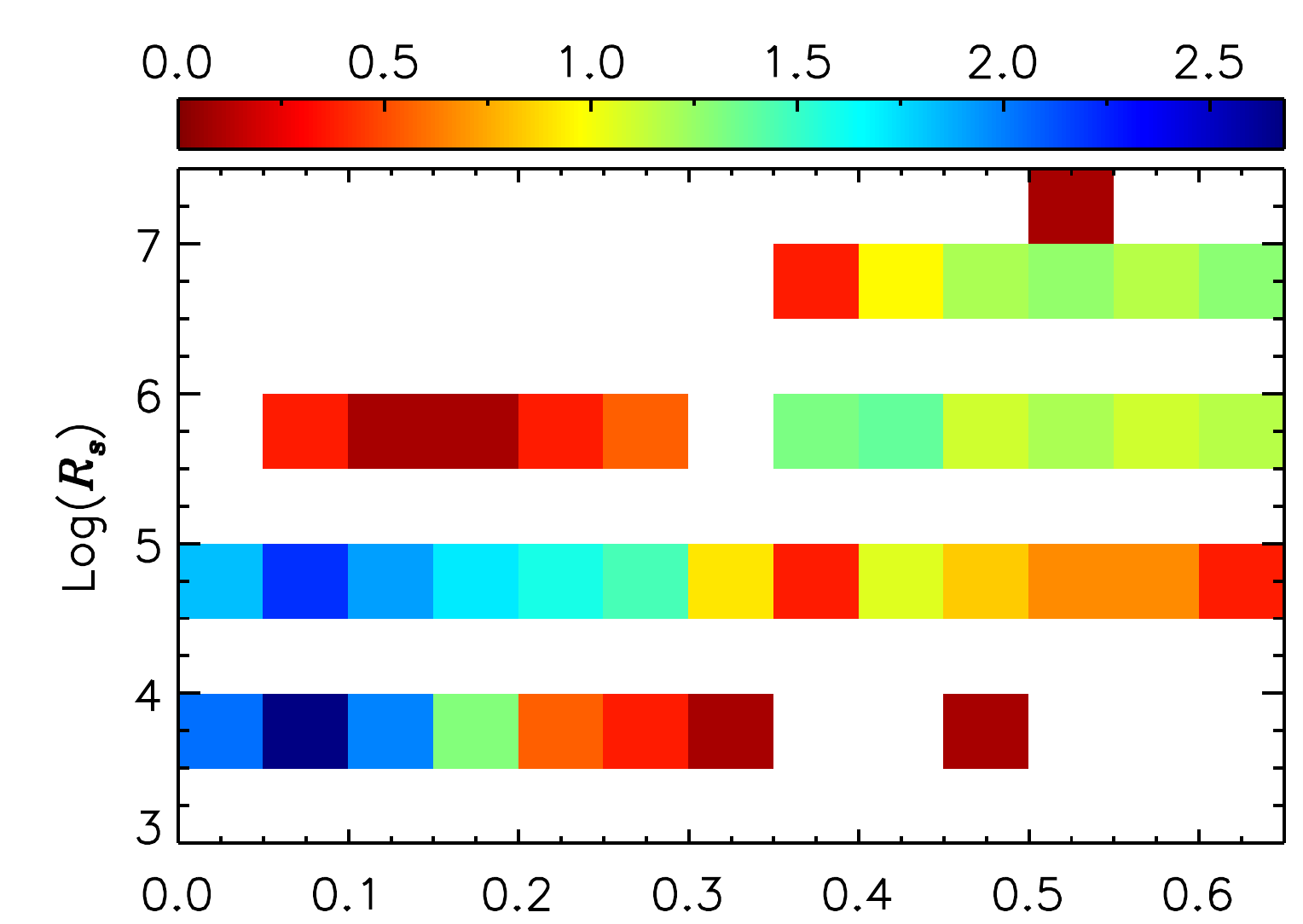}%
\includegraphics[clip]{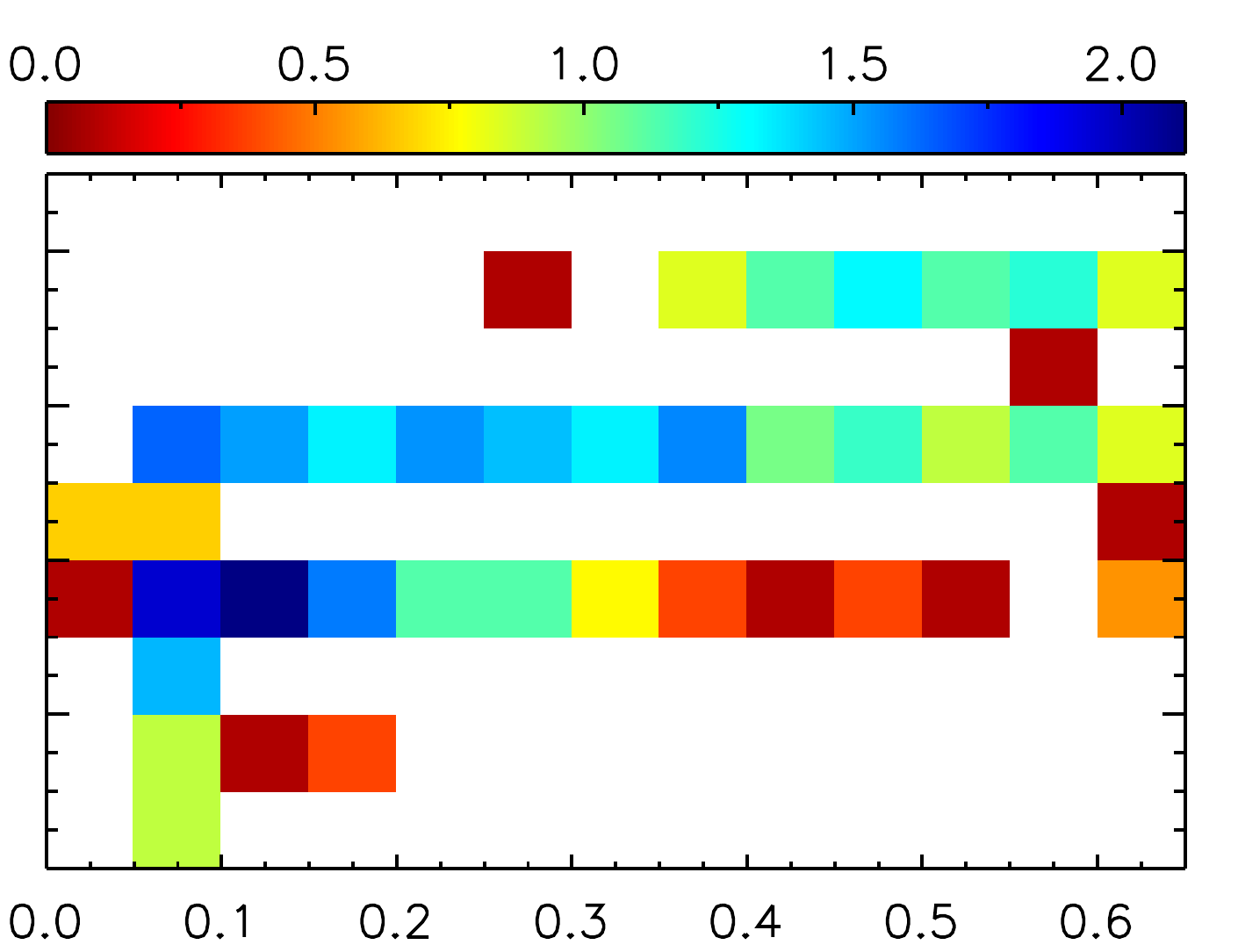}}
\resizebox{\linewidth}{!}{%
\includegraphics[clip]{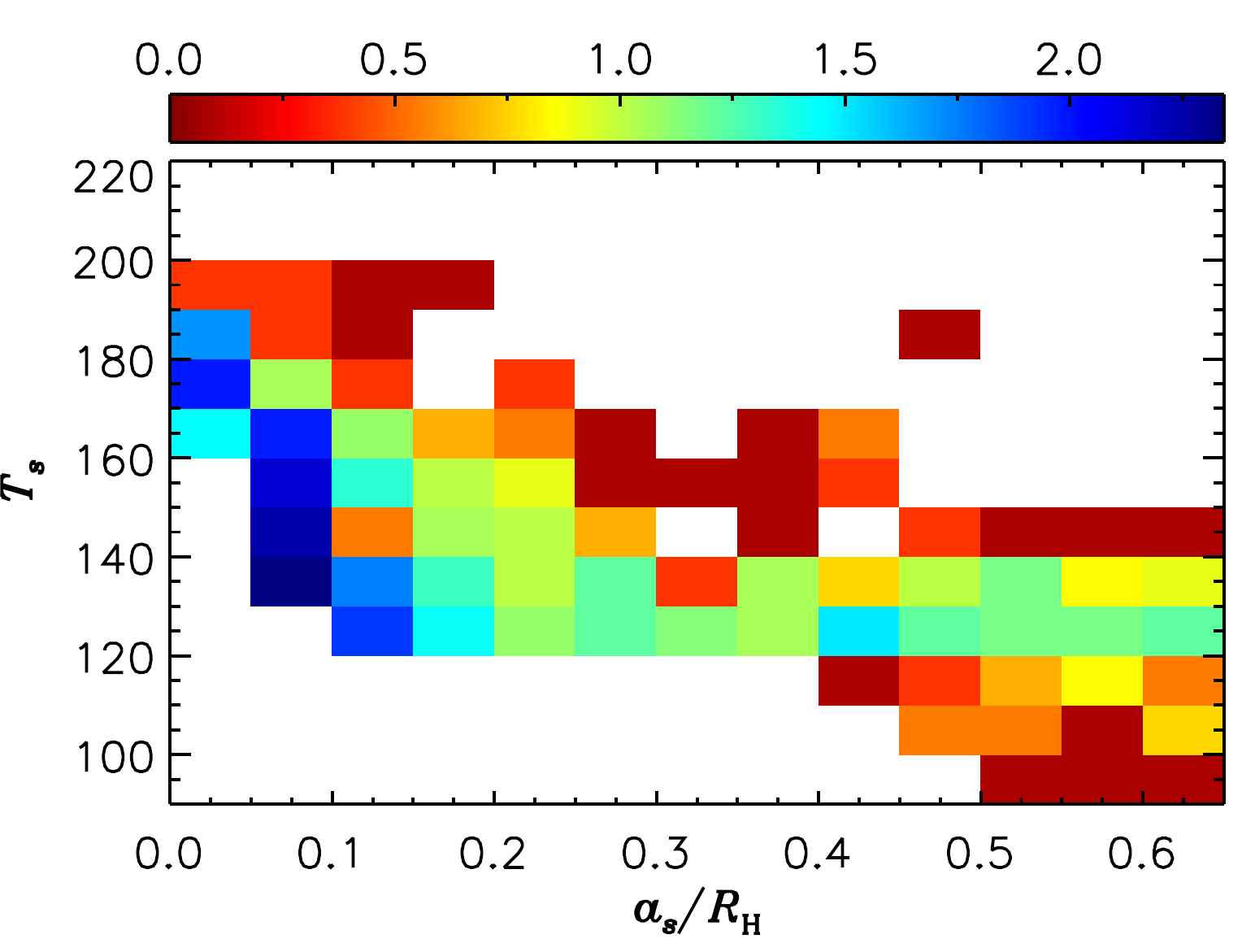}%
\includegraphics[clip]{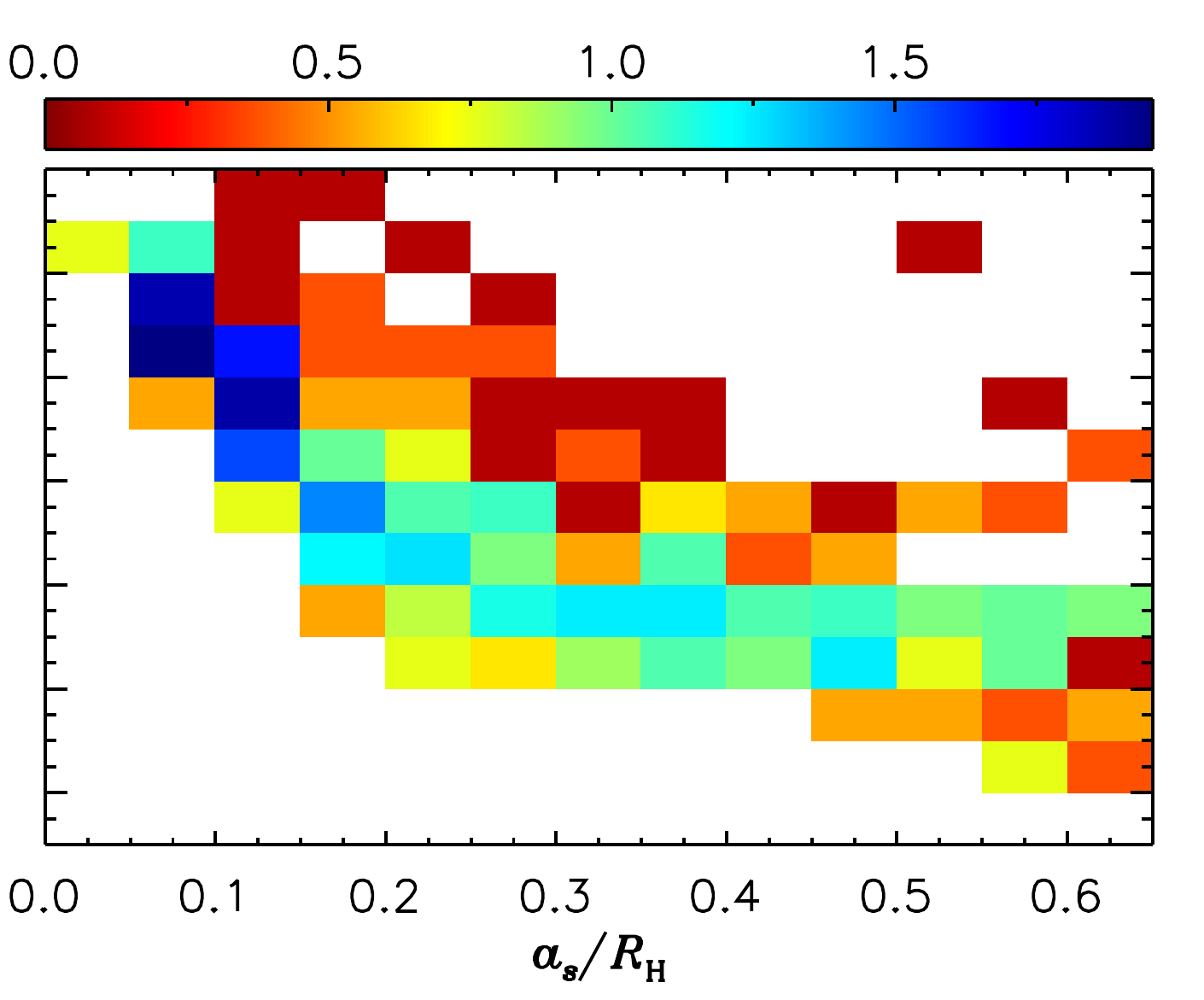}}
\caption{%
              Distribution of icy and mixed-composition planetesimals bound to
              the planet versus semi-major axis and versus orbital eccentricity, 
              orbital inclination, radius, and temperature, as indicated on the 
              vertical axes. Osculating orbital elements are computed from
              positions and velocities relative to the planet.
              Left and right panels refer, respectively, to
              $\rho_{0}=10^{-12}$ and $10^{-11}\,\mathrm{g\,cm}^{-3}$. 
              The color
              scale indicates the logarithm of the number of particles.
              Units are as in Figure~\ref{fig:disk_2dh}.
              }
\label{fig:cpd_2dh}
\end{figure}

Consider the reference frame $\{\mathcal{O}^{\prime}; x,y,z\}$, introduced 
in Section~\ref{sec:CPD}, whose origin is fixed to the planet. Based on 
relative positions and velocities, we derive the osculating orbital elements 
in this frame of the planetesimals bound to the planet. 
Relative trajectories that are not representable by osculating ellipses 
are disregarded, although they may belong to bodies that will accrete 
on the planet. 
Notice that we maintain the same notations as in previous sections,
even though the orbits are relative to the planet.
Since planetesimals are initially placed on orbits about the star well outside 
the Roche lobe region, these are all objects captured by the planet's
gravity. Capture is aided by dissipation of kinetic energy through gas drag.
Figure~\ref{fig:h_cpd} shows semi-major axes (left), eccentricities 
(center), and inclinations (right) of these orbits, including both icy and 
mixed-composition bodies. The top/bottom panels refer to the lower/higher 
background density, $\rho_{0}$. The histograms include only those objects 
whose (planetocentric) orbit has a semi-major axis $\le 0.6\,\Rhill$.
Inclination distributions indicate the presence of retrograde orbits
($i_{s}>90^{\circ}$, see also Figure~\ref{fig:cpd_2dh}). 
These objects are among those initially moving
in the corotation region and in the region exterior of the planet's orbit, 
although the absence of retrograde objects originating from the inner
disk region may be a result of small-number statistics.
As explained below, since the orbital elements are based on osculating
ellipses, not all planetesimals represented in Figure~\ref{fig:h_cpd} are 
permanently captured (or accreted) by the planet.

The histograms in the top row of Figure~\ref{fig:h_cpd} include
about $1.8$ times as many bodies as the histograms in the bottom row.
Yet, the distributions for the reference density 
$\rho_{0}=10^{-11}\,\mathrm{g\,cm}^{-3}$ contain about $25$\% as much 
mass (see distributions of $R_{s}$ in Figure~\ref{fig:cpd_2dh}).
Most captured bodies have radii $\lesssim 10\,\mathrm{km}$, but
most of the mass is carried by $R_{s}\sim 100\,\mathrm{km}$ planetesimals.

Average mass fractions on the order of several times $10^{-4}$ 
are captured from the planetesimals initially placed in the inner disk and 
somewhat larger fractions, $\sim 10^{-3}$, are captured from bodies initially 
deployed in the region beyond the planet's orbit. Much smaller 
mass fractions are instead captured from the corotation region.
The fractions are computed as the ratio of the mass of planetesimals 
moving inside $\tilde{r}\le 0.6\,\Rhill$ to the total available mass in solids.
Masses are averaged over the last $\approx 50$ orbits of the planet.
The mass in solids accreted by the planet is not taken into account.
This fractional mass may be considered as an ``equilibrium'' mass 
between the supply of solids from the circumstellar disk and 
the mass loss due to ejection, ablation, break-up, and accretion 
of planetesimals in the circumplanetary disk.
If results were rescaled so that the surface density of solids between
$0.77\,a_{p}$ and $0.82\,a_{p}$ and between
$1.2\,a_{p}$ and $1.25\,a_{p}$ was $1\,\mathrm{g\,cm}^{-2}$ at the end 
of the calculations, the average mass within $0.6\,\Rhill$ of the planet 
would be $\sim 10^{-3}\,\Mearth$.

Planetesimals that orbit within $\sim 20^{\circ}$ of the equatorial plane
are subject to an increasing drag force, as they approach the planet, 
due to the augmenting gas density (see Figure~\ref{fig:prof_sub}).
Equatorial gas rotation inside $0.1\,\Rhill$ of the planet deviates only 
a few percent from Keplerian rotation $\sqrt{G\Mp/\tilde{r}}$, 
but the relative difference increases with distance, 
becoming $\approx 10$\% at $\tilde{r}\approx 0.2\,\Rhill$ and 
$\approx 40$\% at $\tilde{r}\approx 0.5\,\Rhill$.
The radial velocity of the gas at the equator is much smaller 
in magnitude than the azimuthal velocity.
For near-circular orbits, gas density can be approximated to a constant
and Equation~(\ref{eq:ada}) may be applied, although the term in the
square brackets of Equation~(\ref{eq:rotacu}) should depend on 
$\tilde{r}$ in these cases.
According to Equation~(\ref{eq:ada}), the decay time of orbital 
semi-major axes due to aerodynamics drag at $\tilde{r}\sim 0.5\,\Rhill$
is on the order of a few times $(R_{s}/a_{s})(\rho_{s}/\rho_{g})$, 
in units of the local orbital period around the planet.
We recall that the length $a_{s}$ here represents the semi-major axis 
of the planetocentric orbit. 
The impact of gas drag on higher inclination orbits 
($40^{\circ} \lesssim i_{s}\lesssim 140^{\circ}$) is probably somewhat smaller, 
since high densities are mostly encountered when these orbits are close 
the disk's equatorial plane (see top panels of Figure~\ref{fig:img_sub}).

\begin{figure}
\centering%
\resizebox{\linewidth}{!}{\includegraphics[clip]{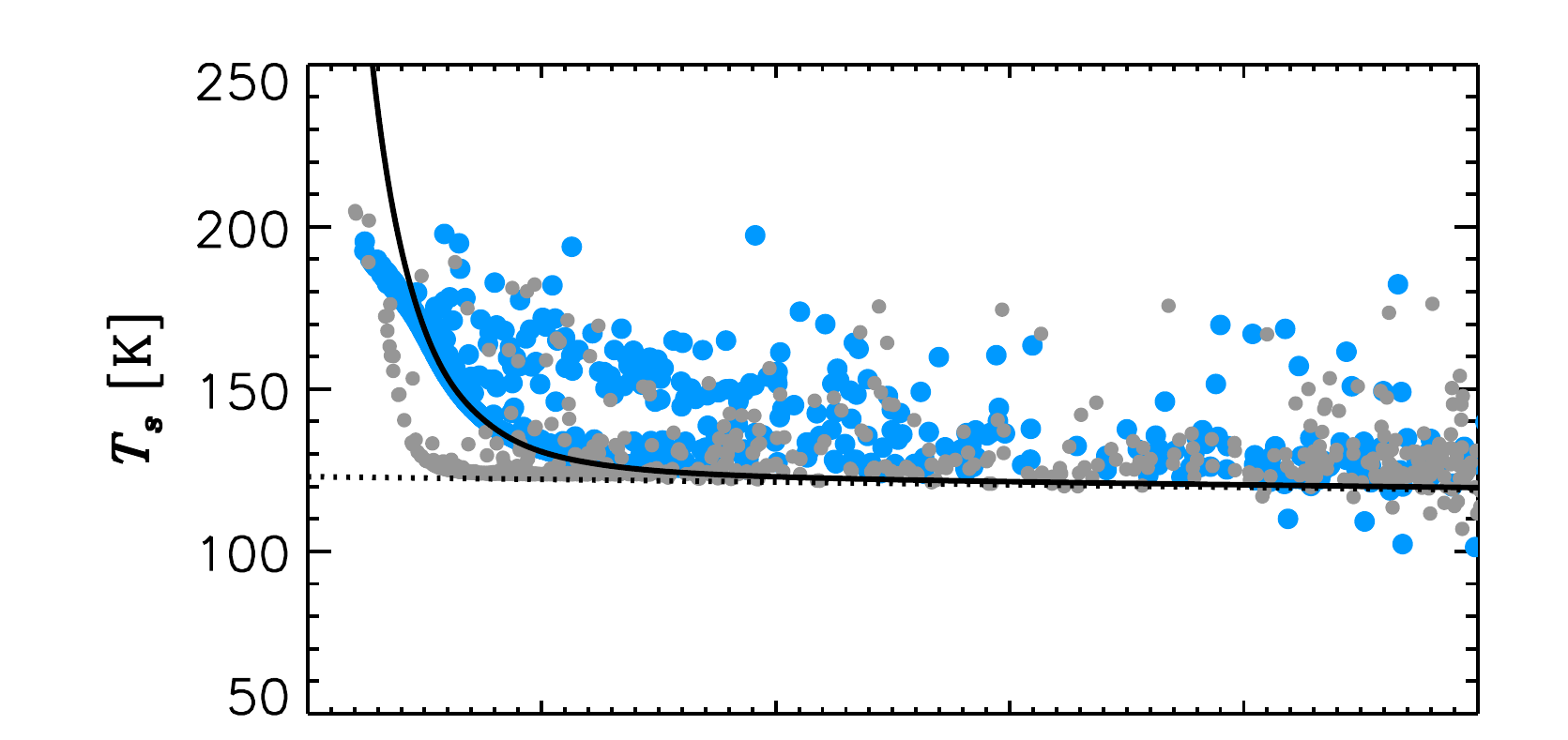}}
\resizebox{\linewidth}{!}{\includegraphics[clip]{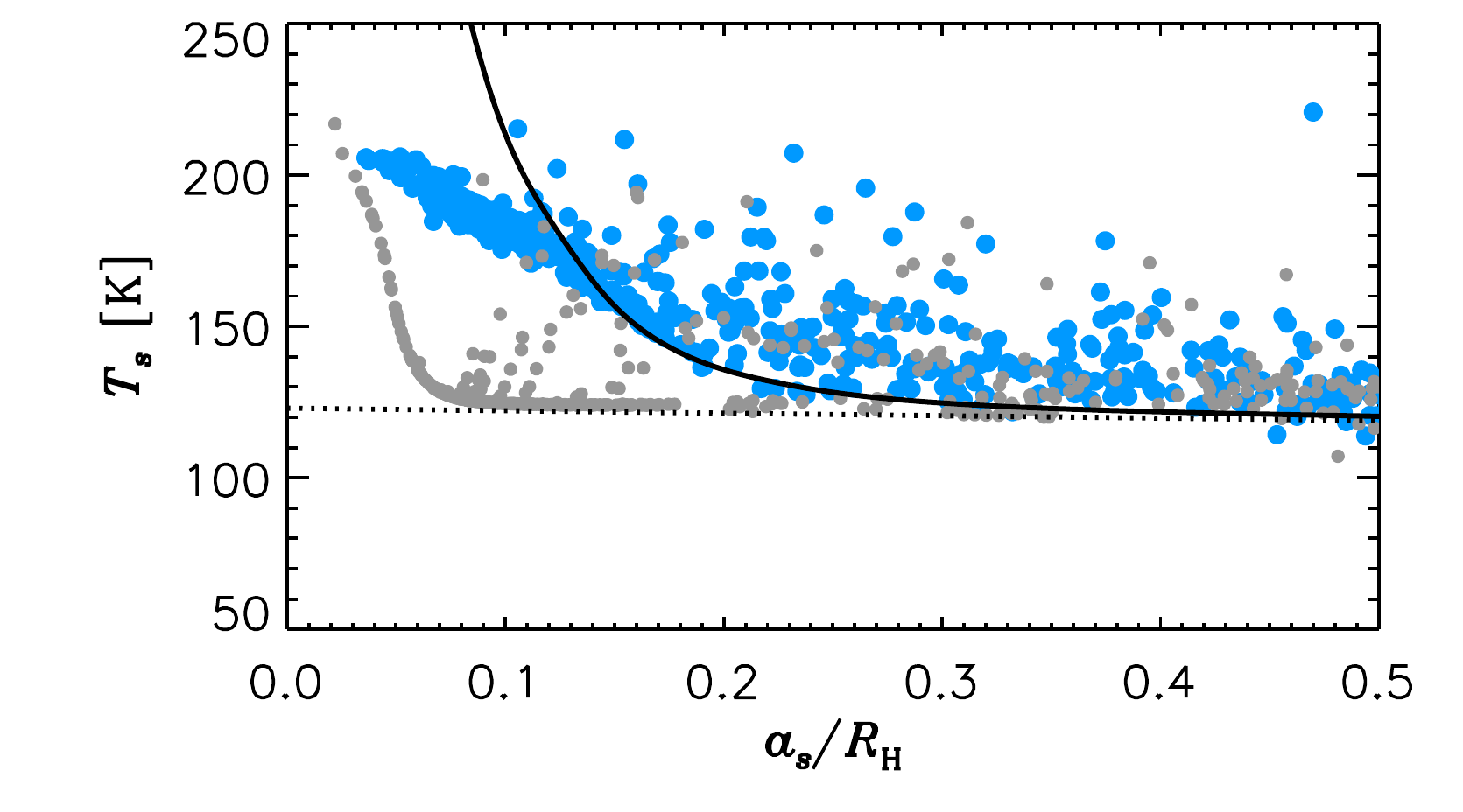}}
\caption{%
             Temperature versus semi-major axis of some planetesimals bound
             to the planet (blue circles) for the reference gas density
             $\rho_{0}=10^{-12}$ (top) and $10^{-11}\,\mathrm{g\,cm}^{-3}$ (bottom).
             The light gray circles indicate the temperature of planetesimals in the
             isothermal circumplanetary disk calculations (see Section~\ref{sec:PCPiso}).
             The solid lines represent the gas temperature $T_{g}$ in 
             Equation~(\ref{eq:Tcp}), also plotted in  Figure~\ref{fig:tcp}. 
             The dotted line is the gas temperature $T_{n}$ in Equation~(\ref{eq:Tg})
             that, in the ``isothermal'' calculations,  is applied everywhere in the disk.
             }
\label{fig:tcp_sp}
\end{figure}
Figure~\ref{fig:cpd_2dh} shows two-dimensional distributions analogous to 
those in Figure~\ref{fig:disk_2dh}, but for planetesimals bound to the planet.
Note that counts of $1$, shown in Figure~\ref{fig:cpd_2dh}, are not visible 
in histograms of Figure~\ref{fig:h_cpd}, as they lie on the horizontal axis.
The figure indicates that planetesimal temperatures are typically
$T_{s}\lesssim 200$--$220\,\K$ (see also Figure~\ref{fig:tcp_sp}). 
As argued above,  the ablation timescale rapidly decreases at 
higher temperatures. A finer sampling of the semi-major axis 
histograms reveals that number densities start to decline inward of 
$\tilde{r}\approx 0.05\,\Rhill$ and $\approx 0.08\,\Rhill$ for 
$\rho_{0}=10^{-12}$ and $10^{-11}\,\mathrm{g\,cm}^{-3}$, respectively. 
Figure~\ref{fig:tcp_sp} shows that particle temperatures (blue circles)
become consistently lower than gas temperature (solid line) when 
substantial ablation begins, as expected from Equation~(\ref{eq:dTsdt})
for slowly varying frictional heating and radiative gain/loss energy terms. 
The gray circles belong to isothermal circumplanetary disk calculations
discussed in Section~\ref{sec:PCPiso}.
The large ablation rates can enrich these circumplanetary disk regions 
with ice and rock, locally increasing the solid-to-gas mass ratio.

It is safe to assume that, over the course of the calculations, solid material 
is practically only ablated either inward of $2.8\,\AU$ or in close proximity 
of the planet. By separating these two contributions, we estimate that 
$\sim 10^{-7}\,\Mearth\,\mathrm{yr}^{-1}$ worth of ice and silicates would be 
released in the gas close to the planet, if the initial surface density 
of planetesimals between $0.77\,a_{p}$ and $0.82\,a_{p}$ and between 
$1.2\,a_{p}$ and $1.25\,a_{p}$ was $1\,\mathrm{g\,cm}^{-2}$.
At this production rate, the average metallicity of the circumplanetary disk 
would become 
$\sim 0.01\,(10^{-12}\,\mathrm{g\,cm}^{-3}/\rho_{0})$ 
in $\sim 100$ planet's periods.
There are only marginal differences in the amounts of material ablated 
from icy and mixed-composition bodies ($\sim 10$\%, roughly consistent 
with the predictions from Equation~(\ref{eq:tabla}) if equal temperature 
bodies are assumed) and similar differences are obtained for the two 
values of $\rho_{0}$.

\begin{figure}
\centering%
\resizebox{\linewidth}{!}{\includegraphics[clip]{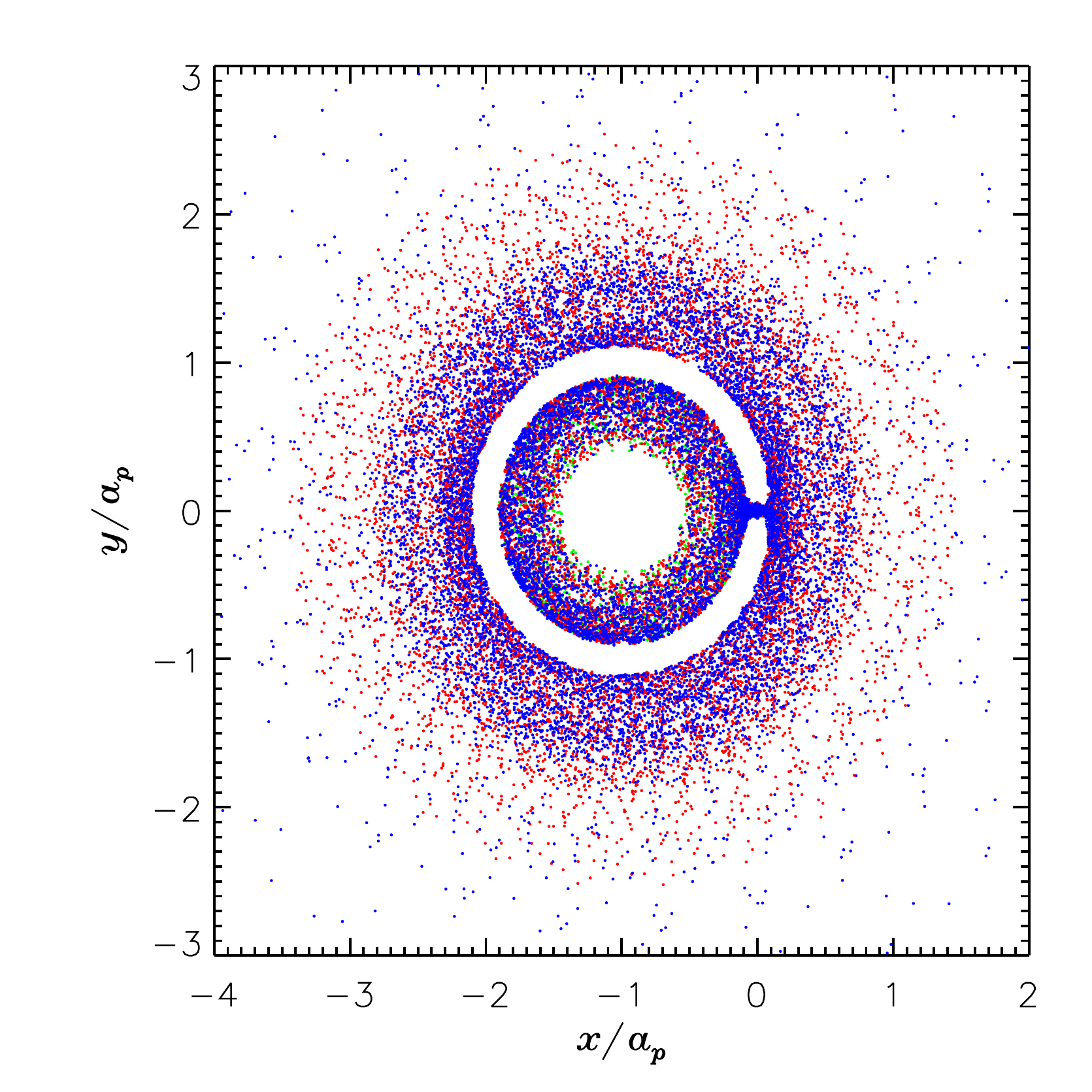}}
\caption{%
             Positions along $18$ trajectories, in a reference frame
             fixed to the planet (Section~\ref{sec:CPD}), of icy planetesimals
             scattered out of the planet's Roche lobe. The star is located
             at $(x,y)=(-a_{p},0)$. The reference density
             in the disk is $\rho_{0}=10^{-12}\,\mathrm{g\,cm}^{-3}$. The 
             trajectories are integrated for about $57$ orbital periods of
             the planet. The color scale of the dots indicates the
             planetesimals radius (which increases from lighter to darker
             dots). Except for one planetesimal ($R_{s}\sim 1\,\mathrm{km}$,
             scattered inward), about equal numbers of bodies have radii 
             $\sim 10$ and $\sim 100\,\mathrm{km}$.
             }
\label{fig:cpd_scat}
\end{figure}
In the calculations, scattering of planetesimals out of the Roche lobe occurs 
through interaction with the planet, following one or more close encounters. 
In Figures~\ref{fig:h_cpd} and \ref{fig:cpd_2dh}, 
both low and high eccentricity orbits may eventually lead 
to scattering events. Of the objects plotted in the upper panels of
Figure~\ref{fig:h_cpd}, by following the subsequent evolution of a sample, 
we estimate that the ratio of accreted to scattered (out of the planet's Roche lobe)
objects is about $1/7$. 
The $R_{s}\sim 100\,\mathrm{km}$ planetesimals, which are the least affected
by gas drag, are those most easily ejected.
Bodies are scattered both inward and outward of the planet's orbit. 
In this sample, whose initial positions and
velocities are the same as those used to draw the distributions in 
Figure~\ref{fig:h_cpd}, the majority of scattered bodies with a starting semi-major 
axis $\gtrsim 0.3\,\Rhill$ have eccentricities $\gtrsim 0.6$.
To give an idea of the spatial distribution of scattered objects,
Figure~\ref{fig:cpd_scat} shows the positions along the trajectories of $18$ 
such bodies, of radius $\sim 10$ and $\sim 100\,\mathrm{km}$ (see figure 
caption for details), extracted from said sample. 
In particular, the planetesimals in the figure are originally placed exterior 
of the planet's orbit, before being diverted toward the circumplanetary disk
(but all trajectories illustrated in the figure start from within $0.6\,\Rhill$ 
of the planet).

\begin{deluxetable*}{ccccccc}
\tablecolumns{7}
\tablewidth{0pc}
\tablecaption{Accretion Rates of Planetesimals on the Planet\tablenotemark{a}\label{table:pac}}
\tablehead{\colhead{}&\colhead{}&%
\multicolumn{2}{c}{Icy Bodies}&\colhead{}&%
\multicolumn{2}{c}{Mixed Bodies}\\%
\cline{3-4} \cline{6-7}\\[-5pt]
\colhead{Zone}&\colhead{}&\colhead{$\rho_{0}=10^{-12}$}&\colhead{$10^{-11}$}%
                        &\colhead{}&\colhead{$\rho_{0}=10^{-12}$}&\colhead{$10^{-11}$}%
}
\startdata
Interior%
&       & $1.2\times 10^{-6}$ & $1.0\times 10^{-6}$ & & $1.2\times 10^{-6}$ & $1.5\times 10^{-6}$\\
Corotat.%
&       & $8.7\times 10^{-7}$ & $8.2\times 10^{-7}$ & & $8.8\times 10^{-7}$ & $7.1\times 10^{-7}$\\
Exterior%
&       & $2.7\times 10^{-5}$ & $2.6\times 10^{-5}$ & & $2.4\times 10^{-5}$ & $2.7\times 10^{-5}$
\enddata
\tablenotetext{a}{In units of $\Mearth\mathrm{yr}^{-1}$ and scaled to $\Sigma_{s}=1\,\mathrm{g\,cm}^{-2}$.}
\end{deluxetable*}
Table~\ref{table:pac} lists the accretion rates of planetesimals on the planet,
estimated over the last $50$--$70$ planet's revolutions. The accretion rates
are rescaled so that the average surface density of solids in the regions
$0.77\le r/a_{p} \le 0.82$, $0.965\le r/a_{p} \le 1.035$, and 
$1.2\le r/a_{p} \le 1.25$ 
is $1\,\mathrm{g\,cm}^{-2}$. Contributions from each of the three
regions are listed separately. There are relatively small differences 
($\lesssim 10$\%) between the values obtained from calculations using 
different reference gas densities, $\rho_{0}$, and different material compositions. 
The accretion of solids arises almost entirely from the regions interior and exterior 
of the planet's orbit, the former region contributing around $5$\% of the total.
The corotation region provides only a minimal fraction of the solids' accretion.
Adding up the regional contributions and averaging out the results, we have
\begin{equation}
\langle\dMp\rangle_{s}=2.8\times 10^{-5}\left(\frac{\Sigma_{s}}{1\,\mathrm{g\,cm}^{-2}}\right)%
\Mearth\,\mathrm{yr}^{-1},
\label{eq:mzdot}
\end{equation}
where $\Sigma_{s}$ is the surface density of solids in the three radial
regions mentioned above. This result applies for initial planetesimal
populations with equal numbers of bodies per size bin.
Additionally, as stressed in the previous section, 
planetesimal-planetesimal interactions are neglected and, therefore,
Equation~(\ref{eq:mzdot}) is valid as long as the effects of collisions
and encounters among solids can be neglected, 
i.e., for low enough values of $\Sigma_{s}$.
But in absence of a mechanism (like collisions and gravitational stirring) 
to replenish the exterior disk region with solids (resupply via gas drag 
would not be effective at the gas density levels considered here due to 
long orbital decay times, see Section~\ref{sec:PCD}), 
a total mass in solids of order
$0.2\,\Sigma_{s}/(1\,\mathrm{g\,cm}^{-2})\,\Mearth$ can be 
delivered to the planet. A somewhat smaller mass would be 
accessible from the interior disk, but over a much longer timescale.
This amount of solids may represent only a relatively small addition 
to the heavy element content of the planet.

\begin{figure}
\centering%
\resizebox{\linewidth}{!}{\includegraphics[clip]{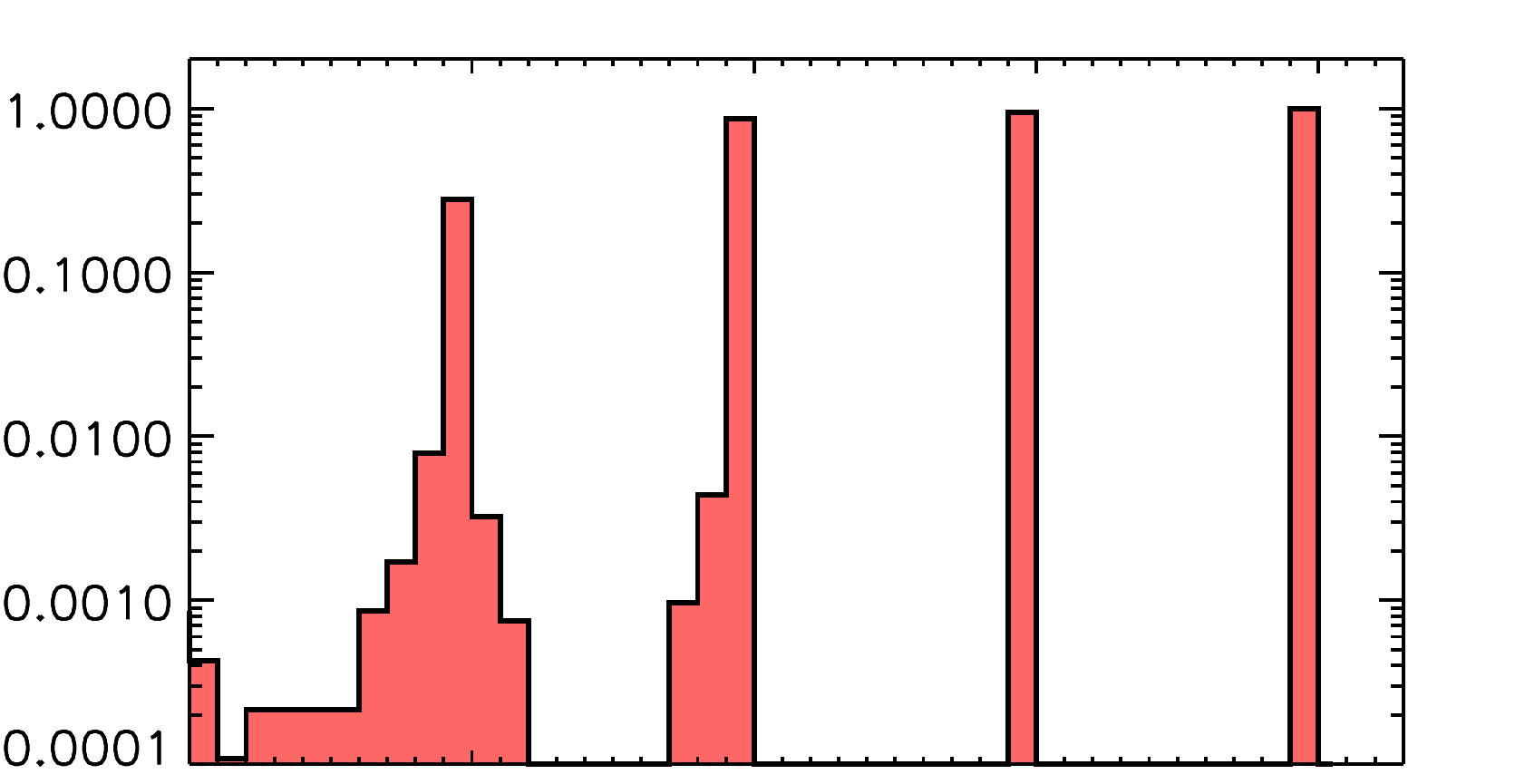}}
\resizebox{\linewidth}{!}{\includegraphics[clip]{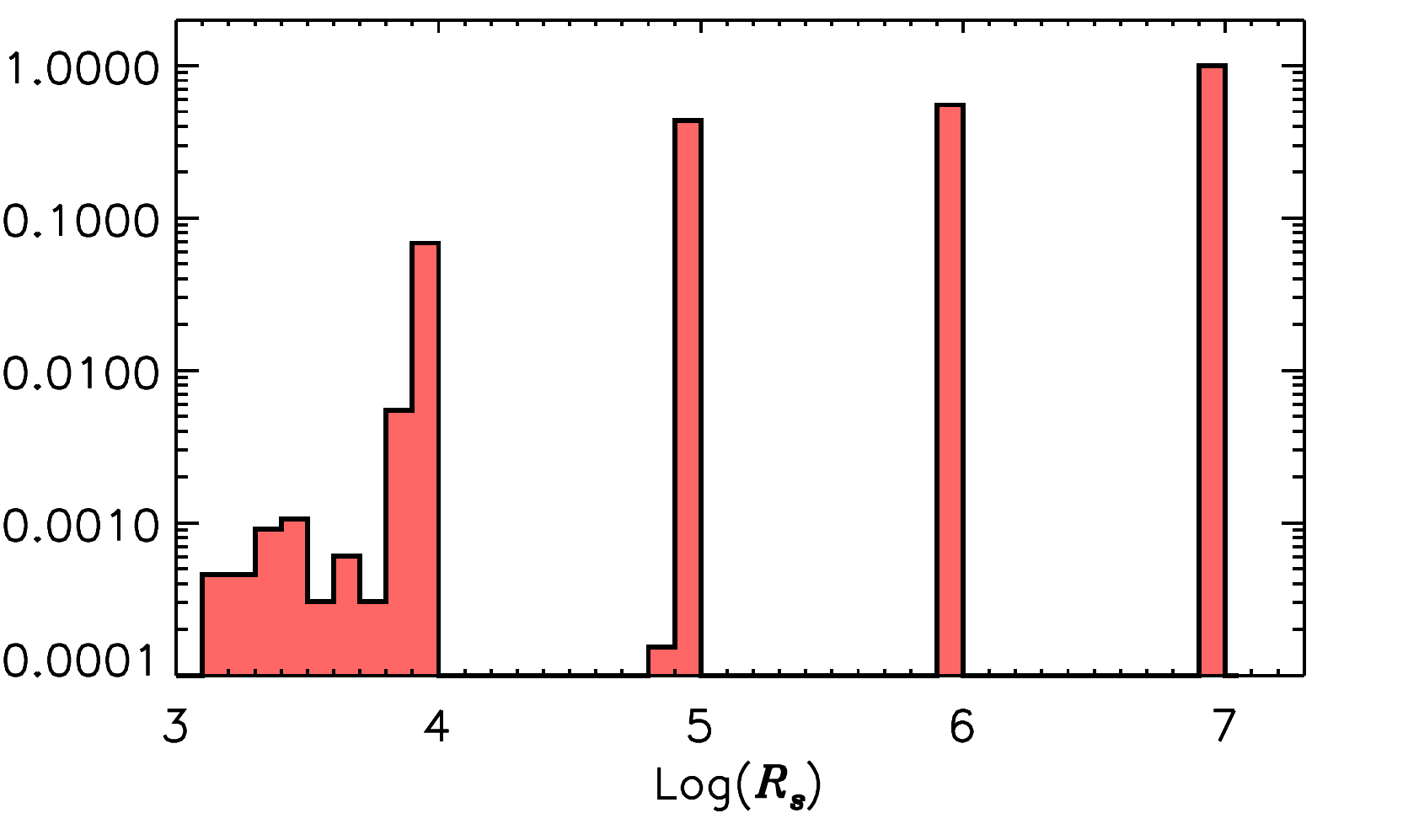}}
\caption{%
             Distributions of the radii of planetesimals accreted by the planet 
             during the course of the calculations. 
             The histograms include bodies of both compositions, initially placed 
             in all three disk regions. The reference gas density is
             $\rho_{0}=10^{-12}$ (top) and $10^{-11}\,\mathrm{g\,cm}^{-3}$ (bottom).
             The radii are in $\mathrm{cm}$.
             Histograms are normalized so that the maximum is $1$.
             }
\label{fig:mzdot}
\end{figure}
The histograms in Figure~\ref{fig:mzdot} show the radii in $\mathrm{cm}$
of accreted planetesimals for $\rho_{0}=10^{-12}$ (top) and 
$10^{-11}\,\mathrm{g\,cm}^{-3}$ (bottom). Histograms are rescaled
so that the maximum count is $1$. 
For the lower reference density case, there is an almost equal probability, 
within $\lesssim 15$\%, of accreting bodies in the size range 
$1\,\mathrm{km}\lesssim R_{s}\lesssim 100\,\mathrm{km}$.
The difference increases for the higher reference density case, in which 
the probability of accreting $R_{s}\approx 1\,\mathrm{km}$ bodies is about 
half that of accreting $R_{s}\approx 100\,\mathrm{km}$ bodies.
(As discussed below, one reason for the different accretion probability lies
in the fact that smaller bodies are more likely to break up.)
Consequently, 
our choice of starting the planetesimal populations with equal numbers 
of objects in different size bins determines the result that the accretion 
rate in solids is basically supplied by the largest planetesimals.
In fact, the accretion rate in Equation~(\ref{eq:mzdot}) is basically
that of the largest planetesimals and $\Sigma_{s}$ is the surface 
density of the largest bodies. 
Therefore, Figure~\ref{fig:mzdot} implies that, within factors of order unity, 
Equation~(\ref{eq:mzdot}) provides the accretion rates of a mono-size 
population of planetesimals (whose radius belongs to the approximate 
range $1$--$100\,\mathrm{km}$) with $\Sigma_{s}$ being the surface 
density of that population.
Figure~\ref{fig:mzdot} includes all planetesimals accreted during the
course of the calculations, hence it contains some bias due to the choice
of the initial distributions of planetesimals. For example, the probability 
of delivering to the circumplanetary disk, and thus of accreting, 
$R_{s}\sim 0.1\,\mathrm{km}$ bodies declines over time because 
of the widening gap in solids of this size (see Figure~\ref{fig:solid_gap}).

\begin{figure*}[t!]
\centering%
\resizebox{\figlew}{!}{\includegraphics[clip]{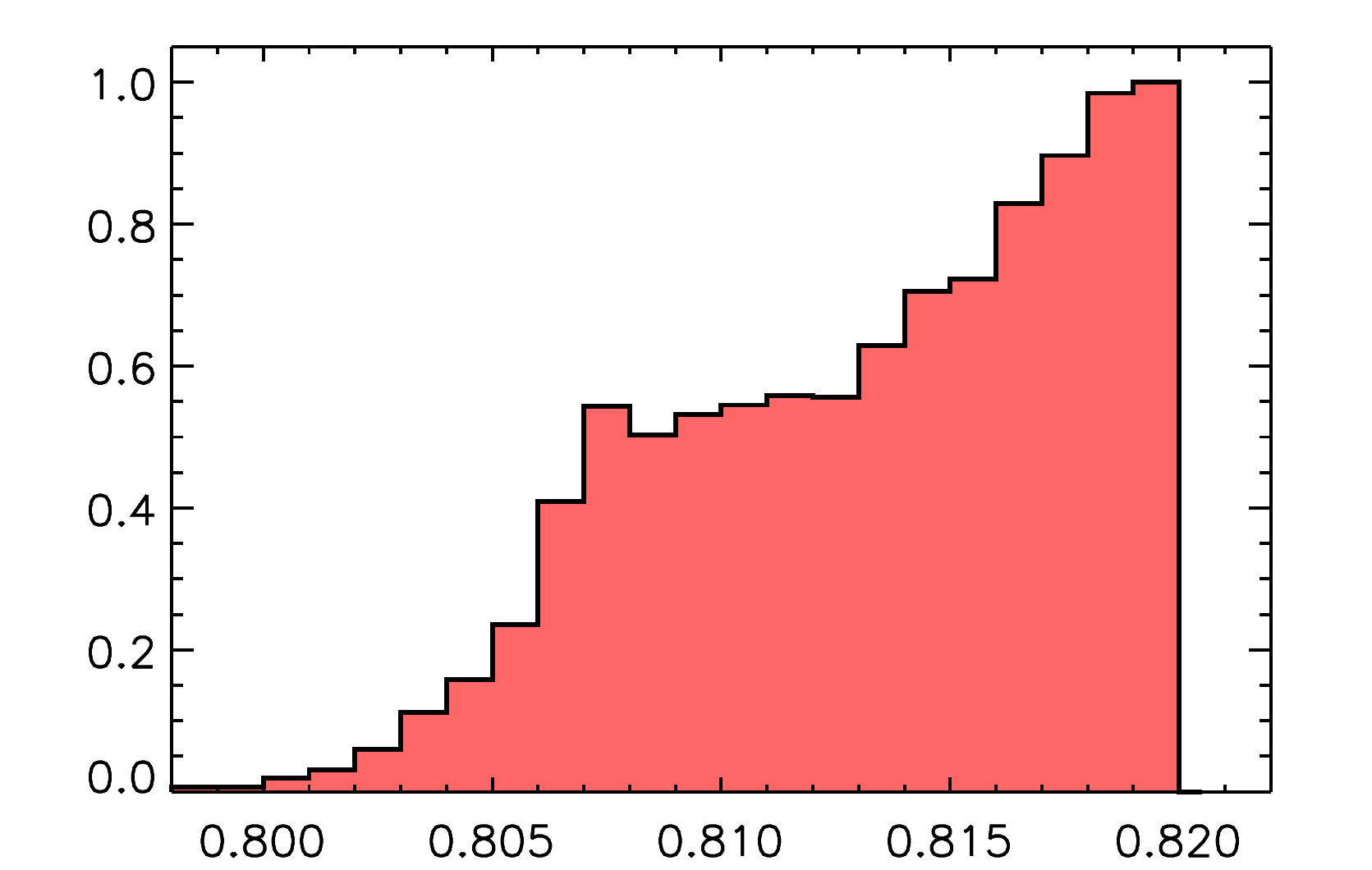}%
                                 \includegraphics[clip]{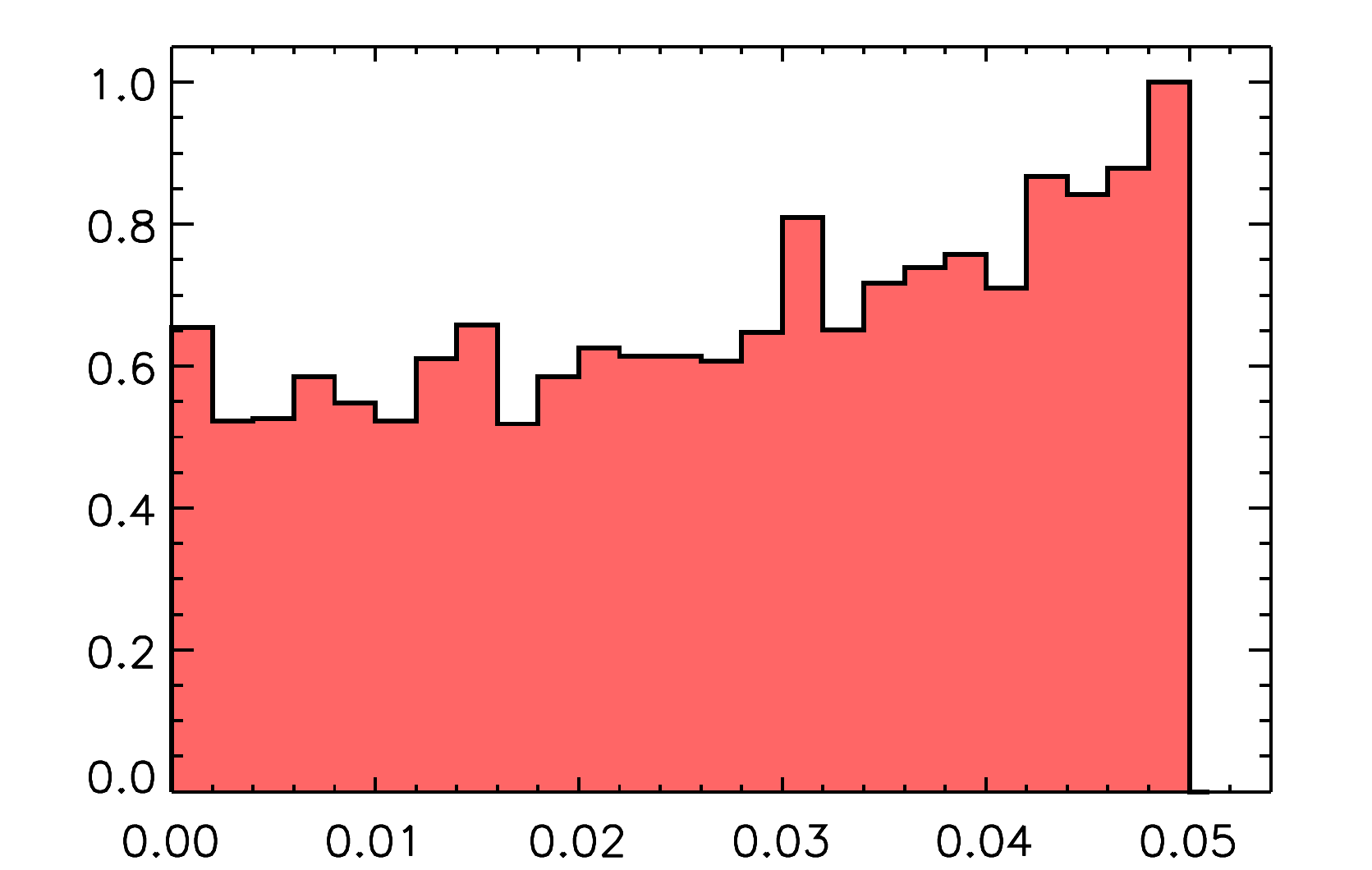}%
                                 \includegraphics[clip]{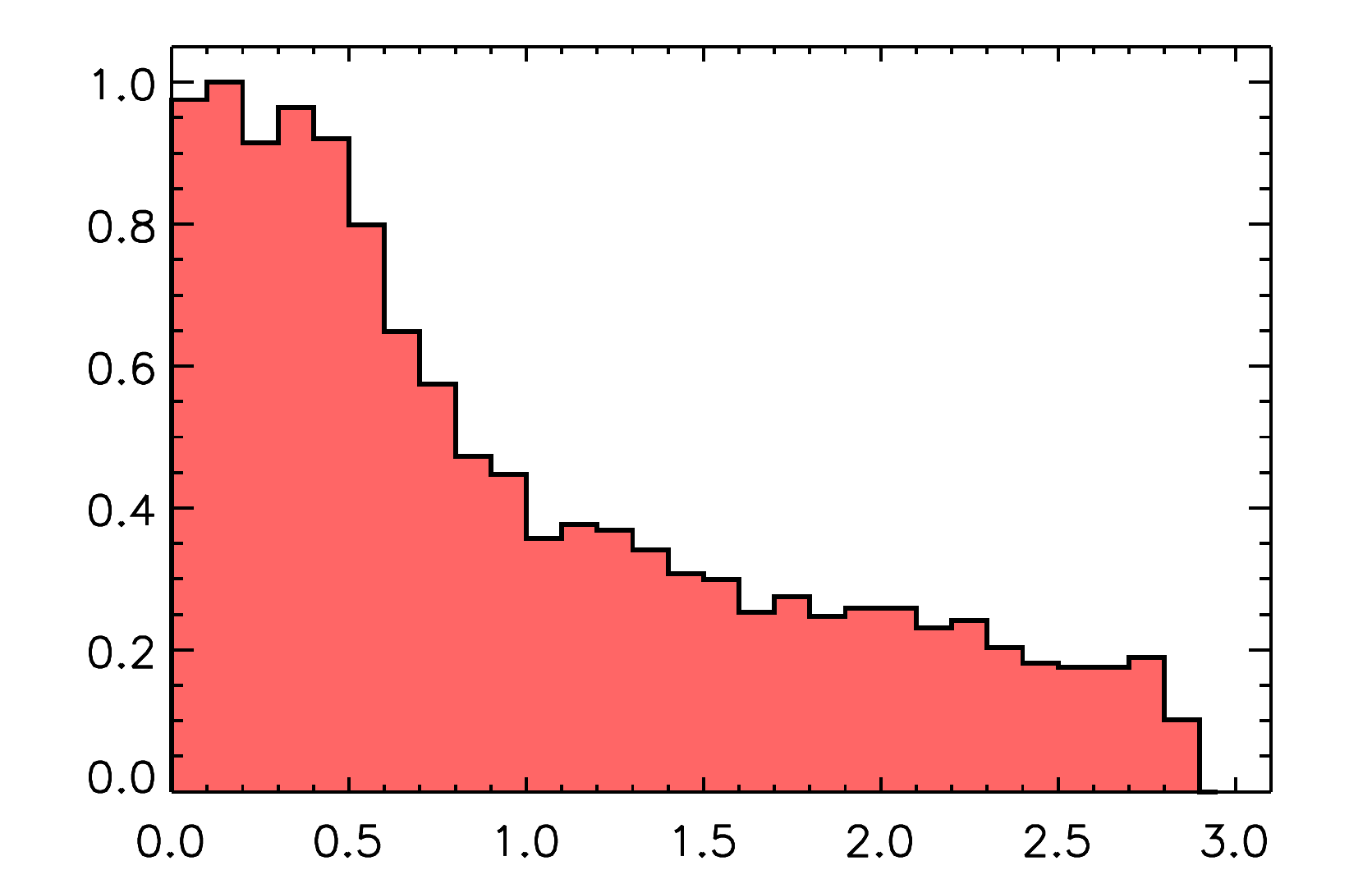}}
\resizebox{\figlew}{!}{\includegraphics[clip]{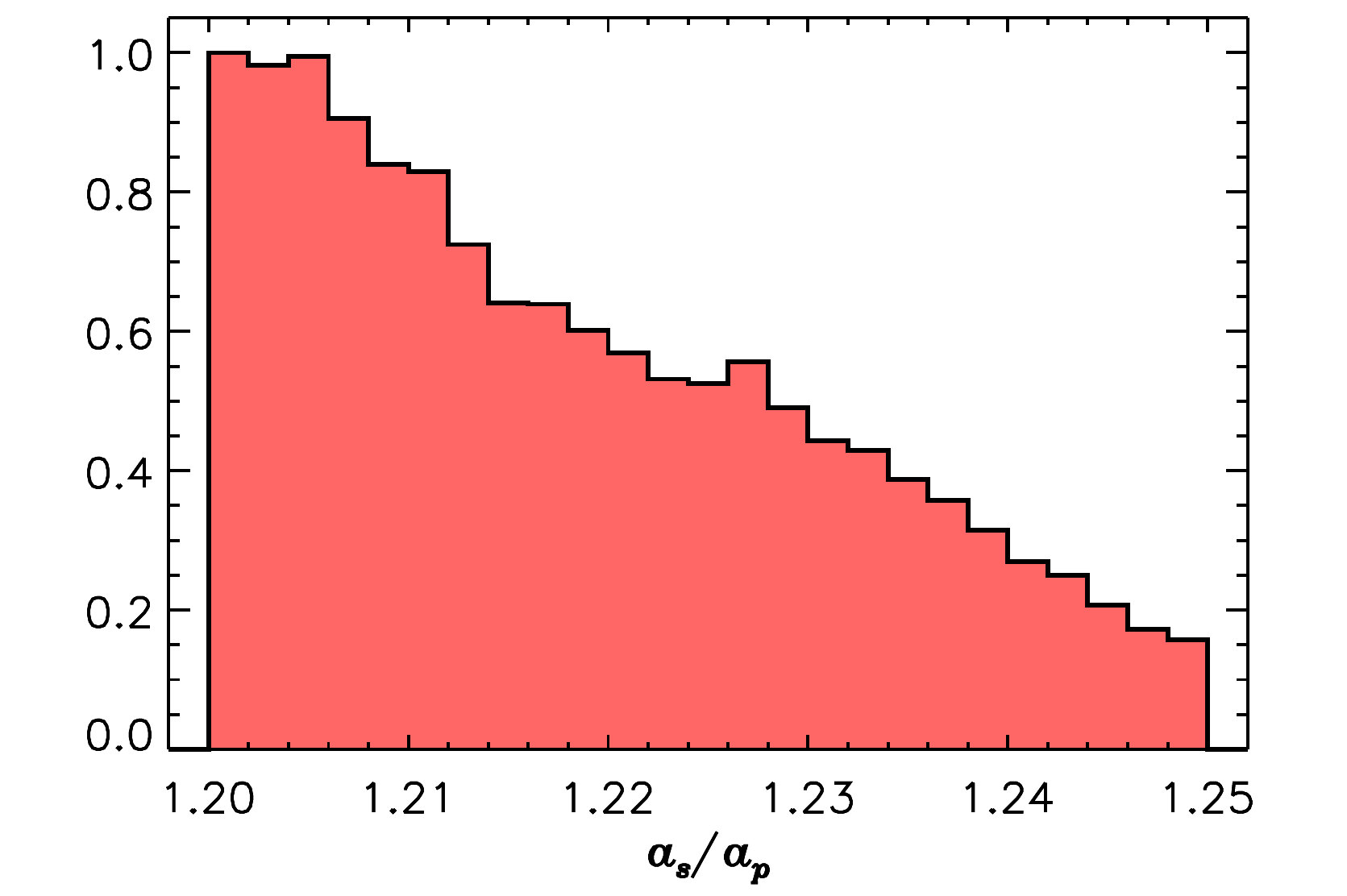}%
                                 \includegraphics[clip]{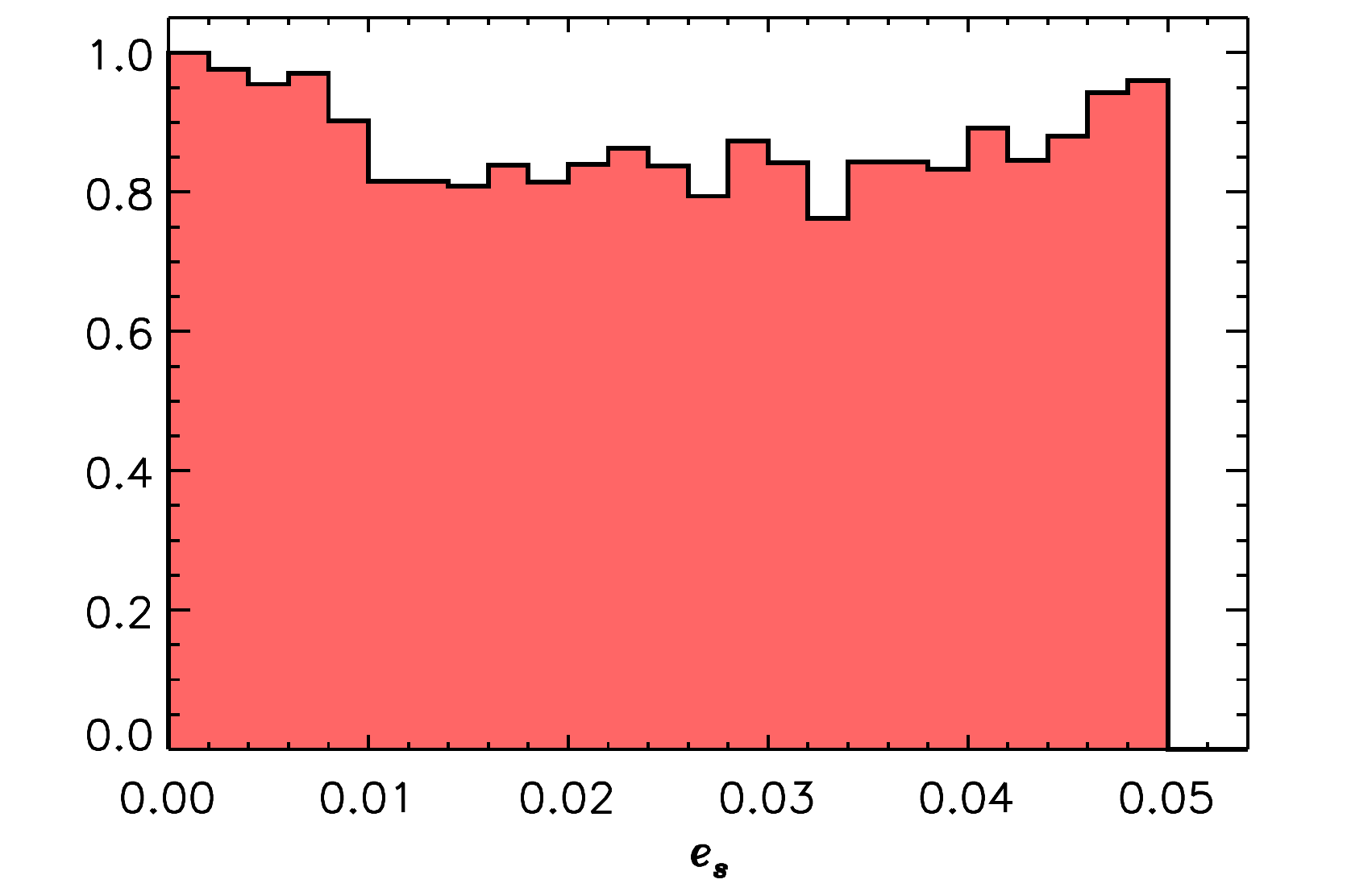}%
                                 \includegraphics[clip]{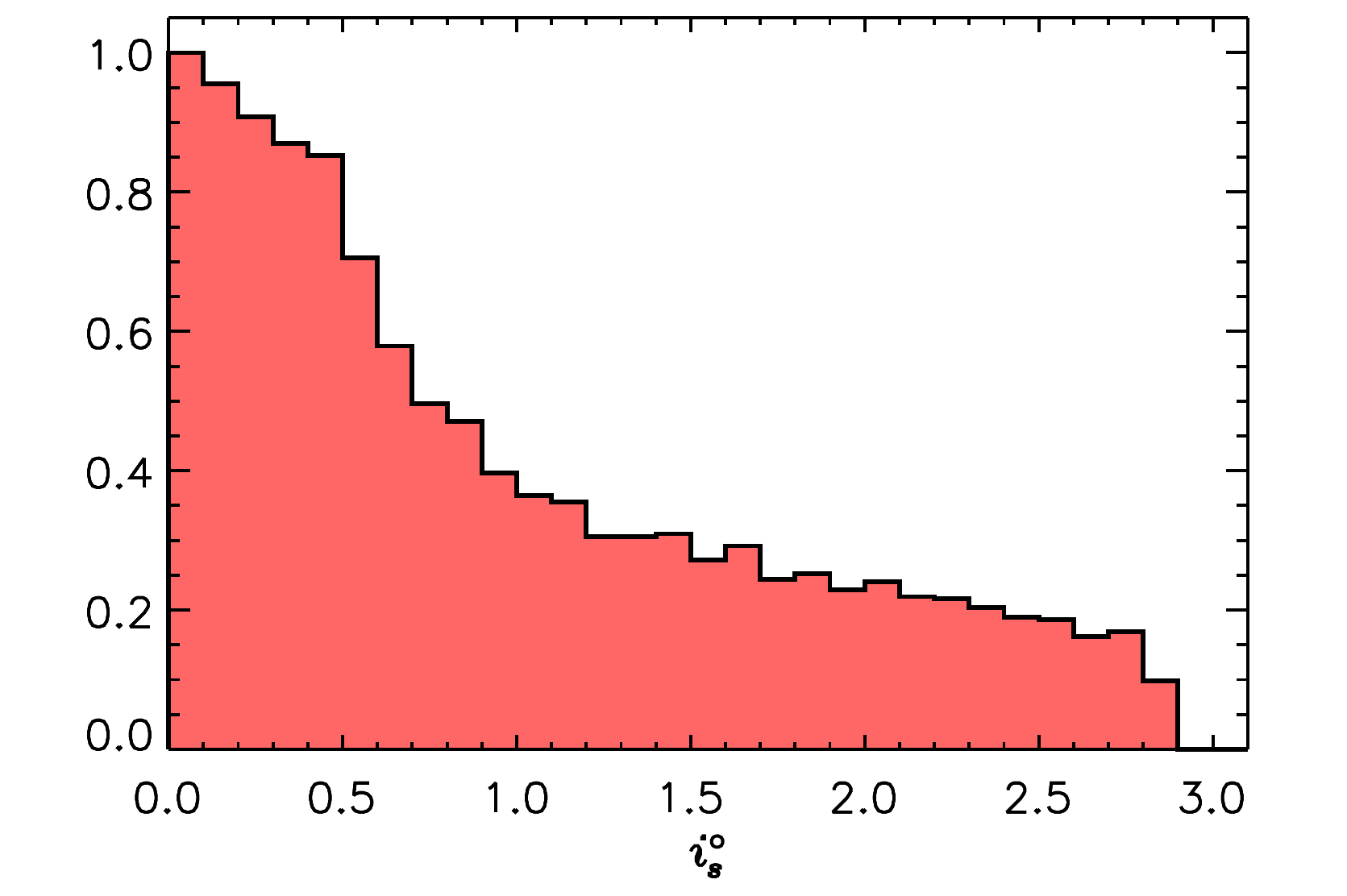}}
\resizebox{\figlew}{!}{\includegraphics[clip]{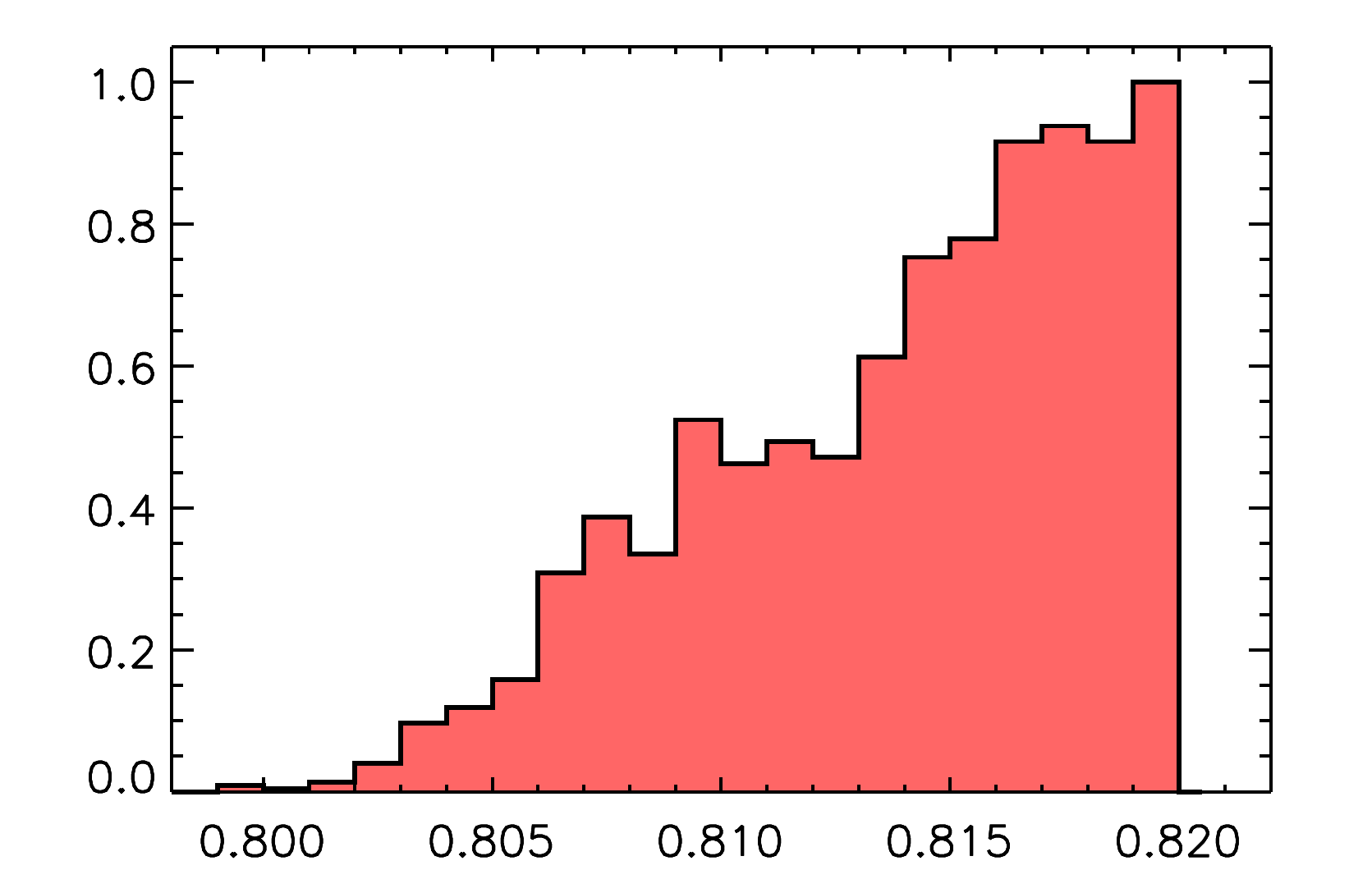}%
                                 \includegraphics[clip]{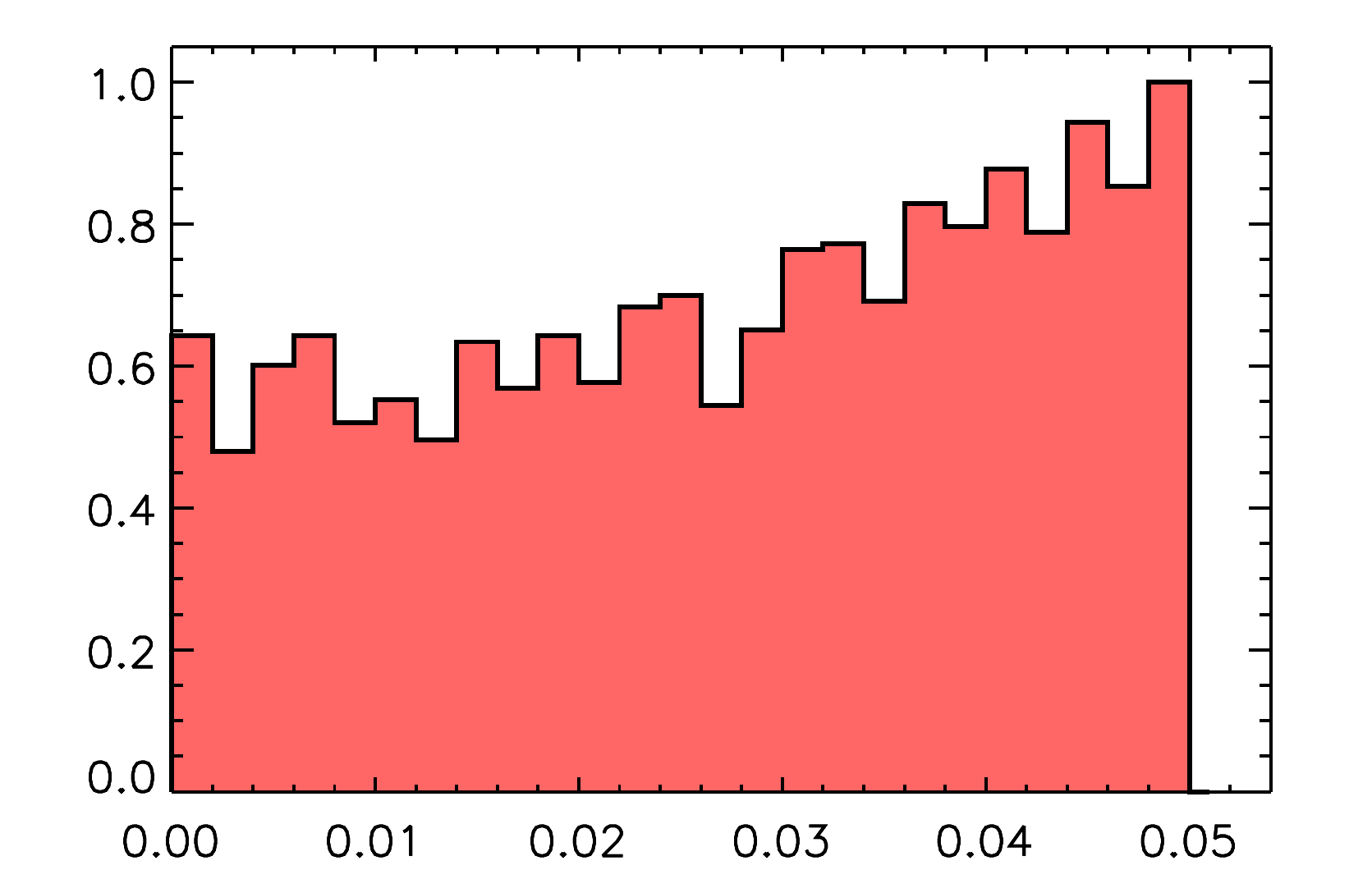}%
                                 \includegraphics[clip]{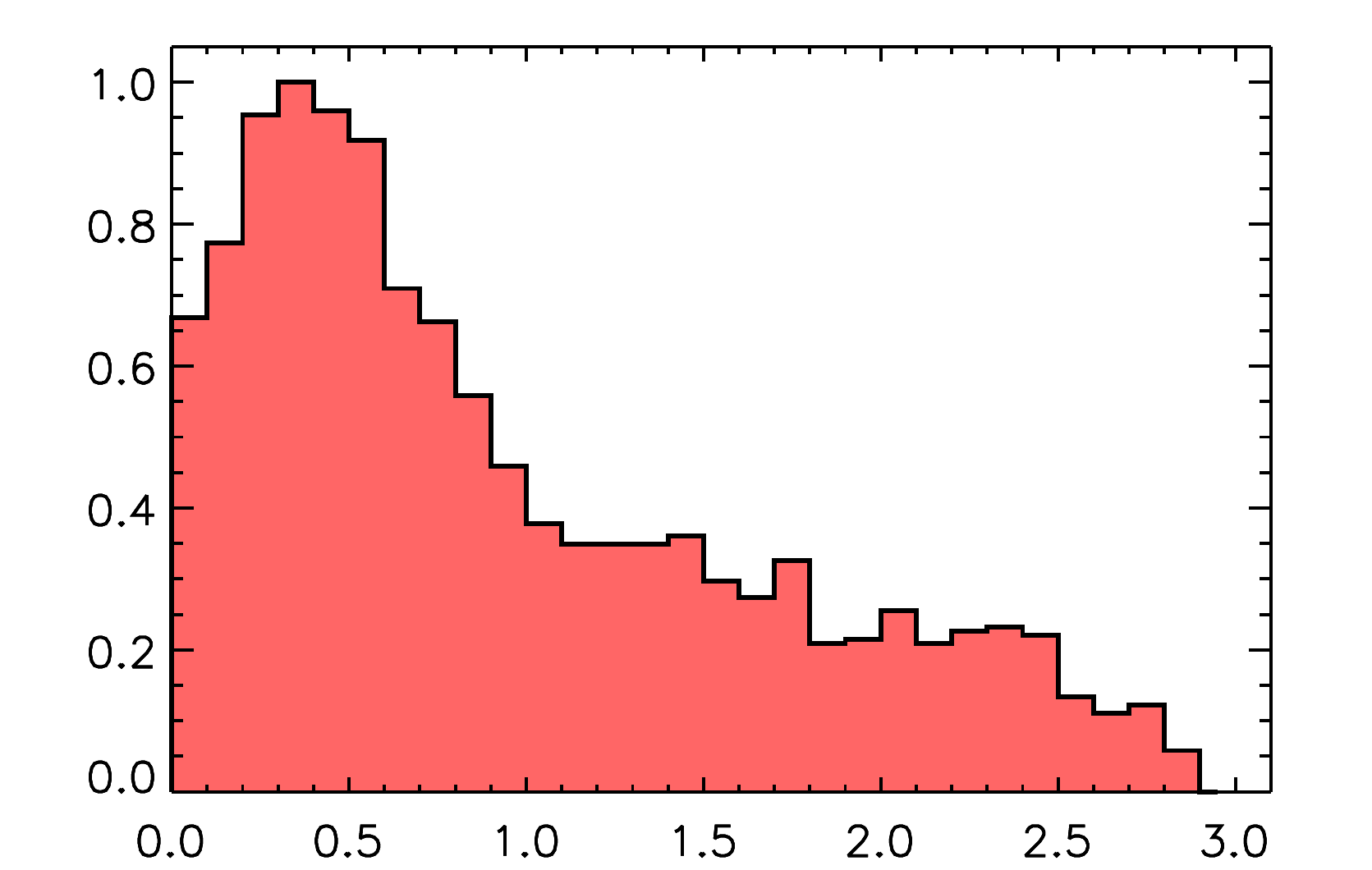}}
\resizebox{\figlew}{!}{\includegraphics[clip]{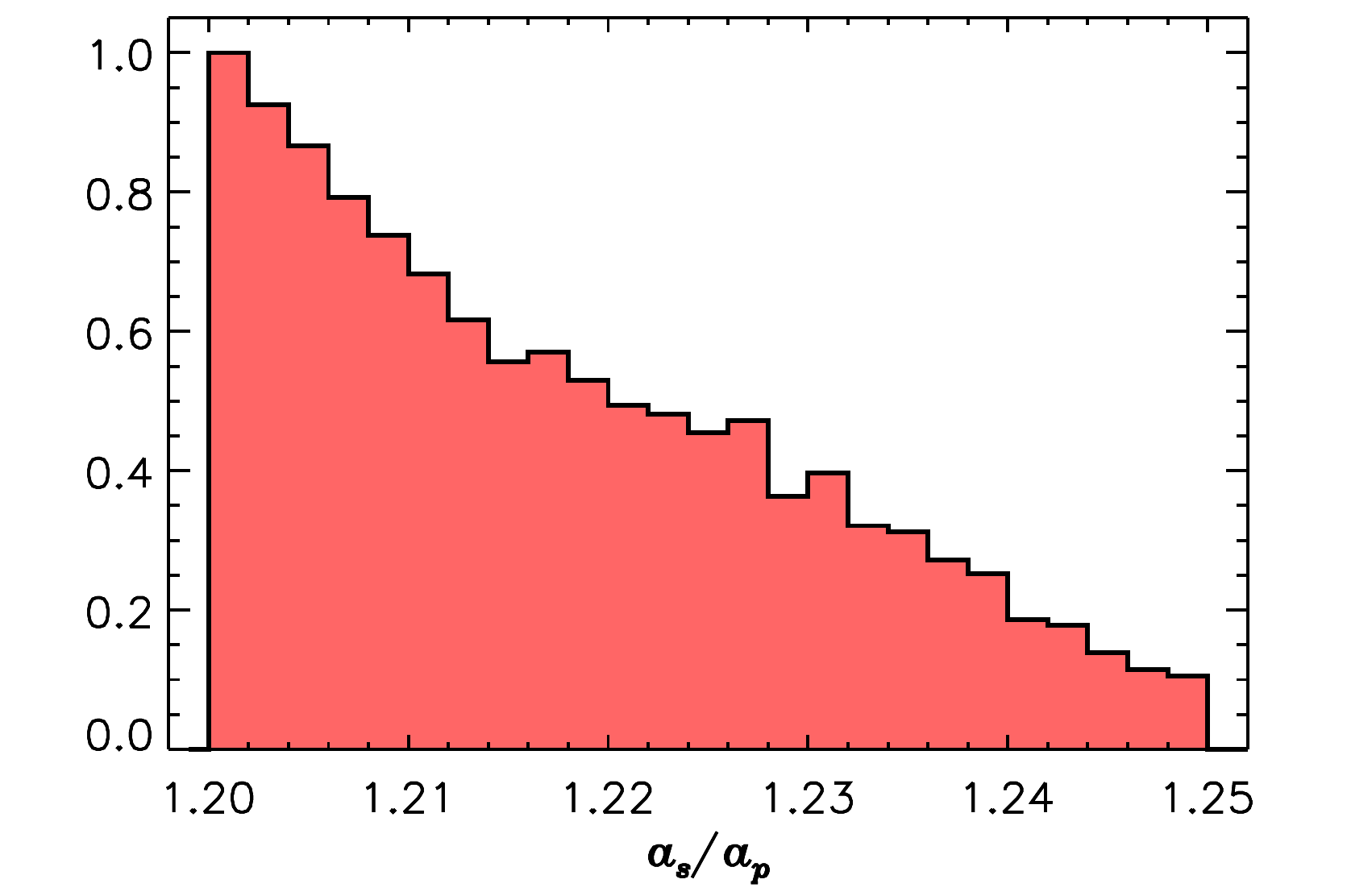}%
                                 \includegraphics[clip]{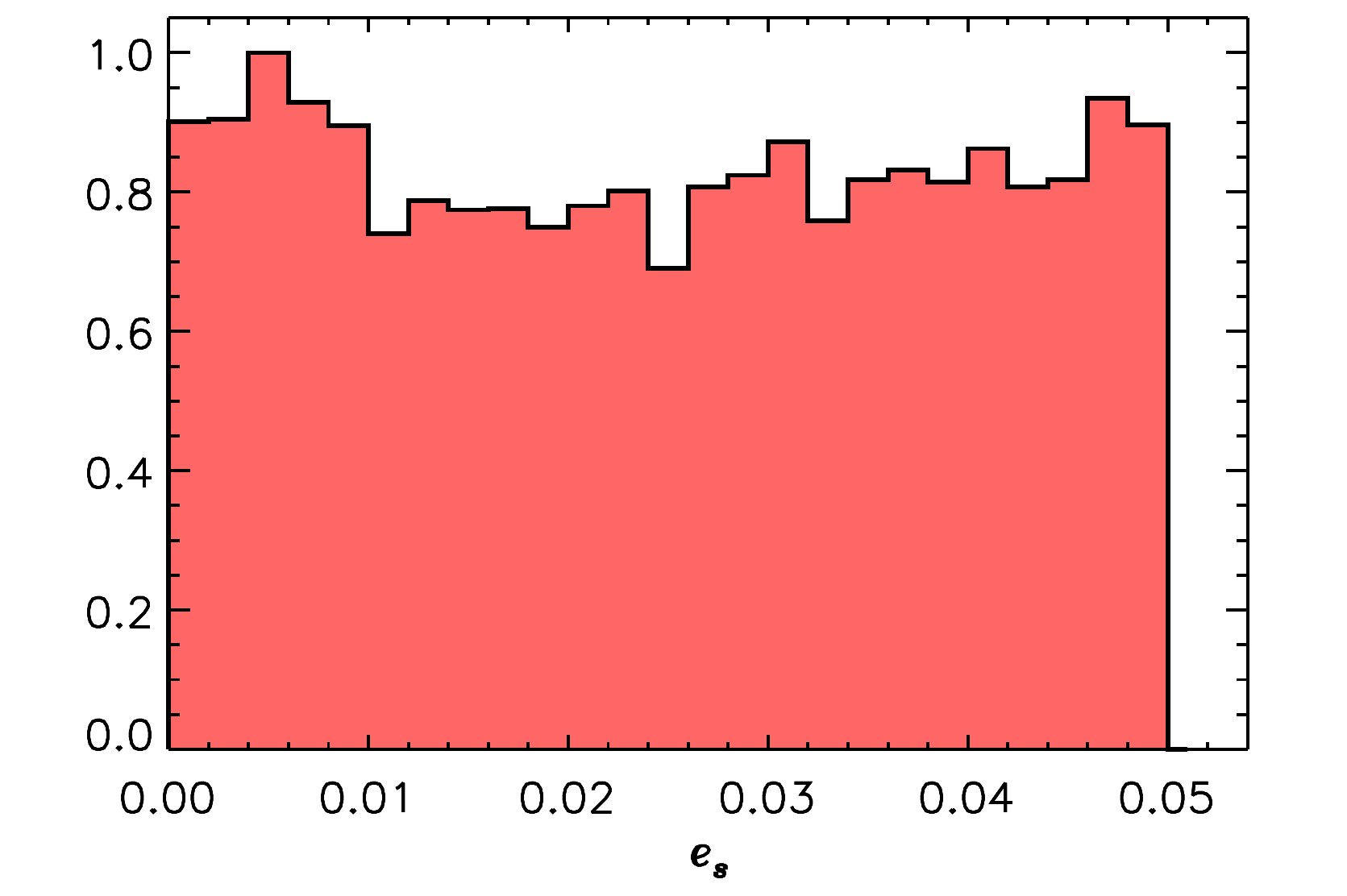}%
                                 \includegraphics[clip]{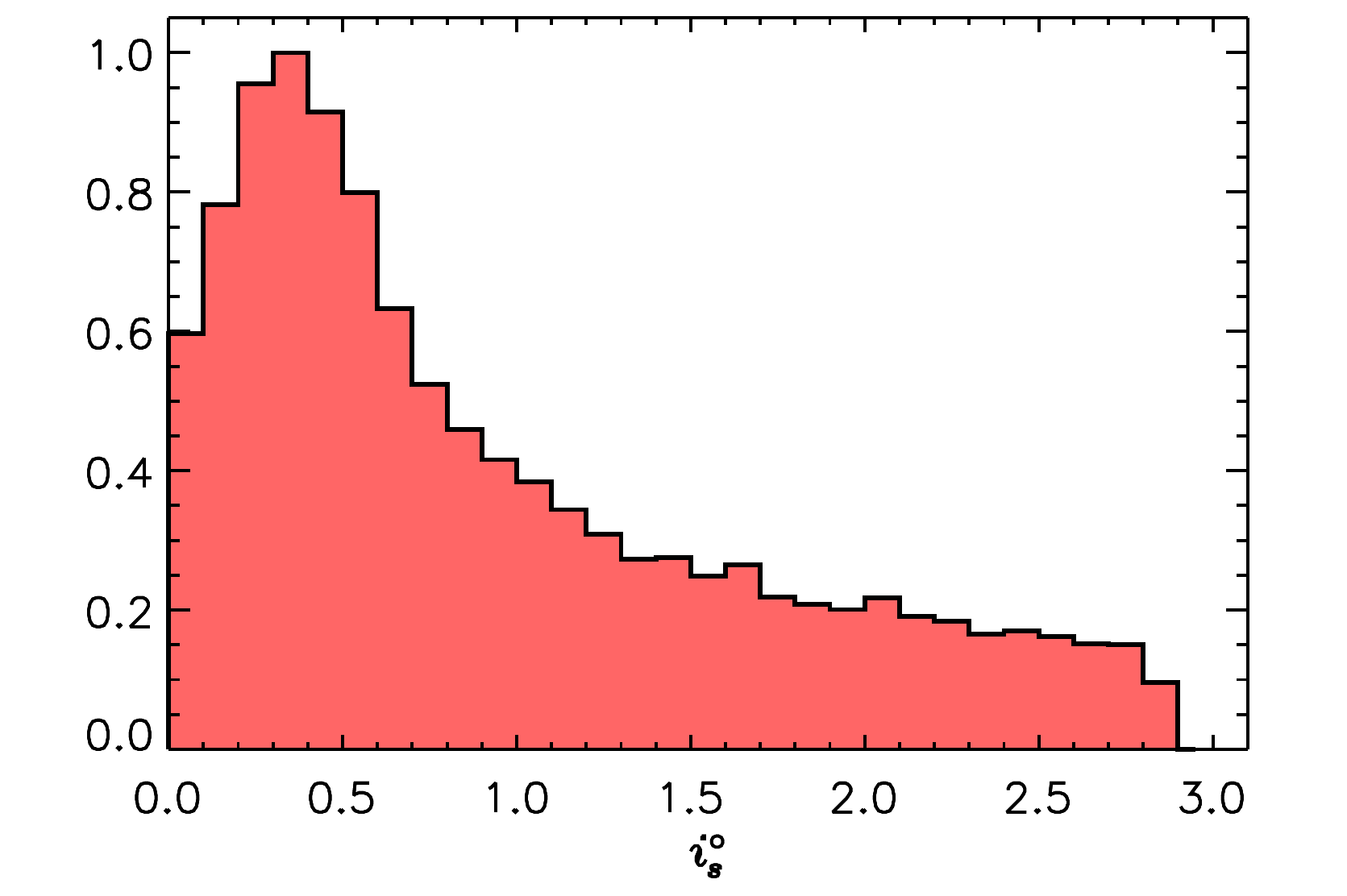}}
\caption{%
             Initial distributions of semi-major axis (left), eccentricity (center), and inclination
             (right) of accreted planetesimals
             for $\rho_{0}=10^{-12}$ (upper pair of rows) and $10^{-11}\,\mathrm{g\,cm}^{-3}$ 
             (lower pair of rows). The histograms refer to the initial distributions of 
             both icy and mixed-composition bodies.
             Histograms are normalized so that the maximum is $1$.
             }
\label{fig:h_iacc}
\end{figure*}
The initial semi-major axis, eccentricity, and inclinations of accreted 
planetesimals are plotted in Figure~\ref{fig:h_iacc} (see figure caption 
for details). The results indicate that planetesimals deployed in proximity
of the edge closer to the planet's orbit are more likely to be accreted than 
are more distant bodies (see left columns).
For bodies released in the cororation region (not shown in the figure), the
trend is opposite, as expected, due to the stability of tadpole orbits.
The probability of accreting planetesimals deployed exterior of the planet's
orbit is fairly independent of the initial eccentricity, whereas (in the range
$0$--$0.05$) larger initial eccentricities favor accretion of bodies deployed
interior of planet's orbit (see center columns).
Initially co-planar orbits lead to accretion more easily than do inclined orbits
(see right columns).
The smaller accretion rates provided by the interior disk (see 
Table~\ref{table:pac}) can be explained by observing that the peak number
densities around the inner edge of the solids' gap are at $\approx 0.8\,a_{p}$ 
(see Figure~\ref{fig:solid_gap}), where the probability of accretion is 
relatively low (see Figure~\ref{fig:h_iacc}, left-even panels).

As mentioned in the previous section, break-up of planetesimals may 
occur when they encounter the dense gas of the circumplanetary disk, 
inside $\approx 0.2\,\Rhill$ of the planet (see Figure~\ref{fig:prof_sub}). 
Assuming that planetesimal fragments
can quickly drift toward the planet, the mass of disrupted bodies would 
contribute to the accretion rate of solids on the planet. However, 
this addition would amount to $\sim 0.01$\% or $\sim 0.1$\% 
(depending on $\rho_{0}$, see above) of the values reported in 
Table~\ref{table:pac}.
This small contribution largely depends on the fact that accretion of solids 
is dominated by $R_{s}\approx 100\,\mathrm{km}$ planetesimals,
which do not tend to break up. However, smaller bodies break up more easily.
In fact, for $\rho_{0}=10^{-11}\,\mathrm{g\,cm}^{-3}$, 
the mass of shattered planetesimals with radii
$1\,\mathrm{km}\lesssim R_{s}\lesssim 10\,\mathrm{km}$
is comparable to (and sometimes larger than) the accreted mass
contributed by bodies of these sizes.
Thus, if fragments ought to be considered as accreted material,
the accretion rate in this size range (e.g., given by 
Equation~(\ref{eq:mzdot}) applied to a mono-size population) 
may be higher by a factor of up to a few.
For the reference density $\rho_{0}=10^{-12}\,\mathrm{g\,cm}^{-3}$,
break-up of planetesimals is less relevant, and it would contribute 
$\sim 10$\% to the accretion of $R_{s}\sim 10\,\mathrm{km}$ bodies 
and even less to the accretion of smaller planetesimals.

\subsection{Isothermal Circumplanetary Disk Calculations}
\label{sec:PCPiso}

As anticipated in previous sections, we also consider models that apply 
the temperature $T_{n}$ in Equation~(\ref{eq:Tg}) also in the circumplanetary disk,
which then becomes nearly isothermal with a temperature of $\approx 120\,\K$
(see Figure~\ref{fig:tcp} at $\tilde{r}\approx \Rhill$ and the dotted line in
Figure~\ref{fig:tcp_sp}). 
A smaller number of bodies is released in these calculations.
Since the gas density is the same as in the models discussed above, 
differences may be expected especially in the thermal evolution and ablation 
of planetesimals moving in close proximity of the planet. 
Nonetheless, we find that there are not large differences between the two approaches. 

As above, average fractions of order $10^{-3}$ are captured within $0.6\,\Rhill$ 
of the planet from the available mass of planetesimals. More bodies are captured 
at the lower reference density, $\rho_{0}$, but a somewhat larger mass is 
retained at the higher value of $\rho_{0}$. Most of the captured bodies have 
radii $\lesssim 10\,\mathrm{km}$, but most mass is carried by 
the $100\,\mathrm{km}$-radius planetesimals.

The gray circles in Figure~\ref{fig:tcp_sp} show the planetesimal temperatures
as a function of the planetocentric semi-major axis, while the dotted line
represents the local gas temperature (see Equation~(\ref{eq:Tg})).
In the calculations with reference density $\rho_{0}=10^{-12}\,\mathrm{g\,cm}^{-3}$,
$T_{s}$ is generally comparable with the planetesimals' temperature 
obtained from the calculations discussed above (top panel, darker circles), 
which use the gas temperature in Equation~(\ref{eq:Tcp}).
For the higher reference density, the discrepancy is larger, but only
when $0.03\,\Rhill\lesssim a_{s}\lesssim 0.2\,\Rhill$ (see bottom panel).
Body temperatures become comparable again ($T_{s}\gtrsim 200\,\K$) 
when planetesimals get close to the planet ($a_{s}\lesssim 0.03\,\Rhill$), 
where ablation is most vigorous.
It is possibly for this reason that the amounts of ablated material are similar
in the two sets of calculations.

The accretion rates of planetesimals on the planet are comparable to those
given by Equation~(\ref{eq:mzdot}). 
Relative differences between isothermal and non-isothermal calculations 
of accretion rates versus $R_{s}$ are also small, $\lesssim 20\%$.

The thermal distribution of the gas does not directly affect the fragmentation 
of planetesimals, as neither the dynamical pressure, $P_{\mathrm{dy}}$,
nor the material compressive strength, $\sigma_{s}\sqrt{1\,\mathrm{km}/R_{s}}$, 
is explicitly dependent on $T_{g}$ ($T_{n}$) or $T_{s}$. Indirect effects
may nonetheless occur, e.g., because of different ablation histories.
Consistently with the results presented above,
break-up of planetesimals occurs in the proximity of the planet, and mostly
in the size range $1\,\mathrm{km}\lesssim R_{s}\lesssim 10\,\mathrm{km}$.
Overall, solid material made available through break-up would only contribute
negligibly to the accreted mass due to the fact that most mass is
delivered to the planet in the form of $R_{s}\sim 100\,\mathrm{km}$ bodies, 
which rarely break up. But this effect depends on body size and gas density. 
As above, at $\rho_{0}=10^{-11}\,\mathrm{g\,cm}^{-3}$ (but not at the lower 
$\rho_{0}$), disruption of 
$1\,\mathrm{km}\lesssim R_{s}\lesssim 10\,\mathrm{km}$ planetesimals 
would significantly contribute to the accretion of solids, if fragments efficiently
accreted on the planet.

\section{Discussion and Conclusions}
\label{sec:DC}

The results presented here show how planetesimals of various sizes,
initially orbiting in three narrow radial regions 
($\Delta r \approx 0.26$--$ 0.36\,\AU$) 
around a star and in proximity of a Jupiter-mass planet, are scattered 
through the circumstellar disk and toward the planet. 
Scattering is dominated by three-body interactions (star-planet-planetesimal) 
while gas drag typically operates as a perturbing force. 
In some case, for $R_{s}\lesssim 0.1\,\mathrm{km}$, gas drag does
determine planetesimal dynamics.
As expected, the evolution of mixed-composition (ice/quartz with a $60$\% 
ice mass fraction) and icy planetesimals is similar in most thermodynamical 
aspects. 

The surface temperature of planetesimals, $T_{s}$ (defined in 
Section~\ref{sec:PT}), evolves toward an equilibrium value 
($dT_{s}/dt\approx 0$), which in absence of significant mass loss is
such that (see Equation~(\ref{eq:dTsdt}))
\begin{equation}
(T^{\mathrm{eq}}_{s})^{4}\approx T^{4}_{g}+
                                  \frac{C_{D}}{32\sigma_{\mathrm{SB}}}\frac{\rho_{g}}{\epsilon_{s}}\,
                                  |\mathbf{v}_{g}-\mathbf{v}_{s}|^{3}.
\label{eq:Ts_eq}
\end{equation}
By using the approximation 
$|\mathbf{v}_{s}-\mathbf{v}_{g}|\approx\xi^{2}a_{s}\Omega_{\mathrm{K}}/2$
with $\xi\approx H/r$ (which holds far from the planet's orbit, see 
Section~\ref{sec:DDD} and Appendix~\ref{sec:DT}), the second term 
on the right-hand side of Equation~(\ref{eq:Ts_eq}) becomes
$(\xi/2)^{6}C_{D}/(4\sigma_{\mathrm{SB}})(\rho_{g}/\epsilon_{s})%
(G\Ms/a_{s})^{3/2}$. In the calculations described here, this term is
usually  small 
compared to $T^{4}_{g}$ and thus $T^{\mathrm{eq}}_{s}\approx T_{g}$.
This conclusion is supported by the calculations.
Vigorous mass loss tends to lower $T_{s}$ relative to $T_{g}$ (see 
Figure~\ref{fig:tcp_sp}). The inverse of the timescale required to reach 
equilibrium, neglecting both frictional heating and cooling via latent heat 
release, is
\begin{equation}
\frac{1}{T_{s}}\left|\frac{dT_{s}}{dt}\right|\approx\frac{\epsilon_{s}}{\delta_{s}}%
                                                        \frac{\sigma_{\mathrm{SB}}T^{3}_{s}}{\rho_{s}C_{s}}
                                                        \left|1-\left(\frac{T_{g}}{T_{s}}\right)^{4}\right|.
\label{eq:time_eq}
\end{equation}
For a given deviation from equilibrium ($T_{g}$), this timescale is 
$\propto \delta_{s}/T^{3}_{s}\propto \lambda_{s}/T^{6}_{s}$ 
(see Equation~(\ref{eq:diso})) and increases with approaching equilibrium.
Some numerical examples on the evolution of $T_{s}$ toward equilibrium 
are shown in Appendix~\ref{sec:TT}. 
The situation is more complex for eccentric orbit
bodies, as they experience a varying gas temperature along their orbit.
Nonetheless, an equilibrium temperature can be reached if 
$|dT_{s}/dt|/T_{s}\ll \Omega_{\mathrm{K}}$.

Results concerning the distributions of solids can be rescaled to an arbitrary 
surface density (in the radial regions of initial deployment), $\Sigma_{s}$, 
provided that collisions and encounters among bodies do not significantly 
alter their dynamics, i.e., that $\Sigma_{s}$ remains relatively low, 
as it may be the case at the late epochs of giant planet formation
(see Section~\ref{sec:ECR}).
The ejection rate of mass out of the disk domain ($2\,\AU\lesssim r\lesssim 20\,\AU$)
is 
$\sim 2\times 10^{-5}\,(\Sigma_{s}/1\,\mathrm{g\,cm}^{-2})\,\Mearth\,\mathrm{yr}^{-1}$.
In reality, most of these bodies are on orbits bound to the star when they cross 
the boundaries, and ``ejection'' generally classifies orbits whose perihelia (aphelia) 
are inside (outside) of $r_{\mathrm{mn}}$ ($r_{\mathrm{mx}}$).
Planetesimals are scattered from the interior to the exterior of the planet's
orbit at a rate of
$\sim 5\times 10^{-6}\,(\Sigma_{s}/1\,\mathrm{g\,cm}^{-2})\,\Mearth\,\mathrm{yr}^{-1}$,
and in the opposite direction at a rate of
$\sim 2\times 10^{-5}\,(\Sigma_{s}/1\,\mathrm{g\,cm}^{-2})\,\Mearth\,\mathrm{yr}^{-1}$.
These rates refer to scattered objects that have elliptical orbits about the star.
Arguably, these scattering rates  may only apply for a limited period
of time, if solids are not replenished via collisions or some other mechanism
(e.g., gravitational stirring or scattering by other planets).
Therefore, the minimum masses that can be scattered out of boundaries, 
inside and outside the planet's orbit are 
$\sim 0.25\,(\Sigma_{s}/1\,\mathrm{g\,cm}^{-2})\,\Mearth$,
$\sim 0.3\,(\Sigma_{s}/1\,\mathrm{g\,cm}^{-2})\,\Mearth$, and
$\sim 0.1\,(\Sigma_{s}/1\,\mathrm{g\,cm}^{-2})\,\Mearth$, respectively
(neglecting contributions from the mass in the corotation region).

For the disk temperatures applied here, both icy and mixed-composition 
bodies would be ablated inside $r\approx 2.8\,\AU$ ($T_{g}=T_{n}\approx 220\,\K$). 
However, disk models \citep[e..g,][]{dalessio2005} suggest lower temperatures
after a few to several million years, hence ice-rich planetesimals may survive
at radii $r\lesssim 2\,\AU$. Regardless, scattering by a Jupiter-mass planet
provides an important source of hydrated planetesimals to inner disk regions.
A mass equal to the current mass of the main asteroid belt \citep{krasinsky2002}
would be delivered in $\sim 50\,(1\,\mathrm{g\,cm}^{-2}/\Sigma_{s})\,\mathrm{yr}$. 
These bodies would still orbit in a relatively dense gas, but the orbital decay time
of $R_{s}\gtrsim 1\,\mathrm{km}$ planetesimals around $2.5\,\AU$
would be $\gtrsim 5\times 10^{5}$ local orbital periods for 
$\rho_{0}\approx 10^{-11}\,\mathrm{g\,cm}^{-3}$. 
Orbital eccentricities and inclinations would be damped on timescales shorter 
by factors of $\sim 200$.

The planetesimals orbiting in the corotation region $a_{p}\pm\Rhill$ are removed 
at an average rate of
$\sim 6\times 10^{-6}\,(\Sigma_{s}/1\,\mathrm{g\,cm}^{-2})\,\Mearth\,\mathrm{yr}^{-1}$.
However, the tadpole orbits around the L$_{4}$ and L$_{5}$ points are very stable,
due to low gas densities (see Figure~\ref{fig:img_disk}),
and number densities do not drop around these points.
Since the rate of capture in the corotation region appears to be much smaller than 
the removal rate, the local density may bear information about dynamical and physical 
conditions at earlier times, before the giant planet acquires its massive envelope 
\citep[see also][]{peale1993}.

The steep radial pressure gradient induced by the planet at the edges 
of the gap in the gas density profile (as function of $r$, see 
Figure~\ref{fig:vphiu}) can partially prevent small planetesimals 
($R_{s}\sim 0.1\,\mathrm{km}$) from crossing the planet's orbit, 
but this effect reduces as gas dissipates (see Figure~\ref{fig:solid_gap}). 
The size range most affected is determined by the strength of the drag 
acceleration (Equation~(\ref{eq:aD})), and hence by $\partial P_{g}/\partial r$ 
at the gap's outer edge.  
A condition for gap formation derived from the balance of viscous and 
tidal torques \citep[and references therein]{gennaro2012} is
\begin{equation}
\frac{1}{\sqrt{3\pi\alpha_{g}}}\left(\frac{\Mp}{\Ms}\right)%
\left(\frac{a_{p}}{H}\right)\left(\frac{a_{p}}{\widetilde{\Delta}}\right)^{3/2}\gtrsim 1,
\label{eq:gap_con}
\end{equation}
where $\widetilde{\Delta}=\max{(H,\Rhill)}$.
The value of the left-hand side of this inequality is $\approx 5.6$ for the 
parameters adopted here.
A similar number is obtained for a Saturn-mass planet and somewhat smaller
values for $H/r$ and $\alpha_{g}$, which suggests that a reduction in the inward 
flux of $R_{s}\sim 0.1\,\mathrm{km}$ (and smaller) planetesimals 
may begin prior to reaching the current mass of Jupiter.

Experiments conducted on $1\,\mathrm{cm}\le R_{s}\le 10\,\mathrm{m}$ 
bodies, initially released exterior of the planet's orbit ($1.2<a_{s}/a_{p}<1.25$), 
show that these particles remain segregated. 
After $\sim 600$ orbital periods of the planet, results indicate that 
the amount of solids delivered to the interior disk is negligible
(for $\rho_{0}=10^{-12}\,\mathrm{g\,cm}^{-3}$)
or virtually zero (for $\rho_{0}=10^{-11}\,\mathrm{g\,cm}^{-3}$). 
The radial position of the swarm's inner edge is between 
$r\approx 1.27\,a_{p}$ and $1.4\,a_{p}$ for 
$1\,\mathrm{cm}\lesssim R_{s}\lesssim 10\,\mathrm{cm}$, 
and between $r\approx 1.36\,a_{p}$ and $1.4\,a_{p}$ for 
$1\,\mathrm{m}\lesssim R_{s}\lesssim 10\,\mathrm{m}$ (the position also 
depends on $\rho_{0}$).
For $\rho_{0}=10^{-11}\,\mathrm{g\,cm}^{-3}$, particles of $1\,\mathrm{cm}$ 
in radius are halted at $r\approx 1.18\,a_{p}$, close to the peak of
super-Keplerian rotation in Figure~\ref{fig:vphiu}.
Segregation also leads to negligible fluxes of these bodies toward the
circumplanetary disk. When $\rho_{0}=10^{-12}\,\mathrm{g\,cm}^{-3}$, 
we do find that $10\,\mathrm{m}$-bodies can be scattered toward 
the planet and the inner disk when $a_{s}\approx 1.2\,a_{p}$ and 
$e_{s}\approx i_{s}\approx 0$. However, this scattering event lasts 
only briefly at the beginning of the calculation, before the swarm 
recedes. Therefore, this is likely a transient effect induced by the
choice of the initial distributions. Nonetheless, if $R_{s}\sim 10\,\mathrm{m}$ 
bodies are produced via collisional comminution of planetesimals 
around $r=1.2\,a_{p}$, part of them may be scattered inward.

Planetesimals can be captured in the circumplanetary disk with 
a wide range of (planetocentric) orbits, including retrograde ones
(see Figures~\ref{fig:h_cpd} and \ref{fig:cpd_2dh}) as also found 
by other recent studies \citep[e.g.,][]{fujita2013,tanigawa2014}.
The ensemble of bodies with retrograde orbits comprises planetesimals 
with radii $0.1\,\mathrm{km} \lesssim R_{s}\lesssim 100\,\mathrm{km}$,
although most of those coming from the exterior disk have
$R_{s}\gtrsim 10\,\mathrm{km}$.
Capture of planetesimals provides the circumplanetary disk with 
a time-averaged solids' reservoir of 
$\sim 10^{-3}\,(\Sigma_{s}/1\,\mathrm{g\,cm}^{-2})\,\Mearth$, which may 
be considered as a balance between the external supply and the loss 
due to ejection, ablation, break-up, and accretion on the planet.
This amount of solids would account for a relatively low surface density
(although planetesimals are continuously supplied), $\sim 0.3\,\Sigma_{s}$,
which may indicate relatively long times,
$\sim 10^{7}\,(1\,\mathrm{g\,cm}^{-2}/\Sigma_{s})$ local orbital periods
about the planet, for the formation of $\sim 10^{3}\,\mathrm{km}$-radius 
satellites. 
Applied to the Galilean satellites, these formation times appear compatible 
with Callisto, which is partially undifferentiated \citep{stevenson1986}, and 
may suggest post-formation differentiation of the inner three satellites 
\citep{schubert2004}.
Type~I migration due to tidal interactions \citep[e.g.,][]{tanaka2002}
with the thick circumplanetary disks considered here (see 
Figures~\ref{fig:img_sub} and \ref{fig:prof_sub}) would lead 
to timescales for the orbital decay, at $\tilde{r}\sim 0.04\,\Rhill$, of
$\sim 10^{9}\,(10^{-12}\,\mathrm{g\,cm}^{-3}/\rho_{0})$ 
local orbital periods, longer than formation timescales. 

Sustained ablation close to the planet ($\tilde{r}\lesssim 0.1\,\Rhill$) releases
heavy elements in the gas at a rate of
$\sim 10^{-7}\,(\Sigma_{s}/1\,\mathrm{g\,cm}^{-2})\,\Mearth\,\mathrm{yr}^{-1}$,
which is large enough to significantly alter the gas metallicity over relatively
short timescales, possibly leading to a dust laden system. 
Disruption of planetesimals may also contribute to the solids' content 
of the circumplanetary disk.

Equation~(\ref{eq:mzdot}) approximates the accretion of solids on the planet
supplied by a mono-size swarm of planetesimals, with radius in the range from
$\sim 1\,\mathrm{km}$ to $\sim 100\,\mathrm{km}$, where $\Sigma_{s}$ is 
the solids' surface interior and exterior of the planet's orbit.
Figure~\ref{fig:h_iacc} indicates that the efficiency of accretion of planetesimals
declines with increasing distance from the planet's orbit. If the edges of the gap 
in the solids' distribution are eroded, because of lack of supply or because 
they recede due to gas drag torques, the accretion rate is expected to decrease.
Probably, late accretion of solids only represents a minor addition to
the heavy element content of a giant planet 
(unless $\Sigma_{s}$ is still quite large).

We estimate the mean accretion energy per unit mass, 
$\langle\Delta E_{\mathrm{acc}}/\Delta M_{s}\rangle$, 
delivered to the planet by accreted planetesimals during the course of 
the calculations. The quantity $\Delta E_{\mathrm{acc}}$ contains both
kinetic and gravitational energy. We assume that all this energy is 
delivered close to the planet surface. 
The energy per unit time produced by accretion of solids is then 
\begin{equation}
\langle\Delta E_{\mathrm{acc}}/\Delta M_{s}\rangle\langle\dMp\rangle_{s}\sim%
10^{-5}\left(\frac{\Sigma_{s}}{1\,\mathrm{g\,cm}^{-2}}\right)L_{\odot}.
\label{eq:ezdot}
\end{equation}
This accretion power can be compared to the planet's luminosity due 
to envelope contraction, between 
$\sim 10^{-6}\,L_{\odot}$ and $\sim 10^{-4}\,L_{\odot}$ \citep{lissauer2009}.
In reality, Equation~(\ref{eq:ezdot}) gives only a lower limit to 
the accretion power since additional energy is released as planetesimals 
sink into the planet. In fact, while ice dissolves at relatively shallow depths, 
where the temperature is $\lesssim T_{\mathrm{cr}}\sim 650\,\K$, rock
(which makes $40$\% of the mass of mixed-composition planetesimals)
can sink to much deeper layers on account of the higher critical temperature 
($T_{\mathrm{cr}}=4500\,\K$ for quartz). 

In some instances ($1\,\mathrm{km}\lesssim R_{s}\lesssim 10\,\mathrm{km}$ 
and $\rho_{0}\sim 10^{-11}\,\mathrm{g\,cm}^{-3}$), planetesimal break-up in 
the circumplanetary disk produces significant amounts of solids that may 
increase $\langle\dMp\rangle_{s}$ by factors of up to a few, if debris is accreted 
before being completely ablated (the dissolution timescale is $\propto R_{s}$, 
see Equation~(\ref{eq:tabla})). 
However, if break-up produces large ($\sim 0.1\,\mathrm{km}$) fragments,
ablation appears to dominate over accretion by a large margin: 
while $\sim 50$\% of their mass is ablated, only $\sim 0.01$\% 
is accreted.
To examine more in detail the fate of smaller fragments, 
we present tests that use $1\,\mathrm{cm}\le R_{s}\le 10\,\mathrm{m}$ 
bodies as a proxy.
These are released on circular orbits around the planet, between 
$\tilde{r}\approx 0.1\,\Rhill$ and $\approx 0.6\,\Rhill$, at the disk's equator.
We find zero or negligible accretion.
For radii $1\,\mathrm{m}\lesssim R_{s}\lesssim 10\,\mathrm{m}$, planetesimals
migrate inward very quickly, but not enough to overcome ablation. 
Essentially, they are all ablated.
At the reference density $\rho_{0}\sim 10^{-12}\,\mathrm{g\,cm}^{-3}$, 
disruption of planetesimals releases less mass, but fragments can 
still be produced via collisional comminution.
The same tests reveal similar conclusions.
Bodies with radii $1\,\mathrm{m}\lesssim R_{s}\lesssim 10\,\mathrm{m}$ are
almost entirely ablated without any significant amount being accreted. 
For comparison, $R_{s}\sim 0.1\,\mathrm{km}$ planetesimals shed in the gas 
via ablation $\approx 30$ times the mass they deliver to the planet via accretion.

Solids in the range $1\,\mathrm{cm}\lesssim R_{s}\lesssim 10\,\mathrm{cm}$,
for both values of $\rho_{0}$, are also much more prone to ablation than 
they are to accretion, if they move toward the planet. In the tests, none
of these particles is accreted.
In either case, the mass that is not ablated remains beyond 
$\tilde{r}\approx 0.1\,\Rhill$ for the duration of the calculations, possibly 
because these small solids are more efficiently coupled to the gas than 
are larger particles and the gas radial velocity 
$\mathbf{\tilde{r}}\mathbf{\cdot}\mathbf{v}_{g}/\tilde{r}$
(where $\mathbf{v}_{g}$ is relative to the planet)
is positive at the equator, i.e., directed away from the planet
\citep[see also][]{tanigawa2012}.
The conclusion is that small fragments resulting from disruption should 
not significantly contribute to accretion, but should rather contribute to 
the local reservoir of solids and to enriching the gas with heavy elements.

The main limitation of this study is the lack of planetesimal-planetesimal 
interactions, especially in the circumplanetary disk (see Section~\ref{sec:ECR}), 
which could affect the distribution of solids but which allows 
us to rescale the outcomes of the calculations to different values of 
the initial surface density of solids. 
The relatively short time span covered by the models is also a limiting 
factor.
Another limitation is obviously the ``discrete'' approach, i.e., that of treating
each particle as an individual body, which prevents from dealing with more
realistic swarms of planetesimals, in terms of both number densities and
size distributions. However, this approach allows us to model the evolution 
of the thermodynamical properties of single planetesimals at levels of detail 
not accessible to other, e.g., statistical or hybrid, approaches.
Therefore, the method applied here can complement other techniques by
providing detailed information on restricted populations of planetesimals 
at selected epochs of evolution.

\acknowledgments

We wish to express our gratitude to Jack Lissauer and Peter Bodenheimer
for their valuable feedback. 
We thank an anonymous referee for prompt and constructive comments.
G.D.\ acknowledges support from NASA Outer Planets Research Program
grant 202844.02.02.01.75 and from NASA Origins of Solar Systems Program 
grants NNX11AD20G, NNX11AK54G, and NNX14AG92G.
Resources supporting this work were provided by the NASA High-End
Computing (HEC) Program through the NASA Advanced Supercomputing
(NAS) Division at Ames Research Center.

\appendix

\section{The Drag Coefficient}
\label{sec:CD}

\begin{figure*}[t!]
\centering%
\resizebox{\figlen}{!}{%
\includegraphics[clip]{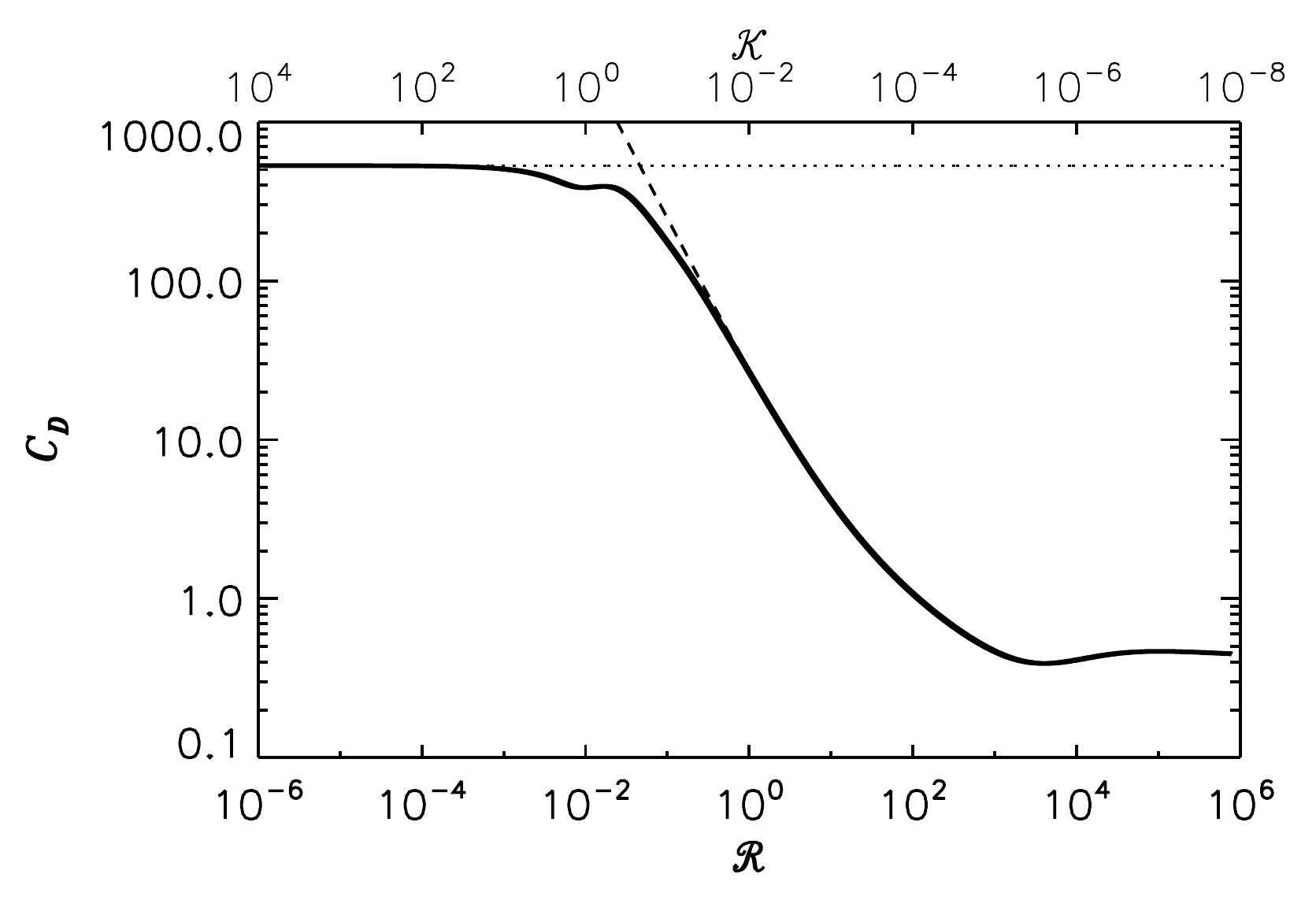}%
\includegraphics[clip]{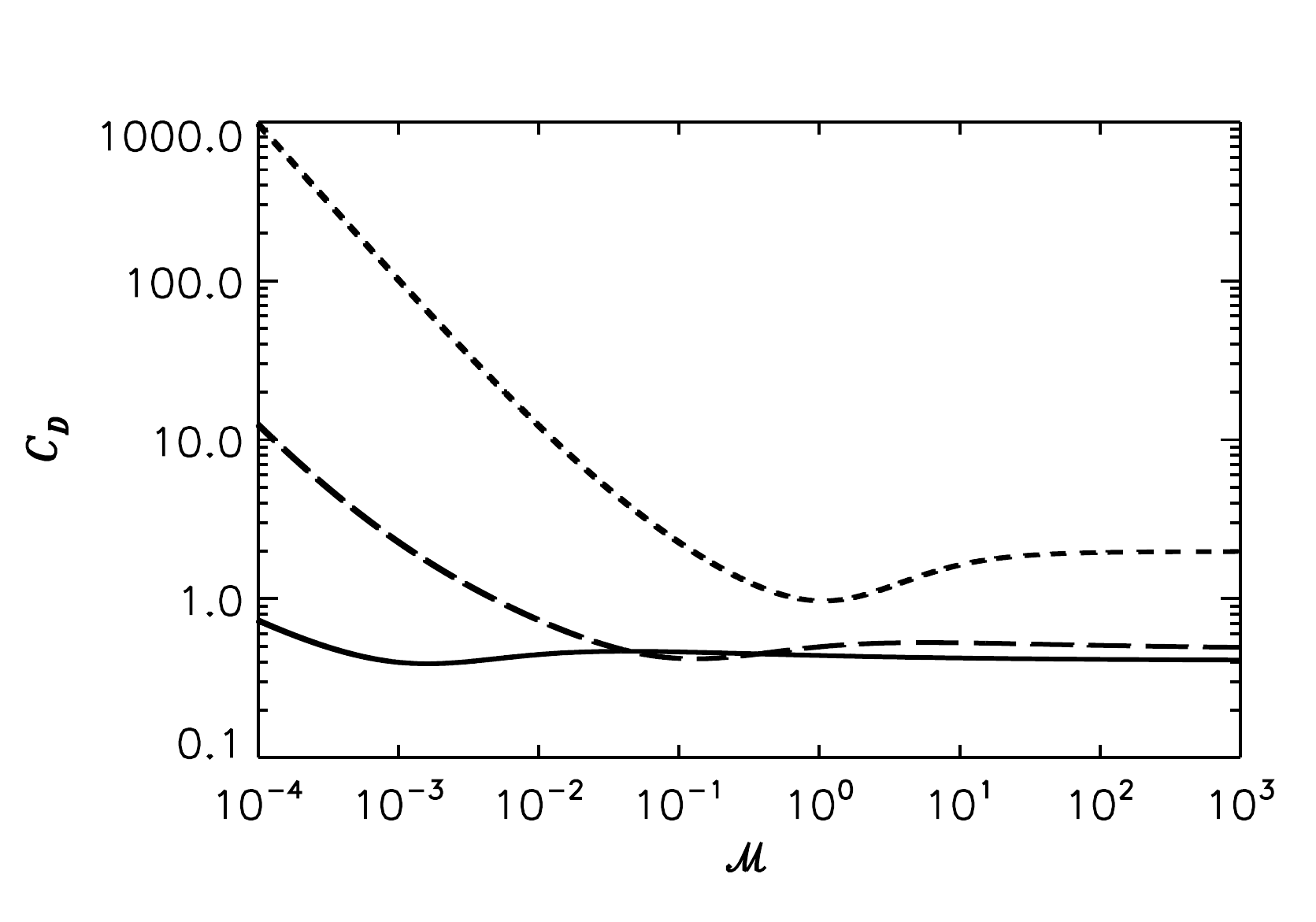}}
\resizebox{\figlen}{!}{%
\includegraphics[clip]{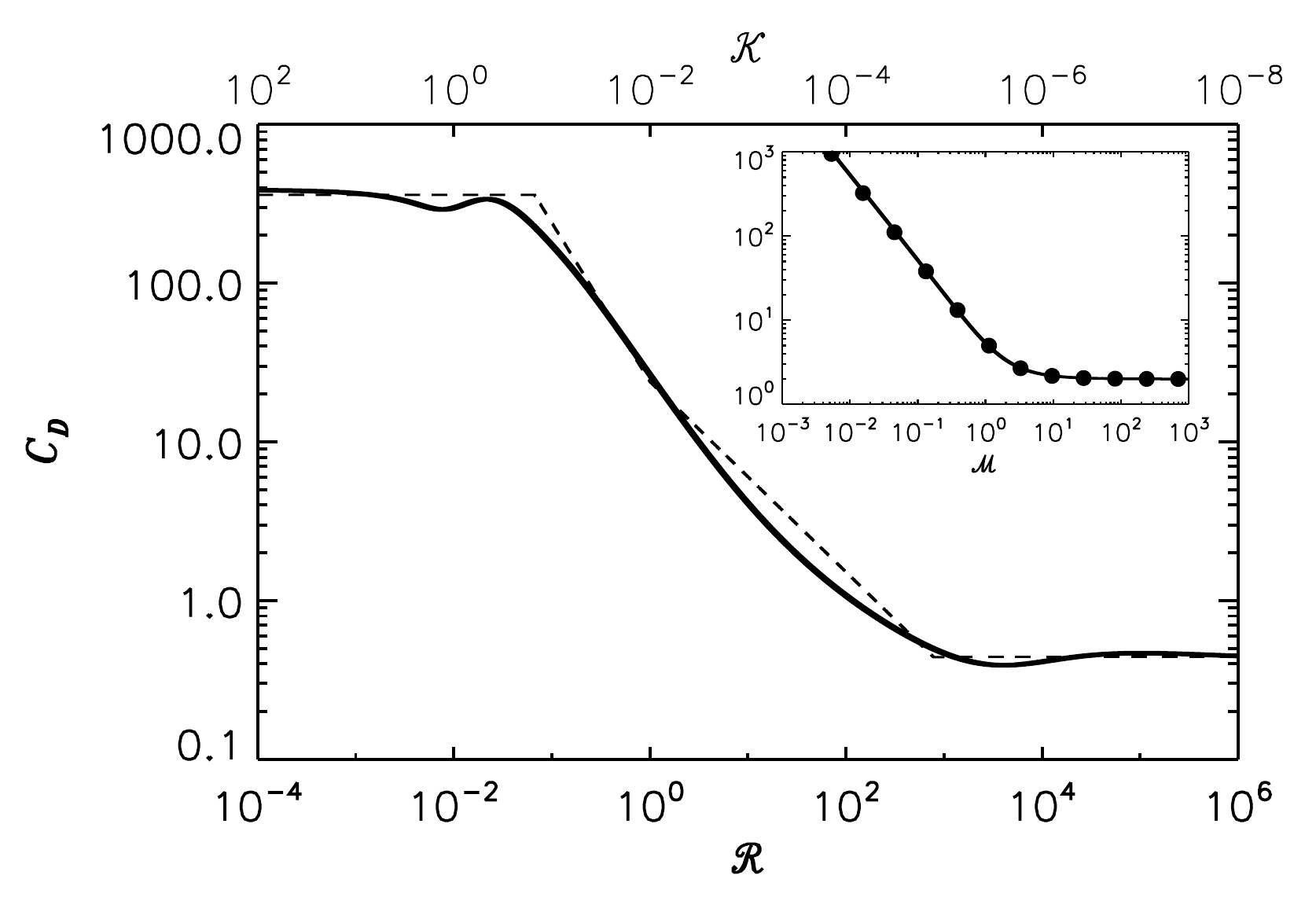}%
\includegraphics[clip]{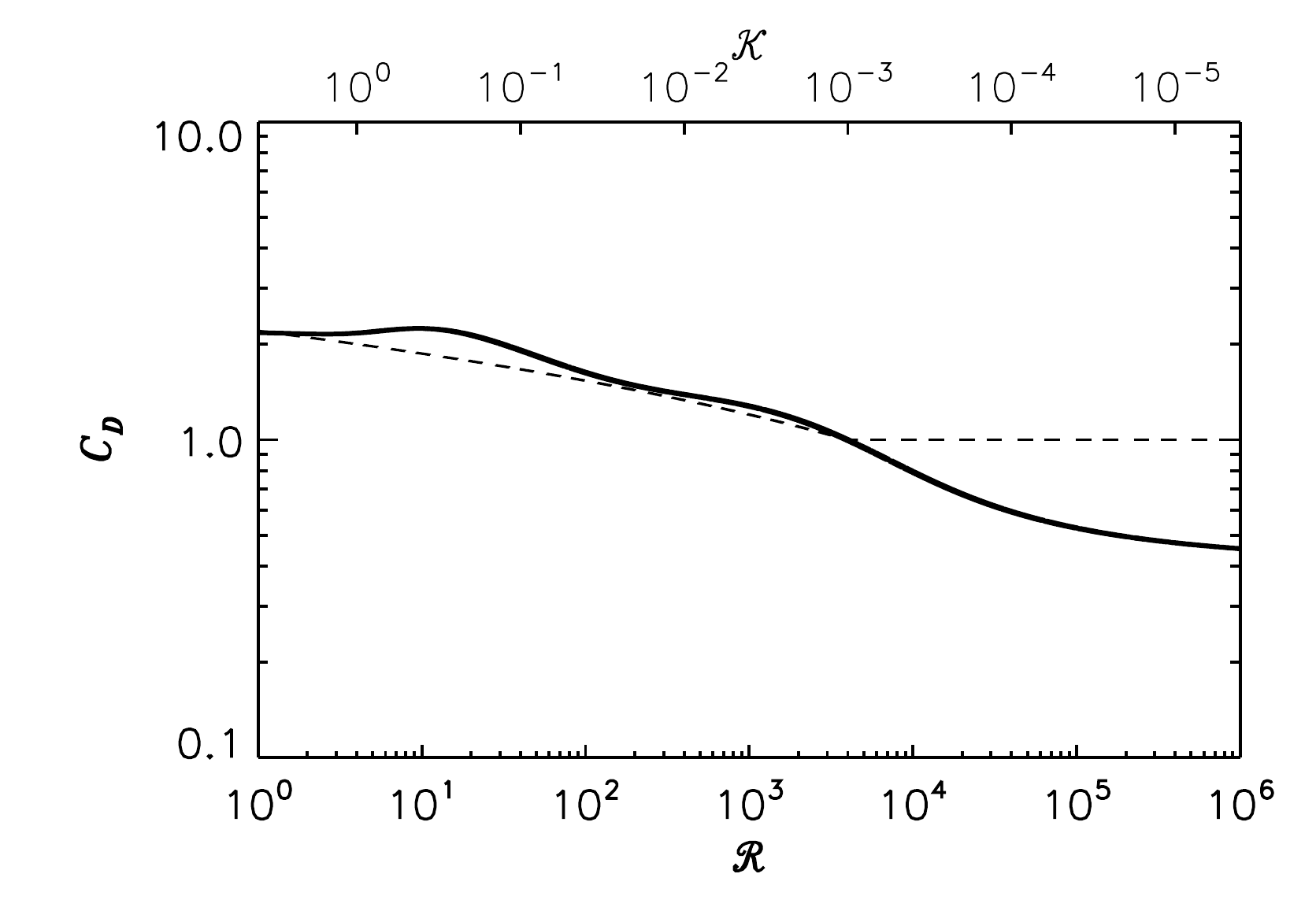}}
\caption{%
             Top-left: the drag coefficient, $C_{D}$, given in
             Equation~(\ref{eq:CD}), versus the
             Reynolds number, $\mathcal{R}$ (bottom axis), and 
             $\mathcal{K}=\mathcal{M}/\mathcal{R}$ (top axis), for
             $T_{s}=T_{g}$ and a Mach number $\mathcal{M}=0.01$. 
             The dotted line is the limit for $\mathcal{K}\gg 1$,
             given in Equation~(\ref{eq:CEC}).
             The dashed line is the limit for $\mathcal{K}\ll 1$, given in 
             Equation~(\ref{eq:CS}).
             Top-right: $C_{D}$ as a function of the Mach number for
             particle of different radius: $10^{7}\,\mathrm{cm}$ (solid line),
             $10^{5}\,\mathrm{cm}$ (long-dashed line), and
             $10^{3}\,\mathrm{cm}$ (short-dashed line).
             Bottom-left: The solid line is $C_{D}$ from Equation~(\ref{eq:CD})
             while the dashed line is the drag coefficient used by
             \citet{whipple1973} and \citet{weidenschilling1977b}
             ($T_{s}$ is set to zero in Equation~(\ref{eq:CD}), 
             as discussed in the text). The inset shows a comparison
             with the drag coefficient used, among others, by
             \citet{stalder1951,probstein1968,hood1991,tedeschi1999,liffman2000}
             (solid circles), which applies for $\mathcal{K}\gg 1$.
             Bottom-right: as in the bottom-left panel, but the dashed line is 
             the coefficient for $\mathcal{R}>1$ and $\mathcal{M}>1$ 
             used by \citetalias{podolak1988}.
             }
\label{fig:cda1}
\end{figure*}

\citet{melosh2008} performed an extensive study of the literature
on existing gas drag experiments 
\citep[see, e.g.,][and references therein]{walsh1976}. 
They derived an expression for
the drag coefficient $C_{D}$, in Equation~(\ref{eq:FD}) and 
(\ref{eq:aD}), as a function of the Mach number $\mathcal{M}$ 
(see Equation~(\ref{eq:Mn})), the Reynolds number $\mathcal{R}$ 
(see Equation~(\ref{eq:Rn})), and their ratio 
$\mathcal{K}=\mathcal{M}/\mathcal{R}$. The function is continuous 
and extends over the entire (plausible) ranges of $\mathcal{M}$ and 
$\mathcal{R}$.

Let us introduce the adiabatic gas sound speed
\begin{equation}
c_{g}=\sqrt{\gamma_{g}\frac{k_{\mathrm{B}}T_{g}}{\mu_{g}m_{\mathrm{H}}}},
\label{eq:cg}
\end{equation}
where $\gamma_{g}$ is the adiabatic index of the gas (the isothermal 
sound speed is obtained for $\gamma_{g}=1$) and the mean thermal 
velocity of the gas \citep[e.g.,][]{m&m}
\begin{equation}
\bar{\mathcal{V}}_{g}=\sqrt{\frac{8}{\pi}%
                             \frac{k_{\mathrm{B}}T_{g}}{\mu_{g}m_{\mathrm{H}}}},
\label{eq:vth}
\end{equation}
which is the equivalent of Equation~(\ref{eq:vvap}) for the gas 
constituents (atoms and/or molecules). Let us now define the magnitude 
of the relative velocity between the gas and a solid particle as 
$u=|\mathbf{v}_{g}-\mathbf{v}_{s}|$, then the (relative) Mach number 
can be written as
\begin{equation}
\mathcal{M}=\sqrt{\frac{8}{\pi\gamma_{g}}}%
                      \left(\frac{u}{\bar{\mathcal{V}}_{g}}\right).
\label{eq:Mn1}
\end{equation}

The definition of the relative Reynolds number in Equation~(\ref{eq:Rn})
involves the dynamical molecular viscosity of the gas. If interactions
among gas atoms/molecules can be described as collisions between
two rigid elastic spheres, an approximation of the dynamical molecular 
viscosity is \citep{m&m}
\begin{equation}
\eta_{g}=\frac{5\sqrt{2}}{64}\left(\frac{m_{\mathrm{H}}}{d^{2}_{\mathrm{H}}}\right)%
        \mu_{g}\bar{\mathcal{V}}_{g},
\label{eq:eta}
\end{equation}
where  $m_{\mathrm{H}}$ is the hydrogen mass and
$d_{\mathrm{H}}$ is the typical diameter of the gas constituents.
This length is $d_{\mathrm{H}}=2.71\times 10^{-8}\,\mathrm{cm}$
for hydrogen molecules and $2.15\times 10^{-8}\,\mathrm{cm}$ 
for helium \citep{CRC92}.
Although the interaction model based on the rigid sphere representation
of gas constituents, which interact only upon ``contact'', is rather simple
\citep[see discussion in][]{m&m}, Equation~(\ref{eq:eta}) agrees 
within $25$\% with molecular hydrogen viscosity data (and $20$\% with 
helium) in the temperature range from $100$ to $600\,\K$ \citep{CRC92}.

By substituting Equations~(\ref{eq:eta}) and (\ref{eq:Mn1}) into 
Equation~(\ref{eq:Rn}), one finds that the (relative) Reynolds 
number can be cast into the following form
\begin{equation}
\mathcal{R}=\frac{32\sqrt{\pi}}{5}\left(\frac{d^{2}_{\mathrm{H}}}{m_{\mathrm{H}}}\right)%
\left(\frac{\sqrt{\gamma_{g}}}{\mu_{g}}\right) \rho_{g}R_{s}\mathcal{M}.
\label{eq:Rn1}
\end{equation}
The ratio $\mathcal{K}$ of the Mach number to the Reynolds number 
can also be written as
\begin{equation}
\mathcal{K}=\frac{5}{32\sqrt{\pi}}\left(\frac{m_{\mathrm{H}}}{d^{2}_{\mathrm{H}}}\right)%
\left(\frac{\mu_{g}}{ \rho_{g}R_{s}\sqrt{\gamma_{g}}}\right).
\label{eq:Kn1}
\end{equation}
It is important to note that $\mathcal{K}$ is proportional 
to the Knudsen number, which is defined as the ratio between 
the mean-free path of a gas atom/molecule and the particle 
diameter. Therefore, $\mathcal{K}$ may be regarded as 
a modified Knudsen number.
The proportionality factor depends on the form 
adopted for the dynamical viscosity 
$\eta_{g}$ \citepalias[see][]{podolak1988}. 
In our case, Equation~(\ref{eq:eta}) yields a proportionality factor 
equal to 
$(16/5)\sqrt{\gamma_{g}/(2\pi)}\simeq 1.28\sqrt{\gamma_{g}}$.

In the derivation of \citet{melosh2008}, the drag coefficient is 
written as
\begin{equation}
C_{D}=2+\left(C_{\mathrm{S}}-2\right)%
               e^{-p_{1}\sqrt{\gamma_{g}}\mathcal{K}G(\mathcal{R})}%
             + C_{\mathrm{E}}\,e^{-1/(2\mathcal{K})},
\label{eq:CD}
\end{equation}
where
\begin{equation}
C_{\mathrm{E}}=
  \frac{1}{\sqrt{\gamma_{g}}\mathcal{M}}%
  \left(\frac{4.6}{1+\mathcal{M}}+1.7\sqrt{\frac{T_{s}}{T_{g}}}\,\right),
\label{eq:CE}
\end{equation}
and the auxiliary function $G(\mathcal{R})$ is such that
\begin{equation}
\log{G}=\frac{2.5\left(\mathcal{R}/312\right)^{p_{2}}}{1+\left(\mathcal{R}/312\right)^{p_{2}}}.
\label{eq:CDG}
\end{equation}
The constants $p_{1}$ in Equation~(\ref{eq:CD}) and $p_{2}$ in
Equation~(\ref{eq:CDG}) are $p_{1}=3.07$ and $p_{2}=0.6688$.
The function $G(\mathcal{R})$ takes limiting values of $1$, for
$\mathcal{R}\rightarrow 0$, and of $10^{2.5}\approx 316.23$, for 
$\mathcal{R}\rightarrow \infty$.

For $\mathcal{K}\gg 1$, when the particle size is much smaller 
than the mean-free path of the gas constituents, a regime referred 
to as free-molecular flow, the drag coefficient takes the value
\begin{equation}
C_{D}\stackrel{\mathcal{K}\gg 1}{\longrightarrow} C_{\mathrm{E}}+2,%
\label{eq:CEC}
\end{equation}
which, for Mach numbers $\ll 1$, becomes 
$(4.6+1.7\sqrt{T_{s}/T_{g}})/\left(\sqrt{\gamma_{g}}\mathcal{M}\right)$,
as in the Epstein regime
\citep[e.g.,][]{whipple1973,weidenschilling1977b}.
We stress here that Equation~(\ref{eq:CE}) ought to be regarded 
as an extension of the Epstein drag coefficient
(see \citealp{hood1991} and discussion in \citealp{liffman2000}).
In fact, the form of Epstein coefficient typically
adopted in the literaure, $(8/3)\sqrt{8/(\pi\gamma_{g})}/\mathcal{M}$ 
\citep[e.g.,][and references therein]{supulver2000,chiang2010},
only applies when there is specular reflection of the gas constituents 
impinging on the particle \citep[see][for details]{epstein1924}, which 
corresponds to assuming $T_{s}=0$ in the limiting  expression above.
\citet[Part~I, Section~7]{epstein1924} also argued that, for 
$\mathcal{K}\gg 1$, particles should be considered as perfect thermal
conductors, i.e., $T_{s}=T_{g}$, and the drag coefficient is then
$(8/3+\pi/3)\sqrt{8/(\pi\gamma_{g})}/\mathcal{M}$, in agreement with
the limiting expression above.
At large Mach numbers, the right-hand side of Equation~(\ref{eq:CEC})
has asymptotic behavior 
$2+(1.7/\mathcal{M})\sqrt{T_{s}/(\gamma_{g}T_{g})}$ 
\citep[e.g.,][]{whipple1950,baker1959}.

For $\mathcal{K}\ll 1$, as happens in the continuum and incompressible 
($\mathcal{M}\ll 1$) flow regimes, the first exponential in Equation~(\ref{eq:CD}) 
tends to $1$ while the second exponential tends to $0$. 
Therefore, the drag coefficient takes the value
\begin{equation}
C_{D}\stackrel{\mathcal{K}\ll 1}{\longrightarrow} C_{\mathrm{S}}.
\label{eq:CSC}
\end{equation}
In this study, for $C_{\mathrm{S}}$, we use a formula suggested by 
\citet{brown2003}
\begin{equation}
C_{\mathrm{S}}=\frac{24}{\mathcal{R}}\left(1+0.15\mathcal{R}^{p_{3}}\right)%
                         +\frac{0.407\mathcal{R}}{\mathcal{R}+8710},
\label{eq:CS}
\end{equation}
in which the constant in the power of $\mathcal{R}$ is $p_{3}=0.681$.
For $\mathcal{R}\lesssim 1$, Equation~(\ref{eq:CS}) becomes the 
classical Stokes drag law $C_{\mathrm{S}}\approx 24/\mathcal{R}$
\citep[e.g.,][]{whipple1973,weidenschilling1977b,brown2003},
whereas, for $\mathcal{R}\gg 1$, we have that 
$C_{\mathrm{S}}\approx 0.407$, sometimes referred to as the
Newtonian drag coefficient \citep[e.g.,][]{whipple1973}.
For non-spherical shapes, e.g., a cube or a short cylinder, this
asymptotic value would be more than twice as large.

In the top panels of Figure~\ref{fig:cda1}, we plot the drag 
coefficient in Equation~(\ref{eq:CD}) versus the Reynolds number 
(Equation~(\ref{eq:Rn1})) and the modified Knudsen number
(Equation~(\ref{eq:Kn1})), and also display the two limiting 
cases in Equations~(\ref{eq:CEC}) and (\ref{eq:CS}). In the
right panel, $C_{D}$ is plotted for three different particle radii
versus the Mach number.
In the bottom panels, we make comparisons with drag 
coefficients used in previous studies (see figure caption for details), 
including the widely used coefficient for free-molecular flows
of \citet[their Equations~(A15) and (A17)]{stalder1951}.

\section{Tests on Solutions of the Particle Evolution}
\label{sec:TPE}

In this Appendix, we present tests of the ordinary differential equation 
solver applied to the system of eight Equations~(\ref{eq:rtpeq}),
(\ref{eq:momeqr}), (\ref{eq:momeqtet}), (\ref{eq:momeqphi}), 
(\ref{eq:dTsdt}), and (\ref{eq:dmsdt}) or (\ref{eq:dmsdt_crit}).
In order to make comparisons with compact analytic solutions,
in the various tests we solve a reduced system and discuss separately
dynamical problems (Equations~(\ref{eq:rtpeq}) (\ref{eq:momeqr}), 
(\ref{eq:momeqtet}), and (\ref{eq:momeqphi})) 
in Appendix~\ref{sec:DT} and thermodynamical problems 
(Equations~(\ref{eq:dTsdt}) and (\ref{eq:dmsdt}) or (\ref{eq:dmsdt_crit})) 
in Appendix~\ref{sec:TT}.

\subsection{Dynamics Tests}
\label{sec:DT}

\begin{figure*}[t!]
\centering%
\resizebox{\figlew}{!}{%
\includegraphics[clip]{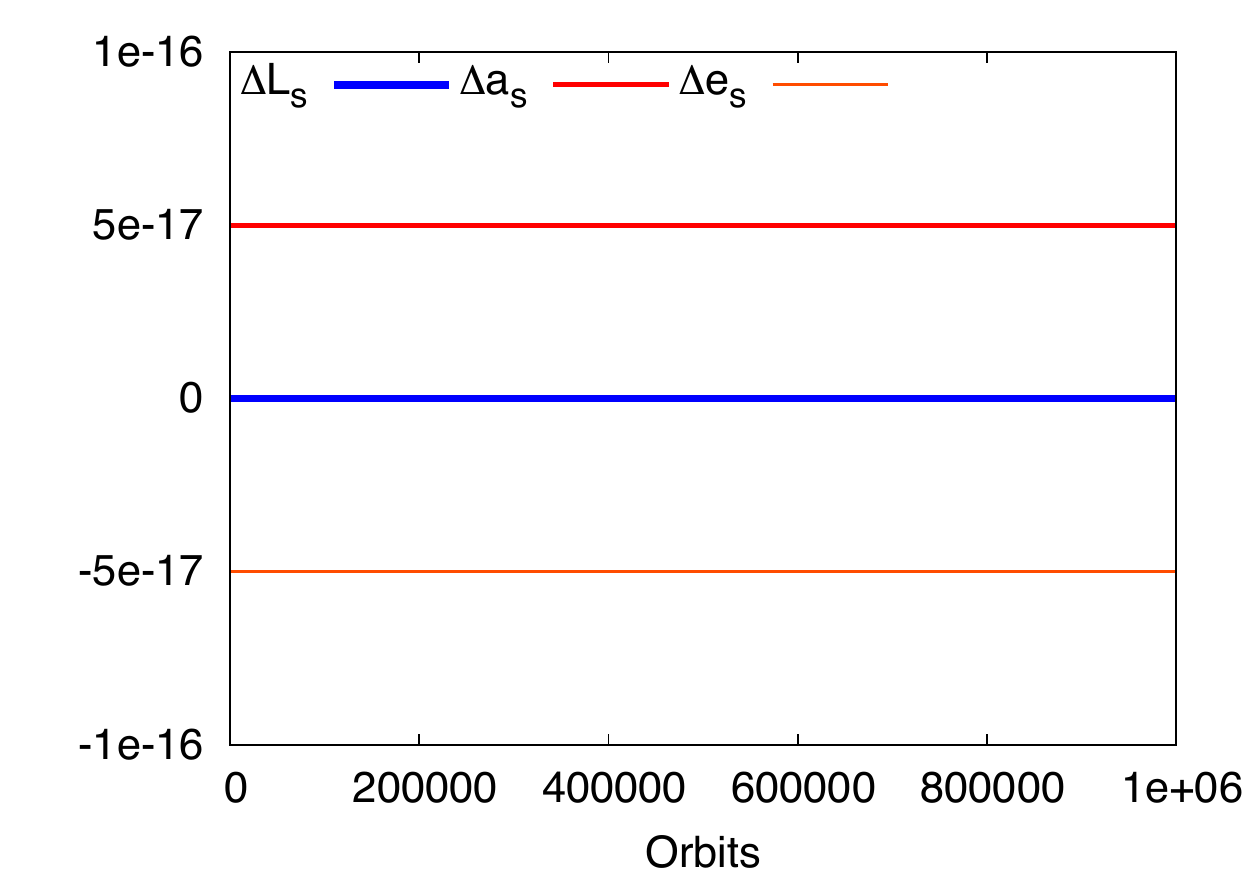}%
\includegraphics[clip]{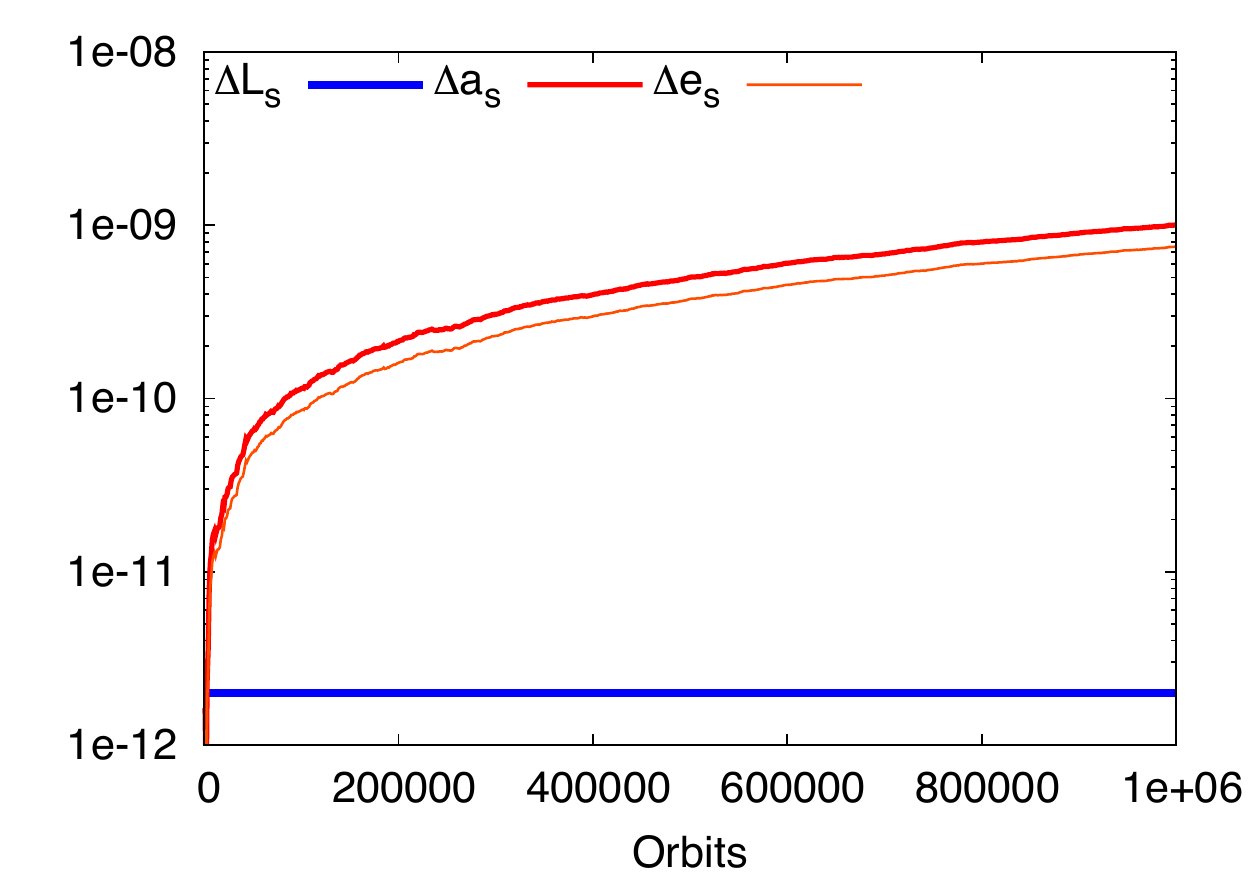}%
\includegraphics[clip]{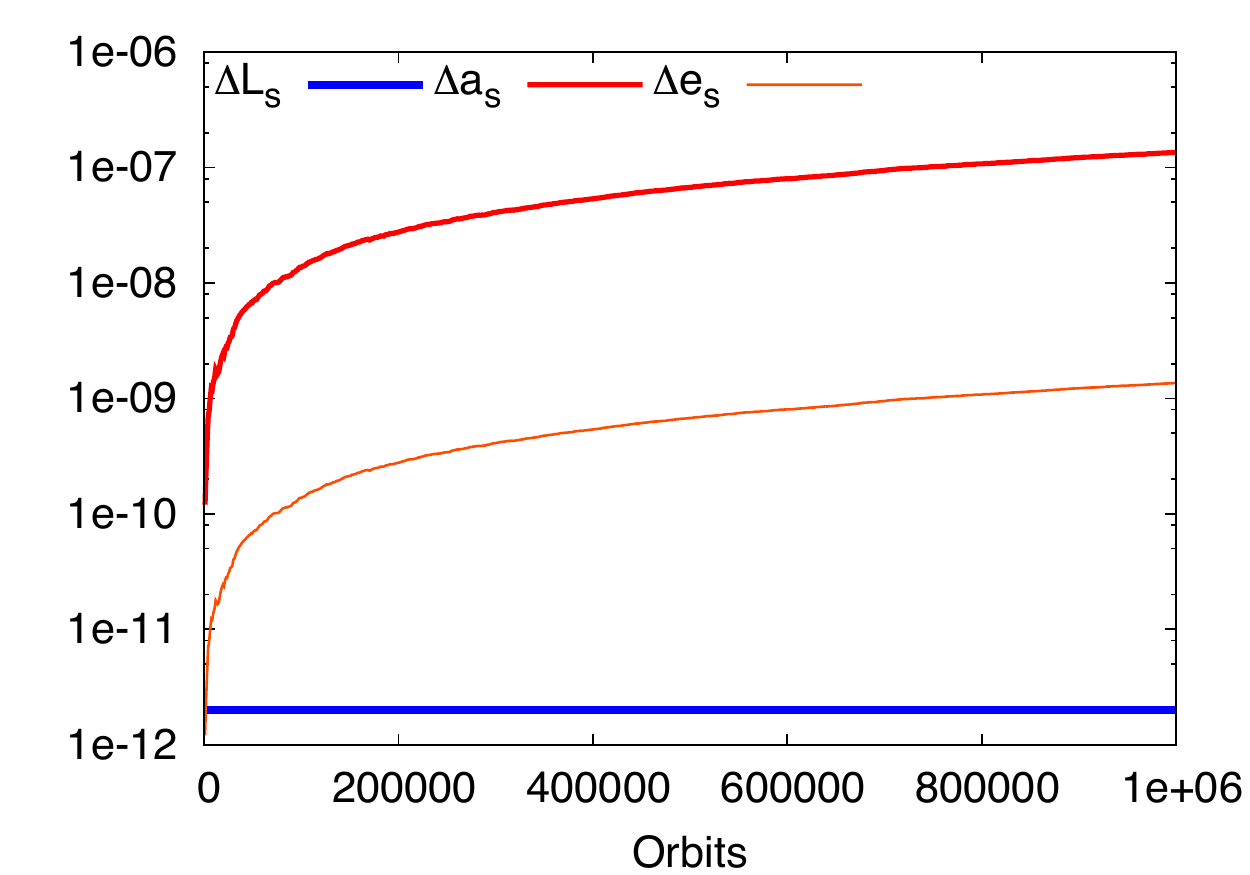}}
\caption{%
             Variations of orbital angular momentum, semi-major axis and
             eccentricity in two-body problems with initial eccentricites
             $0$ (left), $0.5$ (center), and $0.99$ (right). In the left panel,
             to separate the three curves, $\Delta a_{s}$ and $\Delta e_{s}$ 
             are shifted by an amount equal to $5\times 10^{-17}$. 
             In the center and right panels, $\Delta L_{s}$ is zero within 
             machine precision, hence it is shifted by $2\times 10^{-12}$ 
             to appear in the plot.
             }
\label{fig:test_tb}
\end{figure*}

The solver is first tested against standard two-body problems, in which
the particle orbits the star. Orbital energy and angular momentum are
expected to be conserved in these problems and the extent to which
this requirement is fulfilled provides an indication of accuracy.

Let us indicate with $\Ms$ the stellar mass, with $M_{s}$ the
particle mass, and with $a_{s}$ and $e_{s}$ the particle's semi-major 
axis and eccentricity. 
In a two-body problem, the orbital energy and angular momentum 
per unit mass are, respectively,
$E_{s}=-G\left(\Ms+M_{s}\right)/(2 a_{s})$ 
and
\begin{equation}
L_{s}=\sqrt{G\left(\Ms+M_{s}\right) a_{s} \left(1-e^{2}_{s}\right)},
\label{eq:Ls}
\end{equation}
where $G$ is the gravitational constant (here $L_{s}$ should not be confused
with the specific vaporization energy). Conservation of energy and
angular momentum  translates into constancy of $a_{s}$ and $e_{s}$ 
or of  $a_{s}$ and $L_{s}$. In fact, taking the differential of
Equation~(\ref{eq:Ls}) and dividing by $L^{2}_{s}$, we have
\begin{equation}
\frac{d L_{s}}{L_{s}}=\frac{1}{2}\frac{d a_{s}}{a_{s}}%
                                       -\left(\frac{e^{2}_{s}}{1-e^{2}_{s}}\right)%
                                       \frac{d e_{s}}{e_{s}},
\label{eq:dLs}
\end{equation}
which connects the relative variations of $\Delta L_{s}/L_{s}$,
$\Delta a_{s}/a_{s}$, and $\Delta e_{s}/e_{s}$
\citep[e.g.,][]{beutler2005}.

Experiments indicate that an advantage of integrating the equation 
of motion in terms spherical polar coordinates and angular momenta, 
in place of the usual cartesian positions and velocities, is a substantial 
improvement in conservation of angular momenta and, typically, 
of energy. In the calculations reported in Figure~\ref{fig:test_tb},
we consider orbits with eccentricities $e_{s}=0$ (left), $0.5$ (center), 
and $0.99$ (right). In all cases $L_{s}$ is conserved to machine precision,
whereas the expected error in energy for the most eccentric orbit is 
one part in $10^{4}$ over a period of $1\,\mathrm{Gyr}$, as the
asymptotic error is linear in time in that case \citep[e.g.,][]{calvo1993a}. 
If necessary, better conservtion can be obtained by constraining 
the internal time step of the solver at the expense of an increased 
run time.

\begin{figure*}
\centering%
\resizebox{\figlew}{!}{%
\includegraphics[clip]{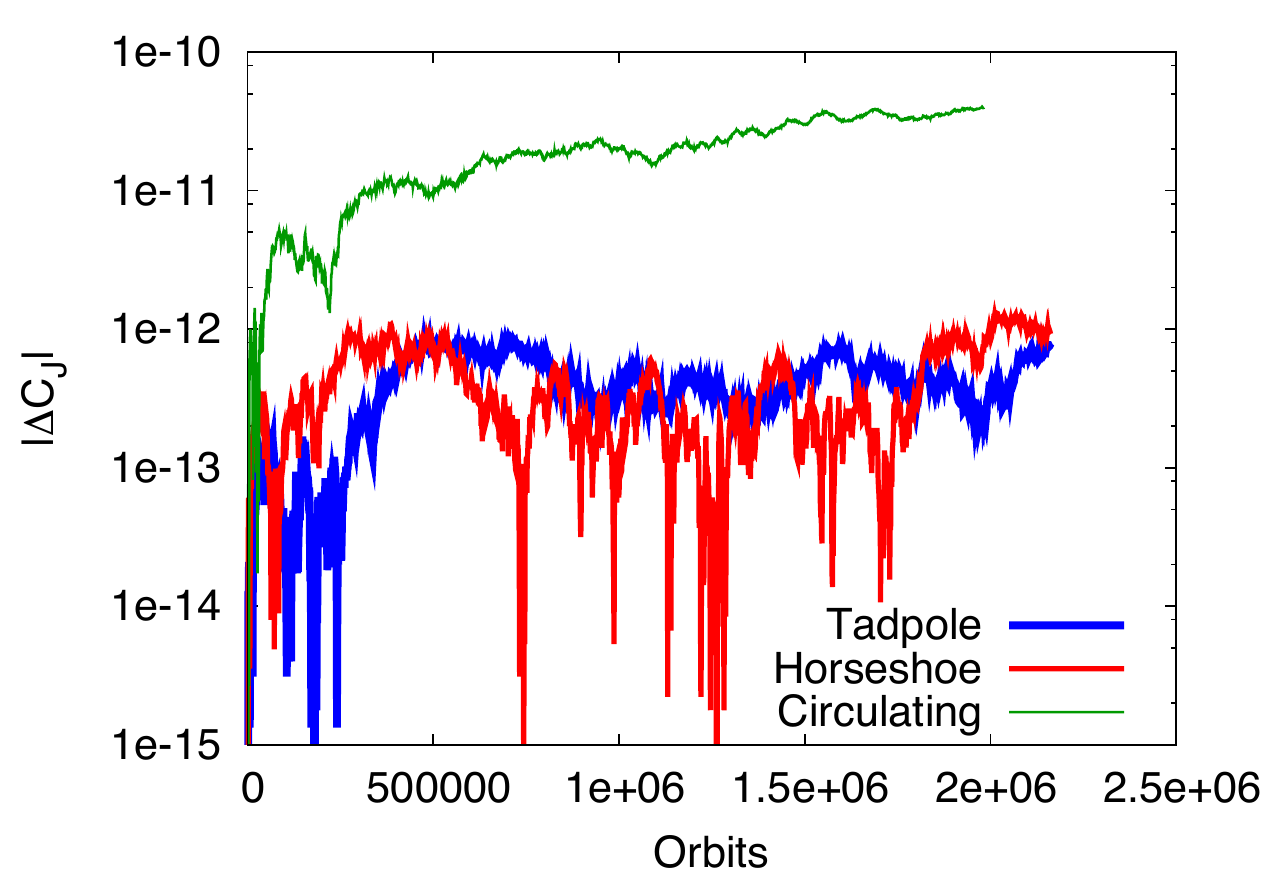}%
\includegraphics[clip]{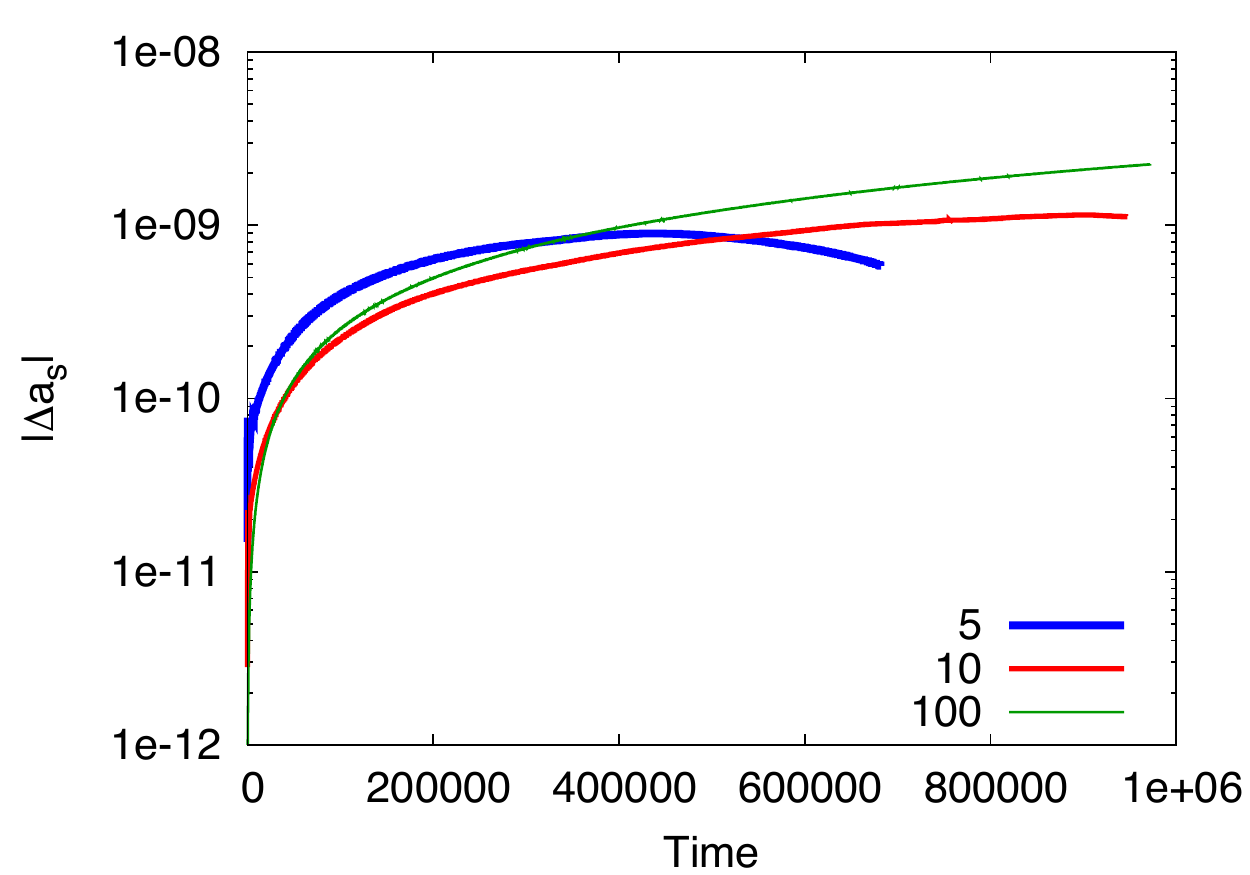}%
\includegraphics[clip]{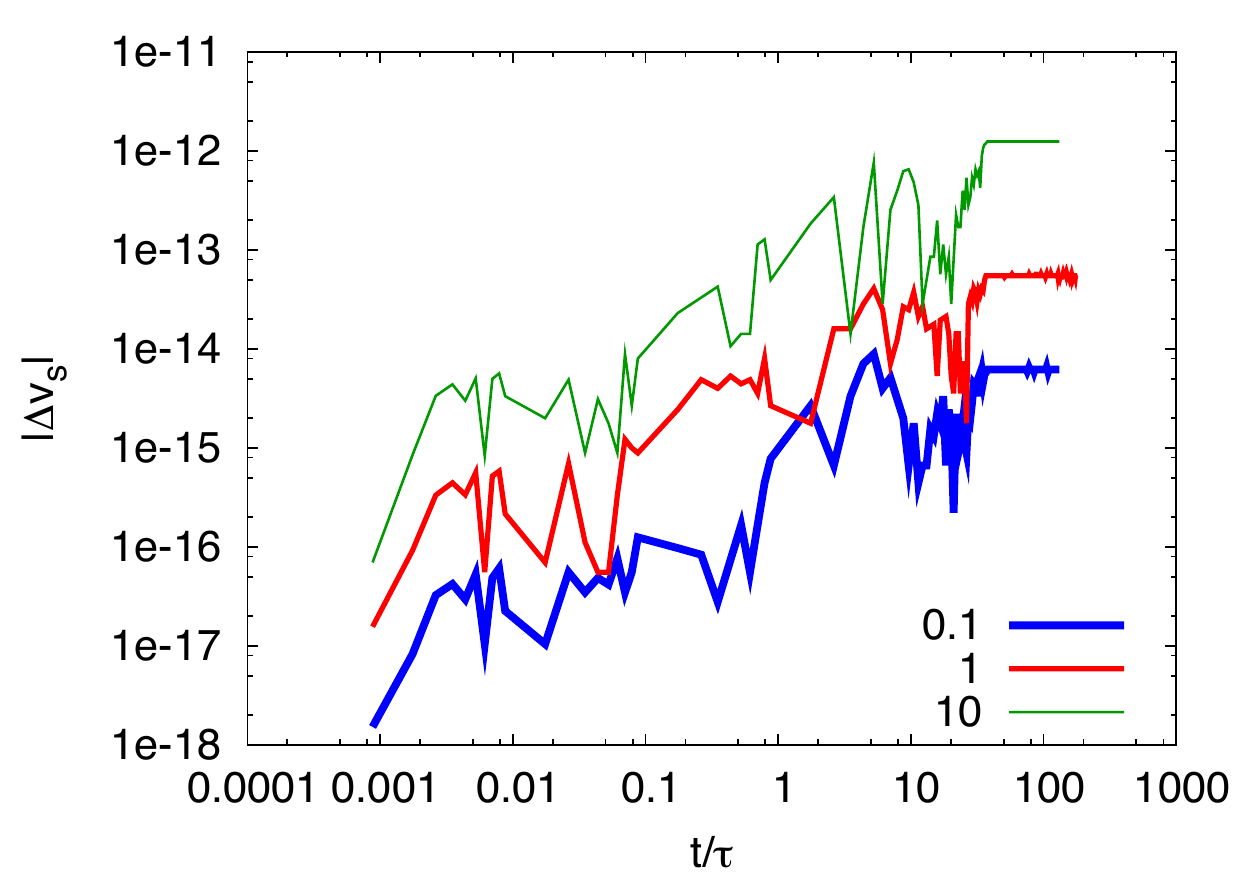}}
\caption{%
             Left. Variation of the Jacobi's integral (Equation~(\ref{eq:CJ})) 
             versus the orbital period of the massive bodies. 
             The three curves refer to particles on different types of orbits: 
             tadpole (thicker line) horseshoe, and circulating (thinner line).
             Center. Difference $|\Delta a_{s}|$ between calculated
             and predicted positions of a particle subject to gas drag
             versus time in units of $2\pi/\Omega_{0}$.
             The predicted position is given by Equation~(\ref{eq:aspeale})
             with parameter $b=0$. The three curves refer to different
             values of the constant $\Omega_{0}\tau$, as indicated in the
             legend ($\tau$ is defined after Equation~(\ref{eq:dasdtpeale})). 
             Right. Difference $|\Delta v_{s}|$ between calculated
             and predicted (Equation~(\ref{eq:vsff})) velocities of a free-falling 
             particle. The curves correspond to three values of the transient 
             time $\tau$ (defined after Equation~(\ref{eq:dvsdt})) in units of
             $1/\Omega_{0}$, as indicated in the legend.
             }
\label{fig:test_crtb}
\end{figure*}
Another test we discuss is a circular restricted three-body problem, 
constituted by two massive bodies, whose masses are $M_{1}$ and
$M_{2}$, and a massless particle. All bodies orbit in the same plane and 
the radius of the massive bodies' orbit is $a$.
The Jacobi's integral of motion for such system is \citep[e.g.,][]{murray2000}
\begin{equation}
C_{\mathrm{J}}=\left[\frac{G\left(M_{1}+M_{2}\right)}{a^{3}}\right] r^{2}%
                        +2\left(\frac{GM_{1}}{r_{1}}+\frac{GM_{2}}{r_{2}}\right)%
                        -v^{2},
\label{eq:CJ}
\end{equation}
where $r$ and $v$ are the distance and velocity of the particle relative
to the center of mass of the massive bodies, and $r_{1}$ and $r_{2}$
are the distances relative to these bodies.

In Figure~\ref{fig:test_crtb} (left), we set $M_{1}+M_{2}=1$ and 
$M_{2}/M_{1}=0.001$. The Jacobi's integral is plotted as a function 
of the orbital period of $M_{2}$ around $M_{1}$, for particles on 
tadpole (thicker curve), horseshoe, and circulating (thinner curve) 
orbits (see figure caption for details). 
While the error in the circulating orbit test displays a typical
asymptotic linear behavior (usually due to truncation errors in 
the algorithm), no systematic errors appear in the solutions for 
the tadpole and horseshoe orbits.

In order to test the solver in the presence of  drag, we follow the 
approach of \citet{peale1993}. Consider a particle orbiting a star
in gaseous disk. The rate of change of the particle's orbital energy
is equal to the work done on it in the inertial frame, that is
$M_{s}dE_{s}/dt=\mathbf{F}_{D}\mathbf{\cdot}\mathbf{v}_{s}$,
where $\mathbf{F}_{D}$ is the drag force given by Equation~(\ref{eq:FD}).
Differentiating the specific orbital energy $E_{s}$ (see above), 
we have
\begin{eqnarray}
\left[\frac{G\left(\Ms+M_{s}\right)}{2 a^{2}_{s}}\right]\left(\frac{d a_{s}}{dt}\right) &=&%
\frac{3}{8}\frac{C_{D}}{R_{s}}\left(\frac{\rho_{g}}{\rho_{s}}\right)%
|\mathbf{v}_{g}-\mathbf{v}_{s}|\nonumber\\%
& &\times%
\left(\mathbf{v}_{g}\mathbf{\cdot}\mathbf{v}_{s}-|\mathbf{v}_{s}|^{2}\right).
\label{eq:dedtpeale}
\end{eqnarray}
For the sake of simplicity, the drag coefficient is taken to be constant,
the disk's gas velocity is approximated as sub-Keplerian (due to support
provided by the pressure gradient) with no radial component, and 
$|\mathbf{v}_{s}|\approx a_{s}\Omega_{\mathrm{K}}$, where 
$\Omega^{2}_{\mathrm{K}}=G\left(\Ms+M_{s}\right)/a^{3}_{s}$. 
Hence, we have that
$|\mathbf{v}_{s}-\mathbf{v}_{g}|=%
a_{s}\Omega_{\mathrm{K}}(1-\sqrt{1-\xi^{2}})$.
The quantity $\xi$ is connected to the gradients of temperature and
surface density of the disk's gas, as well as to the disk's local 
thickness, $H/r$ \citep[see, e.g.,][]{peale1993,takeuchi2002,tanaka2002}, 
and is assumed to be constant. 
Under typical disk conditions, one finds that 
$\xi\sim H/r$ (see also Equation~(\ref{eq:rotacu})). 
If we indicate with $a_{0}$ and $\Omega_{0}$ the
initial values of $a_{s}$ and $\Omega_{\mathrm{K}}$, 
Equation~(\ref{eq:dedtpeale}) can be written as
\begin{equation}
\frac{d}{dt}\left(\frac{a_{s}}{a_{0}}\right)=%
-\frac{1}{\tau}\left(1-\sqrt{1-\xi^{2}}\right)^{2}%
\left(\frac{a_{0}}{a_{s}}\right)^{b}
\sqrt{\frac{a_{s}}{a_{0}}},
\label{eq:dasdtpeale}
\end{equation}
in which 
$1/\tau=(3/4)C_{D}(\rho_{g0}/\rho_{s})(a_{0}/R_{s})\Omega_{0}$
and $\rho_{g}=\rho_{g0}\left(a_{0}/a_{s}\right)^{b}$ with $b\ge 0$.
The solution of Equation~(\ref{eq:dasdtpeale}) is
\begin{equation}
\frac{a_{s}}{a_{0}}=%
\left[%
1-\left(\frac{1+2b}{2}\right)\left(1-\sqrt{1-\xi^{2}}\right)^{2}\left(\frac{t}{\tau}\right)%
\right]^{2/(1+2b)}.
\label{eq:aspeale}
\end{equation}

Our working assumptions imply that
$|da_{s}/dt|\ll a_{s}\Omega_{\mathrm{K}}(1-\sqrt{1-\xi^{2}})$
which, by using Equation~(\ref{eq:dasdtpeale}), becomes
$(a_{0}/\tau)(1-\sqrt{1-\xi^{2}})(a_{0}/a_{s})^{b-1/2}\ll%
 a_{s}\Omega_{\mathrm{K}}$, or equivalently
\begin{equation}
\frac{3}{4}C_{D}\left(\frac{\rho_{g}}{\rho_{s}}\right)%
\left(\frac{a_{0}}{R_{s}}\right)\left(1-\sqrt{1-\xi^{2}}\right)\left(\frac{a_{0}}{a_{s}}\right)^{b-1}\ll 1.
\label{eq:peale_con}
\end{equation}
If $\xi^{2}\ll 1$ and $0\le b\lesssim 1$, the inequality~(\ref{eq:peale_con}) becomes
$(3/8)C_{D}(\rho_{g}/\rho_{s})(a_{0}/R_{s})\xi^{2}\ll 1$.

In the center panel of Figure~\ref{fig:test_crtb}, we show results for
the orbital evolution of a particle initially moving on a circular orbit and 
subject to gas drag. The difference $\Delta a_{s}$ between the calculated 
position and that predicted by Equation~(\ref{eq:aspeale}) is illustrated 
for three values of $\Omega_{0}\tau$, assuming a radially
constant gas density $\rho_{g}$ (i.e., $b=0$).

We also present a test conducted on a classical free-fall problem.
Consider a particle at some height above the equatorial 
plane of the disk and suppose that it is subject to a constant gravitational 
acceleration, $g$, directed toward the equatorial 
plane, and to gas drag (Equation~(\ref{eq:aD})). Moreover,
suppose that the particle's velocity, $\mathbf{v}_{s}$, is perpendicular 
to the disk's equatorial plane and that there is no vertical motion of the 
gas. The scalar acceleration of the particle is then
\begin{equation}
\frac{d v_{s}}{dt}=-g%
                            -\frac{3}{8}\left(\frac{C_{D}}{R_{s}}\frac{\rho_{g}}{\rho_{s}}\right)%
                            |v_{s}| v_{s}.
\label{eq:dvsdt0}
\end{equation}
If $v_{s}>0$, the acceleration is always negative, and
eventually the velocity becomes first zero and then negative
($dv_{s}/dt<0$ if $v_{s}=0$).
If $v_{s}\le0$, Equation~(\ref{eq:dvsdt0}) can be written as
\begin{equation}
\frac{d v_{s}}{dt}=-g%
                            +\left(\frac{1}{4 g \tau^{2}}\right) v^{2}_{s},
\label{eq:dvsdt}
\end{equation}
where $1/(g\tau^{2})=(3/2)(C_{D}/R_{s})(\rho_{g}/\rho_{s})$ and
$\tau$ is a timescale.
For $v_{s}<-2g\tau$, the acceleration is positive and negative
otherwise. Thus, the particle will always approach the velocity $-2g\tau$,
which is referred to as terminal or asymptotic velocity of the
free-fall problem. The solution to Equation~(\ref{eq:dvsdt}) is 
\begin{equation}
v_{s}=2g\tau%
          \left(\frac{1\mp B e^{t/\tau}}{1\pm B e^{t/\tau}}\right).
\label{eq:vsff}
\end{equation}
Quantity $B$ is a \emph{positive} integration constant determined 
through the initial condition. The top (bottom) signs in front of $B$ 
apply if $v^{2}_{s}$ is smaller (larger) than $(2g\tau)^{2}$.
In the limit $t\rightarrow \infty$, either solution tends to the terminal 
velocity.

In the right panel of Figure~\ref{fig:test_crtb}, we plot the
difference $\Delta v_{s}$ between the calculated free-fall velocity 
of a particle (with zero initial velocity) and that predicted by 
Equation~(\ref{eq:vsff}) as a function of the normalized time $t/\tau$.
A velocity within $1$\% of the terminal velocity is attained for
$t/\tau\gtrsim 5$.

\subsection{Thermodynamics Tests}
\label{sec:TT}

\begin{figure*}
\centering%
\resizebox{\figlew}{!}{%
\includegraphics[clip]{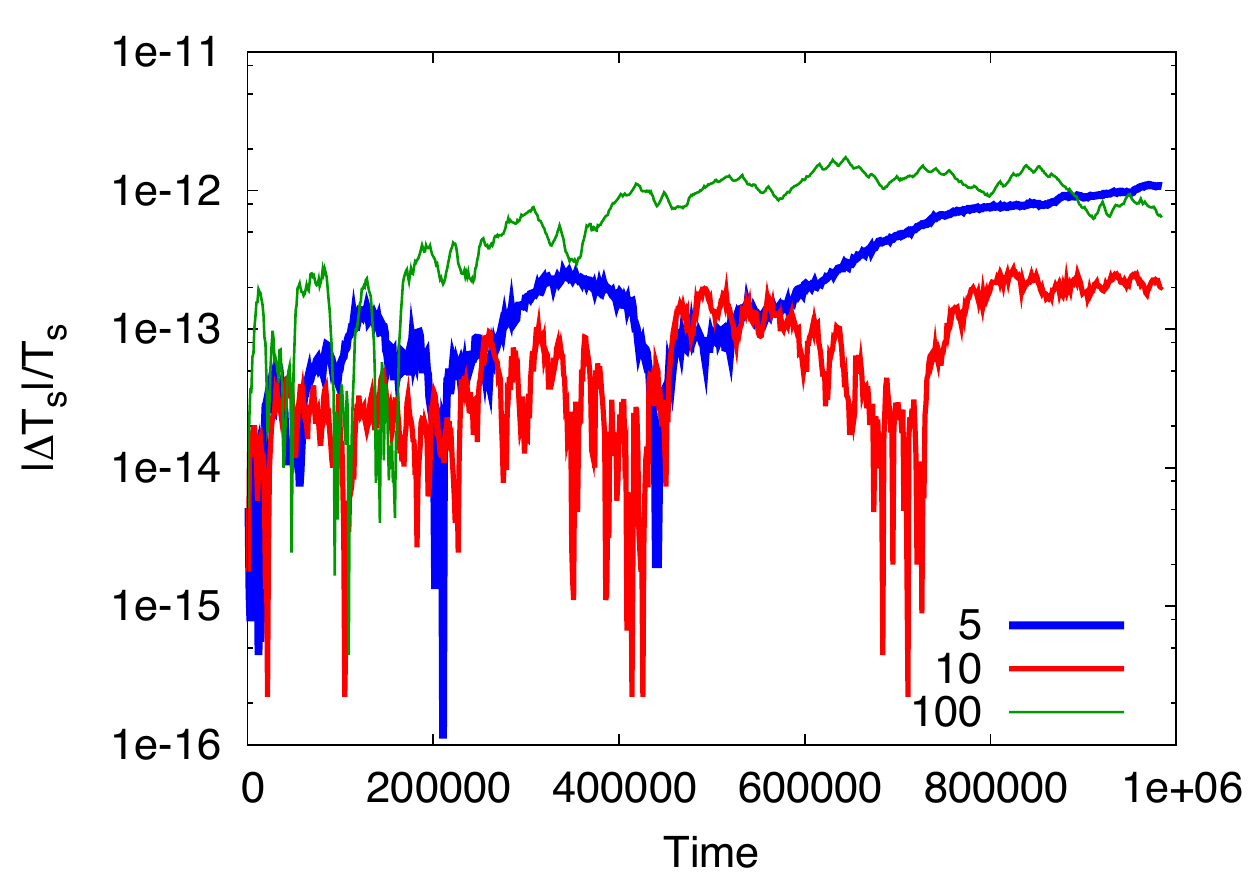}%
\includegraphics[clip]{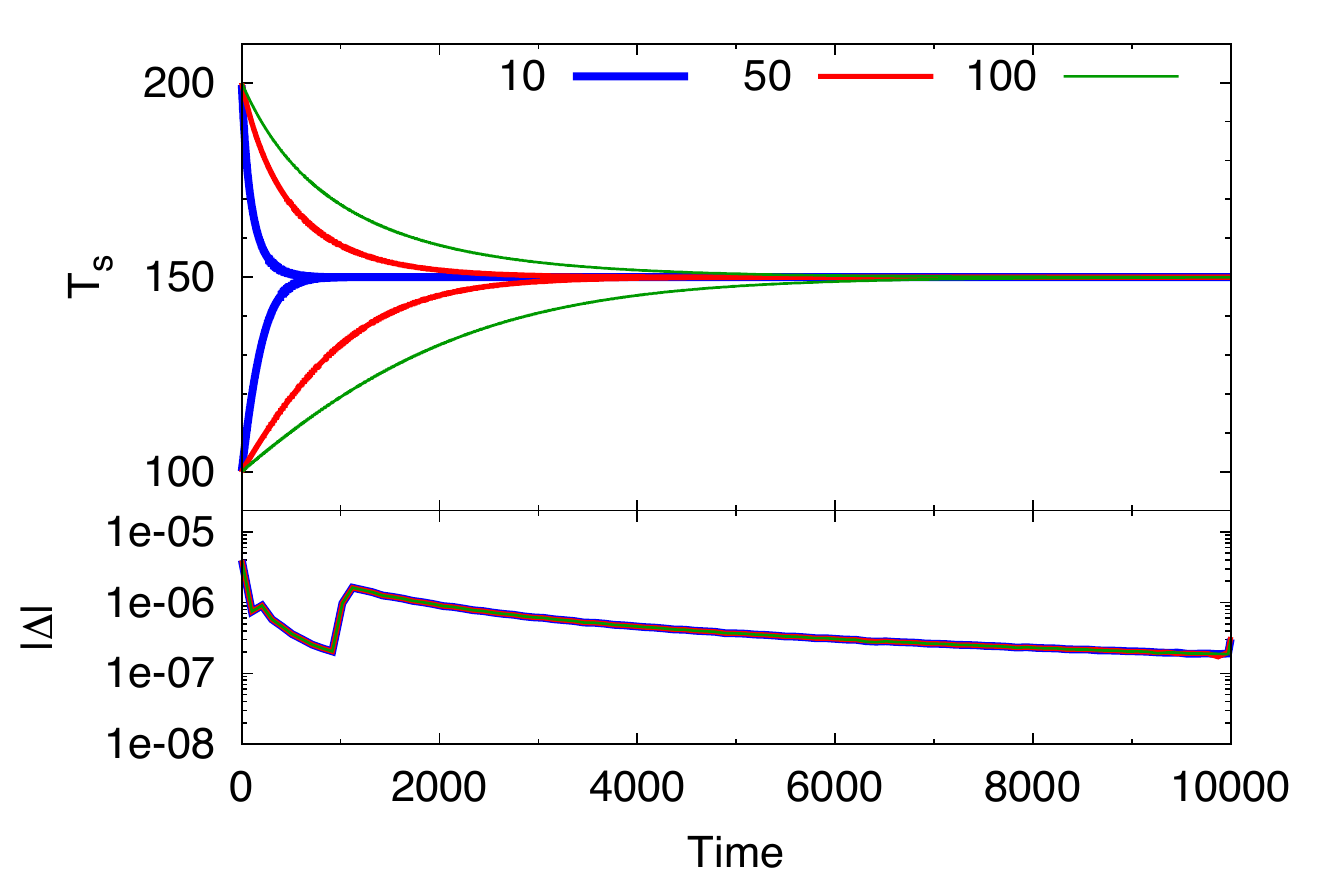}%
\includegraphics[clip]{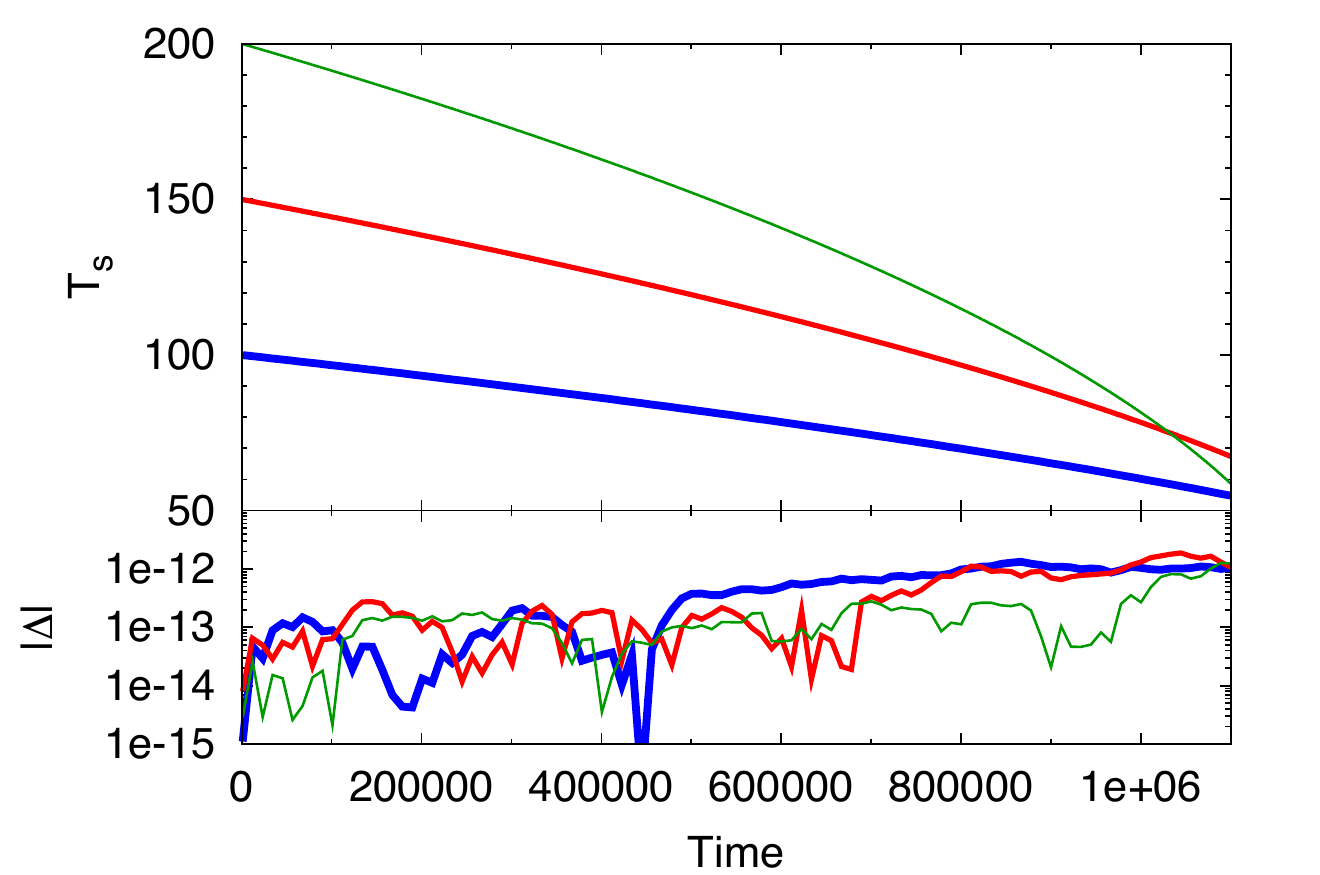}}
\caption{%
             Left. Difference $|\Delta T_{s}|$ between calculated
             and predicted temperature of planetesimals, divided
             by the predicted temperature. Planetesimals are only 
             heated via gas friction and cannot cool.
             The predicted temperature is given by Equation~(\ref{eq:Ts_CD}). 
             The curves refer to different values of the constant $\Omega_{0}\tau$, 
             as indicated, and $\tau$ is defined after Equation~(\ref{eq:dasdtpeale}).
             See text for further details.
             Center. Temperature evolution of planetesimals heated by
             the gas radiation field and losing energy via radiative cooling.
             The gas temperature is fixed at $T_{g}=150\,\K$.
             The initial temperature of the planetesimals is $100\,\K$
             and $200\,\K$ and their radius in km is indicated in the legend. 
             The lower panel shows the normalized difference
             between the numerical and the analytical solution 
             (Equation~(\ref{eq:Ts_HC})).
             Right. Thermal evolution of planetesimals that lose energy due
             to ablation. The temperature is shown on the top while the
             normalized difference between numerical and analytical
             temperature (Equation~(\ref{eq:Ts_L})) is shown on the
             bottom. In all panels, the time units are $2\pi/\Omega_{0}$. 
             }
\label{fig:test_thte}
\end{figure*}

In this section, we test the numerical solution of reduced forms 
of Equation~(\ref{eq:dTsdt0}) against analytical solutions. 
We shall assume that heating and cooling processes affect 
the entire volume of the particle, which thus has a uniform 
temperature throughout. This assumption basically implies 
that the thermal conductivity of the body, $\lambda_{s}$, tends 
to infinity (see discussion in Section~\ref{sec:PT}). 
We adopt this simplified approach here, instead of solving 
Equation~(\ref{eq:dTsdt}), because it helps in searching for 
analytical solutions of the reduced equations.

Consider the situation in which the particle mass, $M_{s}$, is constant 
and the temperature of the gas, $T_{g}$, is always equal to $T_{s}$, 
the particle temperature. Hence, if the particle moves through gas
around a star, it is constantly heated via friction so that its temperature 
changes in time according to
\begin{equation}
\frac{dT_{s}}{dt}=\frac{3}{32}%
\left(\frac{C_{D}\rho_{g}}{R_{s}\rho_{s} C_{s}}\right)%
|\mathbf{v}_{g}-\mathbf{v}_{s}|^{3}.
\label{eq:dTsdt_CD0}
\end{equation}
All quantities in parenthesis  on the right-hand side are taken
as constants. Following one of the problems in Section~\ref{sec:TT},
we assume that the gas is partially supported by pressure and has 
no radial velocity component, then
$|\mathbf{v}_{s}-\mathbf{v}_{g}|=a_{s}\Omega_{\mathrm{K}}(1-\sqrt{1-\xi^{2}})$,
where the azimuthal velocity of the particle is equal to $a_{s}\Omega_{\mathrm{K}}$.
If the radial position of the particle is given by Equation~(\ref{eq:aspeale}) with
$b=0$, Equation~(\ref{eq:dTsdt_CD0}) becomes
\begin{equation}
\frac{dT_{s}}{dt}=\frac{3}{32}%
\left(\frac{C_{D}\rho_{g}}{R_{s}\rho_{s} C_{s}}\right)%
\left(\frac{a_{0}\Omega_{0}\sqrt{c\tau}}{1-ct/2}\right)^{3}
\label{eq:dTsdt_CD}
\end{equation}
where $\tau$ is defined beneath Equation~(\ref{eq:dasdtpeale}) and
$c=(1-\sqrt{1-\xi^{2}})^{2}/\tau$. As above, $a_{0}$ and 
$\Omega_{0}$ are the initial values of $a_{s}$ and 
$\Omega_{\mathrm{K}}$.
Note that condition~(\ref{eq:peale_con}) applies since we are using
Equation~(\ref{eq:aspeale}). The solution of Equation~(\ref{eq:dTsdt_CD})
is
\begin{eqnarray}
T_{s} = T_{s}(0) &+& \frac{3}{32}%
\left(\frac{C_{D}\rho_{g}}{R_{s}\rho_{s} C_{s}}\right)%
\left(a_{0}\Omega_{0}\right)^{3}%
\left(c\tau\right)^{3/2} \nonumber \\%
                          &  &\times%
\left[\frac{1}{c(1-c t/2)^{2}}-\frac{1}{c}\right].
\label{eq:Ts_CD}
\end{eqnarray}
A Taylor expansion around $t=0$ of the function in square brackets 
on the right-hand side gives $t$. The temperature diverges as
$t\rightarrow 2/c$, i.e., as $a_{s}\rightarrow 0$, since the
relative velocity $|\mathbf{v}_{s}-\mathbf{v}_{g}|$ diverges.

We solve Equation~(\ref{eq:dTsdt_CD}) for different values 
of the constant $\Omega_{0}\tau$ and initial temperature
of $100\,\K$. In Figure~\ref{fig:test_thte} (left), we plot
the difference $|\Delta T_{s}|$, between numerical and
analytic solution (Equation~(\ref{eq:Ts_CD})), divided by
the analytic solution. The rise in temperature is limited to
a few degrees in case with longest $\tau$ and to over
$50\,\K$ in the opposite case.

Consider another situation in which the particle has again a constant
mass but its temperature differs from the gas temperature. Hence, 
the particle experiences heating by gas-emitted photons and cooling 
via back-body emission.
Suppose also that the frictional heating is negligible or, otherwise stated,  
that the quantity in parenthesis  on the right-hand side of 
Equation~(\ref{eq:dTsdt_CD0}) is vanishingly small. Thus,
the temperature variation of the particle is governed by
\begin{equation}
\frac{dT_{s}}{dt}=3%
\left(\frac{\epsilon_{s} \sigma_{\mathrm{SB}}}{R_{s}\rho_{s} C_{s}}\right)%
\left(T^{4}_{g}-T^{4}_{s}\right).
\label{eq:dTsdt_HC}
\end{equation}
Equation~(\ref{eq:dTsdt_HC}) implies that the particle temperature evolves
towards $T_{g}$, which is assumed to be constant. All quantities in first set
of parenthesis on the right-hand side are also supposed to be constants. 
A solution to the equation is
\begin{equation}
\left(\frac{\epsilon_{s} \sigma_{\mathrm{SB}}}{R_{s}\rho_{s} C_{s}}\right) t%
=I(T_{s})-I(T_{s0}),
\label{eq:Ts_HC}
\end{equation}
where $T_{s0}$ is the initial particle temperature. 
For $T_{s}>T_{g}$, we have
\begin{equation}
I(T_{s})=\frac{1}{4T^{3}_{g}}\left[%
              \ln{\left(\frac{T_{s}+T_{g}}{T_{s}-T_{g}}\right)}%
             +2\arctan{\left(\frac{T_{s}}{T_{g}}\right)}%
                                             \right]
\label{eq:IBYgt}
\end{equation}
whereas for $T_{s}<T_{g}$, the function is
\begin{equation}
I(T_{s})=\frac{1}{4T^{3}_{g}}\left[%
              \ln{\left(\frac{T_{g}+T_{s}}{T_{g}-T_{s}}\right)}%
             +2\arctan{\left(\frac{T_{s}}{T_{g}}\right)}%
                                             \right].
\label{eq:IBYlt}
\end{equation}
Equation~(\ref{eq:Ts_HC}) defines implicitly $T_{s}$ as a function of time.

Equation~(\ref{eq:dTsdt_HC}) is solved numerically for three different 
radii (see figure caption) of planetesimals, orbiting a star with a period
$2\pi/\Omega_{0}$.
Two values of the initial planetesimal temperatures are applied: $100\,\K$
and $200\,\K$, so that the bodies will either heat up or cool down toward
the gas temperature of $150\,\K$. The results are shown 
in the center panel of Figure~\ref{fig:test_thte}, which illustrates $T_{s}$
(top) and the normalized difference between numerical
and analytical solutions (bottom) versus time.

Finally, we discuss a test in which a particle is neither subject to
frictional heating (as in the first test of this section) nor to radiative heating 
and cooling (as in the second test).
We assume that a particle loses mass, due to ablation, releasing 
vaporization energy in the process. For simplicity, the variation
of the particle mass is such that  $dR_{s}/dt$ is constant, so that
$dM_{s}/dt\propto R^{2}_{s}$. Thus, the energy budget
reduces to
\begin{equation}
C_{s}(T_{s})\frac{dT_{s}}{dt}=%
\left(\frac{L_{s}}{M_{s}}\right)\frac{dM_{s}}{dt},
\label{eq:dTsdt_L}
\end{equation}
where, $L_{s}$, the specific energy of vaporization, is constant
\citepalias[see][]{podolak1988}.
The specific heat has the form $C_{s}=c T^{b}$ ($b\approx 1$), 
which is an approximation to the specific heat of ice between 
$30\,\K$ and $273\,\K$ \citep{CRC92}. 
The solution to Equation~(\ref{eq:dTsdt_L}) is 
 \begin{equation}
T_{s}=\left[T_{s0}^{b+1}%
        +3\left(\frac{b+1}{c}\right)L_{s}\ln{\left(\frac{R_{s}}{R_{s0}}\right)}\right]^{1/(b+1)},
\label{eq:Ts_L}
\end{equation}
where $T_{s0}$ and $R_{s0}$ are the initial temperature and radius
of the particle.

The right panel of Figure~\ref{fig:test_thte} shows the thermal evolution
of particles that ablate and lose energy. Again, it is assumed that the
particle orbits the star with a period $2\pi/\Omega_{0}$. Tests are performed
for various values $T_{s0}$ and $R_{s0}$. The temperature is illustrated
on top while the normalized difference between the computed and the 
analytical temperature in Equation~(\ref{eq:Ts_L}) is shown on the
bottom of the figure.




\end{document}